\documentclass[letterpaper,twoside,12pt]{book}
\usepackage[latin2]{inputenc}
\usepackage{appendix}
\usepackage{amsmath,amssymb}
\usepackage{graphicx,multicol}
\usepackage[a4paper,margin=1in]{geometry}
\usepackage{fancyhdr}
\usepackage{clrscode}
\usepackage{makeidx}
\makeindex

\def\ind{{\mathbb I}}
\def\(({\left(}
\def\)){\right)}                       

\def\ZMM{ZDEB-1}
\def\KM{ZDEB-4}
\def\MM{ZDEB-2}
\def\RS{ZDEB-3}
\def\ZK{ZDEB-5}
\def\KZP{ZDEB-6}
\def\KZJ{ZDEB-7}
\def\MZ{ZDEB-8}
\def\AZ{ZDEB-10}
\def\ZML{ZDEB-9}

\newcommand{\cN}{\mathcal{N}}
\newcommand{\ud}{\mathrm{d}}

\newcommand{\be}{\begin{equation}}
\newcommand{\ee}{\end{equation}}
\newcommand{\bea}{\begin{eqnarray}}
\newcommand{\eea}{\end{eqnarray}}

\newcommand{\da}{\partial a}
\newcommand{\di}{\partial i}

\begin{document}

\begin{titlepage}
 
\begin{minipage}{0.68\linewidth}
 \Large \textbf{Universit\'e Paris-Sud 11} \\
   Facult\'e des sciences d'Orsay
\end{minipage}
\begin{minipage}{0.3\linewidth}
\begin{center}
  \resizebox{4.5cm}{!}{\includegraphics{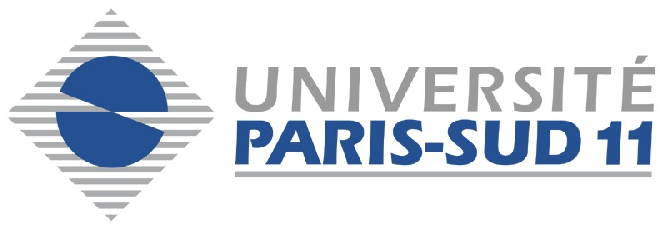} }
\end{center} 
\end{minipage}

\vspace{0.5cm}

\begin{minipage}{0.68\linewidth}
 \Large \textbf{Univerzita Karlova v Praze} \\
  Matematicko-fyzik\'aln\'{i} fakulta
\end{minipage}
\begin{minipage}{0.3\linewidth}
\begin{center}
  \resizebox{3cm}{!}{\includegraphics{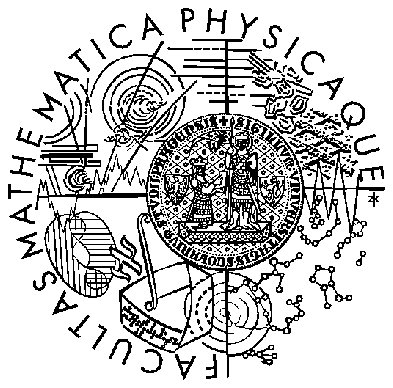} }
\end{center} 
\end{minipage}

\vspace{0.3cm}

\begin{center}
Thesis presented to obtain the degree of

\vspace{0.3cm}

{\bf Doctor of Sciences of the University Paris XI \\ 
Doctor of the Charles University in Prague\\}

\vspace{0.3cm}

Specialization: Theoretical Physics \\

\vspace{0.3cm}

by

\vspace{0.6cm}

{\bf \Large Lenka ZDEBOROV\'A}

\vspace{2.cm}

\resizebox{\linewidth}{!}{\includegraphics{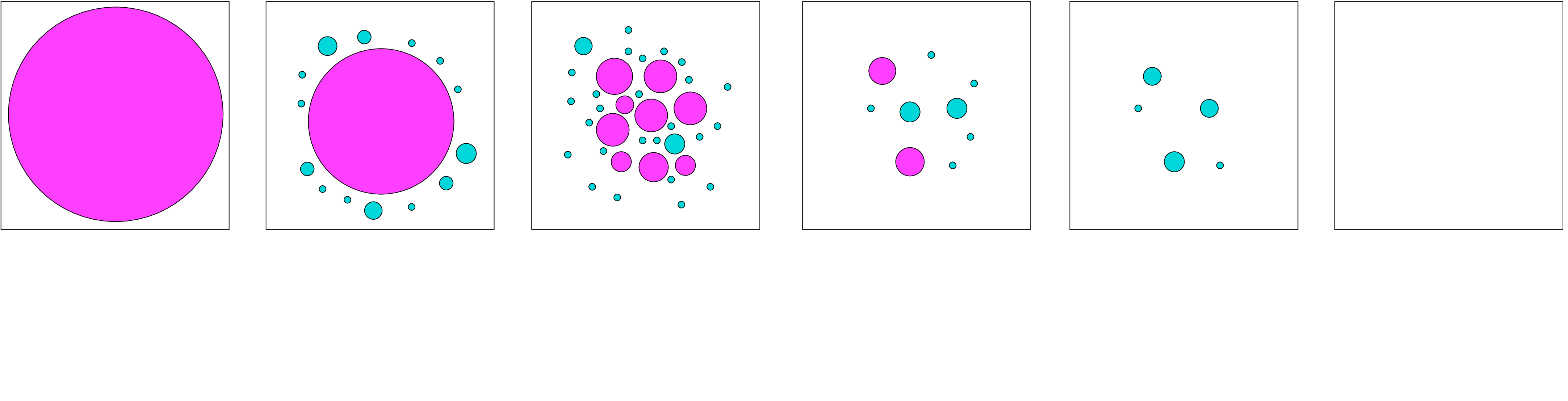} }
\vspace{0.2cm}

\LARGE \textbf{Statistical Physics of \\ Hard Optimization Problems}
\vspace{2.cm} 

\end{center}

{Defended on June 20, 2008, in front of the Thesis Committee:}

{\large
\begin{center}
\begin{tabular}{lll}
 Silvio & FRANZ & \\ 
 V\'aclav & JANI\v S &  thesis advisor (Prague)\\
 Ji\v r\'{i} & LANGER & \\
 Stephan & MERTENS & referee \\
 Marc & MEZARD & thesis advisor (Paris) \\
 Riccardo & ZECCHINA & referee 
\end{tabular}
\end{center} 
}
 
\end{titlepage}

\frontmatter

\newpage
\thispagestyle{plain}

\subsection*{Acknowledgment}
 
\addcontentsline{toc}{chapter}{Acknowledgment}

First of all I would like to express my thanks to my advisor Marc M\'{e}zard from whom I learned a lot. He showed me how to combine enthusiasm, patience, computations, and intuition in the correct proportion to enjoy the delicious taste of the process of discovery.
I thank as well to my advisor V\'{a}clav Jani\v{s} who guided my scientific steps mainly in the earlier stages of my work. 

This work would never be possible without the contact and discussions
with my colleagues and collaborators all over the world. Without them  I would feel lost in the vast world of unknown. I am grateful to all the organizers of workshops, conferences and summer schools where I had participated. I also thank for invitations on visits and seminars which were always very inspiring. 

I am very thankful to the whole Laboratoire de Physique Th\'{e}orique et Mod\`{e}les Statistiques for a very warm reception and to all its members for helping me whenever I needed. I also owe a lot to my professors from the Charles University in Prague, and to my colleagues from the Institute of Physics of the Academy of Sciences in Prague. I also thank the referees of my thesis and other members of the thesis committee for their interest in my work and for accepting this task.

I value profoundly the scholarship granted by the French Government which covered the largest part of my stay in France. Further, I appreciated greatly the subvention from the French ministry for higher education and research "cotutelles internationales de th\'{e}se". I also acknowledge gratefully the support from the FP6 European network EVERGROW.  

My deepest thanks go to my parents for their constant support, encouragement, and love. Finally, thank you Flo for all the items above and many more. You taught me well, and yes I am the Jedi now.

\vspace{3cm}

\begin{flushright}  Lenka  Zdeborov\'a \end{flushright}
\begin{flushright}  Paris, May 12, 2008 \end{flushright}

\resizebox{4.5cm}{!}{\includegraphics{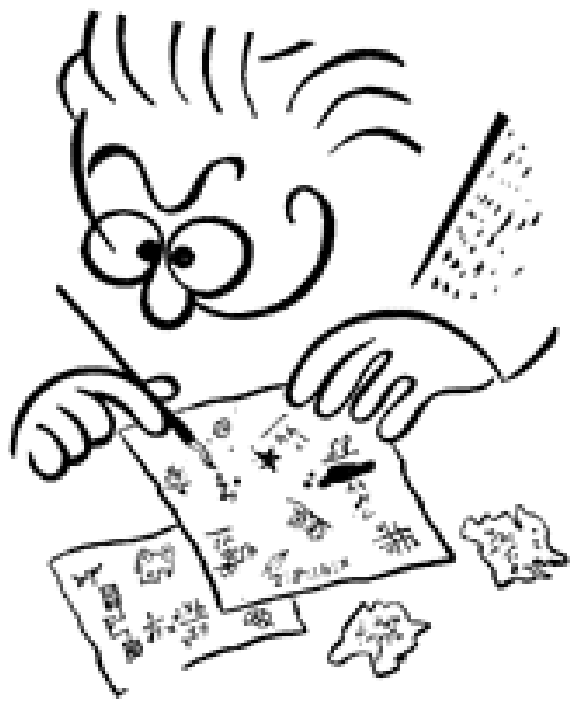} }

\newpage

\chapter*{}

\vspace{5cm}
                 
\begin{flushright}  {\it M\'{y}m rodi\v{c}\r{u}m} \end{flushright}
\begin{flushright}  {\it Josefovi a Bo\v{z}en\v{e}} \end{flushright}

\tableofcontents 

\newpage

\addcontentsline{toc}{chapter}{Abstract}

\newpage

\chapter*{}

\vspace{-3cm}

\thispagestyle{myheadings}
\markboth{ABSTRACT}{ABSTRACT}

\vspace{1cm}

\noindent {\bf Title}: Statistical Physics of Hard Optimization Problems

\noindent {\bf Author}: Lenka Zdeborov\'a

\vspace{1cm}

\noindent {\bf Abstract}: 
Optimization is fundamental in many areas of science, from computer science and information theory to engineering and statistical  physics, as well as to biology or social sciences. It typically involves a large number of variables and a cost function depending on these variables. Optimization problems in the NP-complete class are particularly difficult, it is believed that the number of operations required to minimize the cost function is in the most difficult cases exponential in the system size. However, even in an NP-complete problem the practically arising instances might, in fact, be easy to solve. The principal question we address in this thesis is: How to recognize if an NP-complete constraint satisfaction problem is typically hard and what are the main reasons for this? We adopt approaches from the statistical physics of disordered systems, in particular the cavity method developed originally to describe glassy systems. We describe new properties of the space of solutions in two of the most studied constraint satisfaction problems - random satisfiability and random graph coloring. We suggest a relation between the existence of the so-called frozen variables and the algorithmic hardness of a problem. Based on these insights, we introduce a new class of problems which we named "locked" constraint satisfaction, where the statistical description is easily solvable, but from the algorithmic point of view they are even more challenging than the canonical satisfiability.

\vspace{1cm}

\noindent {\bf Keywords}: Constraint satisfaction problems, combinatorial optimization, random coloring problem, average computational complexity, cavity method, spin glasses, replica symmetry breaking, Bethe approximation, clustering of solutions, phase transitions, message passing, belief propagation, satisfiability threshold, reconstruction on trees.

\newpage

\chapter*{}

\vspace{-3cm}

\thispagestyle{myheadings}
\markboth{RESUME}{RESUME}

\vspace{1cm}

\noindent {\bf Titre}:  Physique statistique des probl\`emes d'optimisation

 \noindent {\bf Autheur}: Lenka Zdeborov\'a

\vspace{1cm}

\noindent {\bf R\'esum\'e}: L'optimisation est un concept fondamental
dans beaucoup de domaines  scientifiques comme l'informatique, la
th\'eorie de l'information, les sciences de l'ing\'enieur et la
physique statistique, ainsi que pour la biologie et les sciences
sociales.  Un probl\`eme d'optimisation met typiquement en jeu un
nombre important de variables et une fonction de co\^{u}t qui d\'epend de
ces variables.  La classe des probl\`emes NP-complets est
particuli\`{e}rement difficile, et il est commun\'{e}ment admis que, dans le
pire des cas, un nombre d'op\'{e}rations exponentiel dans la taille du
probl\`{e}me est n\'{e}cessaire pour minimiser la fonction de co\^{u}t.
Cependant, m\^{e}me  ces probl\`{e}mes peuveut \^{e}tre faciles \`{a}
r\'{e}soudre  en pratique.  La principale question consid\'{e}r\'{e}e dans cette th\`{e}se est
comment reconnaître si un probl\`{e}me de satisfaction de contraintes NP-complet est "typiquement" difficile et
quelles sont les raisons pour cela ? Nous suivons une approche
inspir\'{e}e par la physique statistique des syst\`{e}mes desordonn\'{e}s, en
particulier la m\'{e}thode de la cavit\'{e} d\'{e}velopp\'{e}e originalement pour les
syst\`{e}mes vitreux. Nous d\'{e}crivons les propri\'{e}t\'{e}s de l'espace des
solutions dans deux des probl\`{e}mes de satisfaction les plus \'{e}tudi\'{e}s : la
satisfiabilit\'{e} et le coloriage al\'{e}atoire.  Nous sugg\'{e}rons une relation
entre l'existence  de variables dites "gel\'{e}es" et la difficult\'{e}
algorithmique d'un probl\`{e}me donn\'{e}. Nous introduisons aussi une
nouvelle classe de probl\`{e}mes, que nous appelons "probl\`{e}mes
verrouill\'{e}s", qui pr\'{e}sentent l'avantage d'\^{e}tre \`{a} la fois facilement
r\'{e}soluble analytiquement, du point de vue du comportement moyen, mais
\'{e}galement extr\^{e}mement difficiles du point de vue de la recherche de
solutions dans un cas donn\'{e}.

 \vspace{1cm}

 \noindent {\bf Les mots clefs}:  Probl\`emes d'optimisation de
contraintes, optimisation combinatoire, probl\`emes de coloriage,
complexit\'e de calcul moyenne, m\'ethode de la cavit\'e, verres de
spins, brisure de la sym\'etrie des r\'epliques, approximation de
Bethe, regroupement des solutions en amas,  transitions de phases,
passage de messages, propagation des convictions, seuil de
satisfiabilit\'e, reconstruction sur des arbres.

\newpage 

\chapter*{}

\vspace{-3cm}

\thispagestyle{myheadings}
\markboth{\v{C}ESK\'{Y} ABSTRAKT}{\v{C}ESK\'{Y} ABSTRAKT}

\vspace{1cm}

\noindent {\bf N\'azev}: Statistick\'a fyzika t\v{e}\v{z}k\'{y}ch optimaliza\v{c}n\'{i}ch \'{u}loh 

\noindent {\bf Autor}: Lenka Zdeborov\'a

\vspace{1cm}

\noindent {\bf Abstrakt}: 
Optimalizace je fundament\'{a}ln\'{i} koncept v mnoha v\v{e}dn\'{i}ch oborech, po\v{c}\'{i}naje po\v{c}\'{i}ta\v{c}ovou v\v{e}dou a teori\'{i} informace, p\v{r}es in\v{z}en\'{y}rstv\'{i} a statistickou fyziku, a\v{z} po biologii \v{c}i ekonomii. Optimaliza\v{c}n\'{i} \'{u}loha se typicky skl\'{a}d\'{a} z minimalizace funkce z\'{a}visej\'{i}c\'{i} na  velk\'{e}m mno\v{z}stv\'{i} prom\v{e}nn\'{y}ch. 
Probl\'{e}my z takzvan\'{e} NP-\'{u}pln\'{e} t\v{r}\'{i}dy jsou obzvl\'{a}\v{s}t\v{e} slo\v{z}it\'{e}, v\v{e}\v{r}\'{i} se, \v{z}e po\v{c}et operac\'{i} pot\v{r}ebn\'{y} k nalezen\'{i} \v{r}e\v{s}en\'{i} v tom nejt\v{e}\v{z}\v{s}\'{i}m p\v{r}\'{i}pad\v{e} roste exponenci\'{a}ln\v{e} s po\v{c}tem prom\v{e}n\'{y}ch. Nicm\'{e}n\v{e} i pro NP-\'{u}pln\'{e} \'{u}lohy plat\'{i}, \v{z}e praktick\'{e} p\v{r}\'{i}pady mohou b\'{y}t jednoduch\'{e}. Hlavn\'{i} ot\'{a}zka, kterou se zab\'{y}v\'{a} tato pr\'{a}ce, je: Jak rozpoznat, zda je NP-\'{u}pln\'{y} probl\'{e}m splnitelnosti podm\'{i}nek v typick\'{e}m p\v{r}\'{i}pad\v{e} t\v{e}\v{z}k\'{y} a \v{c}\'{i}m je tato slo\v{z}itost zp\r{u}sobena? K t\'{e}to ot\'{a}zce p\v{r}istupujeme s vyu\v{z}it\'{i}m znalost\'{i} ze statistick\'{e} fyziky neuspo\v{r}\'{a}dan\'{y}ch, a zejm\'{e}na skeln\'{y}ch, syst\'{e}m\r{u}. Pop\'{i}\v{s}eme nov\'{e} vlastnosti prostoru \v{r}e\v{s}en\'{i} ve dvou z nejv\'{i}ce studovan\'{y}ch optimaliza\v{c}n\'{i}ch probl\'{e}m\r{u} -- splnitelnosti n\'{a}hodn\'{y}ch Booleovsk\'{y}ch formul\'{i} a barven\'{i} n\'{a}hodn\'{y}ch graf\r{u}. Navrhneme  existenci vztahu mezi typickou algoritmickou slo\v{z}itost\'{i} a existenc\'{i} takzvan\v{e} zamrzl\'{y}ch prom\v{e}nn\'{y}ch. Na z\'{a}klad\v{e} t\v{e}chto poznatk\r{u} zkonstruujeme novou t\v{r}\'{i}du probl\'{e}m\r{u}, kter\'{e} jsme nazvali "uzamknut\'{e}", zde je statistick\'{y} popis mno\v{z}iny v\v{s}ech \v{r}e\v{s}en\'{i} pom\v{e}rn\v{e} jednoduch\'{y}, ale z algoritmick\'{e}ho pohledu jsou tyto typick\'{e} p\v{r}\'{i}pady t\v{e}chto probl\'{e}m\r{u} je\v{s}t\v{e} te\v{z}\v{s}i ne\v{z} v kanonick\'{e}m probl\'{e}mu splitelnosti Booleovsk\'{y}ch formul\'{i}.

\vspace{1cm}

\noindent {\bf Kl\'{i}\v{c}ov\'{a} slova}: Probl\'{e}my splnitelnosti podm\'{i}nek, kombinatorick\'{a} optimalizace, barven\'{i} n\'{a}hodn\'{y}ch graf\r{u}, pr\r{u}m\v{e}rn\'{a} algoritmick\'{a} slo\v{z}itost, metoda kavity, spinov\'{a} skla, naru\v{s}en\'{i} symetrie replik, Betheho aproximace, shlukov\'{a}n\'{i} \v{r}e\v{s}en\'{i}, f\'{a}zov\'{e} p\v{r}echody, pos\'{i}l\'{a}n\'{i} zpr\'{a}v, propagace domn\v{e}nek, pr\'{a}h splnitelnosti, rekonstrukce na stromech.

\newpage

\chapter*{Foreword}

\markboth{FOREWORD}{FOREWORD}

\addcontentsline{toc}{chapter}{Foreword}

P.-G. de Gennes in his foreword to the book "Stealing the gold -- A celebration of the pioneering physics of Sam Edwards" wrote:

{\it But he \{meaning S. Edwards\} also has another passion, which I \{meaning P.-G. de Gennes\} call "The search for unicorns." To chase unicorns is a delicate enterprise. Medieval Britons practised it with great enthusiasm (and this still holds up to now: read Harry Potter). Sir Samuel Edwards is not far from the gallant knights of the twelfth century. Discovering a strange animal, approaching it without fear, then not necessarily harnessing the creature, but rapidly drawing a plausible sketch of its main features.

One beautiful unicorn prancing in the magic garden of Physics has been names "Spin glass." It is rare: not many pure breeds of Spin glasses have been found in Nature. But we have all watched the unpredictable jumps of this beast. And we have loved its story -- initiated by Edwards and Anderson.} 

\vspace{0.5cm}

Unicorn is a mythical animal, described in the book of Job, together with another strange and fascinating creature which is less peaceful: the leviathan. Leviathans are described as immense terrible monsters, invincible beasts. Most people prefer not to even think about them.

\vspace{0.5cm}

This thesis tells a story about what happens when the fierce and  mysterious beauty of a unicorn meets with the invincibility of a leviathan.

\mainmatter

\chapter{Hard optimization problems}
\label{history}

{\it In this opening chapter we introduce the constraint satisfaction problems and discuss briefly the computer science approach to the computational complexity. We review the studies of the random satisfiability problem in the context of average computational complexity investigations. We describe the connection between spin glasses and random CSPs and highlight the most interesting results coming out from this analogy. We explain the replica symmetric approach to these problems and show its usefulness on the example of counting of matchings [\ZMM]. Then we review the survey propagation approach to constraint satisfaction on an example of 1-in-$K$ satisfiability [\RS]. Finally we summarize the main contributions of the author to the advances in the statistical physics of hard optimization problems, that are elaborated in the rest of the thesis.}

\section{Importance of optimization problems}

Optimization is a common concept in many areas of human activities. 
It typically involves a large number of variables, e.g.
particles, agents, cells or nodes, and a cost function depending on
these variables, such as energy, measure of risk or expenses. The problem consists in finding a state of variables which minimizes the value of the cost function.  

In this thesis we will concentrate on a subset of optimization problems the so-called {\it constraint satisfaction problems} (CSPs). Constraint satisfaction problems are one of the main building blocks of complex systems studied in computer science, information theory and statistical physics. Their wide range of applicability arises from their very general nature: given a set of $N$ discrete variables subject to $M$ constraints, the CSP consists in deciding whether there exists an assignment of variables which satisfies simultaneously all the constraints. And if such an assignment exists then we aim at finding it. 

In computer science, CSPs are at the core of computational complexity studies: the satisfiability of boolean formulas is the canonical example of an intrinsically hard, NP-complete, problem. In information theory, error correcting codes also rely on CSPs. The transmitted information is encoded into a codeword satisfying a set of constraints, so that the information may be retrieved after transmission through a noisy channel, using the knowledge of the constraints satisfied by the codeword. Many other practical problems in scheduling a collection of tasks, in electronic design engineering or artificial intelligence are viewed as CSPs. In statistical physics the interest in CSPs stems from their close relation with the theory of spin glasses. Answering if frustration is avoidable in a system is the first, and sometimes highly nontrivial, step in understanding the low temperature behaviour.

A key point is to understand how difficult it is to solve practical instances of a constraint satisfaction problem. Everyday experience confirms that sometimes it is very hard to find a solution. Many CSPs require a combination of heuristics and combinatorial search methods to be solved in a reasonable time. A key question we address in this thesis is thus {\it why} and {\it when} are some instances of these problems intrinsically hard. Answering this question has, next to its theoretical interest, several practical motivations
\begin{itemize}
   \item Understanding where the hardness comes from helps to push the performance of CSPs solvers to its limit.
   \item Understanding which instances are hard helps to avoid them if the nature of the given practical problem permits.
   \item Finding the very hard problem might be interesting for cryptographic application. 
\end{itemize}
A pivotal step in this direction is the understanding of the onset of hardness in random constraint satisfaction problems. In practice random constraint satisfaction problems are either regarded as extremely hard as there is no obvious structure to be explored or as extremely simple as they permit probabilistic description. Furthermore, random constraint satisfaction models are spin glasses and we shall thus borrow methods from the statistical physics of disordered systems. 

\section{Constraint Satisfaction Problems: Setting}

\subsection{Definition, factor graph representation}

Constraint Satisfaction Problem (CSP)\index{constraint satisfaction problem}: Consider $N$ variables $s_1\dots,s_N$ taking values from the domain $\{0,\dots,q-1\}$, and a set of $M$ constraints\index{constraint}. A constraint $a$ concerns a set of $k_a$ different variables which we call $\partial a$. Constraint $a$ is a function from all possible assignments of the variables $\partial a$ to $\{0,1\}$. If the constraint evaluates to $1$ we say it is satisfied, and if it evaluates to $0$ we say it is violated. The constraint satisfaction problem consists in deciding whether there exists an assignment of variables which satisfies simultaneously all the constraints. We call such an assignment a solution of the CSP. 

In physics, the variables represent $q$-state Potts spins (or Ising spins if $q=2$). The constraints represent very general (non-symmetric) interactions between $k_a$-tuples of spins. In Boolean constraint satisfaction problems ($q=2$) a {\it literal}\index{literal} is a variable or its negation. A {\it clause}\index{clause} is then a disjunction (logical OR) of literals.

\begin{figure}[!ht]
\begin{center}
 \resizebox{0.3\linewidth}{!}{\includegraphics{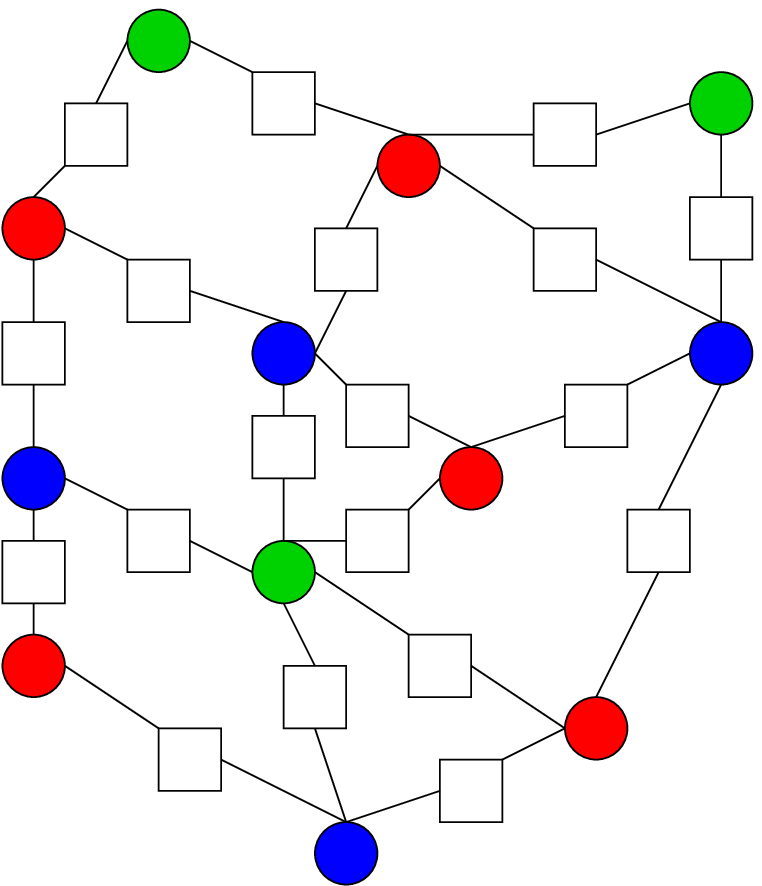}}
    \hspace{0.05\linewidth}
  \resizebox{0.35\linewidth}{!}{  \includegraphics{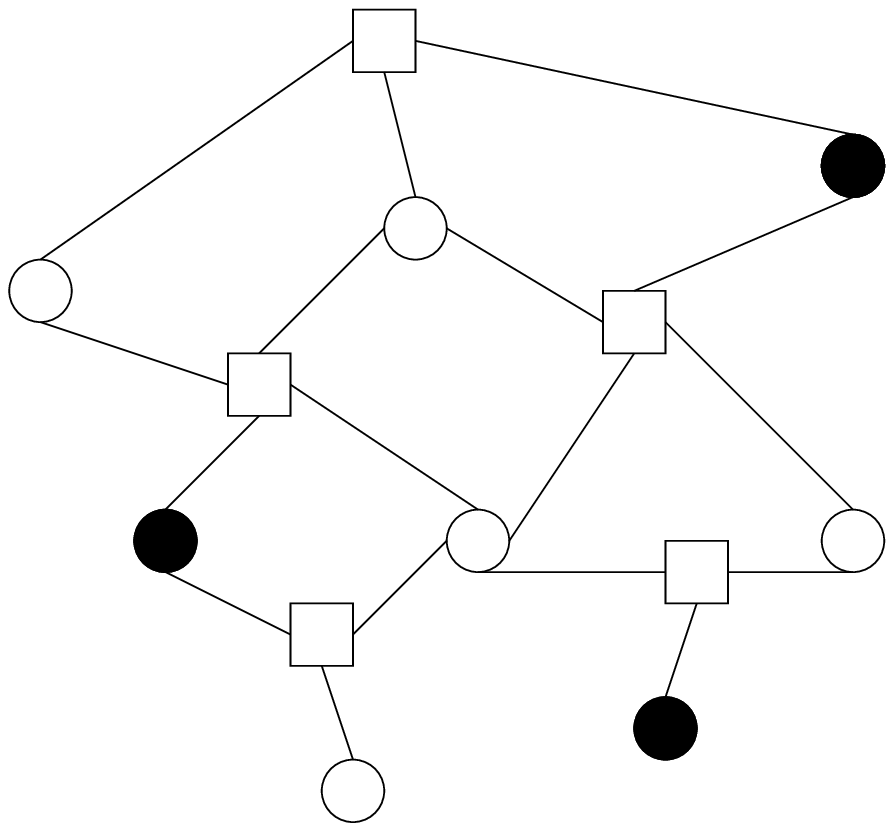}}
 \caption{\label{fig:factor_graphs} Example of a factor graph representation for the coloring (left) and the exact cover (right) problems. The function nodes (squares) in the graph coloring are satisfied if and only if their two neighbours (circles) are in different states (take different colors). The function nodes (squares) in the exact cover problem are satisfied if exactly one variable (circle) around them takes values $1$ (full) and the others $0$ (empty).}
\end{center}
\end{figure}

A handy representation for a CSP is the so-called {\it factor graph}\index{factor graph}, see \cite{KschischangFrey01} for a review. Factor graph is a bipartite graph $G(V,F,E)$ where $V$ is the set of variables (variables nodes, represented by circles\index{variable node}) and $F$ is the set of constraints (function nodes, represented by squares\index{function node}). An edge $(ia)\in E$ is present if the constraint $a\in F$ involves the variable $i\in V$. A constraint $a$ is connected to $k_a$ variables, their set is denoted $\partial a$. A variable $i$ is connected to $l_i$ constraints, their set is denoted $\partial i$. For clarity we specify the factor graph representation for the graph coloring and exact cover problem in fig.~\ref{fig:factor_graphs}, both defined in the following section \ref{sec:def}.

\subsection{List of CSPs discussed in this thesis} 
\label{sec:def}

Here we define constraint satisfaction problems which will be discussed in the following. Most of them are discussed in the classical reference book \cite{GareyJohnson79}. The most studied constraint satisfaction problems are defined over Boolean variables, $q=2$, $s_i\in \{0,1\}$. Sometimes we use equivalently the notation with Ising spins $s_i\in\{-1,+1\}$. CSPs with Boolean variables that we shall discuss in this thesis are:
\begin{itemize}
\item{{\bf Satisfiability (SAT) problem}\index{satisfiability problem}: Constraints are clauses, that is logical disjunctions of literals (i.e., variables or their negations). Example of a satisfiable formula with 3 variables and 4 clauses (constraints) and 10 literals: $(x_1 \vee x_2 \vee \neg x_3)\wedge ( x_2 \vee x_3)\wedge (\neg x_1 \vee \neg x_3) \wedge (x_1 \vee \neg x_2 \vee x_3)$.}
\item{{\bf $K$-SAT}\index{$K$-SAT}: Satisfiability problem where every clause involves $K$ literals, $k_a=K$ for all $a=1,\dots,M$.}
\item{{\bf Not-All-Equal SAT}\index{Not-All-Equal SAT}: Constraints are satisfied everytime except when all the literals they involve are TRUE or all of them are FALSE.}
\item{{\bf Bicoloring}\index{bicoloring}: Constraints are satisfied except when all variables they involve are equal. Bicoloring is Not-All-Equal SAT without negations.}
\item{{\bf XOR-SAT}\index{XOR-SAT}: Constraints are logical XORs of literals.}
\item{{\bf Odd (resp. Even) Parity Checks}\index{parity check}: A constraint is satisfied if the sum of variables it involves is odd (resp. even). Odd parity checks are XORs without negations.}
\item{{\bf 1-in-$K$ SAT}\index{1-in-$K$ SAT}: Constraints are satisfied if exactly one of the $K$ literals they involve is TRUE.}
\item{{\bf Exact Cover, or positive 1-in-$K$ SAT}: Constraints are satisfied if exactly one of the $K$ variables they involve is $1$ (occupied). Exact cover, or positive 1-in-$K$ SAT, is 1-in-$K$ SAT without negations.}
\item{{\bf Perfect matching}\index{matching!perfect matching}:
Nodes of the original graph become constraints, variables are on edges and determine if the edge is or is not in the matching, see fig.~\ref{fig:gr_fact}. Constraints are satisfied if exactly one of the $K$ variables they involve is $1$ (belongs to the matching). Note that perfect matching is just a variant of the Exact Cover}
\item{{\bf Occupation problems} are defined by a binary $(K+1)$ component vector $A$\index{occupation problems}. All constraints involve $K$ variables, and are satisfied if the sum of variables they involve $r=\sum_{\partial a}s_i$ is such that $A_r=1$.}
\item{{\bf Locked Occupation Problems (LOPs)}\index{occupation problems!locked occupation problems}: If the vector $A$ is such that $A_iA_{i+1}=0$ for all $i=0,\dots,K-1$, and all the variables are present in at least two constraints.}
\end{itemize}

We will also consider in a great detail one CSP with $q$-ary variables: The {\bf graph coloring} with $q$ colors\index{coloring}: Every constraint involves two variables and is satisfied if the two variables are not assigned the same value (color). In physics the $q$-ary variables are called Potts spins.

\subsection{Random factor graphs: definition and properties}
\label{sec:random}

Given a constraint satisfaction problem with $N$ variables and $M$ constraints, the {\it constraint density} is defined as $\alpha=M/N$\index{constraint density}. Denote by ${\cal R}(k)$ the probability distribution of the degree of constraints (number of neighbours in the factor graph), and by ${\cal Q}(l)$ the probability distribution of the degree of variables. 
The average connectivity\index{connectivity!average} (degree) of constraints is 
\be
      K = \overline k = \sum_{k=0}^\infty k {\cal R}(k) \, .
\ee
The average connectivity of variables is\index{average connectivity}
\be
      c = \overline l = \sum_{l=0}^\infty l {\cal Q}(l) \, . 
\ee 
The constraint density is then asymptotically
\be
       \alpha = \frac{M}{N}=\frac{\overline l}{\overline k} = \frac{c}{K}\, .
\ee

A random factor graph\index{random graph} with a given $N$ and $M$ is then created as follows: Draw a sequence $\{l_1,\dots,l_N\}$ of $N$ numbers from the distribution ${\cal Q}(l)$. Subsequently, draw a sequence $\{k_1,\dots,k_M\}$ of $M$ numbers from the distribution ${\cal R}(k)$, such that $\sum_{a=1}^M k_i =  \sum_{i=1}^N l_i$. The {\it random factor graph}\index{factor graph!random} is drawn uniformly at random from all the factor graphs with $N$ variables, $M$ constraints and degree sequences $\{l_1,\dots,l_N\}$ and $\{k_1,\dots,k_M\}$.

Another definition leading to a Poissonian degree distribution is used often if the degree of constraints is fixed to $K$ and the number of variables is fixed to $N$. There are ${N \choose K}$ possible positions for a constraint. Each of these positions is taken with probability 
\be
     p = \frac{c N}{K {N \choose K}} \, .
\ee 
The number of constraints is then a Poissonian random variable with average $M=cN/K$. The degree of variables is distributed according to a Poissonian law\index{degree distribution!Poissonian} with average $c$
\be
      {\cal Q}(l) =   e^{-c} \frac{c^l}{l!}\, . \label{eq:Poisson}
\ee
If $K=2$ these are the random Erd\H{o}s-R\'enyi graphs \cite{ErdosRenyi59}\index{random graphs!Erd\H{o}s-R\'enyi}. This definition works also if constraints are changed for variables, that is if the degree of variables and the number of constraints are fixed, as in e.g. the matching problem.

The random factor graphs are called {\it regular}\index{factor graph!regular}\index{random graph!regular} if both the degrees of constraints and variables are fixed, ${\cal R}(k)=\delta(k-K)$ and ${\cal Q}(l)=\delta(l-L)$\index{degree distribution!regular}.
In section \ref{locked} we will also use the {\it truncated Poissonian} degree distribution\index{degree distribution!truncated Poissonian}
\begin{subequations}
\label{eq:Poiss}
\bea
   l\le 1: &&\quad {\cal Q}(l) = 0\, ,\\
   l\ge 2: &&\quad {\cal Q}(l) =  \frac{1}{e^c - (c+1)} \frac{c^l}{l!}  \, .    
\eea
\end{subequations}
The average connectivity for the truncated Poissonian distribution is then 
\be
       \overline l = c \frac{e^c -1}{e^c - (c+1)} \, .  \label{eq:l_aver}
\ee

In the cavity approach, the so-called {\it excess degree distribution}\index{degree distribution!excess} is a crucial quantity. It is defined as follows: Choose an edge $(ij)$ at random and consider  the probability distribution of the number of neighbours of $i$ except $j$. The variables (analogously for constraints) excess degree distribution thus reads
\be
   q(l) = \frac{(l+1) {\cal Q}(l+1)}{\overline l} \, , \quad \quad \quad r(k) = \frac{(k+1) {\cal R}(k+1)}{\overline k} \, . \label{eq:excess}
\ee

We will always deal with factor graphs where $K$ and $c$ are of order one, and $N\to \infty, M\to \infty$. These are called {\it sparse random factor graphs}. Concerning the physical properties of sparse random factor graphs\index{factor graph!sparse}\index{random graph!sparse} the two definitions of a random graph with Poissonian degree distribution are equivalent. Some properties (e.g. the annealed averages) can however depend on the details of the definition. 

\paragraph{The tree-like property of sparse random factor graphs ---} Consider a random variable $i$ in the factor graph. We want to estimate the average length of the shortest cycle going through variable $i$. Consider a diffusion algorithm spreading into all direction but the one it came from. The probability that this diffusion will arrive back to $i$ in $d$ steps reads
\be
     1 - \left(1-\frac{1}{N}\right)^{\sum_{j=1}^d (\gamma_l \gamma_k)^j}\, ,  \label{eq:prob_tree}
\ee
where $\gamma_l = \overline{l^2}/\overline l -1$ and $\gamma_k = \overline{k^2}/\overline k -1$ are the mean values of the excess degree distribution (\ref{eq:excess}). The probability (\ref{eq:prob_tree}) is almost surely zero if 
\be
     d \ll \frac{\log{N}}{\log{\gamma_l \gamma_k}} \, .  \label{eq:distance}
\ee
An important property follows: As long as the degree distributions $ {\cal R}(k)$ and ${\cal Q}(l)$ have a finite variance the sparse random factor graphs are locally trees up to a distance scaling as $\log{N}$ (\ref{eq:distance}). We define this as the {\it tree-like}\index{factor graph!tree-like} property.

In this thesis we consider only degree distributions with a finite variance. A generalization to other cases (e.g. the scale-free networks with long-tail degree distributions) is not straightforward and many of the results which are asymptotically exact on the tree-like structures would be in general only approximative. We observed, see e.g. fig.~\ref{fig:cmplx}, that many of the nontrivial properties predicted asymptotically on the tree-like graphs seems to be reasonably precise even on graphs with about $N=10^2-10^4$ variables. It means that the asymptotic behaviour sets in rather early and does not, in fact, require $\log N\gg 1$.

\section{Computational complexity}

\subsection{The worst case complexity}

Theoretical computer scientists developed the computational complexity theory in order to quantify how hard problems can be in the worst possible case. The most important and discussed complexity classes are the P, NP and NP-complete. 

A problem is in the {\it P (polynomial) class} if there is an algorithm which is able to solve the problem for any input instance of length $N$ in at most $c N^k$ steps, where $k$ and $c$ are constants independent of the input instance. The formal definitions of what is a "problem", its "input instance" and an "algorithm" was formalized in the theory of Turing machines \cite{Papadimitriou94}, where the definition would be: The complexity class P is the set of decision problems that can be solved by a deterministic Turing machine in polynomial time. A simple example of polynomial problem is sorting a list of $N$ real numbers.

A problem is in the {\it NP class} if its instance can be stored in memory of polynomial size and if the correctness of a proposed result can be checked in polynomial time. Formally, the complexity class NP is the set of decision problems that can be solved by a non-deterministic Turing machine in polynomial time \cite{Papadimitriou94}, NP stands for non-deterministic polynomial. Whereas the deterministic Turing machine is basically any of our today computers, the non-deterministic Turing machine can perform unlimited number of parallel computations.  Thus, if for finite $N$ there is a finite number of possible solutions all of them can be checked simultaneously. This class contains many problems that we would like to be able to solve efficiently, including the Boolean satisfiability problem, the traveling salesman problem or the graph coloring. Problems which do not belong to the NP class are for example counting the number of solutions in Boolean satisfiability, or the random energy model \cite{Derrida80,Derrida81}. 

All the polynomial problems are in the NP class. It is not known if all the NP problems are polynomial, and it is  considered by many to be the most challenging problem in theoretical computer science. It is also one of the seven, and one of the six still open, Millennium Prize Problems that were stated by the Clay Mathematics Institute in 2000 (a correct solution to each of these problems results in a \$1,000,000 prize for the author). A majority of computer scientists, however, believes that the negative answer is the correct one \cite{Gasarch02}. 

The concept of NP-complete problems was introduced by Cook in 1971 \cite{Cook71}. All the NP problems can be polynomially reduced to any NP-complete problem, thus if any NP-complete problem would be polynomial then P$=$NP. Cook proved \cite{Cook71} that the Boolean satisfiability problem is NP-complete. Karp soon after added 21 new NP-complete problems to the list \cite{Karp72}. Since then thousands of other problems have been shown to be NP-complete by reductions from other problems previously shown to be NP-complete; many of these are collected in the Garey and Johnson's "Guide to NP-Completeness" \cite{GareyJohnson79}.

Schaefer in 1978 proved a dichotomy theorem for Boolean ($q=2$) constraint satisfaction problems. He showed that if the constraint satisfaction problem has one of the following four properties then it is polynomial, otherwise it is NP-complete. (1) All constraints are such that $s_i=1$ for all $i$ is a solution or $s_i=0$ for all $i$ is a solution. (2) All constraints concern at most two variables (e.g. in 2-SAT). (3) All constraints are linear equations modulo two (e.g. in XOR-SAT). (4) All constraints are the so-called Horn clauses or all of them are the so-called dual Horn clauses. A Horn clause is a disjunction of variables such that at most one variable is not negated. A dual Horn clause is when at most one variable is negated. A similar dichotomy theorem exists for 3-state variables, $q=3$, \cite{Bulatov02}. Generalization for $q>3$ is not known. 

\subsection{The average case hardness}

Given the present knowledge, it is often said that all the polynomial problems are easy and all the NP-complete problems are very hard. But, independently if P$=$NP or not, even polynomial problems might be practically very difficult, and some (or even most) instances of the NP-complete problems might be practically very easy.    

An example of a still difficult polynomial problem is the primality testing, a first polynomial algorithm was discovered by \cite{AgrawalKayal04}. 
But a "proof" of remaining difficulty is the EFF prize \cite{EFFprize} of \$100,000 to the first individual or group who discovers the first prime number with at least 10,000,000 decimal digits.

And how hard are the NP-complete problems? One way to answer is that under restrictions on the structure an NP-complete problem might become polynomial. Maybe the most famous example is 4-coloring of maps (planar factor graphs) which is polynomial. Moreover, it was a long standing conjecture that every map is colorable with 4 colors, proven by Appel and Haken \cite{AppelHaken77,AppelHaken77b}. Interestingly enough 3-coloring of maps is NP-complete \cite{GareyJohnson79}.   

But there are also settings under which the problem stays NP-complete and yet almost every instance can be solved in polynomial time. A historically important example is the Boolean satisfiability where each clause is generated by selecting literals with some fixed probability. Goldberg introduced this random ensemble and showed that the average running time of the Davis-Putnam algorithm \cite{DavisPutnam60,DavisLogemann62} is polynomial for almost all choices of parameter settings \cite{Goldberg79,GoldbergPurdom82}. Thus in the eighties some computer scientist tended to think that all the NP-complete problems are in fact on average easy and it is hard to find the evil instances which makes them NP-complete.     

\begin{figure}[!ht]
\begin{center}
 \resizebox{0.67\linewidth}{!}{\includegraphics{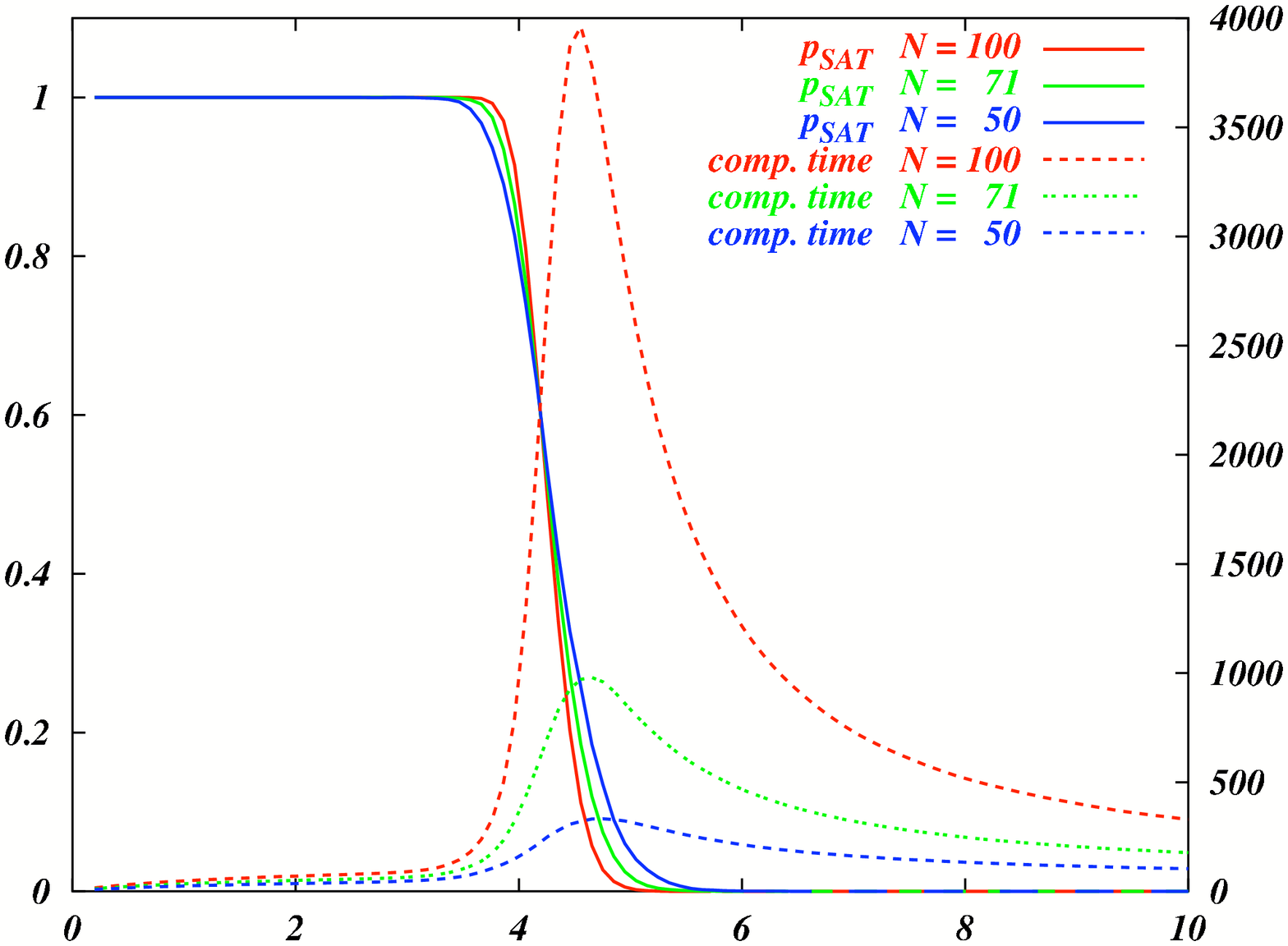}}
 \caption{\label{fig:easy_hard} The easy-hard-easy pattern in the random 3-SAT formulas as the constraint density is changed. Full lines are probabilities that a formula is satisfiable. Dashed lines is the medium running time of the DPLL algorithm. This figure is courtesy of Riccardo Zecchina.}
\end{center}
\end{figure}

The breakthrough came at the beginning of the nineties when Cheeseman, Kanefsky and Taylor asked "Where the {\it really} hard problems are?" in their paper of the same name \cite{CheesemanKanefsky91}. Shortly after Mitchell, Selman and Levesque came up with a similar work \cite{MitchellSelman92}. Both groups simply took a different random ensemble of the satisfiability (in the second case) and coloring (in the first case) instances: the length of clauses is fixed to be $K$ and they are drawn randomly as described in sec.~\ref{sec:random}.  They observed that when the density of clauses $\alpha=M/N$ is small the existence of a solution is very likely and if $\alpha$ is large the existence of a solution is very unlikely. And the {\it really} hard instances were located nearby the critical value originally estimated to be $\alpha_s\approx 4.25$ in the 3-SAT \cite{MitchellSelman92}. The hardness was judged from the median running time of the Davis-Putnam-Logemann-Loveland (DPLL) backtracking-based algorithm \cite{DavisPutnam60,DavisLogemann62}, see fig.~\ref{fig:easy_hard}. This whipped away the thoughts that NP-complete problems might in fact be easy on average. Many other studies and observations followed.
The hard instances of random $K$-satisfiability became very fast important benchmarks for the best algorithms. Moreover, there are some indications that critically constrained instances might appear in real-world applications. One may imagine that in a real world situation the amount of constraints is given by the nature of the problem, and variables usually correspond to something costly, thus the competitive designs contain the smallest possible number of variables.   

\begin{figure}[!ht]
 \resizebox{\linewidth}{!}{
  \includegraphics{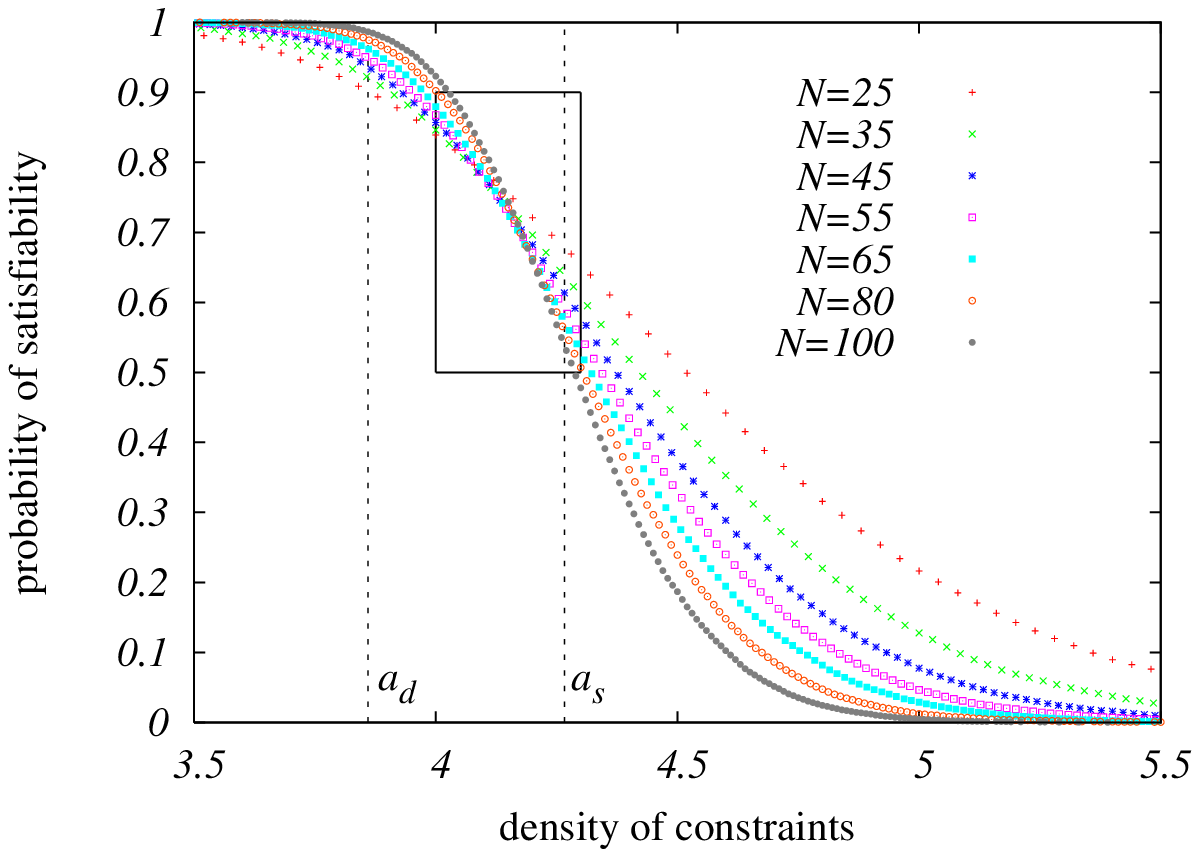}
  \includegraphics{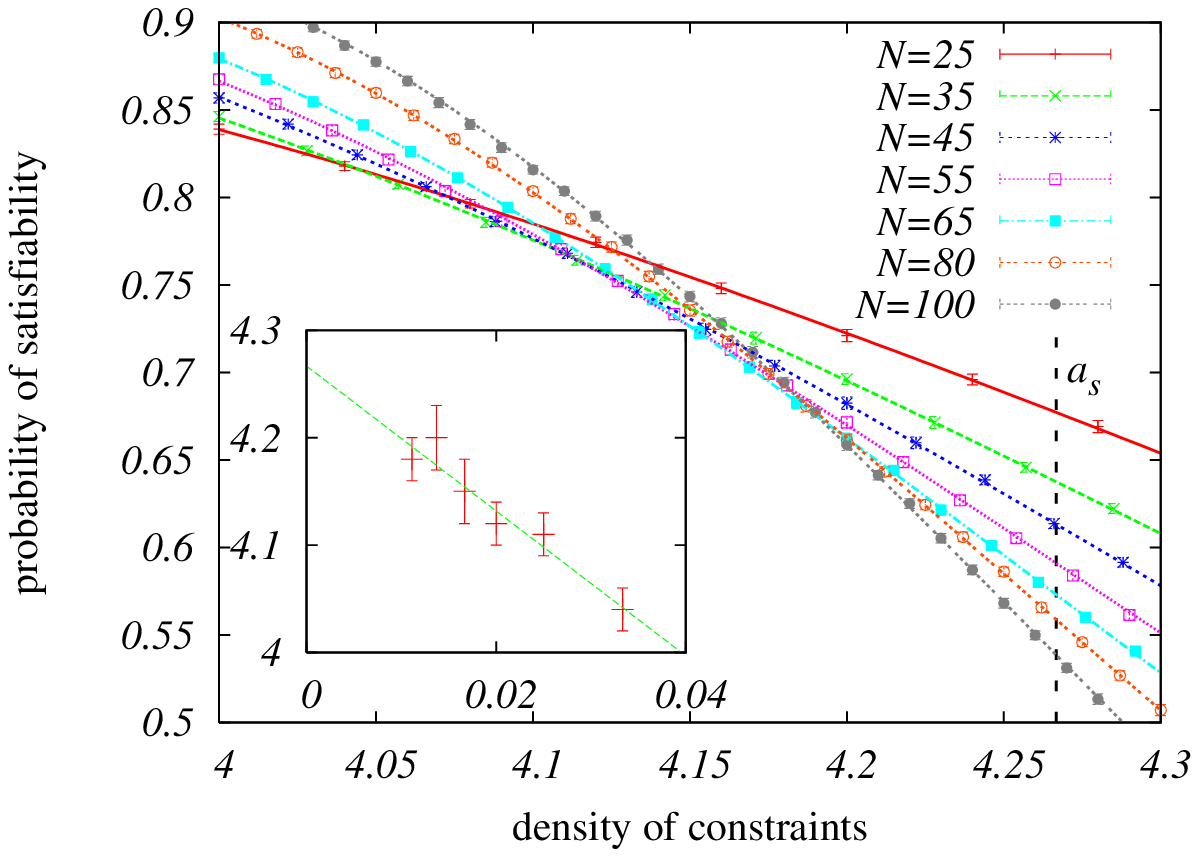}}
 \caption{\label{fig:sat_threshold} Probability that a random 3-SAT formula is satisfiable as a function of the constraint density. In the inset on the left figure is the position of the crossing point between curves corresponding to different sizes as a function of $1/N$. It seems to extrapolate to the analytical value $\alpha_s=4.267$ \cite{MezardZecchina02,MertensMezard06}. This figure should be put in contrast with fig.~\ref{fig:rigidity} where the same plot is presented for the freezing transition with a much smaller size of the inset.}
\end{figure}

Given a random $K$-SAT formula of $N$ variables the probability that it is satisfiable, plotted in fig.~\ref{fig:sat_threshold} for 3-SAT, becomes more and more like a step-function as the size $N$ grows. An analogy with phase transitions in physics cannot be overlooked. The existence and sharpness of the threshold were partially proved \cite{Friedgut99}. The best known probabilistic bounds of the threshold value in 3-SAT are $3.520$ for the lower bound \cite{KaporisKirousis03,HajiaghayiSorkin03} and $4.506$ for the upper bound \cite{DuboisBoufkhad00}. Numerical estimates of the asymptotic value of the threshold are $\alpha_s\approx 4.17$ \cite{KirkpatrickSelman94}, $\alpha_s\approx 4.258$ \cite{CrawfordAuton96}, $\alpha_s\approx 4.27$ \cite{MonassonZecchina99,MonassonZecchina99b}. The finite size scaling of the curves in fig.~\ref{fig:sat_threshold} is quite involved as the crossing point is moving. That is why the early numerical estimates of the threshold were very inaccurate. The work of Wilson \cite{Wilson02}, moreover, showed that the experimental sizes are too small and the asymptotic regime for the critical exponent is not reached in any of the current empirical works. The study of XOR-SAT indeed shows a crossover in the critical exponent at sizes which are not accessible for $K$-SAT \cite{LeoneRicci01}. 

The studies of random $K$-SAT opened up the exciting possibility to connect the hardness with an algorithm-independent property, like the satisfiability phase transition. But what exactly makes the instances near to the threshold hard remained an open question.  

\section{Statistical physics comes to the scene}
\label{sec:stat_mech}

\subsection{Glance on spin glasses}

Spin glass is one of the most interesting puzzles in statistical physics. An example of a spin glass material is a piece of gold with a small fraction of iron impurities. Physicist, on contrary to the rest of the human population, are interested in the behaviour of these iron impurities and not in the piece of gold itself. A new type of a phase transition was observed from the high temperature paramagnetic phase to the low temperature spin glass phase, where the magnetization of each impurity is {\it frozen} to a non-zero value, but there is no long range ordering. More than 30 years ago Edwards and Anderson \cite{EdwardsAnderson75} introduced a lattice model for such magnetic disordered alloys  
\be
       {\cal H} = - \sum_{(ij)} J_{ij} S_i S_j - h \sum_{i} S_i\, , \label{spin_glass}
\ee
where $S_i\in \{-1,+1\}$ are Ising spins on a 3-dimensional lattice, the sum runs over all the nearest neighbours, $h$ is the external magnetic field and the interaction $J_{ij}$ is random (usually Gaussian or randomly $\pm J$). The solution of the Edwards-Anderson model stays a largely open problem even today. 

The mean field version of the Edwards-Anderson model was introduced by Sherrington and Kirkpatrick \cite{SherringtonKirkpatrick75}, the sum in the Hamiltonian (\ref{spin_glass}) then runs over all pairs $(ij)$ as if the underlying lattice would be fully connected.  
Sherrington and Kirkpatrick called their paper "Solvable Model of a Spin-Glass". They were indeed right, but the correct solution came only five years later by Parisi \cite{Parisi80,Parisi80b,Parisi80c}. Parisi's {\it replica symmetry breaking} (RSB) solution of the Sherrington-Kirkpatrick model gave rise to a whole new theory of the spin glass phase and of the ideal glass transition in structural glasses. The exactness of the Parisi's solution was, however, in doubt till 2000 when Talagrand provided its rigorous proof \cite{Talagrand06}. The relevance of the RSB picture for the original Edwards-Anderson model is widely discussed but still unknown. 

A different mean field version of the  Edwards-Anderson model was introduced by Viana and Bray \cite{VianaBray85}, the lattice underlying the Hamiltonian (\ref{spin_glass}) is then a random graph of fixed average connectivity. The complete solution of the Viana-Bray model is also still an open problem.
 
\subsection{First encounter}

The Viana-Bray model of spin glasses can also be viewed as random graph bi-partitioning (or bi-coloring at a finite temperature). The peculiarity of the spin glass phase will surely have some interesting consequences for the optimization problem itself.  Indeed, the close connection between optimization problems and spin glass systems brought forward a whole collection of theoretical tools to analyze the structural properties of the optimization problems. 

All started in 1985 when M\'ezard and Parisi realized that the replica theory can be used to solve the bipartite weighted matching problem \cite{MezardParisi85}. Let us quote from the introduction of this work: {\it "This being a kind of pioneering paper, we have decided to present the method \{meaning the replica method\} on a rather simple problem (a polynomial one) the weighted matching. In this problem one is given $2N$ points $i=1,\dots,2N$, with a matrix of distance $l_{ij}$, and one looks for a matching between the points (a set of $N$ links between two points such that at each point one and only one link arrives) of a minimal length."}  Using the replica symmetric (RS) approach they computed the average minimal length, when the elements of the matrix $l_{ij}$ are random identically distributed independent variables. 

Shortly after Fu and Anderson \cite{FuAnderson86} used the replica method to treat the graph bi-partitioning problem. They were the first to suggest that, possibly, the existence of a phase transition in the average behaviour will affect the actual implementation and performance of local optimization techniques, and that this may also play an important role in the complexity theory. Only later, such a behaviour was indeed discovered empirically by computer scientists \cite{CheesemanKanefsky91,MitchellSelman92}. 

The replica method also served to compute the average minimal cost in the random traveling salesmen problem \cite{MezardParisi86,MezardParisi86b}. Partitioning a dense random graph into more than two groups and the coloring problem of dense random graphs were discussed in \cite{KanterSompolinsky87}.
Later some of the early results were confirmed rigorously, mainly those concerning the matching problem \cite{Aldous01,LinussonWastlund03}. All these early solved models are formulated on dense or even fully connected graph. Thus the replica method and where needed the replica symmetry breaking could be used in its original form. Another example of a "fully connected" optimization problem which was solved with a statistical physics approach is the number partitioning problem \cite{Mertens98,Mertens00}.

And what about our customary random $K$-satisfiability, which is defined on a sparse graph? Monasson and Zecchina worked out the replica symmetric solution in \cite{MonassonZecchina96,MonassonZecchina97}. It was immediately obvious that this solution is not exact as it largely overestimates the satisfiability threshold, the replica symmetry has to be broken in random $K$-SAT. 

An interesting observation was made in \cite{MonassonZecchina99}: They defined the backbone of a formula as the set of variables which take the same value in all the ground-state configurations~\footnote{In CSPs with a discrete symmetry, e.g. graph coloring, this symmetry has to be taken into account in the definition of the backbone.}. No extensive backbone can exist in the satisfiable phase in the limit of large $N$. If it would, then adding an infinitesimal fraction of constraints would almost surely cause a contradiction. At the satisfiability threshold an extensive backbone may appear. The authors of \cite{MonassonZecchina99} suggested that the problem is computationally hard if the backbone appears discontinuously and easy if it appears continuously. They supported this by replica symmetric solution of the SAT problem with mixed 2-clauses and 3-clauses, the so-called $2+p$-SAT. Even if the replica symmetric solution is not correct in random $K$-SAT and even if it overlooks many other important phenomena the concept of backbone is fruitful and we will discuss its generalization in chapter~\ref{freezing}. 

How to deal with the replica symmetry breaking on a sparse tree-like graph was an open question since 1985, when Viana and Bray \cite{VianaBray85} introduced their model. 
The solution came only in 2000 when M\'ezard and Parisi published their paper "Bethe lattice spin glass revisited" \cite{MezardParisi01}. They showed how to treat correctly and without approximations the first step of replica symmetry breaking (1RSB) and described how, in the same way, one can in principal deal with more steps of replica symmetry breaking, this extension is however numerically very difficult. But before explaining the 1RSB method we describe the general replica symmetric solutions. And illustrate its usefulness on the problem of counting matchings in graphs [\ZMM]. Only then we describe the main results of the 1RSB solution and illustrate the method in the 1-in-$K$ SAT problem [\RS]. After we list several "loose ends" which appeared in this approach. Finally we summarize the main contribution of this thesis. This will be the departure point for the following part of this thesis which contains most of the original results.

\section{The replica symmetric solution}

The replica symmetric (RS) solution on a locally tree-like graph consists of two steps:
\begin{itemize}
\item[(1)]{Compute the partition sum and all the other quantities of interest as if the graph would be a tree.}
\item[(2)]{The replica symmetric assumption: Assume that the correlations induced by long loops decay fast enough, such that this tree solution is also correct on the only locally tree-like graph.}
\end{itemize}
Equivalent names used in literature for the replica symmetric solution are Bethe-Peierls approximation (in particular in the earlier physics references) or belief propagation (in computer science or when using the iterative equation as an algorithm to estimate the marginal probabilities - magnetizations in physics). Both these conveniently abbreviate to BP.

\subsection{Statistical physics description}

Let $\phi_a(\partial a)$ be the evaluating function for the constraint $a$ depending on the variables neighbourhooding with $a$ in the factor graph $G(V,F,E)$. A satisfied constraint has $\phi_a(\partial a)=1$ and violated constraint $\phi_a(\partial a)=0$. The Hamiltonian can then be written as
\be
      H_G(\{s\}) = \sum_{a=1}^M   \big[ 1-\phi_a(\partial a) \big]\, . \label{Ham}
\ee
The energy cost is thus one for every violated constraint. The corresponding Boltzmann measure on configurations 
is:
\begin{equation}
   \mu_G(\{s\},\beta) = \frac{1}{Z_G(\beta)} e^{-\beta H_G(\{s\})}\ ,
   \label{Boltzmann}
\end{equation}
where $\beta$ is the inverse temperature and $Z_G(\beta)$ is the partition function. The marginals (magnetizations) $\chi^i_{s_i}$\index{marginals} are defined as the probabilities that the variable $i$ takes value $s_i$ 
\be
        \chi^i_{s_i} =  \frac{1}{Z_G(\beta)}  \sum_{\{s_j\},j=1,\dots,i-1,i+1,\dots,N} e^{-\beta H_G(\{s_j\},s_i)}\, .  \label{eq:marginals}
\ee

The goal is to compute the internal energy $E_G(\beta)$  and the entropy 
$S_G(\beta)$. For $\beta \to \infty$ (zero temperature limit) these two quantities give 
the ground state properties. We are interested in the "thermodynamic" limit of large graphs ($N\to
\infty$), and we shall compute expectations over ensembles of graphs of the
densities of thermodynamical potentials $\epsilon(\beta)=
{\mathbb{E}}[ E_G(\beta) ]/N $ and $s(\beta)=
{\mathbb{E}}[S_G(\beta)]/N$, as well as the average free energy density
\begin{equation}
f(\beta)=\frac{-1}{\beta N}{\mathbb{E}}[\log{Z_G(\beta)} ]= \frac{1}{N}
{\mathbb{E}} [F_G(\beta)]  = \epsilon(\beta)- \frac{1}{\beta} s(\beta)\, .
\end{equation}
The reason for this interest is that, for reasonable 
graph ensembles, $F_G(\beta)$ is self-averaging. This means that 
the distribution of $F_G(\beta)/N$ becomes more and more sharply peaked 
around $f(\beta)$ when $N$ increases.

\subsection{The replica symmetric solution on a single graph}

\begin{figure}[!ht]
\begin{center}
   \resizebox{0.2\linewidth}{!}{\includegraphics{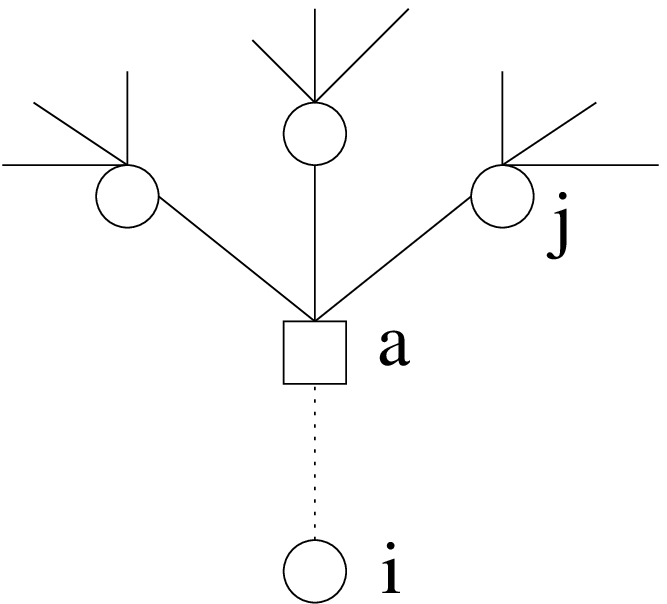}}
 \hspace{0.05\linewidth}
   \resizebox{0.2\linewidth}{!}{\includegraphics{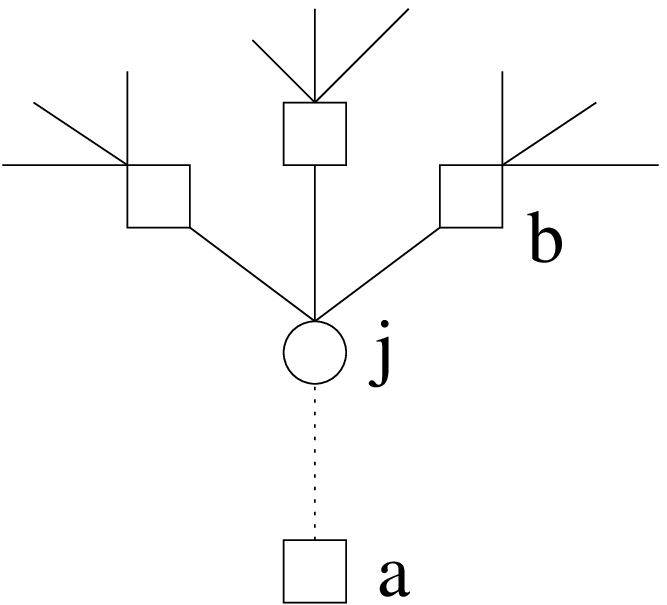}}
\end{center}
\caption{\label{fig:fac_gr} Parts of the factor graph used to compute  $\psi_{s_i}^{a\to i}$ and $\chi_{s_j}^{j\to a}$.}
\end{figure}
  
First suppose that the underlying factor graph is a tree, part of this tree is depicted in fig.~\ref{fig:fac_gr}. We define messages $\psi_{s_i}^{a \to i}$ as the probability that node $i$ takes value $s_i$ on a modified graph where all constraints around $i$ apart $a$ were deleted, and $\chi_{s_j}^{j\to a}$ as the probability that variable $j$ takes value $s_j$ on a modified graph obtained by deleting constraint $a$. On a tree these messages can be computed recursively as 
\begin{subequations}
\label{eq:BP}
\bea
 \psi^{a\to i}_{s_i} &=& \frac{1}{Z^{a\to i}} \sum_{\{s_j\},j\in \partial a-i} \phi_a(\{s_j\},s_i,\beta)  \prod_{j\in \partial a-i} \chi_{s_j}^{j\to a} \equiv {\cal F}_\psi(\{\chi^{j\to a}\}) \, , \label{eq:BP_1}\\
   \chi^{j\to a}_{s_j} &=& \frac{1}{Z^{j\to a}} \prod_{b\in \partial j -a} \psi^{b\to j}_{s_j} \equiv {\cal F}_\chi(\{\psi^{b\to j} \}) \, ,\label{eq:BP_2}
\eea
\end{subequations}
where $Z^{a\to i}$ and $Z^{j\to a}$ are normalization constants, the factor $\phi_a(\{s\},\beta)=1$ if the constraint $a$ is satisfied by the configuration $\{s\}$ and $\phi_a(\{s\},\beta)=e^{-\beta}$ if not. We denote by $\psi^{a\to i}$ the whole vector $(\psi^{a\to i}_0,\dots,\psi^{a\to i}_{q-1})$ and analogically $\chi^{j\to a}=(\chi^{k\to a}_0,\dots,\chi^{j\to a}_{q-1})$.
This is one form of the {\it belief propagation}\index{belief propagation}\index{propagation!belief} (BP) equations 
\cite{KschischangFrey01,Pearl82}, sometimes called sum-product equations.
The probabilities $\psi$, $\chi$ are interpreted as messages (beliefs) living on the edges of the factor graph, with the consistency rules (\ref{eq:BP_1}) and (\ref{eq:BP_2}) on the function and variable nodes. Equations (\ref{eq:BP}) are usually solved by iteration, the name {\it message passing}\index{message passing} is used in this context. 
In the following it will be simpler not to consider the "two-levels" equations (\ref{eq:BP}) but 
\be
   \psi^{a\to i}_{s_i} = \frac{1}{Z^{j\to i}} \sum_{\{s_j\},j\in \partial a-i} \phi_a(\{s_j\},s_i,\beta)  \prod_{j\in \partial a-i}  \prod_{b\in \partial j -a} \psi^{b\to j}_{s_j} \equiv {\cal F}(\{\psi^{b\to j}\})  \label{eq:BP_one}\, ,   
\ee
where $Z^{j \to i} = Z^{a\to i}\prod_{j\in \partial a-i}Z^{j\to a} $. Notice that on simple graphs, i.e., when either $l_i=2$ for all $i=1,\dots,N$ or $k_a=2$ for all $a=1,\dots,M$, the form (\ref{eq:BP_one}) simplifies further. And on constraint satisfaction problems on simple graphs (e.g. the matching or coloring problems) the "two-levels" equations are almost never used.

Assuming that one has found the fixed point of the belief propagation equations (\ref{eq:BP_1}-\ref{eq:BP_2}), one can deduce the various marginal probabilities and the free energy, entropy etc. 
The marginal probability\index{marginals} (\ref{eq:marginals}) of variable $i$ estimated by the BP equations is 
\be
        \chi^i_{s_i} = \frac{1}{Z^i} \prod_{a\in \partial i} \psi_{s_i}^{a\to i} \, .   \label{eq:total}
\ee

To compute the free energy we first define the free energy 
shift $\Delta F^{a+ \partial a}$ after addition of a function 
node $a$ and all the variables $i$ around it, and the free
energy shift $\Delta F^{i}$ after addition of a variable $i$.
These are given in general by:
\begin{subequations}
\label{eq:norms}
\bea
 e^{-\beta \Delta F^{a+\partial a} } &=& Z^{a+\partial a} =  \sum_{\{s_i\},i\in \partial a} \phi_a(\{s_i\},\beta)  \prod_{i\in \partial a} \prod_{b\in \partial i -a} \psi_{s_i}^{b\to i} \, , \label{eq:Zaa}\\
 e^{-\beta \Delta F^{i} } &=& Z^i = \sum_{s_i} \prod_{a\in \partial i} \psi_{s_i}^{a\to i} \, . \label{eq:Zi}
\eea
\end{subequations}
The total free energy is then obtained by summing over all constraints and subtracting the terms counted twice \cite{MezardParisi01,YedidiaFreeman03}:
\begin{equation} 
F_G(\beta)=
\sum_a \Delta F_{a+\partial a} - \sum_i (l_i-1) \Delta F_{i}\, .
  \label{free_given}
\end{equation} 
This form of the free energy is variational, i.e., the derivatives 
$\frac{\partial (\beta F_G(\beta))}{\partial \chi^{i\to a}}$ and $\frac{\partial (\beta F_G(\beta))}{\partial \psi^{a\to i}}$ vanish if and only 
if the probabilities $\chi^{i\to a}$ and $\psi^{a\to i}$ satisfy (\ref{eq:BP_1}-\ref{eq:BP_2}). This allows to compute easily the internal energy as
\be
E_G(\beta) = \frac{\partial \beta F_G(\beta)}{\partial \beta} = - \sum_a \frac{\partial_\beta  Z^{a+\partial a}}{ Z^{a+\partial a}}\, .
\ee
The entropy is then obtained as 
\begin{equation}
S_G(\beta)=\beta[E_G(\beta)-F_G(\beta)]\, .
\label{entropy_G}
\end{equation}

All the equations (\ref{eq:BP})-(\ref{entropy_G}) are exact if the graph $G$ is a tree. The replica symmetric approach consists in assuming that all correlations decay fast enough that application of eqs.~(\ref{eq:BP})-(\ref{entropy_G}) on a large tree-like graph $G$ gives asymptotically exact results. These equations can be used either on a given graph $G$ or to compute the average over the graph (and disorder) ensemble.  

\subsection{Average over the graph ensemble}
\label{sec:average}

We now study the typical instances in an ensemble of graphs. We denote the average over the ensemble by
$\mathbb{E}(\cdot)$. 
We assume that the random factor-graph ensemble is given by a prescribed degree 
distribution ${\cal Q}(l)$ for variables and ${\cal R}(k)$ for constraints.
Let us call ${\cal P}(\psi)$ and ${\cal O}(\chi)$ the distributions of messages $\psi$ and $\chi$ over all the edges of a large typical factor graph from the ensemble. 
They satisfy the following self-consistent  equations
\begin{subequations}
\label{eq:averages}
\bea
   {\cal P}(\psi) = \sum_{l=1}^\infty q(l)
  \int \prod_{i=1}^{l} \left[ {\rm d} \chi^i {\cal O}(\chi^i) \right] 
  \delta\big[\psi - {\cal F}_\psi(\{\chi^i\})\big]\, , \label{eq:aver_P} \\ 
   {\cal O}(\chi) = \sum_{k=1}^\infty r(k)
  \int \prod_{i=1}^{k} \left[ {\rm d} \psi^i {\cal P}(\psi^i) \right] 
  \delta\big[\chi - {\cal F}_\chi(\{\psi^i\})\big] \, , \label{eq:aver_O}
\eea
\end{subequations}
where the functions ${\cal F}_\psi$ and ${\cal F}_\chi$ represent the BP equations (\ref{eq:BP_1}-\ref{eq:BP_2}), $q(l)$ and $r(k)$ are the excess degree distributions defined in (\ref{eq:excess}). If there is a disorder in the interaction terms, as e.g. the negations in $K$-SAT, we average over it at the same place as over the fluctuating degree. 

Solving equations (\ref{eq:aver_P}-\ref{eq:aver_O}) to obtain the distributions ${\cal P}$ and ${\cal O}$ is not straightforward.  In some cases (on regular factor graphs, at zero temperature, etc.) it can be argued that the distributions ${\cal P}$, ${\cal O}$ are sums of Dirac delta functions. Then the solution of eqs.~(\ref{eq:aver_P}-\ref{eq:aver_O}) can be obtained analytically. But in general distributional equations of this type are not solvable analytically. However, a numerical technique called {\it population dynamics} \cite{MezardParisi01} is very efficient for their resolution. In appendix \ref{app:pop_dyn} we give a pseudo-code describing how the population dynamics technique works.

Once the distributions ${\cal P}$ and ${\cal O}$ are known the average of the free energy density can be computed by averaging (\ref{free_given}) over ${\cal P}$. 
This average expression for the free energy  is again in its variational form 
(see \cite{MezardParisi01}), i.e., the functional derivative 
$\frac{\delta f(\beta)}{\delta {\cal P}(h)}$ vanishes if and only if
 ${\cal P}$ satisfies (\ref{eq:aver_m}). The average energy and entropy density
are thus expressed again via the partial derivatives.

\paragraph{Factorized solution ---}\index{solution!factorized} As we mentioned, on the ensemble of random regular factor graphs (without disorder in the interactions) the solution of equations (\ref{eq:averages}) is very simple: ${\cal P}(\psi) = \delta(\psi - \psi^{\rm reg})$,  ${\cal Q}(\chi) = \delta(\chi - \chi^{\rm reg})$, where $\psi^{\rm reg}$ and $\chi^{\rm reg}$ is a self-consistent solution of (\ref{eq:BP}). This is because in the thermodynamical limit an infinite neighbourhood of every variable is exactly identical thus also the marginal probabilities have to be identical in every physical solution.

\subsection{Application for counting matchings}
\label{sec:matching}

To demonstrate how the replica symmetric method works to compute the entropy, that is the logarithm of the number of solutions, we review the results for matching on sparse random graphs [\ZMM].
The reasoning why the replica symmetric solution is exact for the matching problem is done on the level of self-consistency checks in [\ZMM]. And \cite{BayatiNair06} have worked out a rigorous proof for graphs with bounded degree and a large girth (length of the smallest loop).

Consider a graph $G(V,E)$ with $N$ vertices ($N=|V|$) and a set of 
edges $E$. A {\it matching}\index{matching} (dimerization) of $G$ is a subset of edges $M \subseteq E$ such  
that each vertex is incident with at most one edge in $M$. In other words the edges in the matching $M$ do not touch each other. The {\it size of the matching}, $|M|$, is the number of edges in 
$M$. Our goal is to compute the entropy of matchings of a given size on a typical large Erd\H{o}s-R\'enyi random graph.

We describe a matching by the variables $s_{i}=s_{(ab)} \in\{0,1\}$ assigned to each edge $i=(ab)$ of $G$, with $s_{i}=1$ if $i \in M$ and $s_{i}=0$ otherwise. The constraints that two edges 
in a matching cannot touch impose that, on each vertex $a\in V$: $\sum_{b, (ab)\in E} s_{(ab)} \le 1$. To complete our statistical physics description, we define for each given graph $G$ an energy 
(or cost) function which gives, for each matching $M=\{s\}$, the number of unmatched vertices:
\begin{equation}
   H_G(M=\{s\}) = \sum_{a} E_a(\{s\}) = N - 2|M|\ , \label{Ham_matching}
\end{equation}
where $E_a=1-\sum_{\partial b} s_{(ab)}$. 

\begin{figure}[!ht]
\begin{center}
   \resizebox{0.6\linewidth}{!}{
     \includegraphics{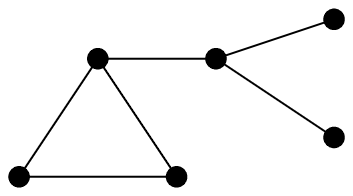}
     \hspace{1cm}
     \includegraphics{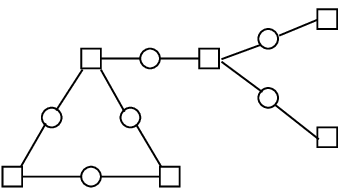}
   }
\end{center}
\caption{\label{fig:gr_fact} On the left, example of a graph with six nodes and six edges. On the right, the corresponding factor graph with six function nodes (squares) and six variable nodes (circles). }
\end{figure}

In the factor graph representation we transform the graph $G$ into a factor graph $F(G)$ as follows (see fig.~\ref{fig:gr_fact}): To each edge of $G$ corresponds a
variable node (circle) in $F(G)$; to each vertex of $G$ corresponds a
function node (square) in $F(G)$. We shall index the variable nodes by
indices $i,j,k,\dots$ and function nodes by $a,b,c,\dots$. The variable $i$
takes value $s_i=1$ if the corresponding edge is in the matching, and $s_i=0$
if it is not. The weight of a function node $a$ is
\begin{equation}
\phi_a(\{\partial a\},\beta)=\ind \left(\sum_{i\in \partial a}s_i\le 1\right) e^{-\beta(1-\sum_{i\in\partial a}s_i)}\ ,
\end{equation}
where $\partial a$ is the set of all the variable nodes which are neighbours of the function node $a$, and the total Boltzmann weight of a configuration
is $\frac{1}{Z_G(\beta)}\prod_a \phi_a(\{\partial a\},\beta)$. 

The belief propagation equation (\ref{eq:BP}) becomes
\begin{eqnarray}
    \chi_{s_i}^{i\to a}&=& \frac{1}{Z^{b\to a}} \sum_{\{s_j\}}
    \ind\left(s_i+\sum_{j\in \partial b - i} s_j  \le 1 \right)
    e^{- \beta(1-s_i-\sum s_j)}\, \prod_{j\in \partial b - i}      
    \chi_{s_j}^{j \to b},  \label{cav_Z}
\end{eqnarray} 
where $Z^{b\to a}$ is a normalization constant.
In statistical physics the more common form of the BP equations uses analog of local magnetic fields instead of probabilities. For every edge between a variable $i$ and a function node $a$, we define a {\it cavity field}\index{cavity field} $ h^{i\to a}$ as 
\begin{equation}
    e^{-\beta h^{i\to a}} \equiv \frac{\chi_{0}^{i\to a} }{
    \chi_{1}^{i\to a} }\ .
    \label{def_cav_h}
\end{equation}
The recursion  relation between cavity fields is then:
\begin{eqnarray}
    h^{i\to a}= - \frac{1}{\beta} \log{\left[e^{-\beta}+ 
        \sum_{j\in \partial b - i} e^{\beta h^{j\to b}} 
    \right]} \label{BP_T}.
\end{eqnarray}
The expectation value (with respect to the Boltzmann distribution) of
the occupation number $s_i$ of a given edge $i=(ab)$ is equal to 
\begin{equation}
\langle s_i\rangle = \frac{1}{1+e^{-\beta(h^{i\to a}+h^{i\to b})}}\, .
\label{marginal}
\end{equation}
 
The free energy shifts needed to compute the total free energy (\ref{free_given}) are
\begin{subequations}
\label{eq:norms_matching}
\begin{eqnarray}
 e^{-\beta \Delta F_{a+i\in \partial a} } &=& e^{-\beta} + 
           \sum_{i \in a}  e^{\beta  h^{i \to a}}\, , \label{df3}\\
 e^{-\beta \Delta F_{i} } &=& 1 +  
            e^{\beta( h^{i \to a}+ h^{i \to b})} \, . \label{df4}
\end{eqnarray}
\end{subequations}
The energy, related to the size of the matching via (\ref{Ham_matching}), is then 
\begin{equation}
   E_G(\beta)  = \sum_a \frac{1}{1+\sum_{i \in \partial a}  e^{\beta(1+h^{i\to a})}} .
    \label{energy_G}  
\end{equation}
This is the sum of the probabilities that node $a$ is not matched. 

The distributional equation (\ref{eq:averages}) becomes
\begin{equation}
 {\cal O}(h)= \sum_{k=1}^\infty r(k)  
  \int \prod_{i=1}^{k} \left[ {\rm d} h^i {\cal O}(h^i) \right] 
  \delta{\left[h+ \frac{1}{\beta}
  \log{\left( e^{-\beta} +\sum_{i}  e^{\beta h^{i} } \right)}  \right]}.
  \label{eq:aver_m}
\end{equation}
And the average free energy is explicitly
\begin{eqnarray}
  f(\beta) &=& \frac{{\mathbb{E}}[F_G(\beta)]}{N}  = -\frac{1}{\beta} 
    \sum_{k=0}^\infty {\cal R}(k) 
    \int \prod_{i=1}^k \left[ {\rm d} h^i {\cal O}(h^i)\right] 
    \log{\left( e^{-\beta} +\sum_{i}  e^{\beta h^{i}} \right)} \nonumber \\
    &+& \frac{c}{2\beta} 
    \int {\rm d} h^1 \, {\rm d} h^2\, {\cal O}(h^1)\, 
    {\cal O}(h^2)\, \log{\left(1 + e^{\beta(h^{1}+ h^{2})}\right)}.
    \label{free_en}
\end{eqnarray}
Where ${\cal R}(k)$ is the connectivity distribution of the function nodes, that is the connectivity distribution of the original graph, $c$ is the average connectivity. The distributional equations are solved via the population dynamics method, see appendix \ref{app:pop_dyn}. Fig.~\ref{fig:s_e} then presents the resulting average entropy as a function of size of the matching. 

\begin{figure}[!ht]
\begin{center}
   \resizebox{0.67\linewidth}{!}{\includegraphics{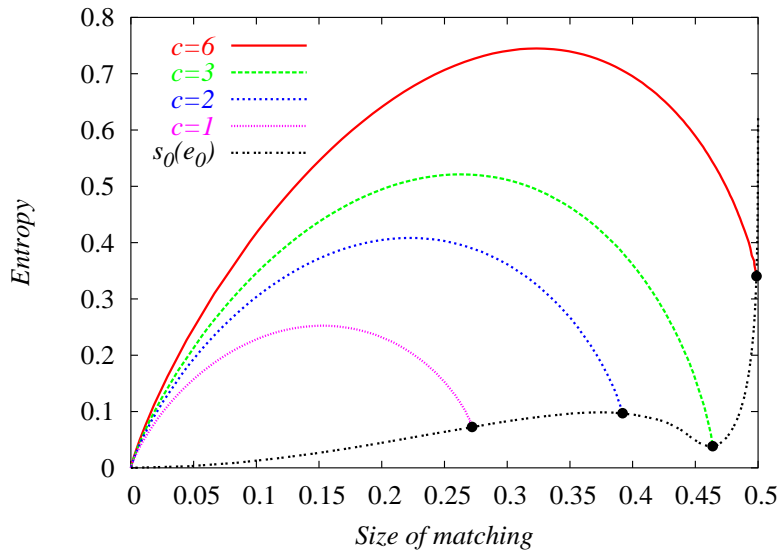}}
\end{center}
\caption{\label{fig:s_e} Entropy density $s(m)$ as a function of relative size of the matching $m=|M|/N$ for Erd\H{o}s-R\'enyi random graphs with mean degrees $c=1, 2, 3, 6$. The lower curve is the ground state entropy density for all mean degrees. The curves are obtained by solving eqs.~(\ref{eq:aver_m})-(\ref{free_en}) with a population dynamics, using a population of sizes $N=2\cdot 10^4$ to $2\cdot 10^5$ and the number of iterations $T=10000$.}
\end{figure}

\section{Clustering and Survey propagation}
\label{sec:energetic}

As we said previously in the random $K$-SAT the replica symmetric solution is not generically correct. 
M\'ezard and Parisi \cite{MezardParisi01} understood how to deal properly and without approximations with the replica symmetry breaking on random sparse graphs, that is how to take into account the correlations induced by long loops. More precisely in their approach only the one-step (at most two-step on the regular graphs) replica symmetry breaking solution is numerically feasible. Anyhow, such a progress opened the door to a better understanding of the optimization problems on sparse graphs. The $K$-satisfiability played again the prominent role. 

To compute the ground state energy within the 1RSB approach we can restrict only to energetic considerations as described in \cite{MezardParisi03}, we call this approach the {\it energetic zero temperature limit}\index{cavity method!energetic}. Applying this method to $K$-satisfiability leads to several outstanding results \cite{MezardParisi02,MezardZecchina02}, we describe the three most remarkable ones. Soon after, analog results were obtained for many other optimization problems, for example graph coloring \cite{MuletPagnani02,BraunsteinMulet03,KrzakalaPagnani04}, vertex cover \cite{Zhou03}, bicoloring of hyper-graphs \cite{CastellaniNapolano03}, XOR-SAT \cite{FranzLeone01,MezardRicci03} or lattice glass models \cite{BiroliMezard02,RivoireBiroli04}.

\paragraph{Clustering ---}
It was known already in the "pre-1RSB-cavity era" that replica symmetry broken solution is needed to solve random $K$-SAT. Such a need is interpreted as the existence of many metastable well-separated states, in the case of highly degenerate ground state this leads to a clustering of solutions in the satisfiable phase \cite{BiroliMonasson00,MezardParisi02,MezardZecchina02}.  The energetic 1RSB cavity method deals with clusters containing frozen variables (clusters with backbones), that is variables which have the same value in all the solutions in the cluster. It predicts how many of such clusters exist at a given energy, the logarithm of this number divided by the system size $N$ defines the complexity function $\Sigma(E)$. According to the energetic cavity method for 3-SAT, clusters exist, $\Sigma(0)\neq 0$, for constraint density $\alpha>\alpha_{\rm SP}=3.92$ \cite{MezardParisi02,MezardZecchina02}. 

It was conjectured \cite{MezardParisi02,MezardZecchina02} that there is a link between clustering, ergodicity breaking, existence of many metastable states and the difficulty of finding a ground state via local algorithms. The critical value $\alpha_{\rm SP}$ was called the {\it dynamical} transition and the region of $\alpha>\alpha_{\rm SP}$ the {\it hard-SAT} phase.  

Clusters were viewed as a kind of pure states, however, in the view of many a good formal definition was missing. It was also often referred to some sort of geometrical separation between different clusters. A particularly popular one is the following: Clusters are connected components in the graph where solutions are the nodes and two solutions are adjacent if they differ in only $d$ variables. Depending on the model and author the value of $d$ is either one of $d$ is a finite number of $d$ is said to be any sub-extensive number. The notion of $x$-satisfiability, the existence of pairs of solutions at a distance $x$, leads to a rigorous proof of existence of exponentially many geometrically separated clusters \cite{MezardMora05,MoraMezard05,AchlioptasRicci06}.

\paragraph{The satisfiability threshold computed ---}
The energetic 1RSB cavity method allows to compute the ground state energy and thus also the satisfiability threshold $\alpha_s$. In 3-SAT its value is $\alpha_s=4.2667$ \cite{MezardParisi02,MezardZecchina02,MertensMezard06}. This value is computed as a solution of a closed distributional equation. This time there is an excellent agreement with the empirical estimations. 
Is the one step of replica symmetry breaking sufficient to locate exactly the satisfiability threshold? The stability of the 1RSB solution was investigated in \cite{MontanariParisi04}, the 1RSB energetic cavity was shown to describe correctly the ground state energy for $4.15<\alpha<4.39$ in 3-SAT. In particular, it yields the conjecture that the location of the satisfiability threshold is actually exact. From a rigorous point of view it was proven that the 1RSB equations give an upper bound on the satisfiability threshold \cite{FranzLeone03,FranzLeone03b,PanchenkoTalagrand04}.

\paragraph{Survey Propagation:  a revolutionary algorithm ---}
The most spectacular result was the development of a new message passing algorithm, the survey propagation \cite{MezardZecchina02,BraunsteinMezard05}. Before the replica and cavity analysis were used to compute the quenched averages of thermodynamical quantities. Using always the self-averaging property that the average of certain (not all) quantities is equal to their value on a large given sample.  
M\'ezard and Zecchina applied the energetic 1RSB cavity equations, later called survey propagation, on a single large graph. This resulted in an algorithm which is arguably still the best known for large instances of random 3-SAT near to the satisfiability threshold. And even more interesting than its performance is the conceptual advance this brought into applications of statistical physics to optimization problems.

\section{Energetic 1RSB solution}
\label{sec:1in3}

In this section we derive the energetic zero-temperature limit of the 1RSB method. When applied to the satisfiability problem this leads, between others, to the calculation of the satisfiability threshold and to the survey propagation equations and algorithm. We illustrate this on the 1-in-3 SAT problem. Before doing so we have to introduce the warning propagation equations, on which the derivation of the survey propagation relies. 

\subsection{Warning Propagation}

In general warning propagation\index{warning propagation}\index{propagation!warning} (min-sum) is a zero temperature, $\beta \to \infty$, limit of the belief propagation (sum-product) equations (\ref{eq:BP_1}-\ref{eq:BP_2}). It can be used to compute the ground state energy (minimal fraction of violated constraints) at the replica symmetric level. A constraint satisfaction problem at a finite temperature gives rise to $\phi_a(\{\partial a\},\beta)=1$ if the constraint $a$ is satisfied by configuration $\{s_{\partial a}\}$, and $\phi_a(\{\partial a\},\beta)=e^{-2\beta}$ if $a$ is not satisfied by $\{s_{\partial a}\}$\footnote{The factor 2 in the Hamiltonian is introduced for convenience and in agreement with the notation of [\RS].}. In a general Boolean CSP, with $N$ variables $s_i\in \{-1,1\}$, the warning propagation can then be obtained from (\ref{eq:BP_1}-\ref{eq:BP_2}) by introducing warnings $u$ and $h$ as
\begin{align}
  e^{2\beta h^{i\to a}} & \equiv \frac{\chi_{1}^{i\to a}}{\chi_{-1}^{i\to a}} \, ,
&
  e^{2\beta u^{a\to i}} & \equiv \frac{\psi_{1}^{a\to i}}{\psi_{-1}^{a\to i}} \, .
\label{def_hu}
\end{align}
This leads in the limit of zero temperature, $\beta \to \infty$,  to
\begin{subequations}
\label{eq:WP}
\begin{align}
  h^{i \to a} &= \sum_{b \in \di - a} u^{b \to i} \, ,
\\
\begin{split}
  u^{a \to i} &= \frac{1}{2} \Big[
 \max_{\{s_j\}}
   \big( \sum_{j\in \partial a-i} h^{j\to a} s_j -2 E_{a}(\{s_j\},+1) \big) \\
 & \hspace{1.cm}
 - \max_{\{s_j\}}
  \big( \sum_{j\in \partial a-i}  h^{j\to a} s_j -2 E_{a}(\{s_j\},-1) \big)
\Big] \, .
\end{split}
\label{BP_0}
\end{align}
\end{subequations}
where $E_a(\{s_i\})=0$ if the configuration $\{s_i\}$ satisfies the constraint $a$, and $E_a(\{s_i\})=1$ if it does not. The warnings $u$ and $h$ have to be integer numbers, as they can be interpreted as changes in the ground state energy of the cavity subgraphs when the value of variable $i$ is changed from $s_i=0$ to $s_i=1$. Given $E_a\in \{0,1\}$ we have that $h \in \mathbb{Z}$ and $u \in \{-1,0,+1\}$. The correspondence between values of $u$ and $\psi$ are
\begin{subequations}
\label{eq:u}
\bea
       u&=&1     \quad \quad \Leftrightarrow \quad \psi_1=1\, , \quad  \psi_{-1}=0\, ,  \label{eq:u1}\\
       u&=&-1  \,   \quad \Leftrightarrow \quad \psi_1=0\, , \quad  \psi_{-1}=1\, , \label{eq:um1}\\
       u&=&0     \quad \quad \Leftrightarrow \quad \psi_1=\epsilon\, , \quad  \psi_{-1}=1-\epsilon\, , \quad  0<\epsilon<1\, .\label{eq:u0}
\eea
\end{subequations}
The warnings $u$ and $h$ can thus be interpreted in the following way
\begin{center}
\begin{tabular}{rc|cp{13cm}}
 $u^{a\to i}=-1$ &&&
  Constraint $a$ tells to variable $i$: ``I think you should be $-1$.''
\\
 $u^{a\to i}=0$  &&&
  Constraint $a$ tells to variable $i$: ``I can deal with any value you
 take.''
\rule{0pt}{12pt}
\\
 $u^{a\to i}=+1$ &&&
  Constraint $a$ tells to variable $i$: ``I think you should be $+1$.''
\rule{0pt}{12pt}
\end{tabular}
\end{center}

\begin{center}
\begin{tabular}{rc|cp{13cm}}
 $h^{i\to a}<0$ &&&
  Variable $i$ tells to constraint $a$: ``I would prefer to be $-1$.''
\\
 $h^{i\to a}=0$  &&&
  Variable $i$ tells to constraint $a$: ``I don't have any strong preferences.''
\rule{0pt}{12pt}
\\
 $h^{i\to a}>0$ &&&
  Variable $i$ tells to constraint $a$: ``I would prefer to be $+1$.''
\rule{0pt}{12pt}
\end{tabular}
\end{center}

\noindent
Given this interpretation the prescriptions (\ref{eq:WP}) on how to
update the warnings over the graph becomes intuitive. Variable $i$ collects the preferences from all constraints except $a$ and sends the result to $a$. Constraint $a$ then decides which value $i$ should take given the preferences of all its other neighbours. 

\begin{table}[!ht]
\begin{center}
\begin{tabular}{|c|c||c|}\hline
$h^{1\to a}$& $h^{2\to a}$& $u^{a\to 3}$ \\ \hline
+ & + & 0 \\ \hline 
+ & -- & -- \\ \hline 
+ & 0 & -- \\ \hline 
0 & 0 & 0 \\ \hline 
-- & -- & + \\ \hline 
-- & 0 & 0 \\ \hline 
\end{tabular}
\end{center}
\caption{\label{tab:WP}Example of the update (\ref{BP_0}) in the positive 1-in-3 SAT problem, where exactly one variable in the constraint takes value $1$ in order to satisfy the constraint. The first line might seem counter-intuitive, but note that we defined the energy in such a way that configuration $(1,1,1)$ is as bad as $(1,1,-1)$.}
\end{table}
Given the fixed point of the warning propagation (\ref{eq:WP}) the total warning of variable $i$ is
\be
    h^{i} = \sum_{a \in \di } u^{a\to i} \, .   \label{eq:warning}
\ee
The corresponding energy can be computed as
\be 
         E = \sum_a  \Delta E^{a + \da} - \sum_i (l_i-1) \Delta E^{i}\, , \label{eq:WP_ener}
\ee
where $\Delta E^{a + \da}$ is the number of contradictions created when constraint $a$ and all its neighbours are added to the graph, $\Delta E^{i}$ is the number of contradictions created when variables $i$ is added to the graph. The energy shifts can be computed from (\ref{eq:Zaa}-\ref{eq:Zi}) using (\ref{def_hu}) and taking $\beta \to \infty$ they read
\begin{subequations}
\label{eq:WP_ener_s}
\begin{align}
 \Delta E^{a + \da}
&=
 - \max_{\{s_{\partial a}\}}
 \Big[ \sum_{i \in \partial a} h^{i\to a} s_i 
      - E_{a}(\{s_{\partial a}\}) \Big]
 + \sum_{i\in \da} \sum_{b\in \di - a} |u^{b\to i}|
\, ;
\\ 
 \Delta E^{i}
&=
 - \Big| \sum_{a\in \di} u^{a\to i} \Big|
 + \sum_{a\in \di} |u^{a\to i}|
\, ;
\label{en_s}
\end{align}
\end{subequations}

To summarize, the warning propagation equations neglect every entropic information in the belief propagation (\ref{eq:BP_1}-\ref{eq:BP_2}), thus only the ground state energy can be computed. On the other hand the fact that warnings $u$ and $h$ have a discrete set of possible values simplifies considerably the average over the graph ensemble presented in sec.~\ref{sec:average} as the distribution ${\cal P}$ is a sum of three Dirac function, and can be represented by their weights. Deeper interpretations of warning propagation and its fixed points will be given in chapter~\ref{freezing}. Note that in the literature the value $0$ of warnings is also called $\ast$ or "joker" \cite{BraunsteinMezard02,BraunsteinZecchina04}.

\subsection{Survey Propagation}

Survey propagation (SP)\index{survey propagation}\index{propagation!survey} \cite{MezardParisi02,MezardZecchina02} is a form of belief propagation which aims to count the logarithm of the number of fixed points of warning propagation (\ref{eq:WP}) of a given energy (\ref{eq:WP_ener}). For the sake of simplicity we present the most basic form of SP which aims to count the logarithm of number of fixed points of the warning propagation with zero energy. 

The constraints on values of the warnings assuring that the fixed point of warning propagation corresponds to zero energy are 
\begin{itemize}
   \item For all $i$ and $a\in \partial i$: the warnings $\{ u^{b\to i} \}_{b \in \di - a}$ 
         are all non-negative or all non-positive,
   \item For all $a$ and $i\in \partial a$: the preferred values of all $j\in \partial a -i$ can be realized without violating the constraint $a$.
\end{itemize}
We define probabilities that warnings $u^{a\to i}$ or $h^{i\to a}$ are positive, negative or null. 
\begin{subequations}
\begin{align}
{\cal P}^{a \to i}(u^{a\to i})
&=
  q_-^{a \to i} \delta(u^{a\to i} + 1)
+ q_+^{a \to i} \delta(u^{a\to i} - 1)
+ q_0^{a \to i} \delta(u^{a\to i} )
\, ;
\\ 
{\cal P}^{i \to a}(h^{i\to a})
&=
  p_-^{i \to a} \mu_-(h^{i\to a})
+ p_+^{i \to a} \mu_+(h^{i\to a})
+ p_0^{i \to a} \delta(h^{i\to a})
\, ;
\end{align}
\end{subequations}
where
$q_-^{a \to i} + q_+^{a \to i} + q_0^{a \to i} 
= p_-^{i \to a}+p_+^{i \to a} +p_0^{i \to a}=1$, and
$\mu_{\pm}(h)$ are normalized measures with support over
$\mathbb{Z}^{\pm}$. So, to every oriented edge we associate a message 
$q=(q_-,q_0,q_+)$ or $p=(p_-,p_0,p_+)$
(resp.~if oriented towards the variable or the constraint).
We call these messages surveys, they are analogous to beliefs $\psi^{a\to i}$ and $\chi^{i\to a}$ from (\ref{eq:BP_1}-\ref{eq:BP_2}). And thus, if the factor graph is tree, exact iterative equations for $q$, $p$ can be written.
The update of surveys $p$ given incoming $q$ is common for all Boolean CSPs and reads:
\begin{subequations}
\label{SP_1}
\begin{align}
p_+^{i \to a} + p_0^{i \to a}
  &=
  {\cal N}_{i \to a}^{-1}
  \prod_{b\in \di - a} ( q_+^{a \to i}  + q_0^{a \to i} )
\, ,
\\
p_-^{i \to a} + p_0^{i \to a}
  &=
  {\cal N}_{i \to a}^{-1}
  \prod_{b\in \di - a} ( q_-^{b \to i}  + q_0^{b \to i} )
\, ,
\\
p_0^{i \to a}
  &=
  {\cal N}_{i \to a}^{-1}
  \prod_{b\in \di -  a} q_0^{b \to i}
\, ,
\end{align}
\end{subequations}
where ${\cal N}_{i \to a}$ is the normalization factor. The update of surveys $q$ given the incoming $p$s depends on the details on the constraint functions. For concreteness we write the equation for the positive 1-in-3 SAT problem. The constraints assuring zero energy then forbids that both the warnings incoming to a constraint $a$ have value $+1$.
\begin{subequations}
\label{SP_2}
\begin{align}
q_{+}^{a \to i}
  &=
  {\cal N}_{a \to i}^{-1} \; 
  p^{j \to a}_{-} p^{k \to a}_{-}
\, ,
\\
q_{-}^{a \to i}
  &=
  {\cal N}_{a \to i}^{-1} \; 
  \big[ p_{+}^{j \to a} (1-p_{+}^{k \to a} )
        + (1-p_{+}^{j \to a} ) p_{+}^{k \to a} \big]
\, ,
\\
q_{0}^{a \to i}
  &=
  {\cal N}_{a \to i}^{-1} \;
  \big[ p_{-}^{j \to a} p_{0}^{k \to a} 
      + p_{0}^{j \to a} p_{-}^{k \to a}
      + p_{0}^{j \to a} p_{0}^{k \to a}  \big]
\, ,
\end{align}
\end{subequations}
where ${\cal N}_{a \to i}=1-p^{j \to a}_{+} p^{k \to a}_{+}$ is the normalization factor, $j$ and $k$ are the other two neighbours of $a$.
  
The associated Shannon entropy is called {\it complexity} \cite{Palmer83} (or structural entropy in the context of glasses) and reads \cite{MezardZecchina02}
\begin{equation}
   \Sigma(E=0)
 = \sum_a \log {\cal N}^{a + \da} 
 - \sum_i (l_i-1) \log {\cal N}^{i}  
\, ,
\label{Sigma_0}
\end{equation}
where ${\cal N}^{a+\da}$ is the probability that no contradiction is created when the constraint $a$ and all its neighbours are added, ${\cal N}^i$ is the probability that no contradiction is created when the variable $i$ is added. Remark the exact analogy with (\ref{eq:Zaa}-\ref{eq:Zi}). 
We denote ${\cal P}_0^i \equiv \prod_{a\in \di} q_0^{a \to i}$ and
${\cal P}_{\pm}^i \equiv \prod_{a\in \di} ( q_{\pm}^{a \to i}
   + q_0^{a \to i} )$, then
\begin{subequations}
\label{eq.prob_12}
\begin{align}
{\cal N}^i
&=
{\cal P}_+^i + {\cal P}_-^i - {\cal P}_0^i
\, ,
\label{prob_1}
\\
\begin{split}
{\cal N}^{a + \da}&=
\prod_{i\in \da} ({\cal P}_{+}^{i\to a} 
   + {\cal P}_{-}^{i\to a} - {\cal P}_0^{i\to a} )
-\prod_{i\in \da} ({\cal P}_{-}^{i\to a} - {\cal P}_0^{i\to a} )
-\prod_{i\in \da} ({\cal P}_{+}^{i\to a} - {\cal P}_0^{i\to a} )
\\
& \quad
-\sum_{i \in \da}
{\cal P}_{-}^{i \to a}
\prod_{j \in \da - i} 
      ({\cal P}_{+}^{j \to a} - {\cal P}_0^{j \to a} )
\, .
\end{split}
\label{prob_2}  
\end{align}
\end{subequations}
The second equation collects the contributions from
all combinations of arriving surveys except the ``contradictory'' ones
$(+,+,+)$, $(-,-,-)$, $(+,+,0)$ and $(+,+,-)$ (plus permutations of
the latter). 

The survey propagation equations (\ref{SP_1}-\ref{SP_2}) and the expression for the complexity function (\ref{Sigma_0}) are exact on tree graphs. In the spirit of the Bethe approximation, we will assume sufficient decay of correlations and use these equations on a random graph~\footnote{The fact that on a given tree with given boundary conditions the warning propagation has a unique fixed point might seem puzzling at this point. Clarification will be made in the chapter~\ref{clustering}.}. To average over the ensemble of random graphs we adopt the same equations as we did for the belief propagation in sec.~\ref{sec:average}.

\subsection{Application to the exact cover (positive 1-in-3 SAT)}

\begin{figure}[!ht]
\begin{center}
  \resizebox{0.67\linewidth}{!}{\includegraphics{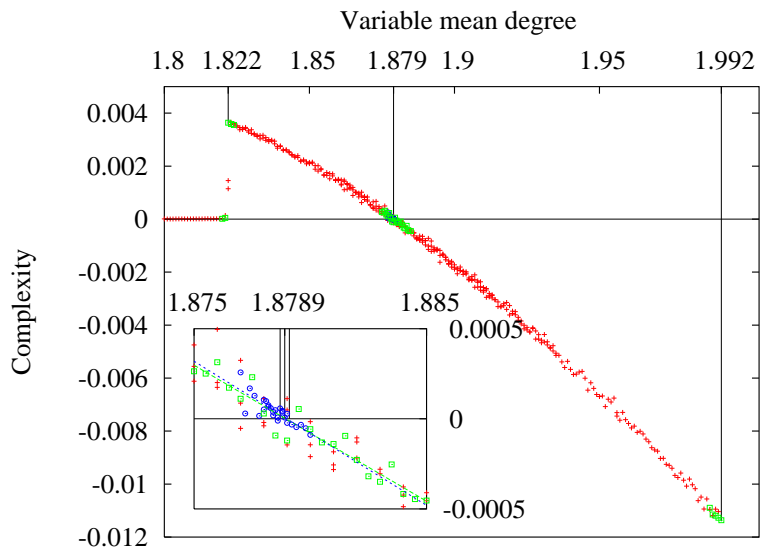}}
\end{center}
\caption{\label{fig:sigma_1in3} Average complexity density (logarithm of number of states divided by the number of variables) as a function of the mean degree $c$ for the positive 1-in-3 SAT problem. At $c_{\rm SP}=1.822$ a nontrivial solution of the survey propagation equations appears, with positive complexity. At $c_s=1.8789\pm 0.0002$ the complexity becomes negative: this is the satisfiability transition. At $c_p=1.992$ the solution at zero energy ceases to exist. The inset magnifies the region where the complexity crosses zero, together with the error bar for the satisfiability 
transition. Crosses represent results of a population dynamics with $N=0.5\cdot 10^5$ elements, squares of $N=1\cdot 10^5$, and circles $N=2\cdot 10^5$. }
\end{figure}

The 1-in-3 SAT problem (with probability of negating a variable equal to one-half) is a rare example of an NP-complete problem which is on average algorithmically easy and where the threshold can be computed rigorously  \cite{AchlioptasChtcherba01}. In particular it was shown that for $\alpha\neq 1$ an instance of the problem can be solved in polynomial time with probability going to one as $N\to \infty$. This result was generalized into random 1-in-3 SAT where the probability of negating a variable is $p\neq 1/2$ [\RS]. In particular we showed that for all $0.273<p<0.718$ the RS solution is correct and almost every instance can be solved in polynomial time if the constraint density $\alpha\neq 1/[4p(1-p)]$. When, however, $p<0.273$ the phase diagram is more complicated, see [\RS]. For $p=0$ the solution of the positive 1-in-3 SAT (exact cover) problem becomes very similar to the one of 3-SAT \cite{MezardZecchina02}. The result for the complexity (\ref{Sigma_0}) in the positive 1-in-3 SAT obtained from the population dynamics method is plotted in fig.~\ref{fig:sigma_1in3}. For more detailed discussion of how the phase diagram changes from the almost-always-easy to the very-hard pattern see [\RS].

Up to certain average connectivity of variables $c_{\rm SP}=1.822$ the only iterative fixed point of the population dynamics gives $q_0^{a\to i}=p_0^{i\to a}=1$ for all $(ia)$. The associated complexity function is zero. In an interval $(c_{\rm SP},c_s)=(1.822,1.879)$ there exist a nontrivial solution giving positive complexity function. There are thus exponentially many different fixed points of the warning propagation. Asymptotically, almost every warning propagation fixed point is associated to a cluster of solutions\footnote{There might exist fixed points of the warning propagation which are not compatible with any solution, thus do not correspond to a cluster. Such "fake" fixed points are negligible if the 1RSB approach is correct.}. Above $c_s=1.879$ there is a nontrivial solution to the SP equations giving a negative complexity function. There are thus almost surely no nontrivial fixed points of warning propagation at zero energy. 

Before interpreting the survey propagation results, we should check that its application on tree-like random graphs is justified. The method to do this self-consistency check has been developed in \cite{MontanariParisi04} and is discussed in appendix \ref{app:1RSB_stab}. For 1-in-3 SAT the result in that SP is stable, thus the results are believed to be correct, for $c\in (1.838,1.948)$ [\RS]. The point $c_s$ belongs to this interval, thus we can interpret it safely as the satisfiability threshold. However, the point $c_{\rm SP}$ has no physical meaning, and some statements that are suggested by its existence are wrong. For example it is not true that there is not exponentially many fixed points of the warning propagation, thus no clustering, for $c<c_{\rm SP}$. This has been remarked in [\KM] and a part of chapter~\ref{clustering} will be devoted to understanding this. 

\section{Loose ends}
\label{sec:loose}

We could summarize the understanding of the subject three years ago in the following way: The 1RSB cavity method was able to compute the satisfiability threshold. The clustered phase was predicted and its existence partially proven. The conjecture that clustering is a key element in understanding of the computational hardness was accepted. The survey propagation inspired decimation algorithm was breath-taking, and the computer science community was getting gradually more and more interested in the concepts which lead to its derivation. It might have seemed that a real progress can be made only on the mathematical side of the theory, in the analytical analysis of the performance of the message passing algorithms, or in new applications. But several loose ends hanged in the air and the opinions on their resolution were diverse. I will list three of them which I consider to be the most obtruding ones.

\paragraph{(A) The "no man's land'', RS unstable but SP trivial ---}

The energetic 1RSB cavity method (survey propagation) predicts the clustering in 3-SAT at $\alpha_{\rm SP}=3.92$. But the replica symmetric solution is unstable at already $\alpha_{\rm RS}=3.86$, at this point the spin glass susceptibility diverges and equivalently the belief propagation algorithm stops to converge on a single graph, see appendix~\ref{app:RS_stab}. What is the solution in the "no man's land" between $\alpha_{\rm RS}$ and $\alpha_{\rm SP}$? The values are even more significant for the 3-coloring or Erd\H{o}s-R\'enyi graphs where the corresponding average connectivities are $c_{\rm RS}=4$ and $c_{\rm SP}=4.42$. 

\paragraph{(B) No solutions with nontrivial whitening cores ---}

An iterative procedure called \textit{whitening of a solution}\index{whitening} is defined as iteration of the warning propagation equations initialized from a solution. {\it Whitening core} is the corresponding fixed point. We call white those variables which are assigned the "I do not care" state in the whitening core. A crucial asymptotic property is that if the 1RSB solution is correct then the
whitening core of all solutions from one cluster is the same and the non-white variables are the frozen ones in that cluster. Consequently, knowing a solution, the whitening may be used to tell if the solution was or was not in a frozen cluster.

Survey propagation uses information only about frozen cluster. It might seem that every cluster is uniquely described by its whitening core, that is by the set and values of the frozen variables.

Yet, the solutions found by survey propagation have always a trivial, all white, whitening core. This paradox was pointed out in \cite{ManevaMossel05} and observed also by the authors of \cite{BraunsteinZecchina04}. It was suggested that the concept of whitening might be meaningful only in the thermodynamical limit. But that was not a satisfactory explanation. 

\paragraph{(C) Where do the simple local algorithms actually fail ---}

The clustered phase, baptized "Hard" in \cite{MezardZecchina02} does not seem to be that hard. There is no local algorithm which would perform well exactly up to $\alpha_{\rm SP}=3.92$. For a while it was thought that the 1RSB stability point $\alpha_{II}=4.15$, see appendix \ref{app:1RSB_stab} , is a better alternative. It was argued that the full-RSB states are more "transparent" for the dynamics than the 1RSB states which should be well defined and separated. Moreover there was at least one empirical result which suggested that the Walk-SAT algorithm stops to work in linear time at that point \cite{AurellGordon04}. But other version of Walk-SAT stopped before or even after, as for example the ASAT which was argued in \cite{ArdeliusAurell06} to work in linear time at least up to $\alpha=4.21$. 

\section{Summary of my contributions to the field}
\label{sec:summary}

In my first works [\ZMM, \MM, \RS] I applied the replica symmetric and the energetic 1RSB method to the matching and the 1-in-$K$ SAT problems. This is why I used these two problems to illustrate the methods in sec.~\ref{sec:matching} and \ref{sec:1in3}.

The problem of matching on graphs is a common playground for algorithmic and methodological development. I studied the
problem of counting maximum matchings in a random graph in [\ZMM]. Finding a maximum matching is a well known polynomial problem, while their approximative counting is a much more difficult task. We showed, that the entropy of maximum matchings can be computed using the belief propagation algorithm, a result which was later on partially proved rigorously \cite{BayatiNair06}.

My interest in the 1-in-$K$ SAT problem stemmed from the work \cite{AchlioptasChtcherba01} where the authors computed rigorously the satisfiability threshold and showed that the NP-complete problem is in fact on average algorithmically easy. In [\MM, \RS] we studied the random 1-in-3 SAT in two-parameter space. One parameter is the classical constraint density, the other is the probability $p$ of negating a variable in a constraint ($p=1/2$ in \cite{AchlioptasChtcherba01}). We showed that for $0.2627<p<0.7373$ the problem is on average easy and the satisfiability threshold can be computed rigorously. On the other hand for $p<0.07$ the problem is qualitatively similar to the 3-SAT. We computed the threshold from the energetic 1RSB approach. In the intermediate region the 1RSB approach is not stable, thus it stays an open question how exactly does the problem evolve from an on average easy case to a 3-SAT like case. Qualitatively similar phase diagram was described in the $2+p$ SAT problem \cite{MonassonZecchina99b,AchlioptasKirousis01}.
We also found an interesting region of the parameter space in the 1-in-3 SAT where the unit clause algorithm provably finds solutions despite the replica symmetric solution being not correct (unstable).

The rest of my works [\KM, \ZK, \KZP, \KZJ, \MZ, \AZ, \ZML] tied up the loose ends from the previous section and mainly addressed the original question of this thesis: Why are some constraint satisfaction problems intrinsically hard on average and what causes this hardness?

I used the entropic zero temperature 1RSB approach, introduced in \cite{MezardPalassini05}, to study the structure of solutions in random CSPs. In [\KM, \ZK] we discovered that the true clustering (dynamical) transition does not correspond to the onset of a nontrivial solution of the survey propagation equations. We gave a proper definition of the clustering transition and formulated it in terms of extremality of the uniform measure over solutions. The clustering transition happens always before or at the same time as the replica symmetric solution ceases to be stable. This tied up the loose end (A), as in the "no man's land" the energetic 1RSB solution was simply incomplete. 

We showed that in general there exist {\it two} distinct clustered phases below the satisfiable threshold. In the first, {\it dynamic clustered phase}, an exponentially large number of pure
states is needed to cover almost all solutions. However, average
properties (such as total entropy) still
behave as if the splitting of the measure did not count. In
particular, a simple algorithm such as belief propagation gives
asymptotically correct estimates of the marginal probabilities. However, the measure over solutions
is not extremal and, more importantly, the Monte Carlo equilibration
time diverges, thus making the {\it sampling of solutions} a hard
problem.  The second kind of clustered phase is the {\it condensed clustered phase} where a finite
number of pure states is sufficient to cover almost all solutions. A number of nontrivial predictions follows: for instance the total entropy has a non-analyticity at the transition to this phase, 
the marginal probabilities are non-self-averaging and not given anymore by the belief propagation algorithm.

In the context of the coloring problem, i.e. anti-ferromagnetic Potts glass, I also addressed related questions of what does the 1RSB solution predict for the finite temperature phase diagram and when is the 1RSB solutions correct (stable) [\ZK]. We give the full phase diagram for this model and argue that in the colorable phase for at least 4 colors the 1RSB solutions is stable, and thus believed to be exact. 

In order to clarify and substantiate this heuristic picture, we introduced the random subcubes model in [\MZ], a generalization of the random energy model.
The random subcubes model is exactly solvable and reproduces the sequence of phase transitions in the real CSPs (clustering, condensation, satisfiability threshold). Its, perhaps, most remarkable property is that it reproduces quantitatively the behaviour of random $q$-coloring and random $K$-SAT in the limit of large $q$ and $K$. We showed that the random subcubes model can also be used as a simple playground for the studies of dynamics in glassy systems. 

An important and quite novel phenomena I investigated in [\ZK, \KZJ] is the freezing of variables. A variable is frozen when in all the solutions belonging to one cluster it takes the same value. I discovered that the fraction of such frozen variables undergoes a first order phase transition when the size of states is varied. I introduced the notion of the {\it rigidity transition} as the point where almost all the dominating clusters become frozen and the {\it freezing transition} as the point where all the clusters become frozen. The solutions belonging to the frozen clusters can be recognized via the whitening procedure. 

We computed the rigidity transition in the random coloring in [\ZK]. And we studied the freezing transition in 3-SAT numerically [\AZ], with the result $\alpha_f=4.254\pm 0.009$ (to be compared to the satisfiability threshold $\alpha_s=4.267$). This study also confirms that the notion of whitening and freezing of variables in meaningful even on relatively small systems.  

This allows us to tie up the loose end (B). The survey propagation algorithm describes the most numerous frozen clusters. The range of connectivities where the SP based algorithms are able to find solutions in 3-SAT lies in the phase where most solutions are in fact unfrozen. It is thus much less surprising that the SP based algorithms always find a solution with a trivial whitening. 

A very natural question cannot be avoided at this point: What happens in the frozen phase where all the solutions are frozen? We know that such a phase exists, this was shown in \cite{AchlioptasRicci06} and numerically in [\AZ]. And we also know from several authors that the known algorithms do not seem to be able to find frozen solutions in polynomial time (that is never for sufficiently large instances). We conjectured in [\ZK] that the freezing is actually a relevant concept for the algorithmical hardness. Thus the answer we suggest to tie up the loose end (C) is that the simple local algorithms stop always before the freezing transition. It is a challenging problem to design an algorithm which would be able to beat this threshold. 

In the coloring and satisfiability problems (at reasonably small $q$ and $K$) the freezing transition is however very near to the satisfiability threshold, see the numbers in [\ZK, \AZ]. It is thus difficult to make strong empirical conclusions about the relation between hardness and freezing. Motivated by the need of problems where the freezing and satisfiability would be well separated I introduced the {\it locked} constraint satisfaction problems where the freezing transition coincides with the clustering one [\ZML]. The locked CSPs are very interesting from several points of view. The clusters in locked CSPs are point-like, this is why the clustering and freezing coincide. This is also connected with a remarkable technical simplification, as these problems can be fully described on the replica symmetric level. 

On the other hand the locked problems are extremely algorithmically challenging. We implemented the best known solvers and showed that they do not find solutions starting very precisely from the clustering (= freezing) transition. At the same time this transition is very well separated from the satisfiability threshold. 

A remarkable point about a subclass of the locked problems which we called {\it balanced} is that the satisfiability threshold can be obtained exactly from the first and second moment calculation. This adds a huge class of constraint satisfaction problems to a handful of other NP-complete CSPs where the threshold is known rigorously. And it also brings the understanding of which properties of the problem introduce fluctuations which make the second moment method fail.  

The numerical work on the 3-SAT problems [\AZ] also addresses another important and almost untouched question: How much are the asymptotic results relevant for systems of practical sizes. We counted the number of clusters in random 3-SAT on instances up to size $N=150$ and compared to the analytical prediction. We saw that the comparison is strikingly good for already so small systems. This should encourage the application of statistical physics methods to the real world problems.

\chapter{Clustering}
\label{clustering}

{\it In this chapter we introduce the concept of clustering of solutions. First we investigate when does the replica symmetric solution fail. Then we derive the one-step replica symmetry breaking equations on trees and give their interpretation on random graphs. We discuss how several geometrical definitions of clusters might be related to the pure states and review the properties of the clustered phase. Finally, we revise how is the clustering related to the algorithmical hardness and conclude that it is considerably less than previously anticipated. The original contributions to this chapter were published in [\KM, \ZK, \AZ].}

\section{Definition of clustering and the 1RSB approach} 
\label{sec:RS_fail}

How to recognize when is the replica symmetric solution correct?
First we have to explain what do we precisely mean by "being correct". 
We obviously require that quantities like the free energy, energy, entropy, marginal probabilities (magnetizations) are asymptotically exact when computed in the replica symmetric approach. But this is not enough, as this is also satisfied in the phase which we will call later the {\it clustered} (dynamical) 1RSB phase. 

A commonly used necessary condition for the validity of the RS solution is referred to as the {\it local stability towards 1RSB}. It consists in checking that the spin glass susceptibility does not diverge, or equivalently that the belief propagation algorithm converges on a large single graph, or in the probability theory this corresponds to the Kesten-Stigum condition \cite{KestenStigum66,KestenStigum66b}. These and other equivalent representations for the replica symmetric stability are discussed in detail in appendix \ref{app:RS_stab}. If the replica symmetric solution is not stable then it predicts wrong free energy, entropy, correlation functions, etc. But the contrary is far from being true: even if stable, the RS solution might be wrong, and even unphysical (predicting negative entropies in discrete models, negative energies in models with strictly non-negative Hamiltonian function, or discontinuities in functions which physically have to be Lipschitzian). 

It is tempting to say: The replica symmetric solution is correct if and only if the assumptions we used when deriving it are correct. In deriving the belief propagation (\ref{eq:BP}) and the RS free energy (\ref{free_given}) we used only one assumption: The neighbours of a variable $i$ are independent random variables, under the Boltzmann measure (\ref{Boltzmann}), when conditioned on the value of $i$. As we will see, this assumption is asymptotically correct also in the dynamical 1RSB phase, and thus the RS marginal probabilities, or the free energy function remain asymptotically exact in that phase.

We thus need a different definition for the "RS correctness" which would determine whether the Boltzmann measure  (\ref{Boltzmann}) can be asymptotically described as a single pure state, and whether the equilibration time of a local dynamics is linear in the system size. At the same time we do not want this definition to refer the RSB solution, because obviously we want to justify the need of the RSB solution by the failure of the RS solution.

A definition satisfying the above requirements appeared only recently \cite{MezardMontanari07,MontanariSemerjian05,MontanariSemerjian06b}, and it can be written in several equivalent ways. From now on we say that the {\it replica symmetric solution is correct}\index{solution!replica symmetric!correctness} if and only if one of the following is true.
\begin{itemize}
    \item[(a)]{The point-to-set correlations decay to zero.}
    \item[(b)]{Reconstruction on the underlying graph in not possible.}
    \item[(d)]{The uniform measure over solutions satisfies the extremality condition.}
    \item[(c)]{The 1RSB equations at $m=1$, initialized in a completely biased configuration, converge to a trivial fixed point.}
\end{itemize}
In the rest of this section we explain these four statements, and show that they are indeed equivalent, and explain how do they correspond to the existence of a nontrivial 1RSB solution. We should mention that in the so-called locked constraint satisfaction problems this definition have to be slightly changed at zero temperature, we will discuss that in sec.~\ref{locked}. The transition from a phase where the RS solution is correct to a phase where it is not is called the {\it clustering} or the {\it dynamical transition}\index{phase transition!clustering}\index{phase transition!dynamical}.

\paragraph{Gibbs measures and why are the sparse random graphs different ---}

Our goal is to describe the structure of the set of solutions of a constraint satisfaction problem with $N$ variables. Let $\phi_a(\partial a)$ be the constraint function depending on variables $s_i\in \partial a$ involved in the constraint $a$, $\phi_a(\partial a)=1$ if the constraint is satisfied, $\phi_a(\partial a)=0$ if not. The uniform measure\index{measure!uniform over solutions} over all solutions can be written as
\be
      \mu(\{s_i\}) = \frac{1}{Z} \prod_{a=1}^M \phi_a(\partial a)\, , \label{uniform} 
\ee 
where $Z$ is the total number of solutions. The uniform measure over solutions is the zero temperature limit, $\beta\to \infty$, of the Boltzmann measure\index{measure!Boltzmann}
\be
      \mu(\{s_i\},\beta) = \frac{1}{Z(\beta)} \prod_{a=1}^M e^{-\beta[1-\phi_a(\partial a)]}\, . \label{Boltzman}
\ee

The above expressions are valid on any given finite factor graph. The theory of Gibbs measures \cite{Georgii88} tries to formally define and describe the limiting object to which (\ref{uniform}-\ref{Boltzman}) converge in the thermodynamical limit, $N\to \infty$. A common way to build this theory is to ask: What is the measure induced in a finite volume $\Lambda$ when the boundary conditions are fixed? Roughly speaking, the good limiting objects, called the {\it Gibbs measures} or the {\it pure states}, are such that boundaries taken from the Gibbs measure induce the same measure inside the finite large volume $\Lambda$\index{measure!Gibbs}. 

The Ising model on a 2D lattice gives an excellent example of how a phase transition is seen via Gibbs measures. Whereas in the high temperature paramagnetic phase the Gibbs measure is unique, in the ferromagnetic phase there are two extremal measures, one corresponding to the positive average magnetization, the other to the negative average magnetization. Indeed, if a boundary condition is chosen from one of these two then the correct magnetization will be induced in the bulk. In general the bulk in equilibrium can be described by a linear combination of these two {\it extremal} objects. 

In the disordered models the situation might be much more complicated. Indeed the proper definition of the Gibbs measure in the Edwards-Anderson model (\ref{spin_glass}) and other glassy models is a widely discussed but still an open problem \cite{Bovier06,Talagrand03,NewmanStein92}. 

The locally tree-like lattices, we are interested in here, are also peculiar from this point of view. The main difference is that in any reasonable definition of the boundary variables, the boundary has volume comparable to the volume of the interior. Thus again the usual theory of Gibbs measure implies very little. On the other hand the tree structure makes some considerations simpler. We will try to understand what sort of long range correlations might appear on the tree-like graphs by studying the tree graphs with general boundary conditions.

\subsection{Properties and equations on trees}
\label{trees}

It is a well known fact that on arbitrary tree, with arbitrary boundary conditions, the belief propagation equations and the Bethe free energy are exact (the thermodynamical limit is not even needed here) \cite{Pearl88,KschischangFrey01,YedidiaFreeman00}. 

But what if the boundary conditions are chosen from a complicated measure? Then very little (if anything) is known in general. However, there is a way how to choose the boundary conditions such that the tree is then described by the one-step replica symmetry breaking equations. This is closely linked to the problem of reconstruction on trees, studied in mathematics \cite{EvansKenyon00,Mossel01,Mossel04}. The link with 1RSB was discovered by M\'ezard and Montanari \cite{MezardMontanari06}. We chose to present the 1RSB equations in this new way, because it opens the door to further mathematical developments. For the original statistical physics derivation we refer to \cite{MezardParisi00}. Another recent computer science-like derivation, which is based on the construction of a decorated constraint satisfaction problem and writing belief propagation on such a problem, in presented in \cite{MezardMontanari07,Mora07}.  

\paragraph{Reconstruction\index{reconstruction} on trees ---}

We explain the concept of reconstruction on trees \cite{Mossel04}. For simplicity we consider  $q$-coloring on a rooted tree with constant branching factor $\gamma$ (sometimes also called the Cayley tree). A more general situation (with disorder, in the interaction or in the branching factor) is described in appendix~\ref{app:m1}.

\begin{figure}[!ht]
\begin{center}
  \hspace{2cm}
  \resizebox{0.67\linewidth}{!}{\includegraphics[angle=90]{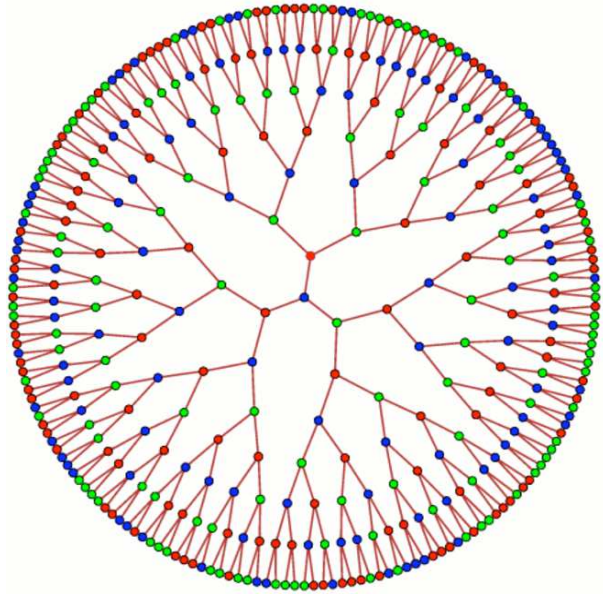}}
 \caption{\label{fig:rec} Illustration of the broadcasting of colors on a binary tree ($\gamma=2$) for the reconstruction problem.}
\end{center}
\end{figure}

Create a rooted tree with branching $\gamma$ and with $L$ generations. An example of $\gamma=2$ and $L=8$ is in fig.~\ref{fig:rec}. Assign a color $s_0$ to the root and broadcast over the edges towards the leaves of the tree in such a way that if a parent node $i$ was assigned color $s_i$ then each of its ancestors is assigned random one of the remaining $q-1$ colors. 

At the end of this broadcasting, every node in the tree is assigned a color, and this assignment corresponds to a proper coloring (neighbours have different colors). Now in an imaginary experiment we forget the colors everywhere but on the leaves. The problem of reconstruction consists in deciding if there is any information left in the values on the leaves (and their correlation) about the original color $s_0$ of the root in the limit of infinite tree $L\to \infty$. If the answer is yes then we say that the reconstruction is possible, if the answer is no then the reconstruction is not possible. 

Call $\{s\}_l$ the assignment of colors in the $l^{\rm th}$ generation of the tree. Consider formally the probability $\psi_{s_0}(\{s\}_l)$ that a broadcasting process which finished at the configuration $\{s\}_l$ started from the color $s_0$ at the root. In other words, in what fraction of assignments in the interior of the tree  (compatible with the boundary conditions $\{s\}_l$) is the color of the root $s_0$?
Reconstruction is possible if and only if 
\be
   \lim_{l\to \infty} \sum_{r=1}^q  \psi_r(\{s\}_l)  \log{\big[ q \psi_r(\{s\}_l)\big]} > 0\, .
\ee

Intuitively when the branching $\gamma$ is small and the number of colors large the information about the root will be lost very fast. If, on the contrary, the branching is large compared to the number of colors some information remains. A simple exercise is to analyze the so-called {\it naive reconstruction}\index{reconstruction!naive} algorithm \cite{Semerjian07}. The naive reconstruction is possible if the probability that the leaves determine uniquely the root does not go to zero as the number of generation goes to infinity. 
We compute the probability $\eta$ that the far-away boundary is compatible with only one value of the root. Denote $\eta_l$ the probability that a variable in the $l^{\rm th}$ generation is directly implied conditioned on the value of its parent. The probability $\eta_{l-1}$ can be computed recursively as
\bea
 \eta_{\, l-1} &=&  1 - (q-1) \big(1 - \frac{1}{q-1} \eta_{\, l} \big)^\gamma + \frac{(q-1)(q-2)}{2} \big(1 - \frac{2}{q-1} \eta_{\, l} \big)^\gamma - \dots \nonumber \\
  &=& \sum_{r=0}^{q-1} (-1)^r {q-1 \choose r} \Big(1 - \frac{r}{q-1} \eta_{\, l} \Big)^\gamma \, . \label{eq:hard}
\eea
The terms in this telescopic sum come from probabilities that number $r$ out of the $q-1$ colors are not present in the $\gamma$ descendants.
In the last generation we know the colors by definition of the problem, thus $\eta_{\infty}=1$. If the iterative fixed point of (\ref{eq:hard}) is positive then the reconstruction is possible. 

This simple upper bound on the branching $\gamma$ for which the reconstruction is possible is actually quite nontrivial and in the limit of large number of colors it coincides with the true threshold at least in the first two orders, see [\ZK] and \cite{Semerjian07,Sly08}. This upper bound is connected to the presence of frozen variables and will be discussed in a greater detail in chapter~\ref{freezing}.  

\paragraph{Self-consistent iterative equations for the reconstruction ---}
The iterative equations for the reconstruction problem are equivalent to the one-step replica symmetry breaking equations with Parisi parameter $m$\index{Parisi parameter!m}, $m=1$ will apply to the original question of reconstructibility. This was first derived by M\'ezard and Montanari \cite{MezardMontanari06} and it has some deep consequences for the understanding of the RSB solution. We now explain this derivation, still for the coloring problem with a fixed branching $\gamma$ and $q$ colors. A more general form is presented in appendix~\ref{app:m1}. 

For given boundary conditions $\{s\}_l$, constructed as described above, we compute the probability $\psi^{i\to j}_{s_i}$ (over all broadcasting experiments leading to these boundary conditions) that a variables $i$ had color $s_i$, where $j$ is the parent of $i$ and the edge $(ij)$ has been cut. Given the probabilities on the descendants of $i$, which are indexed by $k=1,\dots,\gamma$, we can write
\be
    \psi^{i\to j}_{s_i} = \frac{1}{Z^{i\to j}}  \prod_{k=1}^\gamma (1- \psi^{k\to i}_{s_i}) \equiv {\cal F}_{s_i}(\{ \psi^{k\to i}\})\, , \label{eq:col_BP}
\ee
because the descendants can take any other color but $s_i$. The $Z^{i\to j}$ is a normalization constant. It should be noticed that this is in fact the belief propagation equation (\ref{eq:BP}) for the graph coloring.
This equation can also be derived by counting how many assignments are consistent with the boundary conditions $\{s\}_l$.
This gives a natural interpretation to $Z^{i\to j}$
\be
       Z^{i\to j} = \frac{ Z^{(i)} }{\prod_{k=1}^\gamma Z^{(k)}} \, .\label{eq:rew}
\ee
where $Z^{(i)}$ is the total number of solutions consistent with $\{s\}_l$ if $i$ were the root. Thus $Z^{i\to j}$ is a change in the number of solutions compatible with the boundary conditions when the $\gamma$ branches are merged.

Now we consider the distribution over all possible boundary conditions which are achievable by the broadcasting process defined above. We have to specify the probability distribution on the boundary conditions. We consider that the probability of every boundary conditions $\{s\}_l$ is proportional to the power $m$ of the number of ways by which we could create $\{s\}_l$, denote this number $Z(\{s\}_l)$. In other words, the probability of a given boundary condition is proportional to the power $m$ of the number of possible assignments in the bulk of the tree. 
\be
      \mu(\{s\}_l) = \frac{ \big[ Z(\{s\}_l) \big]^m  }{{\cal Z}(m)  } \, ,  \quad {\rm where} \quad {\cal Z}(m) = \sum_{\{s\}_l} \big[ Z(\{s\}_l)\big]^m \, .   \label{eq:mu_bound}
\ee
The value of $m=1$ is natural for the original question of reconstruction, because every realization of the broadcasting experiment is then counted in a equiprobable way. We, however, introduced a general power $m$. The parameter $m$ will play a role of the Legendre parameter, changing its value focuses on boundary conditions compatible with a given number of assignments inside the tree. 

Denote $P^{i\to j}(\psi^{i\to j} )$ the distribution of $  \psi^{i\to j} $, over the measure on the boundary conditions (\ref{eq:mu_bound}) 
\be
    P^{i\to j}(  \psi^{i\to j} ) \equiv \sum_{\{s\}_l}   \mathbb{I}(\{s\}_l\,  {\rm induce}\,    \psi^{i\to j})   \frac{\big[ Z^{(i)}(\{s\}_l) \big]^m}{{\cal Z}^{(i)}(m)  }\, .\label{eq:1RSB_tree}
\ee
Where $Z^{(i)}(\{s\}_l)$ is the number of solutions induced on the subtree rooted in vertex $i$, ${\cal Z}^{(i)}(m)$ is the corresponding normalization. To express the probability distribution $P^{i\to j}(  \psi^{i\to j} )$ as a function of $P^{k\to i}(  \psi^{k\to i}  )$ we need that $  \psi^{i\to j} = {\cal F}(\{  \psi^{k\to i}\})$, eq.~(\ref{eq:col_BP}). Moreover, $Z^{i\to j}$ is the increase in the total number of solutions after merging the branches rooted at $k=1,\dots,\gamma$ into one branch rooted at $i$. The distributional equation for $P$ is then
\be
    P^{i\to j}(  \psi^{i\to j} ) = \frac{1}{{\cal Z}^{i\to j}} \int \prod_{k=1}^\gamma {\rm d}P^{k\to i}(  \psi^{k\to i}) \,   (Z^{i\to j})^m \,   \delta\big[  \psi^{i\to j} - {\cal F}(\{  \psi^{k\to i}\})\big] \, ,\label{eq:1RSB_m1_tree}
\ee
where $  {\cal F}$ and $Z^{i\to j}$ are defined in (\ref{eq:col_BP}), and ${\cal Z}^{i\to j}$ is a normalization constant equal to
\be
     {\cal Z}^{i\to j} = \frac{{\cal Z}^{(i)}(m)}{\prod_{k=1}^\gamma {\cal Z}^{(k)}(m)} =  \int \prod_{k=1}^\gamma {\rm d}P^{k\to i}(  \psi^{k\to i}) \,   (Z^{i\to j})^m \, .  \label{eq:1RSB_Z}
\ee
where ${\cal Z}^{(i)}$ is the normalization from (\ref{eq:1RSB_tree}) if $i$ were the root. 
Notice that if we start from boundary conditions which are not compatible with any solution then the re-weighting $Z^{i\to j}=0$ at the merging where a contradiction is unavoidable. Initially at the leaves the colors of nodes are known. Call $\delta_r$ the $q$-component vector $\psi_{s_i}^{i\to j} = \delta(s_i,r)$, then the initial distribution is just a sum of singletons 
\be
      P^{\rm init}(  \psi) = \frac{1}{q}\sum_{r=1}^q \delta\big( \psi - \delta_r \big) \, . \label{eq:init}
\ee
Denote $P_0(  \psi )$ the distribution created from (\ref{eq:init}) after many iteration of (\ref{eq:1RSB_m1_tree}) with $m=1$. The reconstruction is possible if and only if $P_0(  \psi )$ is nontrivial, that is different from singleton on $\psi_{s_i}=1/q\, , \,  \forall s_i$.  We define the critical branching factor $\gamma_d$ in such a way that for $\gamma<\gamma_d$ the reconstruction is not possible\index{reconstruction!not possible}, and for $\gamma\ge \gamma_d$ the reconstruction is possible\index{reconstruction!possible}. The critical values  $\gamma_d=c_d-1$ for the coloring problem are reviewed in tab.~\ref{tab:regular}.

\paragraph{What are clusters on a tree?}

If the reconstruction is not possible, then almost all (with respect to (\ref{eq:mu_bound}) at $m=1$) boundary conditions do not contain any information about the original color of the root. However, for rare boundary conditions this might be different. Obviously as long as $\gamma \ge q-1$ one can always construct boundary conditions which determine uniquely the value of the root (by assigning every of the $q-1$ colors to the descendants of every node). If $\gamma<q-1$ then this is no longer possible. And it was proven in \cite{Jonasson02} that for $\gamma<q-1$ every boundary conditions lead to an expectation $1/q$ for every color on the root. If the reconstruction is possible, then different boundary conditions may lead to different expectations on the root. 

The basic idea of the definition of clusters on a tree is the same as in the classical definition of a Gibbs measure \cite{Georgii88}\index{cluster!on a tree}. However, some more work is needed to make the following considerations rigorous. Define a $d$-neighbourhood of the root as all the nodes up to $d^{\rm th}$ generation, consider $1 \ll d \ll l$. Consider the set ${\cal S}$ (resp. ${\cal S}'$) of all assignments on the $d$-neighbourhood compatible with a given boundary condition $\{s\}_l$ (resp. $\{s'\}_l$). Define two boundary conditions $\{s\}_l$ and $\{s'\}_l$ as equivalent if the fraction of elements in which the two sets ${\cal S}$ and ${\cal S}'$ differ goes to zero as $l,d\to \infty$. Clusters are then the equivalence classes in the limit $l\to \infty$, $d\to \infty$, $d\ll l$. The requirement $d\ll l$ comes from the fact that in $l-d$ iterations the equation (\ref{eq:1RSB_tree}) should converge to its iterative fixed point. 

As we explained, more than one cluster exists as soon as the branching factor $\gamma\ge q-1$, but as long as the iterative fixed point of eq.~(\ref{eq:1RSB_m1_tree}) at $m=1$ is trivial all but one clusters are negligible because they contain an exponentially small fraction of solutions. Indeed, if the reconstruction is not possible it means that the information about the $d$-neighbourhood is almost surely lost at the $l^{\rm th}$ generation. Thus almost every broadcasting will lead to a boundary condition from the only relevant giant cluster. 

Only for $\gamma\ge \gamma_d$, when the reconstruction start to be possible, the total weight of all solutions will be split into many clusters. In every of them the set of expectation values (beliefs) $\psi^{i\to j}$ will be different. This is related to another derivation of the 1RSB equations where the clusters of solutions on a given graph are identified with fixed points of the belief propagation equations \cite{MezardMontanari07,Mora07}.  There are exponentially many (in the total number of variables $N$) initial conditions, it is also reasonable to expect that the number of clusters will be exponentially large in $N$.

\paragraph{The complexity function ---}

The number of solutions compatible with a boundary condition $(\{s\}_l)$ was denoted $Z(\{s\}_l)$ in eq.~(\ref{eq:mu_bound}). The associated entropy is then, due to interpretation of $Z^{i\to j}$ (\ref{eq:rew})
\be
   S(\{s\}_l) \equiv \log{\big[Z(\{s\}_l)\big]} = \sum_{i} \log{Z^{i\to j}} \, ,\label{eq:ent_tree}
\ee
where the sum is over all the vertices $i$ in the tree, if $i$ is a leaf then $Z^{i\to j}=1$, if $i$ is the root that $j$ is a imaginary parent of the root. An intuition about this formula is the following: $\log Z^{i\to j}$ is the change in the entropy when the node $i$ and all edges $(ki)$, where $k$ are descendants of $i$, are added. Summing over all $i$ then creates the whole tree. 

More commonly, we introduce also messages going from the parents to the descendants and write the expression for the entropy (\ref{eq:ent_tree}) in the equivalent Bethe form \cite{YedidiaFreeman03} 
\be
   S(\{s\}_l) =  \sum_i \log{ Z^{i+\partial i}} - \sum_{ij}  \log{Z^{ij}} \, ,  \label{eq:col_entr}
\ee
where  
\be
   Z^{i+\partial i} = \sum_{r=1}^q  \prod_{k \in \partial i} \big[ 1 - \psi_r^{k\to i}\big]\, ,\quad  \quad
  Z^{ij} = 1 - \sum_{r=1}^q  \psi_r^{i\to j}  \psi_r^{j\to i}\, , \label{eq:ZZ}   
\ee
where $\partial i$ are all the neighbours (descendants and the parent) of node $i$. The first sum in (\ref{eq:col_entr}) goes over all the nodes in the tree, the root included, leaves have only one allowed color, thus eq.~(\ref{eq:ZZ}) changes correspondingly. Again the meaning of $\log{ Z^{i+\partial i}}$ is the change in the entropy when node $i$ and his neighbourhooding edges are added, each edge is then counted twice, thus the shift in the entropy when an edge $(ij)$ is added, $\log{Z^{ij}}$, have to be subtracted.

We denote $\Phi(m)\equiv\log{{\cal Z}(m)}$ the thermodynamical potential associated to the measure (\ref{eq:mu_bound}). To avoid confusion with the real free energy, associated to the uniform measure over solutions (\ref{uniform}), we call it the {\it replicated free entropy}\index{replicated free entropy}. If a nonzero temperature is involved then $-\Phi(m)/(\beta m)$ is called the {\it replicated free energy}\index{replicated free energy}. The replicated free entropy on a tree can be expressed in totally analogous way as the entropy. From (\ref{eq:1RSB_Z}) we derive 
\be
 \Phi(m) \equiv \log{{\cal Z}(m)} = \sum_{i} \log{{\cal Z}^{i\to j}} \label{eq:free_ent} \, ,
\ee
which is usually written in the equivalent way
\be
  \Phi(m) =  \sum_i \log{ {\cal Z}^{i+\partial i}} - \sum_{ij}  \log{{\cal Z}^{ij}}\, ,  \label{eq:1RSB_entr_tree}
\ee
where we introduced 
\begin{subequations}
\label{eq:1RSB_entropies_tree}
\bea
   {\cal Z}^{i+\partial i} &=&  \int \prod_{k\in \partial i} {\rm d}P^{k\to i}(  \psi^{k\to i})  \big( Z^{i+\partial i} \big)^m  \, ,\\
  {\cal Z}^{ij} &=& \int  {\rm d}P^{i\to j}(  \psi^{i\to j})  {\rm d}P^{j\to i}(  \psi^{j\to i})  \big( Z^{ij} \big)^m    \, .
\eea
\end{subequations}

We denote $\Sigma(m)$ the Shannon entropy corresponding to measure on the boundary conditions (\ref{eq:mu_bound}), and we call it the {\it complexity}\index{complexity function!on trees} function.
\be
     \Sigma(m) \equiv -\sum_{\{s\}_l} \mu(\{s\}_l) \log{\mu(\{s\}_l)} = -m  S(m)  + \Phi(m)\, , \label{eq:1RSB_Sigma_tree}
\ee
where $S$ is the entropy averaged with respect to $\mu(\{s\}_l)$ 
\be
     S(m)  = \sum_{\{s\}_l}  \frac{ \big[ Z(\{s\}_l) \big]^m  }{{\cal Z}(m)  }   \log{Z(\{s\}_l)}   = \frac{\partial \Phi(m)}{\partial m}\, . \label{eq:1RSB_entropy_tree}
\ee
Thus the complexity can also be written as a function of the internal entropy via the Legendre transform of the replicated free entropy $\Phi(m)$ 
\be
     \Sigma(S)  = -m  S  + \Phi(m) \quad {\rm with} \quad  \frac{\partial \Sigma(S)}{\partial S} = -m\, .  \label{eq:Legendre_tree}
\ee
The reader familiar with the cavity approach surely recognized eqs.~(\ref{eq:1RSB_m1_tree}) and (\ref{eq:free_ent}--\ref{eq:Legendre_tree}) as the 1RSB equations.

\paragraph{Interpretation of the complexity function ---}

In the cavity method \cite{MezardParisi01} the exponential of the complexity function $\Sigma(m)$ (\ref{eq:1RSB_Sigma_tree}) counts the number of clusters  corresponding to a given value of the parameter $m$, that is of a given entropy $S$ (\ref{eq:1RSB_entropy_tree}). Complexity defined on the full tree is never negative, as it is a Shannon entropy of a discrete random variable. The same is, of course true, about the entropy (\ref{eq:ent_tree}).

It is more interesting to consider the complexity (or the entropy) function $\Sigma_d(m)$ on the $d$-neighbourhood of the root. If the total number of generations of the tree is $l$ we take $1\ll d\ll l$. And moreover we require $l-d$ to be large enough, such that the distributional iterative equation (\ref{eq:1RSB_m1_tree}) converges to its fixed point in less than $l-d$ iterations. The average complexity function on the $d$-neighbourhood can then be computed from this fixed point. And it can be both positive or negative. Its negative value then means that the number of clusters is decreasing as we are getting nearer to the root. Two important critical connectivities can be defined 
\begin{itemize}
    \item{$\gamma_c$: at which the complexity of the "natural" clusters $\Sigma_d(m=1)$ becomes negative.}
    \item{$\gamma_s$: at which the maximum of the complexity  $\Sigma_d(m=0)$ becomes negative.}
\end{itemize}
The connectivity $\gamma_s$ is the tree-analog of the satisfiability threshold\index{phase transition!satisfiability on trees}. The connectivity $\gamma_c$ is the tree-analog of the condensation transition on random graphs, see chapter~\ref{condensation}\index{phase transition!condensation on trees}.  

Strictly speaking, it is not known how to justify the interpretation of the complexity function as the counter of clusters in the derivation we just presented. In the original cavity derivation \cite{MezardParisi01} or in the later derivations  \cite{MezardMontanari07,Mora07} this point is well justified. We, however, find the purely tree derivation appealing for further progress on the mathematical side of the theory and that is why we have chosen to present this approach despite this current incompleteness.

\subsection{Back to the sparse random graphs}

We stress that the equations, derived in the previous section, are all exact on a given (even finite) tree and that we have not use any approximation. We were just describing boundary conditions correlated via (\ref{eq:mu_bound}). These, in nature recursive, equations are solved via the population dynamics technique, see the appendix \ref{app:pop_dyn}.

To come back to the sparse random graphs, which are only locally tree-like, we can consider equations (\ref{eq:1RSB_m1_tree}-\ref{eq:Legendre}) as an approximation on arbitrary graphs, just as we did with belief propagation. This leads to the one-step replica symmetry breaking (1RSB) approach\index{solution!1RSB}. Note that on random graphs we will always speak about densities of the entropy, complexity or free-entropy etc. Thus on random graphs: instead of the entropy $S$ defined in (\ref{eq:ent_tree}) we consider $s=S/N$. The replicated free entropy $\Phi$ (\ref{eq:free_ent}) and complexity $\Sigma$ (\ref{eq:1RSB_Sigma_tree}) are also divided by the number of variables. We, however, denote them by the same symbol, as confusion is not possible. 
 
Let us discuss once again, now from the random graph perspective, what are the correlations which make the replica symmetric approach fail. This will finally explain the definition of the {\it replica symmetric solution being correct} given at the beginning of this section.

\paragraph{Point-to-set correlations ---}

The concept of the point-to-set correlations is common in the theory of glassy systems. Usually it is considered in the phenomenology of the real glassy systems on finite-dimensional lattices, see for example \cite{BouchaudBiroli04} and references therein. Here we restrict the discussion to properties relevant for the tree-like lattices. 

Call $\overline B_d(i)$ all vertices of the graph which are at distance at least $d$ from $i$, define {\it point-to-set}\index{correlation point-to-set} correlation function as
\be
     C_d(i)=  ||\mu(i,\overline B_d(i)) - \mu(i) \mu(\overline B_d(i))||_{\rm TV}\, ,
\ee
where $\mu(\cdot)$ is the uniform measure over solutions (\ref{uniform}), and the total variation distance of two probability distributions is defined as $||q-p||_{\rm TV}=\sum_x |q(x)-p(x)|/2$. The average point-to-set correlation is
\be
 C_d= \frac{1}{N} \sum_{i=1}^N C_d(i) \, . \label{eq:pts}
\ee
The reconstruction on graphs\index{reconstruction!on graphs} is then defined via the decay of this correlation function. The reconstruction on tree-like graphs is not in general equivalent to the reconstruction on trees. Roughly said, it is not equivalent in the ferromagnetic models, e.g. the ferromagnetic Ising model, which spontaneously break some of the discrete symmetries. On the other hand on most of the frustrated models they are equivalent. A general condition, which might be very nontrivial to check, is given in \cite{GerschenfeldMontanari07}. 

If the point-to-set correlation function decays to zero, $\lim_{d\to \infty} C_d =0$, then almost every variable is independent of its far away neighbours. The replica symmetric approach then has to be asymptotically correct on locally tree-like lattices. 

On the other hand if the point-to-set correlations do not decay to zero, then the far-away neighbours influence the value of the variable
$i$.
And the replica symmetric solution fails to give the correct picture of the properties of the model. The lack of decay of the point-to-set correlations is equivalent to the reconstruction on graphs, and is also equivalent to the existence of a nontrivial solution of the 1RSB equation (\ref{eq:1RSB_m1_tree}) at $m=1$. This is also equivalent to the extremality\index{measure!extremality} condition for the uniform measure (\ref{uniform}), which was used in definition of [\KM] and reads
\be
      \mathbb{E} \Big[ \sum_{\overline B_d(i)} \mu(\overline B_d(i)) \,  || \mu(i|\overline B_d(i)) - \mu(i)||_{\rm TV} \Big] \mathop{\rightarrow}_{d \to \infty}  0 \, ,
\ee
where the external average is over quenched disorder (in interactions or connectivities).

The point-to-set correlations do not decay to zero for example in the low temperature phase of the ferromagnetic Ising model on a random graph. There it is sufficient to introduce the pure state "up" and the pure state "down" and within these pure states the point-to-set correlations will decay to zero again. On the frustrated models the situation is more complicated but the idea of the resolution is the same: If we manage to split the set of solutions into clusters (pure states) such that within each cluster the point-to-set correlations again decay, the situation is fixed. A statistical description of the properties of clusters can be obtained using the {\it one-step replica symmetry breaking} (1RSB) equations, derived in the previous section \ref{trees} and summarized in the next section \ref{sec:1RSB_equations}. 

However, the correlations might be more complicated and might not be captured fully by the 1RSB approach. In particular the 1RSB approach is correct if and only if the point-to-set correlation decay to zero within clusters and if the replica symmetric statistical description of clusters is correct. In appendix \ref{app:1RSB_stab} we will discuss a necessary condition for the 1RSB approach being correct. In case the 1RSB approach does not fully describe the system further steps of replica symmetry breaking might provide a better approximation (that means splitting clusters into sub-clusters or aggregation of clusters) \cite{MezardParisi00}. However, on the tree-like lattices, the exact solutions is not known in such cases.

\paragraph{Relation with equilibration time ---}
In glasses, the clustering transition is usually studied at finite temperature and is called the dynamical transition. The clustered phase\index{phase!clustered} with $\Sigma(m=1)>0$ is called the {\it dynamical} 1RSB phase\index{phase!dynamical 1RSB}. This phase, where most of the static properties do not differ from the replica symmetric (liquid) ones, was first described and discuss in \cite{KirkpatrickThirumalai87a,KirkpatrickThirumalai87b}.
The dynamical transition is associated with a critical slowing down of the dynamical properties, e.g. the equilibration time is expected to diverge at this point. Note that such a purely dynamical phase transition is typical for mean-field models. In the finite dimensional glassy systems the barriers between a metastable and an equilibrium state are finite (independent of the system size). This is because the nucleation length might be large but have to be finite. Thus instead of a sharp dynamical transition in finite dimensional systems we observe only a crossover.

However, even at the mean field level, the exact dynamical description is known only in a few toy models, e.g. the spherical $p$-spin model \cite{CugliandoloKurchan93} or the random subcubes model [\MZ]. 
In general, the dynamical solution is only approximative, still many very interesting results were obtained. For a review see \cite{BouchaudCugliandolo98}. In the models on sparse random lattices even the approximation schemes are rather poor, see e.g. \cite{SemerjianWeigt04}. Thus the exact general relation between dynamics and the dynamical (clustering) transition is not known. 

An important contribution in establishing the link between dynamics and the static solution on random graphs is \cite{MontanariSemerjian05,MontanariSemerjian06,MontanariSemerjian06b} where the divergence of the point-to-set correlation length is linked with divergence of the equilibration time of the Glauber dynamics. 
This suggests that beyond the clustering transition the Monte Carlo sampling (or maybe even sampling in general) will be a hard task.

Note also that in the mathematical literature the Glauber dynamics is often studied. Many results exist about the so-called {\it rapid mixing}\index{rapid mixing} of the associated Markov chain \cite{Sinclair93}. But the rapid mixing questions equilibration in polynomial time, whereas in physics the relevant time scale is linear. Moreover rapid mixing is defined as convergence to the equilibrium measure from any possible initial conditions, whereas in physics of glasses the notion of a typical initial condition should be used instead.

\subsection{Compendium of the 1RSB cavity equations}
\label{sec:1RSB_equations}

We review the 1RSB equations on a general CSP. The order parameter is a probability distribution of the cavity field (BP message) $\psi^{a\to i}=(\psi^{a\to i}_0,\dots,\psi^{a\to i}_{q-1})$. The self-consistent equation for $P^{a\to i}$ reads
\be
    P^{a\to i}(\psi^{a\to i} ) = \frac{1}{{\cal Z}^{j\to i}} \int \prod_{j\in \partial a -i}\prod_{b\in \partial j -a} \big[{\rm d}P^{b\to j}(\psi^{b\to j})\big] \,   (Z^{j\to i})^m \,   \delta\big[ \psi^{a\to i} -{\cal F}(\{\psi^{b\to j}\})\big]\, , \label{eq:1RSB_m1}
\ee
where the function ${\cal F}(\{\psi^{b\to j}\})$ and the term $Z^{j\to i}$ are defined by the BP equation (\ref{eq:BP_one}), ${\cal Z}^{j\to i}$ is a normalization constant. 

The associated thermodynamical potential (\ref{eq:free_ent}) is computed as
\begin{subequations}
\label{eq:1RSB_entropies}
\bea
  \Phi(m) &=&  \frac{1}{N} \Big[ \sum_a \log{ {\cal Z}^{a+\partial a}} - \sum_{i} (l_i-1) \log{{\cal Z}^{i}} \Big] \, ,\label{eq:1RSB_entr} \\
   {\cal Z}^{a+\partial a} &=&  \int \prod_{i\in \partial a} \prod_{b\in \partial i-a}\big[  {\rm d}P^{b\to i}(\psi^{b\to i}) \big] \big( Z^{a+\partial a} \big)^m  \, ,\\
   {\cal Z}^{i} &=& \int  \prod_{a\in \partial i} \big[{\rm d}P^{a\to i}(  \psi^{a\to i}) \big]  \big( Z^{i} \big)^m   \, ,
\eea
\end{subequations}
where the terms $Z^{a+\partial a}$ and $Z^i$ are the partition sum contributions defined in (\ref{eq:norms}).

The logarithm of the number of states divided by the system size defines the complexity function $\Sigma$\index{complexity function}. Inversely the number of states is $e^{N\Sigma}$. At finite temperature the complexity of states with a given internal free energy is a Legendre transformation of the potential $\Phi(m)$ 
\be
    \Phi(m) = -\beta m f + \Sigma(f)\, ,   \label{eq:Legendre}  
\ee
Useful relations between the free energy, complexity and potential $\Phi$ are
\be
\partial_f \Sigma(f) = \beta m \, ,  \quad  \quad   \partial_m \Phi(m) = -\beta f\, ,    \quad  \quad   m^2  \partial_m \frac{\Phi(m)}{m} = - \Sigma\, .  
\ee
At zero energy, $E=0$, and zero temperature, $\beta \to \infty$, the free energy becomes entropy $-\beta f \to s$.  Then the complexity is a function of the internal entropy of states and (\ref{eq:Legendre}) becomes
\be
     \Phi(m) = m s + \Sigma(s)   \, ,   \label{eq:1RSB_Sigma} 
\ee
with
\be
 \partial_s \Sigma(s) = - m\, ,   \quad  \quad \partial_m \Phi(m) = s\, ,    \quad  \quad   m^2\partial_m \frac{\Phi(m)}{m} = - \Sigma \, .
\ee
This is called the {\it entropic}\index{cavity method!entropic} zero temperature limit.
The internal entropy is expressed as
\be
       s =  \frac{1}{N} \Big( \sum_a  {\Delta S}^{a+\partial a} - \sum_{i} (l_i-1) {\Delta S}^{i} \Big)\, , \label{eq:1RSB_entropy}
\ee
where $\Delta S^{a+\partial a}$ ($\Delta S^i$ resp.) is an internal entropy shift when the constraint $a$ and all its neighbour (the variable $i$ resp.) are added to the graph.  
\begin{subequations}
\label{eq:entr_shift}
\bea
   {\Delta S}^{a+\partial a} &=&  \frac{\int \prod_{i\in \partial a} \prod_{b\in \partial i-a}  \big[{\rm d}P^{b\to i}(\psi^{b\to i})\big] \big( Z^{a+\partial a} \big)^m  \log{ Z^{a+\partial a} } }{\int \prod_{i\in \partial a} \prod_{b\in \partial i-a} \big[ {\rm d}P^{b\to i}(\psi^{b\to i})\big] \big( Z^{a+\partial a} \big)^m } \, ,\\
   {\Delta S}^{i} &=& \frac{\int  \prod_{a\in \partial i} \big[{\rm d}P^{a\to i}(  \psi^{a\to i})\big]  \big( Z^{i} \big)^m  \log{Z^i}}{\int  \prod_{a\in \partial i} \big[{\rm d}P^{a\to i}(  \psi^{a\to i})\big]  \big( Z^{i} \big)^m }  \, .
\eea
\end{subequations}
In the energetic zero temperature limit\index{cavity method!energetic}, described in sec.~\ref{sec:energetic} for zero energy, the Parisi parameter $y=\beta m$\index{Parisi parameter!y} is kept constant, thus $m \to 0$. The free energy then converges to the energy, and (\ref{eq:Legendre}) becomes
\be
    \Phi(y) = -y e + \Sigma(e) \label{eq:Legendre_ener}  \, ,
\ee
where the complexity is this time a function of the energy density $e$. The survey propagation equations generalized to nonzero $y$ are called the SP-$y$ equations\index{survey propagation!SP-$y$}. 

Equations (\ref{eq:1RSB_m1}-\ref{eq:entr_shift}) are defined on a single instance of the constraint satisfaction problem. Averages  ${\cal P}$ over the graph ensemble are obtained in a similar manner as in sec.~\ref{sec:average} for the replica symmetric solution.
\be
   {\cal P}\big[P(\psi)\big] = \sum_{\{l_i\}} \Big[\prod_{l_i} {\cal Q}_1(l_i) \Big] \int \prod_{i=1}^{K-1} \prod_{j_i=1}^{l_i} \Big\{{\rm d}{\cal P}\big[P^{j_i}(\psi^{j_i})  \big] \Big\} \, \delta\big[ P(\psi) -{\cal F}_2(\{P^{j_i}(\psi^{j_i})\})\big]\, , \label{eq:1RSB_aver}
\ee
where in the sum over $\{l_i\}$, $i\in \{1,\dots,K-1\}$, and the functional ${\cal F}_2$ is defined by (\ref{eq:1RSB_m1}). Analogical expression holds for the average of the complexity or internal entropy. A general method to solve the equation (\ref{eq:1RSB_aver}) is the population of populations described in appendix \ref{sec:pop_pop}.

\section{Geometrical definitions of clusters}
\label{sec:geom}

Up to now we were describing clusters, i.e., partitions of the space of solutions, in a very abstract way which was defined only in the thermodynamical limit. We showed how to compute the number of clusters of a given size (internal entropy) (\ref{eq:1RSB_Sigma}), and we argued that the description makes sense if the point-to-set correlation (\ref{eq:pts}) decays to zero within almost every cluster of that size. In this last sense clusters are what we would call in statistical physics pure equilibrium states.

On a very intuitive level, cluster are groups of nearby solutions which are in some sense separated from each other. Several geometrical definitions are used in the literature, we want to review the most common ones and state their relation to the definition above. We want to stress that it is not know whether any of the geometric definitions is equivalent to the description given above and used usually in the statistical physics literature.

\paragraph{Strong geometrical separation, $x$-satisfiability ---}

First rigorous proofs of existence of an exponential number of clusters of solutions in the random $K$-SAT were based on the concept of $x$-satisfiability. Two solutions are at distance $x$ if they differ in exactly $xN$ variables. A formula is said $x$-satisfiable if there is a pair of solutions at distance $x$, and $x$-unsatisfiable if there is not. 

Mora, M\'ezard and Zecchina \cite{MezardMora05,MoraMezard05} managed to prove that for $K\ge 8$ and a constraint density $\alpha$ near enough to the satisfiability threshold the formulas are almost surely $x$-satisfiable for $x<x_0$, almost surely $x$-unsatisfiable for $x_1<x<x_2$, and almost surely $x$-satisfiable at $x_3<x<x_4$, where obviously $0<x_0<x_1<x_2<x_3<x_4<1$. This means that at least two well separated clusters of solutions exist. Proving that there is an exponentially smaller number of pairs of solutions at distances $x<x_1$ than at distances $x>x_2$ leads to the conclusion that an exponential number of well geometrically separated clusters exists \cite{AchlioptasRicci06}.  

However, the $x$-satisfiability gives too strong conditions of separability. This is illustrated for example in the XOR-SAT problem \cite{MoraMezard06}. It is still an open question if there is or not a gap in the $x$-satisfiability in the random 3-SAT near to the satisfiability threshold.

\paragraph{Connected-components clusters ---}

Another popular choice of a geometrical definition of clusters is that clusters are connected components in a graph where every solution is a vertex and solutions which differ in $d$ or less variables are connected. The distance $d$ is often said to be any sub-extensive (in the number of variables $N$) distance, that is $d=o(N)$. However, such a rule is not very practical for numerical investigations.\index{cluster!connected-components} 

\begin{figure}[!ht]
 \resizebox{\linewidth}{!}{
  \includegraphics{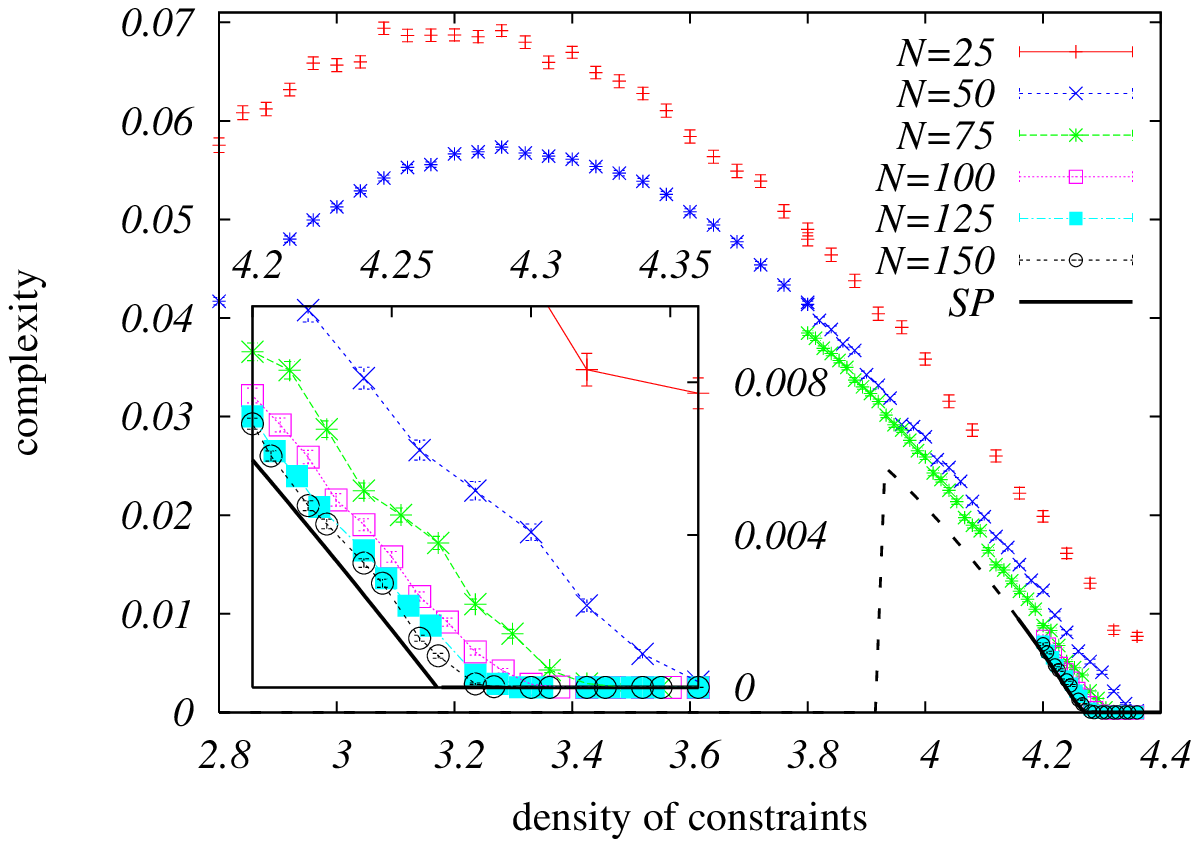}
  \includegraphics{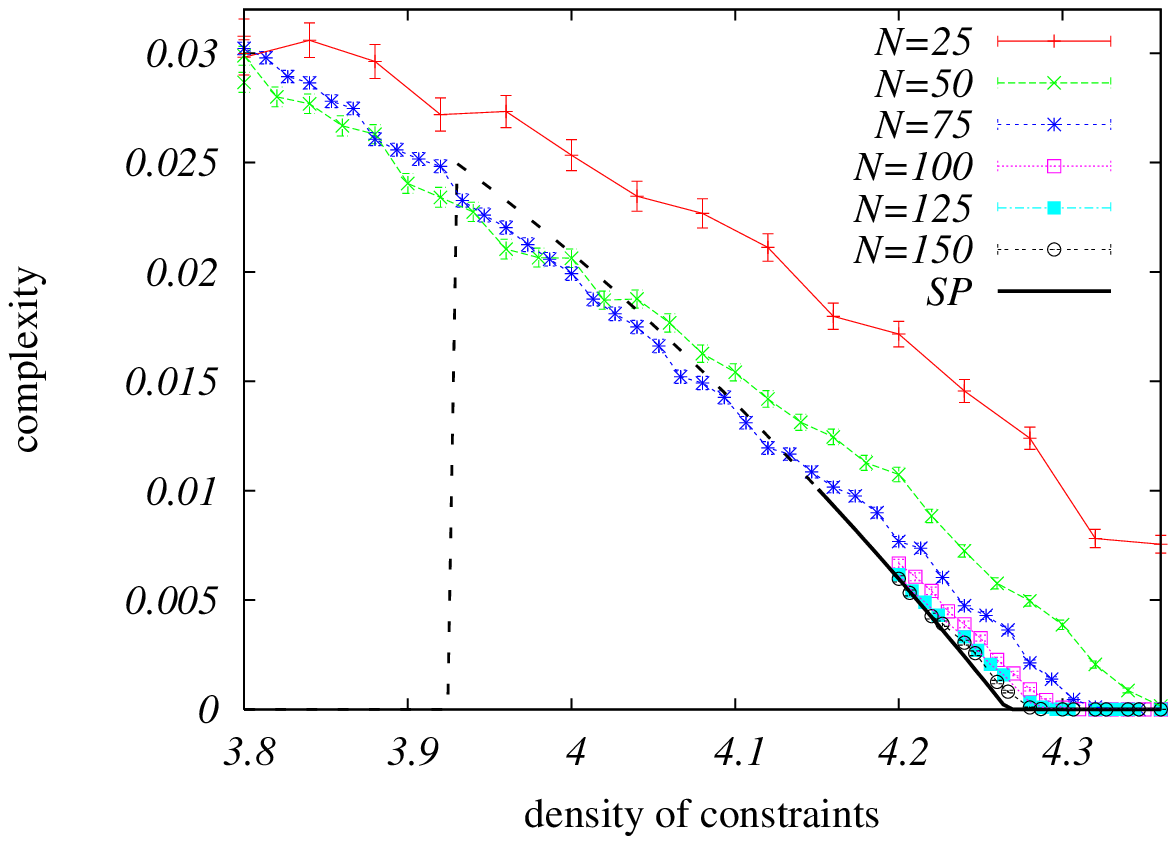}}
 \caption{\label{fig:cmplx} Right: Complexity of the connected-components clusters. Left: Complexity of the whitening-core clusters. Both compared to the complexity computed from the survey propagation equations. The data for the SP complexity are courtesy of Stephan Mertens, from \cite{MertensMezard06}.
}
\end{figure}

In $K$-SAT, in fact, $d=1$ seems to be a more reasonable choice. There are two reasons: First, clusters defined via $d=1$ have correct "whitening" properties as we explain in the next paragraph. Second, we numerically investigated the complexity of $d=1$ connected-components clusters, fig.~\ref{fig:cmplx} right, and the agreement with the total number of clusters computed from (\ref{eq:1RSB_Sigma}) at $m=0$ is strikingly good. In particular, near to the satisfiability threshold $\alpha>4.15$, where the 1RSB result for the total complexity function is believed to be correct (stable) \cite{MontanariParisi04}. 

Formally, connected-components clusters have no reason to be equivalent to the notion of pure states. They are not able to reproduce purely entropic separation between clusters, which might exist in models like 3-SAT. However, fig.~\ref{fig:cmplx} suggests that there is more in this definition than it might seem at a first glance.  

\paragraph{Whitening-core clusters ---}

We define the {\it whitening of a solution}\index{whitening!of a solution} as iterations of the warning propagation equations (\ref{eq:WP}) initialized in the solution. The fixed point is then called the {\it whitening core}\index{whitening!core}. Note, that the whitening core is well defined in the sense that the fixed point of the warning propagation initialized in a solution does not depend on the order in which the messages were updated. A whitening core is called trivial if all the warning messages are $0$, that is "I do not care".\index{cluster!whitening-core} 

The 1RSB equations at $m=0$, which give the total complexity function, can be derived as belief propagation counting of all possible whitening cores \cite{MezardZecchina02,BraunsteinZecchina04,ManevaMossel05}. Thus another reasonable definition of clusters is that two solutions belong to the same cluster if and only if their whitening core is identical.  In fig.~\ref{fig:cmplx} left we plot numerically computed  complexity of the whitening-core clusters compared to the complexity computed from (\ref{eq:1RSB_Sigma}) at $m=0$. The agreement is again good, in particular near to the satisfiability threshold, $\alpha>4.15$, where the SP gives a correct result. 

The $d=1$ connected-components clusters share the property that all the solution from one clusters have the same whitening core. Proof: If this would not be true then there have to exist a pair of solutions which do not have the same whitening core but differ in only one variable, this is not possible because then the whitening could be started in that variable. 

Note, however, that the definition of whitening-core clusters put all the solutions with a trivial whitening core into one cluster. This is not correct as, at least near to the clustering threshold, there are many pure states with a trivial whitening core. This is closely connected to the properties of frozen variables which will be discussed in chapter~\ref{freezing}. 

\paragraph{Enumeration of clusters in 3-SAT: the numerical method ---}

In order to obtain the data in fig.~\ref{fig:cmplx} we generate instances of the random 3-SAT problem with $N$ variables and $M$ clauses, constraint density is then $\alpha=M/N$. We count number of solutions in $A=999$ random instances and choose the median one where we count the number of connected-components and whitening-core clusters ${\cal S}$. This is repeated $B=1000$ times. The average complexity is then computed as $\Sigma=\sum_{i=1}^B\log{\cal S}_i/(B N)$, if the median instance was unsatisfiable then we count zero to the average, that is if all the $B$ instances are unsatisfiable then the complexity is zero. We do such a non-traditional sampling to avoid rare instances with very many solutions, which we would not be able to cluster.

\section{Physical properties of the clustered phase}

Let us give a summary of the properties of the clustered phase, also called the dynamical 1RSB phase. We describe only the situation when $\Sigma(m=1)>0$ (\ref{eq:1RSB_Sigma}), when the opposite is true the properties are completely different as we will discuss in the next chapter \ref{condensation}. 

The complexity function computed from (\ref{eq:1RSB_Sigma}) is the log-number of clusters of a given internal entropy. If a solution is chosen uniformly at random it will almost surely belong to a cluster with entropy $s^*$ such that $\Sigma(s)+s$ is maximized in $s^*$, $\partial_s \Sigma(s^*)=-1$, that is $m=1$. At $m=1$ the total entropy $\Sigma(s^*)+s^*=\Phi(m=1)$. The replicated free entropy at $\Phi(m=1)$ is equal to the replica symmetric entropy. Thus the total entropy in the dynamical 1RSB phase is equal to the RS entropy. Also the marginal probabilities at $m=1$ are equal to the replica symmetric ones
\be
       \int {\rm d} P^{i\to j}(  \psi^{i\to j})\,   \psi^{i\to j}_{s_i}  = (\psi_{\rm RS})^{i\to j}_{s_i}   \quad {\rm if} \quad m=1\, .
\ee
Thus the clustering transition is not a phase transition in the Ehrenfest sense, because the thermodynamical potential, entropy in our case, in analytical at the transition. 

The overlap (or here distance) distribution, which is often used to describe the spin glass phase, is also trivial and equal to the replica symmetric one in the dynamical 1RSB phase. Indeed, if exponentially many clusters are needed to cover almost all solutions, then the probability that two solutions happen to belong to the same cluster is zero. 

The correlation function between two variables at a distance (shortest path in the graph) $d$ is defined as $\langle s_i s_j \rangle_c=||\mu(s_i,s_j)-\mu(s_i)\mu(s_j)||_{\rm TV}$. The variance of the overlap distribution, which is negligible compared to $1$  as we explained, can be expressed as $\sum_{i,j}\langle s_i s_j \rangle^2_c /N^2$, and thus the two-point correlation have to decay faster with distance than the number on vertices at that distance is growing.  This means in particular that two neighbours of a node $i$ are independent if we condition on the value of $i$, this is again consistent with the fact that the belief propagation equations predict correct total entropy and marginal probabilities.

So far nothing is different form the replica symmetric phase. It is thus not straightforward to recognize the dynamical 1RSB phase based on the original replica computation. Presence of this phase was discovered and discussed in \cite{KirkpatrickThirumalai87a,KirkpatrickThirumalai87b}. Later purely static methods were developed to identify this phase. The most remarkable is perhaps the $\epsilon$-coupling and the "potential" of \cite{FranzParisi95,FranzParisi97}. 

In our setting the main difference between the replica symmetric phase and the dynamical 1RSB phase is that in the later the point-to-set correlations do not decay to zero.  Consequently the equilibration time of the local Monte Carlo dynamics diverges and Monte Carlo sampling becomes difficult \cite{MontanariSemerjian06}.

\section{Is the clustered phase algorithmically hard?}

Clustering has important implications for the dynamical behaviour. It slows down the equilibration and thus uniform sampling of solutions via local single spin flip Monte Carlo is not possible, or exponentially slow, beyond the dynamical threshold. But finding one solution is a much simpler problem than sampling. 

\paragraph{Analytic arguments ---}
In the 3-coloring of Erd\H{o}s-R\'enyi graphs the clustering threshold is $c_d=4$, as at this point the spin glass susceptibility diverges, see appendix \ref{app:RS_stab}. In the terms of the reconstruction problem the Kesten-Stigum \cite{KestenStigum66,KestenStigum66b} bound is sharp. On the other hand Achlioptas and Moore \cite{AchlioptasMoore03} proved that a simple heuristic algorithm is able to find a solution in average polynomial time up to at least $c=4.03$. This shows that the RSB phase is not necessarily hard. 

A similar observation was made in the 1-in-3 SAT problem in [\RS]. There is a region in the values of the average density of constraints and the probability of negating a variable in a clause in which the replica symmetric solution is unstable and yet the unit clause propagation algorithm with the short clause heuristics was proven to find a solution in polynomial average time. 

We should mention a common contra-argument; which is that in the above mentioned regions the 1RSB approach might not be correct, and the presumably full-RSB phase \cite{Parisi80} is more "transparent" for the dynamics of algorithms, see e.g.~\cite{MontanariRicci04}. However, at least in the 3-coloring, the 1RSB approach seems to be correct in the  interval in question, as we argue in appendix \ref{app:1RSB_stab}. 

\begin{figure}[!ht]
 \resizebox{\linewidth}{!}{
  \includegraphics{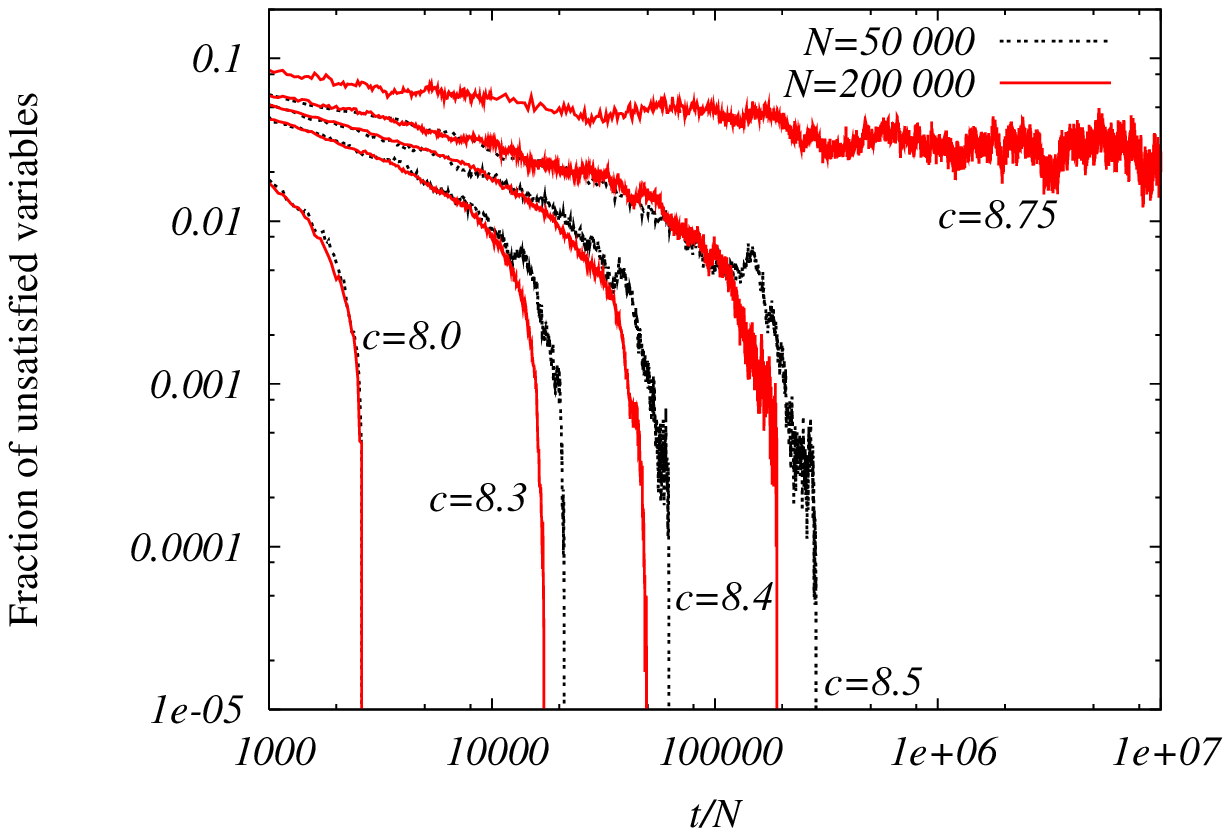}
  \includegraphics{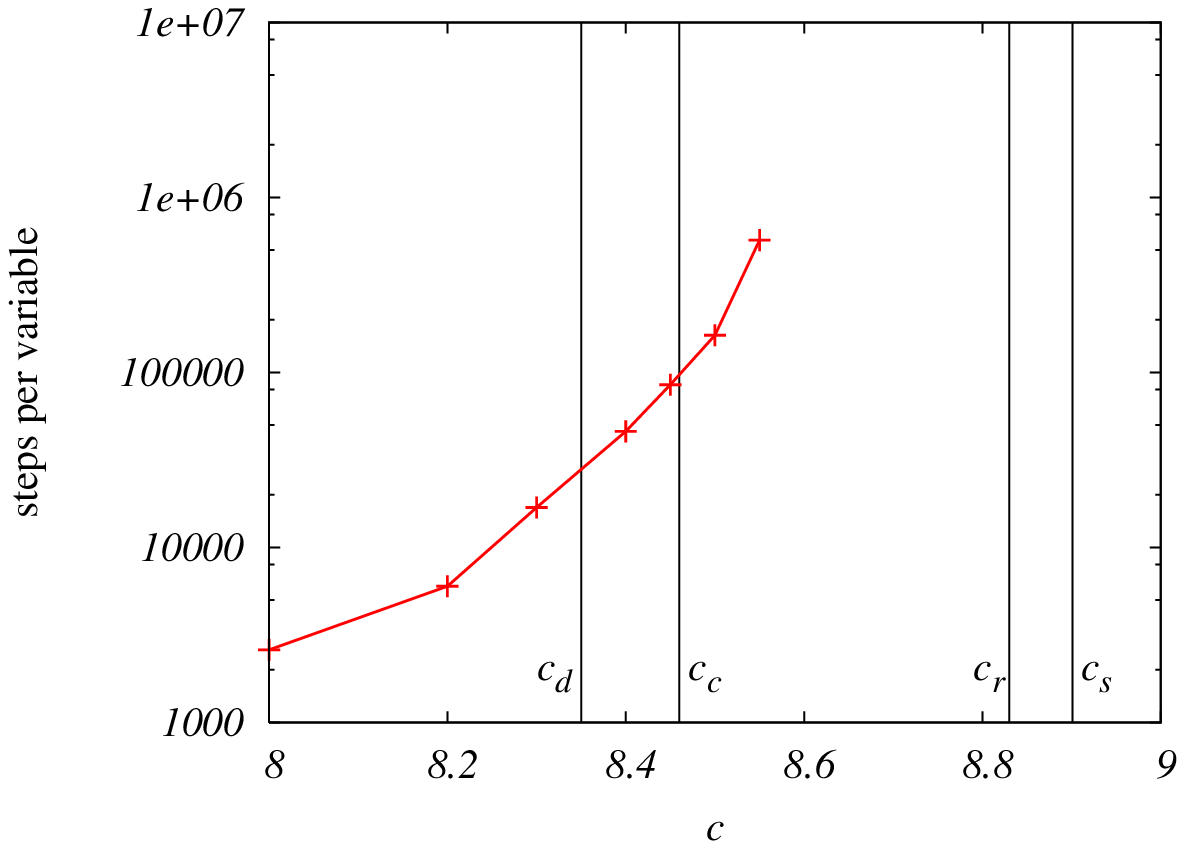}}
 \caption{\label{fig:WalkCOL} The performance of the ASAT algorithm in  the 4-coloring of random Erd\H{o}s-R\'enyi graphs. Left: The energy density plotted against the number of steps per variable. Right: The average running time (per variable) as a function of the connectivity. The time does not diverge at the clustering transition $c_d$, but beyond it. The other phase transitions marked are the condensation transition $c_c$ (chap.~\ref{condensation}) the rigidity transition $c_r$ (chap.~\ref{freezing}) and the colorability threshold $c_s$}
\end{figure}

\paragraph{Stochastic local search ---}
There is a lot of numerical evidence that relatively simple single spin flip stochastic local search algorithms are able to find solutions in linear time deep in the clustered region. Examples of works where performance of such algorithms was analyzed are \cite{KrzakalaKurchan07,SeitzAlava05,ArdeliusAurell06,AlavaArdelius07}. In fig.~\ref{fig:WalkCOL} we give an example of performance of the ASAT\index{algorithms!ASAT} algorithm \cite{ArdeliusAurell06} in 4-coloring of Erd\H{o}s-R\'enyi random graphs [\ZK]. The algorithm is described in appendix \ref{sec:SLS}. In the 4-coloring ASAT is able to find solutions in linear time beyond the clustering transition $c_d=8.35$

\paragraph{Simulated annealing ---}
There is no paradox in the observations above. Quantitative statements are, however, difficult to make. Let us describe on an intuitive level the behaviour of an algorithm (dynamics) which satisfies the detailed balance condition and thus in infinite time samples uniformly from the uniform measure (\ref{uniform}). We think for example about the simulated annealing \cite{KirkpatrickGelatt83}. Above the dynamical temperature $T_d$ corresponding to an energy $E_d$ the point-to-set correlation function (\ref{eq:pts}) decay fast and thus simulated annealing is able to reach the equilibrium. Below temperature $T_d$ this is not the case anymore and the dynamics is stuck for a very long time in one of the clusters, states. But the bottom of this state $E_{\rm bottom}$ lies lower than $E_d$, thus when lowering the temperature the average energy seen by the simulated annealing also decreases. If $E_{\rm bottom}=0$ then the algorithm will find a solution. It is not known how to compute $E_{\rm bottom}$ in general.  Sometimes, far from the clustering transition, the {\it iso-complexity} approach \cite{MontanariRicci04} gives a lower bound on $E_{\rm bottom}$. But in general, as far as we know, there is no argument saying $E_{\rm bottom}>0$. This picture can be substantiated for several simple models as the spherical $p$-spin model \cite{CugliandoloKurchan93} or the random subcubes model [\MZ]. The connection with the optimization problems was remarked in \cite{KrzakalaKurchan07}.

For the stochastic local search algorithm, which does not satisfy the detailed balanced condition, the situation might be similar. At a point the algorithm is stuck in a cluster, but if this cluster goes down to the zero energy then it might be able to find solutions even in the clustered phase.  

However, the current understanding of the dynamics of the mean field glassy systems is far from complete. More studies are needed to understand better the link between the static clustered phase and the dynamical behaviour.

\chapter{Condensation}
\label{condensation}

{\it In this chapter we will describe the so-called {\it condensed clustered phase}\index{phase!condensed}. Before turning to the models of our interest we present the random subcubes model [\MZ], where the condensation of clusters can be understood on a very elementary probabilistic level. After mentioning that the condensed phase is in fact very well known in spin glasses we describe the Poisson-Dirichlet process which determines the distribution of sizes of clusters in that phase. Further, we discuss general properties of the condensed phase in random CSPs. And finally we address our original question and conclude that the condensation is not much significant for the hardness of finding a solution [\ZK].}

\section{Condensation in a toy model of random subcubes}
\label{sec:subcubes}

The random-subcubes model [\MZ] is defined by its solution space $S\subseteq \{0,1\}^N$; we define $S$ as the union of $\lfloor 2^{(1-\alpha)N}\rfloor$ random clusters (where $\lfloor x \rfloor$ denotes the integer value of $x$). A random cluster $A$ being defined as:
\be \label{eq:defcluster}
A=\{\sigma\ |\ \forall i \in  \{1,\ldots,N\}, \sigma_i\in \pi^A_i\},
\ee
where $\pi^A$ is a random mapping:
\begin{eqnarray}
\pi^A: \{1,\ldots,N\} & \longrightarrow & \{\{0\},\{1\},\{0,1\}\}\, , \\
i & \longmapsto & \pi^A_i\, ,
\end{eqnarray}
such that for each variable $i$, $\pi^A_i=\{0\}$ with probability $p/2$, $\{1\}$ with probability $p/2$, and $\{0,1\}$ with probability $1-p$. 
A cluster is here a random subcube of $\{0,1\}^N$. If $\pi^A_i=\{0\}$ or $\{1\}$, variable $i$ is said ``frozen'' in $A$; otherwise it is said ``free'' in $A$.
In this model one given configuration $\sigma$ might belong to zero, one or several clusters. 

We describe the static properties of the set of solutions $S$ in the random-subcubes model in the thermodynamic limit $N\to \infty$ (the two parameters $0\le \alpha\le 1$ and $0\le p\le 1$ being fixed and independent of $N$). The internal entropy $s$ of a cluster $A$ is defined as $\frac{1}{N}\log_2 |A|$, i.e., the fraction of free variables in $A$. 
The probability ${\cal P}(s)$ that a cluster has internal entropy $s$ 
follows the binomial distribution
\be
  {\cal P}(s) = {N\choose sN} (1-p)^{sN} p^{(1-s)N}\, .     \label{eq:sub_bin}
\ee 
Then the number of clusters of entropy $s$, denoted ${\cal N}(s)$, is with high probability
\be
\lim_{N\to\infty}\frac{1}{N}\log_2\cN(s) =\left\{ \begin{array}{ll}  \Sigma(s)\equiv 1-\alpha-D(s\parallel  1-p ) & \textrm{ if }\Sigma(s)\geq 0,\\
-\infty & \textrm{ otherwise,}\end{array}\right.
\label{sigma}
\ee
where $D(x\parallel  y) \equiv x\log_2{\frac{x}{y}}+(1-x)\log_2{\frac{1-x}{1-y}}$ is the binary Kullback-Leibler divergence. 

We compute the total entropy $s_{\rm tot}=\frac{1}{N}\log_2 |S|$. First note that a random configuration belongs on average to $2^{N(1-\alpha)} (1-\frac{p}{2})^N$ clusters. Therefore, if
\be\label{eq:defalphad}
\alpha< \alpha_d \equiv \log_2{(2-p)},
\ee
then with high probability the total entropy is $s_{\rm tot}=1$.

Now assume $\alpha> \alpha_d$. The total entropy is given by a saddle-point estimation:
\be 
\sum_{A} 2^{s(A)N} = [1+o(1)] N\int_{\Sigma(s)\geq 0} \ud s\ 2^{N[\Sigma(s)+s]},\label{eq:calcstot}
\ee
\be
\textrm{whence}\quad     s_{\rm tot}   =  \max_{s}{\left[\Sigma(s)+s\,|\,\Sigma(s)\ge 0\right]}.
     \label{s_tot}
\ee
We denote by $s^*={\rm argmax}_{s}{[\Sigma(s)+s\,|\,\Sigma(s)\ge 0]}$ the fraction of free variables in the clusters that dominate the sum. Note that our estimation is valid (there is no double counting) since in every cluster the fraction of solutions belonging to more than one cluster is exponentially small as long as $\alpha>\alpha_d$.

Define $\tilde s\equiv 2(1-p)/(2-p)$ such that $\partial_s \Sigma(\tilde s)=-1$.
The complexity of clusters with entropy $\tilde s$ reads:
\be
      \Sigma(\tilde s) = \frac{p}{2-p} + \log_2(2-p) - \alpha .
\ee
$\tilde s$ maximizes eq.~(\ref{s_tot}) as long as $\Sigma(\tilde s)\ge 0$, 
that is if
\be 
\alpha\leq \alpha_c \equiv \frac{p}{(2-p)}+\log_2{(2-p)}.
\ee
Then the total entropy reads
\be
        s_{\rm tot} = 1 - \alpha + \log_2{(2-p)} \, \quad {\rm for} \quad \alpha \le \alpha_c.
\ee
For $\alpha>\alpha_c$, the maximum in (\ref{s_tot}) is realized by the largest possible cluster entropy $s_{\rm max}$, which is given by the largest root of $\Sigma(s)$. Then $s_{\rm tot}=s^*=s_{\rm max}$. We will show in the next section that in such a case almost all solutions belong to only a finite number of largest clusters.  This phase is thus called {\it condensed}\index{phase!condensed}, in the sense that almost all solutions are "condensed" in a small number of clusters.

\begin{figure}[!ht]
\begin{center}
 \resizebox{0.67\linewidth}{!}{\includegraphics{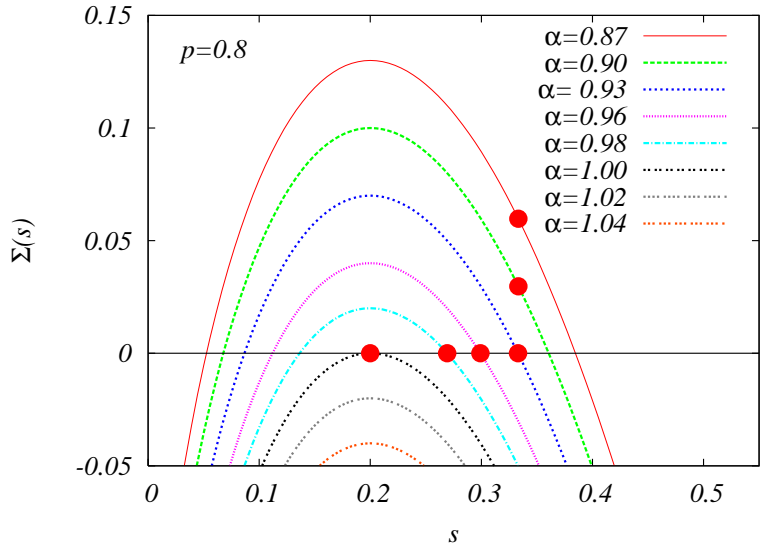}}
 \caption{\label{fig:sub_sigma} The complexity function in the random subcubes model, $\Sigma(s)$ (\ref{sigma}), for $p=0.8$ and several values of $\alpha$. The red dots mark the dominating clusters $s^*,\Sigma(s^*)$. For $p=0.8$ the dynamical transition $\alpha_d\approx 0.263$ is far away from the plotted values, the condensation transition is $\alpha_c\approx 0.930 $, the satisfiability $\alpha_s=1$.}
\end{center}
\end{figure}

In summary, for a fixed value of the parameter $p$, and for increasing values of $\alpha$, four different phases can be distinguished:
\begin{itemize}
   \item[(a)]{Liquid (replica symmetric) phase, $\alpha < \alpha_d$: almost all configurations are solutions.}
   \item[(b)]{Clustered (dynamical 1RSB) phase with many states, $\alpha_d < \alpha < \alpha_c$:
an exponential number of clusters is needed to cover almost all the solutions.}
   \item[(c)]{Condensed clustered phase, $\alpha_c < \alpha < 1$:
a finite number of the biggest clusters covers almost all the solutions.}
   \item[(d)]{Unsatisfiable phase, $\alpha> 1$: no cluster, hence no solution, exists.}
\end{itemize}

\section{New in CSPs, well known in spin glasses}

The complexity function $\Sigma(s)$ (\ref{eq:Legendre}) in random CSPs  is counting the logarithm of the number of clusters per variable which have internal entropy $s$ per variable. We define {\it dominating clusters}\index{cluster!dominating} in the same way as in the random subcubes model, that is clusters of entropy $s^*$ such that
\bea
 s^* = {\rm arg}\max_{s,\Sigma(s)>0} \big[ \Sigma(s) + s \big] \, . \label{eq:s_star}
\eea
In chap.~\ref{clustering} we discussed properties of the dynamical 1RSB phase, that is when $\Sigma(s^*)>0$, in other words when there are exponentially many dominating clusters.

The condensed phase with $\Sigma(s^*)=0$, described in the random subcubes model, exists also in random CSPs. And in the context of constraint satisfaction problems it was first computed and discussed in \cite{MezardPalassini05} and [\KM]. However, historically it was the condensed phase where the 1RSB solution was first worked out \cite{Parisi80}. A very simple example of condensation can also be found in the random energy model \cite{Derrida80,Derrida81}. 
As we discussed in the previous chapter \ref{clustering}, the dynamical 1RSB phase is well hidden within the replica solution --- the total entropy is equal to the replica symmetric entropy, the overlap distribution is trivial and the two-point correlation functions decay to zero etc.  
All this changes in the condensed phase. 

A small digression to the physics of glasses: In structural glasses, the analog of the condensation transition\index{phase transition!condensation} is well known for a long time, its discovery goes back to Kauzmann in 1948 who studied the configurational entropy of glassy materials. Configurational entropy is the difference between the total (experimentally measured) entropy and the entropy of a solid material, this thus corresponds to the complexity function. In the so called fragile structural glasses \cite{Angell95} the extrapolated configurational entropy becomes zero at a positive temperature, nowadays called the Kauzmann temperature. The Kauzmann temperature in the real glasses is, however, only extrapolation. The equilibration time in glasses exceeds the observation time high above the Kauzmann temperature. It is a widely discussed question if there exists a true phase transition at the Kauzmann temperature or not, for a recent discussion see \cite{DebenedettiStillinger01}.

\paragraph{Why does Parisi {\it maximize} the replicated free energy?} As we said, it is the condensed phase which was originally described by Parisi and his one-step replica symmetry breaking solution \cite{Parisi80}. Let us now briefly clarify the relation to the replica solution, similar reasoning first appeared in \cite{Monasson95}. In sec.~\ref{sec:RS_fail} we called the Legendre transform of the complexity function the replicated free entropy $\Phi(m)$ (\ref{eq:Legendre}). In the replica approach the replicated entropy $\Omega(m)=\Phi(m)/m$ is computed. From (\ref{eq:1RSB_Sigma}) follows
\be
       \Omega(m) = s + \frac{\Sigma(s)}{m}   \quad {\rm where} \quad \frac{\partial \Omega(m)}{\partial m} = - \frac{\Sigma(s)}{m^2}\, .
\ee 
Thus, in the condensed phase, computing the largest root of the function $\Sigma(s)$, in order to maximize the total entropy, is equivalent to extremizing the replicated entropy $\Omega(m)$. Moreover, as the function $\Sigma(s)$ is concave and the parameter $m$ is minus its slope this extrema have to be a {\it minima}. Thus in the Parisi's replica solution we have to minimize the replicated entropy function with respect to the parameter $m$. If a temperature is involved then this becomes a maximization of the replicated free energy, this might have seem contra-intuitive in the original solution, but it comes out very naturally in our approach. Other physical interpretation of the maximization was proposed e.g. in \cite{Janis05}.

\section{Relative sizes of clusters in the condensed phase}
\label{sec:PD}

What is the number of dominating clusters in the condensed phase and what are their relative sizes? So far we know that the entropy per variable of the dominating states is $s^*+o(1)$ and that their number is sub-exponential, $\Sigma(s^*)=0$. But much more can be said based on purely probabilistic considerations.

Consider that the total number of clusters ${\cal N}$ is exponentially large in the system size $N$, and that $N\to \infty$. Let the log-number of clusters of a given entropy be distributed according to an analytic function $\Sigma(s)$.  Denote $-m^*=\partial_s \Sigma(s^*)$, in the condensed phase $0 < m^* < 1$.  Denote the size of the $\alpha^{\rm th}$ largest cluster $e^{Ns^*+\Delta_\alpha}$, $\Delta_\alpha=O(1)$. The probability that there is a cluster of size between $e^{Ns^*+\Delta}$ and $e^{Ns^*+\Delta+{\rm d}\Delta}$, $\Delta \gg {\rm d}\Delta$, is $e^{-m^*\Delta}{\rm d}\Delta$, in other words points $\Delta_\alpha$ are constructed from a Poissonian process with rate $e^{-m^*\Delta}$~\footnote{Note that in the random subcubes model the numbers $(Ns^*+\Delta_\alpha)\log(2)$ are integers equal to the number of free variables in the cluster $A_\alpha$. Then $\Delta_\alpha$ are discrete and some of the properties of the resulting process might be different from the Poisson-Dirichlet.}. Relative size of the $\alpha^{\rm th}$ largest cluster is defined as 
\be
         w_\alpha = \frac{e^{\Delta_\alpha}}{\sum_{\gamma=1}^{\cal N} e^{\Delta_\gamma}}\, . \label{eq:weights}
\ee 
Point process $w_\alpha$ which is constructed as described above is in mathematics called the Poisson-Dirichlet process \cite{PitmanYor97}\index{Poisson-Dirichlet process}. 
The connection between this process and the relative weights of states in the mean field models of spin glasses was (on a non-rigorous level) 
understood in~\cite{MezardParisi85b}, for more mathematical review see~\cite{Talagrand03}\footnote{To avoid confusion, note that the Poisson-Dirichlet process we are interested in is the ${\rm PD}(m^*,0)$ in the notation of~\cite{PitmanYor97}. In the mathematical literature, it is often referred to the ${\rm PD}(0,\theta)$ without indexing by the two parameters.}. 

Any moment of any $w_\alpha$ can be computed from the generating function \cite{PitmanYor97}
\begin{equation}
          \mathbb{E} [\exp{(-\lambda/w_\alpha)}] = e^{-\lambda} \phi_{m^*}(\lambda)^{\alpha-1} 
          \psi_{m^*}(\lambda)^{-\alpha} \, ,
\end{equation} 
where $\lambda \ge 0$ and the functions $\phi_{m^*}$ 
and $\psi_{m^*}$ are defined as
\begin{subequations}
\label{eq:generators}
\begin{eqnarray}
     \phi_{m^*}(\lambda) &=& m^* \int_1^\infty  e^{-\lambda x} x^{-1-m^*} {\rm d} x\, , \\
     \psi_{m^*}(\lambda) &=& 1+m^* \int_0^1(1 -e^{-\lambda x}  )  x^{-1-m^*} {\rm d} x\, .
 \end{eqnarray}
\end{subequations}
The second moments can be used to express the average probability $Y$ that two random solutions belong to the same cluster 
\be
    Y =  \mathbb{E}\Big[ \sum_{\alpha=1}^{\cal N} w_\alpha^2 \Big] =1-m^* \label{eq:PD_overlap} \, .
\ee
This was originally derived in \cite{MezardParisi85b}

\begin{figure}[!ht]
\begin{center}
  \resizebox{0.67\linewidth}{!}{\includegraphics{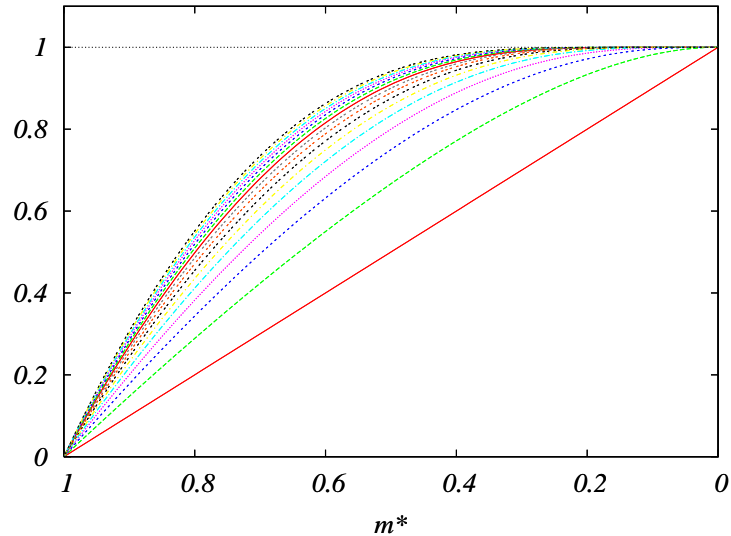}}
\end{center}
\caption{ \label{fig:PD} The fractions of solutions covered by the largest clusters as a function of parameter $m^*$. The lower curve is related to the size of 
the largest clusters as $1/ \mathbb{E}[1/w_1]=1-m^*$. The following curves are 
related to the size of $i$ largest clusters, their distances 
are $ \mathbb{E}[R_\alpha]  \mathbb{E}[R_{\alpha-1}] \dots  \mathbb{E}[R_1] (1-m^*)$.}
\end{figure}

Another useful relation \cite{PitmanYor97} is that the ratio of two consequent points $R_\alpha=w_{\alpha+1}/w_\alpha$,
$\alpha=1,2,\dots,{\cal N}$ is distributed as $\alpha m^* R_\alpha^{\alpha m^*-1}$.  In particular its 
expectation is $ \mathbb{E}[R_\alpha]=\alpha m^*/(1+\alpha m^*)$ and the random variables $R_\alpha$ are mutually independent.
We used these relation to obtain data in figure~\ref{fig:PD}.

From the properties of the Poisson-Dirichlet process, it follows that an arbitrary large fraction of the solutions can be covered by a finite number of clusters. When $m^*$ is near to zero, that is near to the satisfiability threshold, the largest cluster covers a large fraction of solutions. On the other side, when $m^*$ is near to one, that is near to the condensation transition, very many (but finite in $N$) clusters are needed to cover a given fraction of solutions.

\section{Condensed phase in random CSPs}

The total entropy in the condensed phase is strictly smaller than the replica symmetric entropy, $s_{\rm tot}=s^* < s_{\rm RS}$. At the condensation transition $c_c$ the total entropy is non-analytic, it has a discontinuity in the second derivative. This can be seen easily for example from the expressions for the random subcubes model. 
At a finite temperature the discontinuity in the second derivative of the free energy corresponds to a jump in the specific heat.
The parameter $m^*=1$ at the condensation transition and decreases monotonously to $m^*=0$ at the satisfiability threshold.

\paragraph{Concept of self-averaging ---}

In the physics of disordered systems the self-averaging is a crucial concept.
We say that quantity $A$ measured on a system (graph) of $N$ variables is self-averaging if in the limit $N\to \infty$
\be
    \frac{\mathbb{E}(A^2)-\big[\mathbb{E}(A)\big]^2}{\big[\mathbb{E}(A^2)\big]} \to 0\, ,
\ee
where the average $\mathbb{E}\big[\cdot\big]$ is over all the disorder in the system. In other words a quantity is self-averaging if its value on a typical large system is equal to the average value.  By computing the average value we thus describe faithfully the typical large system. And also measuring $A$ on a single large system is enough to represent the whole ensemble. On finite-dimensional lattices and off criticality extensive quantities are always self-averaging. This can be shown by building  the large lattice from smaller blocks, the additivity of an extensive quantity and the central limit theorem then ensures the self-averaging. At the critical point, on a mean field lattice (fully connected or tree-like) or for non-extensive quantities the answer whether $A$ is self-averaging or not becomes nontrivial.

In the condensed phase quantities which involve the weights of clusters are not self-averaging. This arises from the fact that the dominating clusters are different in every realization of the system. Statistical properties of many quantities of interest can be described from the Poisson-Dirichlet process. 

\paragraph{Overlap distribution ---} The overlap between two solutions is defined as one minus the Hamming distance 
\be
q(\{s\},\{s'\}) = \frac{1}{N}\sum_{i=1}^N \delta(s_i,s'_i)\, .
\ee
The overlap between two solutions belonging to two different dominating clusters is $q_0$, and between two solutions belonging to the same dominating cluster $q_1$. Values $q_0$ and $q_1$ are self-averaging.
The distribution of overlaps in the limit $N\to \infty$ can thus be written as 
\be
      P(q) = w\,  \delta(q-q_1) + (1-w)\,  \delta(q-q_0)\, ,
\ee
where the weight $w$ is the probability that two random solutions belong to the same cluster. Thus $w = \sum_{\alpha=1}^{\cal N} w^2_\alpha$, where $w_\alpha$ are weights of the clusters (\ref{eq:weights}) given by the Poisson-Dirichlet process. The weights change from realization to realization, $w$ is thus not a self-averaging quantity, its typical value fluctuates around the mean $\mathbb{E}(w)=1-m^*$ computed in (\ref{eq:PD_overlap}). The distribution of the random variable $w$ is also known \cite{MezardParisi84}.

\paragraph{Two-point correlation functions ---} 

The variance of the overlap distribution is
\be
   {\rm var}\,  q = \int q^2 P(q) \, {\rm d}q - \Big[\int q\,  P(q)\,  {\rm d}q\Big]^2= w(1-w)(q_1-q_0)^2\, .
\ee
At the same time the variance is equal to 
\be
  {\rm var}\,  q = \frac{1}{N^2} \sum_{i,j}  \sum_{s_i,s_j} |\mu(s_i,s_j) - \mu(s_i) \mu(s_j)| \approx  \frac{1}{N} \sum_{i}  \sum_{s_i,s_0} |\mu(s_i,s_0) - \mu(s_i) \mu(s_0)|\, ,
\ee
where $s_0$ is a typical variable in the random graph. 
If we consider that the two-point correlation function is of order one up to a correlation length $\xi$ and zero after that we get 
\be
     {\rm var}\,  q \approx \frac{1}{N}\,  c^\xi  \, ,
\ee
where $c$ is approximately the branching factor. In the condensed phase the variance of the overlap is of order one thus the correlation length has to be of order $\log N$. But the shortest path between two random variables is also of order $\log N$ thus the two-point correlations cannot be neglected in the condensed phase. 


If two-point correlations cannot be neglected then the derivation of belief propagation equations (\ref{eq:BP_1}-\ref{eq:BP_2}) is not valid, because we supposed that the neighbours of a node $i$ are independent when we condition on the value of $i$. It is thus not surprising that the value to which the BP equations converge (if they do), does not correspond to the true marginal probability. Formally, the BP fixed point corresponds to the 1RSB equations at $m=1$, but in the condensed phase $m^*<1$.

In fact, the probability distribution of the true marginal probabilities is another example of a non self-averaging quantity. It again depends on the realization of the Poisson-Dirichlet process.


\section{Is the condensed phase algorithmically hard?}

\begin{figure}[!ht]
  \resizebox{\linewidth}{!}{
  \includegraphics{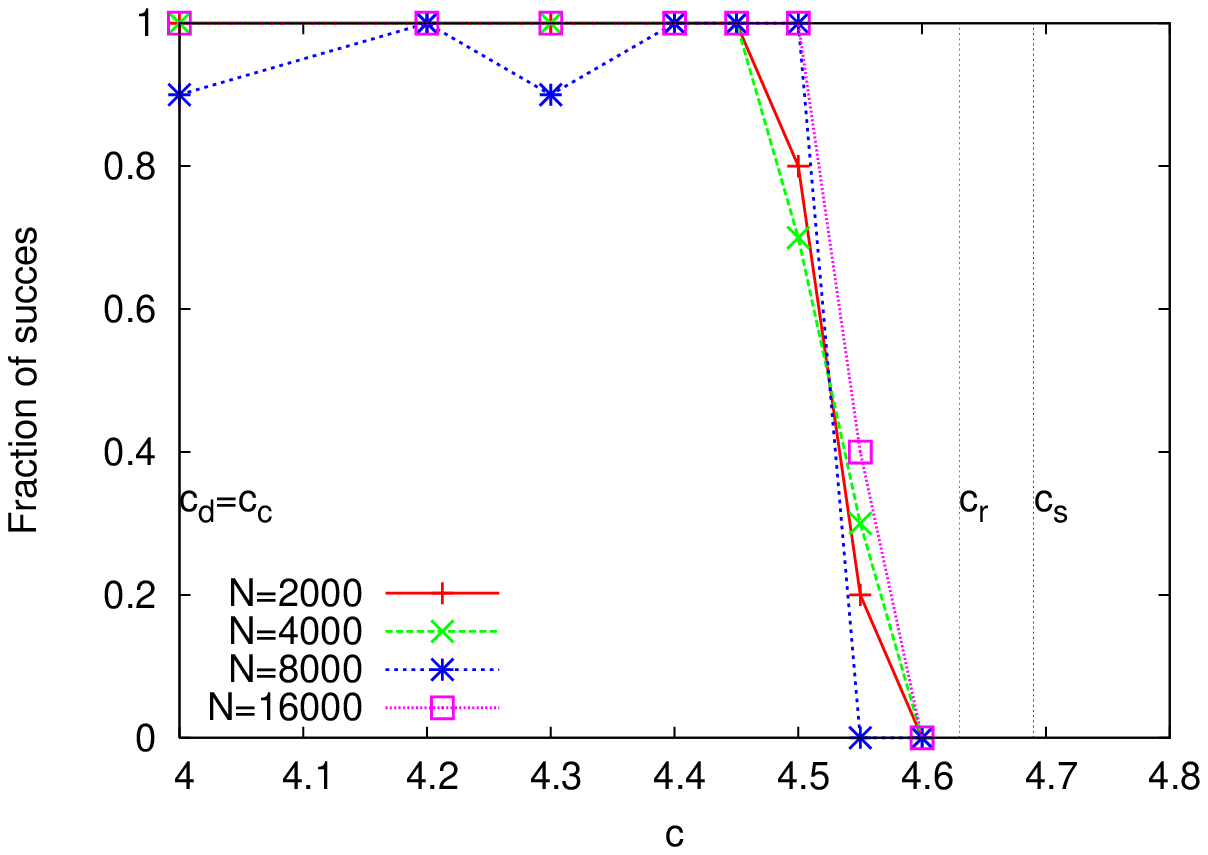}
  \includegraphics{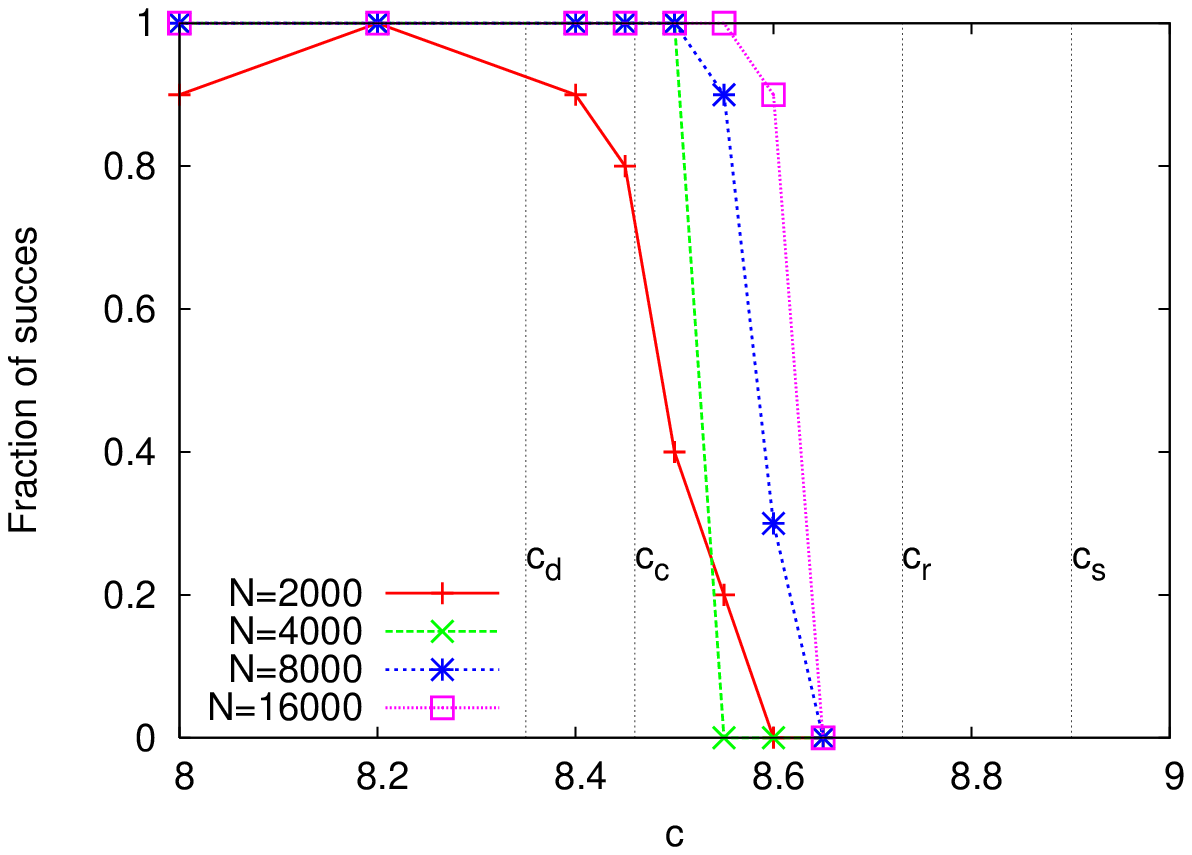}
  }
\caption{ \label{fig:BP_col} The performance of the maximal BP decimation algorithm, described in appendix~\ref{sec:BP_dec}, in the 3-coloring (left) and the 4-coloring (right) of random graphs. This algorithm is able to color random graphs beyond both the clustering $c_d$ and the condensation $c_c$ transitions in 3- and 4-coloring.  }
\end{figure}

From the algorithmic point of view the only important difference between the dynamical 1RSB phase and the condensed phase is that in the condensed phase the belief propagation does not estimate correctly the asymptotic marginal probabilities\index{marginals}. In the condensed phase, the total entropy cannot be estimated from the BP equations either, thus approximative counting and sampling of solutions will probably be even harder than in the dynamical 1RSB phase. 

Concerning the hardness of finding a solution we might expect that the incorrectness of the belief propagation estimates of marginals will play a certain role. However, we used the belief propagation maximal decimation \index{belief propagation!decimation} as described in appendix \ref{sec:BP_dec} in the 3- and 4-coloring, see fig.~\ref{fig:BP_col}. And this algorithm does not seem to have any problem to pass the condensation transition in both these cases. In particular, in the 3-coloring the gap between the condensation threshold $c_c=4$ and the limit of performance of the BP decimation $c\approx 4.55$ is huge. The rigidity transition $c_r$, defined in chapter~\ref{freezing}, and the colorability threshold $c_s$ are also marked for comparison in fig.~\ref{fig:BP_col}.

The condensation transition thus does not seem to play any significant role for the computational hardness of finding a solution.

\chapter{Freezing}
\label{freezing}

\vspace{-0.4cm}

{\it The previous two chapters describe recent contributions to the understanding of the clustering and condensation of solutions in random constraint satisfaction problems. Both these concepts are well known and widely discussed in the mean field theory of glasses and spin glasses for at least a quarter of a century. 

The concept of freezing of variables appeared in the studies of optimization problems, that is systems at zero temperature (or infinite pressure). In this chapter we first define the freezing of variables, clusters and solutions, and discuss its properties both in the thermodynamical limit and on finite-size instances. Then we explain how to describe the frozen variables within the one-step replica symmetry breaking approach and we define several possible phase transition associated to the freezing. To simplify the picture we define and solve the "completely frozen" locked constraint satisfaction problem where every cluster contains only one configuration. Finally we give several arguments about connection between the freezing and the average computational hardness. 
Results of this section are mostly original and were published in
[\ZK, \KZJ, \AZ, \ZML].}

\section{Frozen variables}

Consider a set of solutions ${\cal S}$ of a given instance of a constraint satisfaction problem. Define that a variable $i$ is {\it frozen}\index{variable!frozen} in the set of solutions $A\subset {\cal S}$ if it is assigned the same value in all the solutions in the set. If an extensive number of variables is frozen in the set $A$, then we call $A$ and all the solutions in $A$ {\it frozen}\index{frozen solutions}\index{cluster!frozen}, otherwise $A$ and all the solutions in $A$ are called {\it soft} (unfrozen)\index{variable!soft}.

A first observation is that the set of all solutions ${\cal S}$ is not frozen in the satisfiable phase. If it would be then adding one constraint, i.e., increasing the constraint density by $1/N$, would make the formula unsatisfiable with a finite probability, that would be in a contradiction with the sharpness of the satisfiability threshold. The {\it backbone}\index{backbone} is made of variables frozen in the set of ground states. An extensive backbone can thus exist only in the unsatisfiable phase. Already in \cite{MonassonZecchina99} it was argued that there might be a connection between the backbone and the computational hardness of the problem. The suggestion of \cite{MonassonZecchina99} was that if the fraction of variables covered by the backbone is discontinuous at the satisfiability transition then it is hard to find satisfying assignments on highly constrained but still satisfiable instances. On the other hand if the backbone appears continuously the problem is easy in the satisfiable phase. This was based on the replica symmetric solution of the random $K$-SAT which does not describe fully the phase space, in spite of that the relation between the existence of frozen variables inside clusters and the algorithmical hardness seems to be deep and we will develop it in this chapter. 

\subsection{Whitening: A way to tell if solutions are frozen}
\label{sec:white}

How to recognize if clusters have frozen variables or not. Or how to recognize if a given solution belongs to a frozen cluster or not. An iterative procedure called {\it whitening}\index{whitening} \cite{Parisi02b} gives an answer to these questions. 

Given a formula of a CSP and one of its solutions $\{s_i\}\in \{-1,1\}^N$, $i=1,\dots,N$, the whitening of the solution is defined as iterations of the warning propagation equations (\ref{eq:WP}) initialized on the solution. That is, for a binary CSP
$h^{i\to a}_{\rm init} = s_i$, 
and $u^{a\to i}_{\rm init}$ is computed according to eq.~(\ref{BP_0}). Note that the fixed point of the whitening does not depend on the order in which the warnings are updated. Indeed, during the iterations the only changes in warnings are from non-zero values to zero values. The fixed point is called the {\it whitening core}\index{whitening!core} of the solution. The whitening core is called {\it trivial} if all the warnings are equal to $0$, and {\it nontrivial} otherwise.

In the $K$-SAT problem whitening can be reformulated in a very natural way: Start with the solution $\{s_i\}$, assign iteratively a "$\ast$" (joker) to variables which belong only to clauses which are already satisfied by another variable or already contain a $\ast$ variable. On a general CSP such procedure is not equivalent to the whitening, and the warning propagation definition has to be used instead in order to obtain all the desired properties and relations to the 1RSB solution. 

We now argue that if the 1RSB solution is correct, then frozen variables in the cluster, to which solution $\{s_i\}$ belongs, asymptotically correspond to variables for which in the whitening core the total warning $h^i\neq 0$ (\ref{eq:warning}). Thus whitening can be used to decide if the solution $\{s_i\}$ belongs to a frozen cluster without knowing all the solutions in that cluster. The first step to show this property is, as in sec.~\ref{trees}, to consider the CSP on a tree with given boundary conditions which are compatible with a non-empty set of solutions~${\cal S}$ in the interior of the tree. Starting on the leaves we compute iteratively the warnings (\ref{eq:WP}) down to the root. Variables which have at least one non-zero incoming warning are frozen in the set ${\cal S}$.
The correctness of the 1RSB approach on a tree-like graph means that the picture on a tree captures properly all the asymptotic properties. In particular, the whitening core determines the set of frozen variables on typical large instances of the problem. 
The correctness of the 1RSB solution is an essential assumption for the above statement. Because all the long-range correlations decay within one cluster the warnings $u^{a\to i}$ in the whitening core are independent in the absence of $i$. Thus there truly exist solutions in that cluster in which the variable $i$ takes all the values allowed by the warnings. And on the other hand, if a value is not allowed by the warnings there is no solution where $i$ would be taking this value.
For consistency, all solutions in one cluster have to have the same whitening core. However, two different clusters can have the same whitening core. The most important example are all the soft (not frozen) clusters that all have the trivial whitening core. 

Whitening, as the iterative fixed point of the warning propagation, may be defined not only for a solution but for any configuration. In this way one may find blocking metastable states. For some preliminary numerical considerations see \cite{SeitzAlava05}. 

\subsection{Freezing on finite size instances}

The definition of whitening is applicable to any (non-random, small, etc.) instance. What does then remain from the asymptotic correspondence between frozen variables and whitening cores?

Consider now clusters as connected components in the graph where all solutions are nodes and where edges are between solutions which differ in only one variable, as in sec.~\ref{sec:geom}. Several questions arise about this definition:

\begin{itemize}
\item Do all the solutions in the connected-components cluster have the same whitening core? The answer is yes. If there were two solutions with different whitening cores which can be connected by a chain of single-variable flips, then along this chain there would exist a pair of solutions which differ in only one variable $i$ and have different whitening cores. But this is not possible, as the fixed point of the whitening does not depend on the order in which the warnings were updated, and one could thus start the whitening by setting warnings $h^{i\to a}=0$.  

\item Does the whitening core of a connected-components cluster correspond to the set of frozen variables? The answer is: If in the whitening core $h^i\neq 0$ (\ref{eq:warning}) then the variable $i$ is frozen in the connected-components cluster. Proof: If such a variable $i$ is not frozen, then there have to exist a pair of solutions which differ only in the value of this variable. Then all the constraints around $i$ have to be compatible with both these values, this would be in contradiction with $h^i\neq 0$. On the other hand, if in the whitening core $h^i=0$ then the variable $i$ might still be frozen in the connected-components cluster on a general instance, because correlations which are not considered by the 1RSB solution may play a role. 
\end{itemize}

Consider now clusters as the set of all solutions which share the same whitening core. Whitening-core clusters are aggregations of the connected-components clusters. In particular, all the solutions with a trivial whitening core, which might correspond to exponentially many pure states, are put together.

\begin{itemize}
\item What is the set of frozen variables in the whitening-core clusters? The answer is: Again if in the whitening core $h^i\neq 0$ then the variable $i$ is frozen in the whitening-core cluster. In principle, one whitening-core cluster could be an union of several connected-components cluster, but $i$ is frozen to the same value in each of them. The inverse is not correct in general. On finite size instances some variables with a zero warning $h^i=0$ might be frozen in the whitening-core cluster.

\item Can there be a fixed point of the warning propagation (\ref{eq:WP}) corresponding to zero energy (\ref{eq:WP_ener}) which is not compatible with any solution? The answer is yes. And such fixed points were observed in \cite{BraunsteinZecchina04,ManevaMossel05,KrocSabharwal07}. Again if the 1RSB solution is correct then in the thermodynamical limit these "fake" fixed points are negligible. 
\end{itemize}

\subsection{Freezing transition in 3-SAT - exhaustive enumeration}
\label{sec:3SAT_num}

Before turning to the cavity description of frozen clusters we investigate the \textit{freezing transition}\index{phase! frozen} in the random 3-SAT numerically. We define the freezing transition,  $\alpha_f$, 
as the smallest density of constraints $\alpha$ such that the whitening core of all solutions is nontrivial, i.e., not made only from zero warnings. We use the whitening core in the definition instead of the real set of frozen variables, because it does not depend on the definition of clusters and it has much smaller finite size effects. The existence of such a frozen phase was proven in the thermodynamical limit for $K\ge 9$ of the $K$-SAT near to the satisfiability threshold in \cite{AchlioptasRicci06}. 

\begin{figure}[!ht]
 \resizebox{\linewidth}{!}{
 \includegraphics{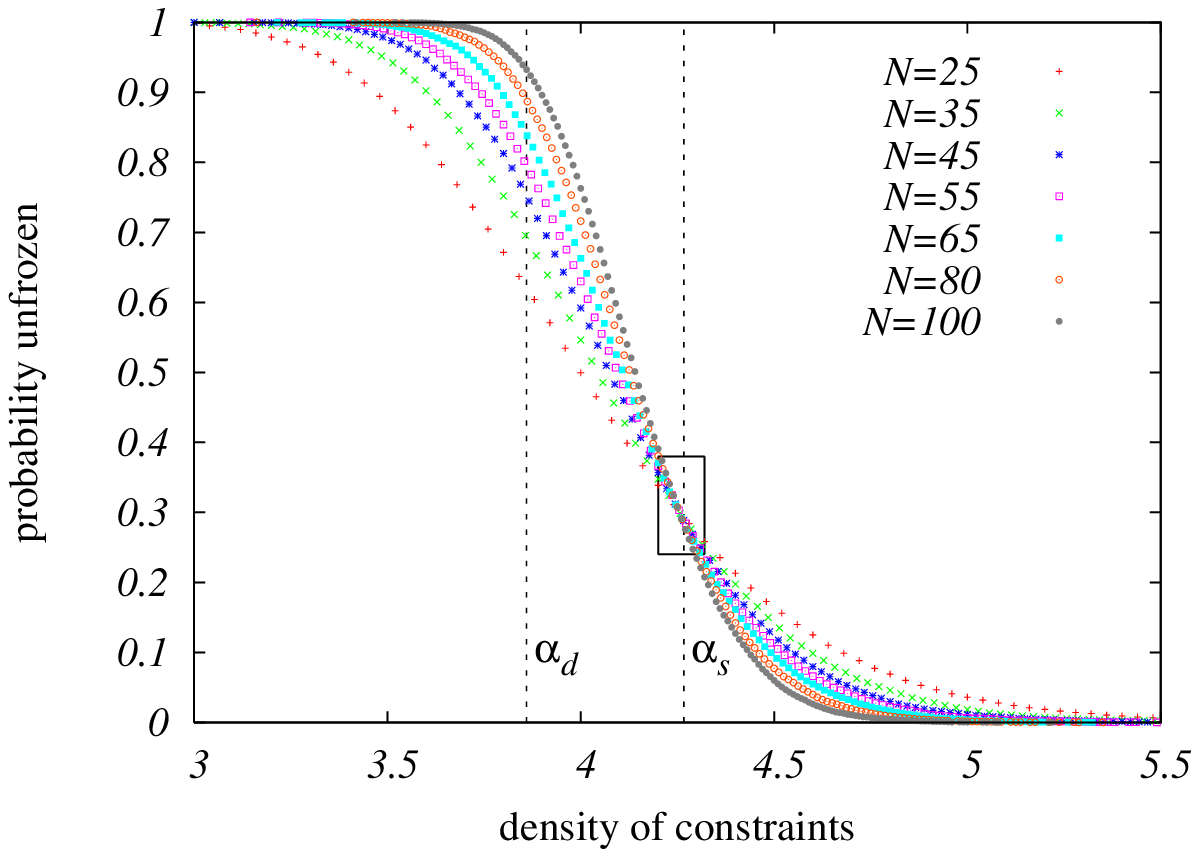}
 \includegraphics{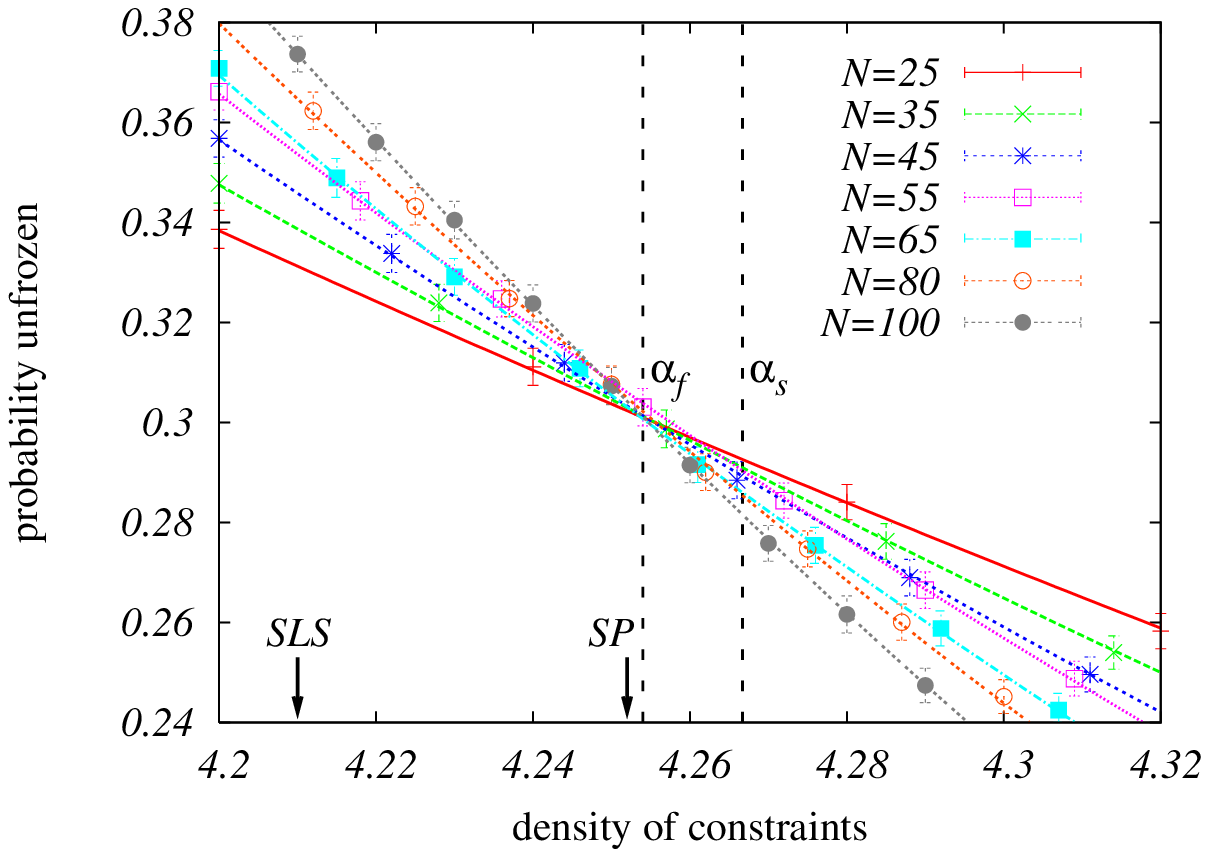}
 }
 \caption{\label{fig:rigidity}Left: Probability that there exists an unfrozen solution as a function of the constraint density $\alpha$ for different system sizes. The clustering [\KM] and satisfiability \cite{MezardParisi02} transitions marked for comparison. Right: A 1:20 zoom on the critical (crossing) point, our estimate for the freezing transition is $\alpha_f=4.254\pm 0.009$. The curves are cubic fits in the interval $\alpha\in (4,4.4)$. The arrows represent estimates of the limits of performance of the best known local search ASAT \cite{ArdeliusAurell06} and survey propagation \cite{Parisi03,ChavasFurtlehner05} algorithms.}
\end{figure}

In order to determine the freezing transition we start with a 3-SAT formula of $N$ variables and all possible clauses, and remove the clauses one by one independently at random\footnote{In practice we do not start with all the clauses, but as many that in all the repetitions of this procedure the initial instance is unsatisfiable.}. We mark the number of clauses $M_s$ where the formula becomes satisfiable as well as the number of clauses $M_f \le M_s$ where at least one solution starts to have a trivial whitening core. We repeat $B$-times ($B=2\cdot10^4$ in fig.~\ref{fig:rigidity}) and compute the probabilities that a formula of $M$ clauses is satisfiable $P_s(\alpha,N)$, and unfrozen $P_f(\alpha,N)$ respectively. 
Due to the memory limitation we could treat only instances which have less than $5\cdot10^7$ solutions which limits us to system sizes $N\le 100$. The results for the satisfiability threshold are shown in fig.~\ref{fig:sat_threshold} and are consistent
with previous studies in \cite{KirkpatrickSelman94,MonassonZecchina99,MonassonZecchina99b}. The probability of being unfrozen, $P_f(\alpha,N)$, is shown in fig.~\ref{fig:rigidity}. 

It is tempting to perform a scaling analysis as has been done in \cite{KirkpatrickSelman94,MonassonZecchina99,MonassonZecchina99b} for the satisfiability threshold. The critical exponent related to the width of the scaling window was defined via rescaling of the constraint density $\alpha$ as $N^{1/\nu_s}[1-\alpha/\alpha_s(N)]$. Note, however, that the estimate $\nu_s=1.5\pm 0.1$ for 3-SAT provided in \cite{MonassonZecchina99b} is not asymptotically correct. It was proven in \cite{Wilson02} that $\nu_s\ge 2$. Indeed, it was shown numerically in \cite{LeoneRicci01} that a crossover exists at sizes of order $N \approx 10^4$ in the related XOR-SAT problem. A similar situation happens for the scaling of the freezing transition, $P_f(\alpha,N)$, as the proof of \cite{Wilson02} applies also here~\footnote{Theorem 1 of \cite{Wilson02} applies to the freezing property where the bystander are clauses containing two leaves.}. It would be interesting to investigate the scaling behaviour on an ensemble of instances where the results of \cite{Wilson02} do not apply (e.g. graphs without leaves). However, we concentrate instead on the estimation of the critical point, which we do not expect to be influenced by the crossover in the scaling. We are in a much more convenient situation for the freezing transition than for the satisfiability one. The crossing point between functions $P_f(\alpha,N)$ for different system sizes seems to depend very little on $N$, while for the satisfiability transition it depends very strongly on $N$, compare the zooms in figs.~\ref{fig:sat_threshold} and \ref{fig:rigidity}.

We determine the value of the freezing transition in random 3-SAT as
\be
\alpha_f=4.254\pm 0.009 \, ,
\ee
which is very near but seems separated from the satisfiability threshold $\alpha_s=4.267$ \cite{MezardZecchina02,MertensMezard06}.
In any case the frozen phase in 3-SAT is very narrow, that is in contrast with the situation in $K\ge 9$ SAT where it covers at least $1/5$ of the large clustered phase \cite{AchlioptasRicci06}.

\section{Cavity approach to frozen variables}
\label{sec:freezing_cav}

In this section we present how to describe the frozen variables within the 1RSB cavity solution. We illustrate the results on an example of the random graph coloring where properties of frozen variables were studied in detail for the first time [\ZK]. 

The energetic 1RSB (survey propagation), sec.~\ref{sec:energetic}-\ref{sec:1in3}, aims to count the total number of frozen clusters. More precisely, it counts the total number of fixed points of the warning propagation (\ref{eq:WP}). It can be used to locate the satisfiability threshold or to design survey propagation based solvers \cite{MezardParisi02,MezardZecchina02}. However, as we understood in chapter \ref{clustering}, by neglecting the soft clusters we cannot locate the clustering transition. In chapter \ref{condensation} we defined the dominant clusters, i.e., those which cover almost all solutions. A natural question arises immediately: Are the dominant clusters frozen or soft? In order to answer the general entropic 1RSB equations (\ref{eq:1RSB_m1},\ref{eq:1RSB_Sigma}) have to be analyzed.

\subsection{Frozen variables in the entropic 1RSB equations}
\label{frozen}

We remind that in the 1RSB solution of the graph coloring problem the components of the messages (called also the cavity fields) $\psi^{i\to j}_{s_i}$ are the probabilities that in a given cluster the node $i$ takes the color $s_i$ when the constraint on the edge $(ij)$ is not present. The belief propagation equations, (\ref{eq:BP}) in general, (\ref{eq:col_BP}) in coloring, then define the consistency rules between the field $\psi^{i\to j}_{s_i}$ and fields incoming to $i$ from the other variables than $j$. In the zero temperature limit we can classify fields $\psi^{i\to j}_{s_i}$  in the following two categories:
\begin{enumerate}
\item[(i)]{ The {\it hard (frozen) field}\index{cavity field!hard field} corresponds to the case when all components of
  $\psi^{i\to j}$ are strictly zero except the one for color $s$. This means that  in the absence of edge $(ij)$, variable $i$ takes color $s$ in {\it all} the solutions from the cluster in question.  }
\item[(ii)]{ The {\it soft field}\index{cavity field!soft field} corresponds to the case when more than one component of $\psi^{i\to j}_{s_i}$ is nonzero. The variable $i$ is thus not frozen in the absence of edge $(ij)$, and the colors of all the nonzero components are allowed.}
\end{enumerate}
This distinction is also meaningful for the full probabilities $\psi^i_{s_i}$ (\ref{eq:total}). By definition, the variable $i$ is frozen in the cluster if and only if $\psi^i_{s_i}$ is a hard field.

It is important to stress that some of the soft fields on a given instance of the problem might be very small. Some of them might even scale like $e^{-N}$. We insist on classifying those as the soft fields because they cannot create real contradictions. This subtle distinction becomes important mainly in the implementation of the population dynamics algorithm, see appendix \ref{app:pop_dyn}.

The distribution of fields over clusters $P^{i\to j}(\psi^{i\to j})$ (\ref{eq:1RSB_m1}), which is the "order parameter" of the 1RSB equation,
can be decomposed into the hard-field part of a weight $\eta_s^{i\to j}$ and the soft-field part $P_{\rm soft}^{i\to j}$ of a weight $\eta^{i\to j}_0=1-\sum_{s=1}^q \eta^{i\to j}_s $
\begin{equation}
    P^{i\to j}(\psi^{i\to j}) = \sum_{s=1}^q \eta^{i\to j}_s \ind(\psi^{i\to j} {\rm frozen\, \, into\, \, } s)+ \eta^{i\to j}_0 
  P_{\rm soft}^{i\to j}(\psi^{i\to j})\, .  \label{eq:frozen}
\end{equation}

\paragraph{Hard fields in the simplest case, $m=0$ ---}
\label{hard_0}

First, we derive equations for the hard fields when the parameter $m=0$ in (\ref{eq:1RSB_m1}). This will, in fact, lead to the survey propagation equations, for coloring originally derived in~\cite{MuletPagnani02,BraunsteinMulet03} from the energetic 1RSB method (\ref{sec:energetic}). For simplicity we write the most general form only for the 3-coloring. 

We plug (\ref{eq:frozen}) into eq.~(\ref{eq:1RSB_m1}). The reweighting factor $(Z^{i\to j})^m$ at $m=0$ is either equal to zero, when the arriving fields are hard and contradictory, or equal to one. This is the origin of a significant simplification. 
The outcoming field $\psi^{i\to j}$ might be frozen in direction $s$ if and only if for every other color $r\neq s$ there is at least one incoming field frozen to the color $r$.  The update of probability $\eta^{i\to j}_s$ that a field is frozen in direction $s$ is for the 3-coloring written as
\begin{equation}
    \eta^{i \to j}_s= \frac{
    \prod_{k\in i - j} (1- \eta^{k\to i}_s) - \sum_{p \neq s} \prod_{k\in i - j} (\eta^{k\to i}_0 + \eta^{k\to i}_p) +  \prod_{k\in i - j} \eta^{k\to i}_0
     }{
  \sum_{p} \prod_{k\in i - j} (1- \eta^{k\to i}_p) -   \sum_{p} \prod_{k\in i - j} (\eta^{k\to i}_0 + \eta^{k\to i}_p) +  \prod_{k\in i - j} \eta^{k\to i}_0
     } \, .\label{SP_color}
\end{equation}
In the numerator there is a telescopic sum counting the probability that color
$s$ and only color $s$ is not forbidden by the incoming fields. In the
denominator there is the normalization, i.e., the telescopic sum counting 
the probability that there is at least one color which is not forbidden. 
The crucial observation is that at $m=0$ the self-consistent equations for $\eta$ do not depend on the soft-fields distribution $P_{\rm soft}^{i\to j}(\psi^{i\to j})$.

If we do not aim at finding of a proper coloring on a single graph but just at computing of the complexity function and similar quantities, we can further simplify eq.~(\ref{SP_color}) by imposing the color symmetry.  Indeed, the probability that in a given cluster a field is frozen in the direction of a color $s$ has to be independent of $s$. Then (\ref{SP_color}) becomes, now for
general number of colors $q$:
\begin{equation}
         \eta^{i \to j} = w(\{ \eta^{k\to i} \}) =
         \frac{ \sum_{l=0}^{q-1} (-1)^l {q-1 \choose l} 
         \prod_{k\in i-j} \left[1-(l+1)\eta^{k\to i} \right]
         }{ \sum_{l=0}^{q-1} (-1)^l {q \choose l+1} 
         \prod_{k\in i-j} \left[ 1-(l+1)\eta^{k\to i} \right] } 
         \label{SP_col}\, .
\end{equation}

We remind that since $\partial \Sigma(s)/\partial s = -m$ (\ref{eq:1RSB_Sigma}), the value 
$m=0$ corresponds to the point $\tilde s$ where the function $\Sigma(s)$ has a zero slope. 
If a nontrivial solution of (\ref{SP_color}) exists, then $\Sigma(\tilde s)|_{m=0}$
is the maximum of the curve $\Sigma(s)$. And if the 1RSB solution for clusters at $m=0$ is correct then it is counting the total
log-number of clusters of size $\tilde s$, which is due to the exponential dependence also the total log-number of all clusters, regardless of their size.

\paragraph{Frozen variables at general $m$, generalized SP ---}
\label{generalized}

Let us compute how the fraction of hard fields $\eta$ evolves after one
iteration of equation (\ref{eq:1RSB_m1}) at a general value of $m$. There are two steps 
in each iteration of (\ref{eq:1RSB_m1}).  
In the first step, $\eta$ iterates via eq.~(\ref{SP_col}). In the
second step we re-weight the fields.
Writing $P^{\rm hard}_m(Z)$ the ---unknown--- distribution of the reweightings $Z^m$  for
the hard fields, one gets
\bea \eta^{i \to j} &=&
\frac{1}{{\cal N}^{i \to j}} \int {\rm d}Z^{i\to j} \, P^{\rm hard}_m(Z^{i\to j}) \,  \big(Z^{i \to j}\big)^m  w(\{ \eta^{k\to i} \})\nonumber \\ 
&=& 
\frac{w(\{ \eta^{k\to i} \})}{{\cal N}^{i\to j}} \int {\rm d}Z^{i\to j} \, P^{\rm hard}_m(Z^{i\to j})\,  \big(Z^{i\to j}\big)^m  =
\frac{ w(\{ \eta^{k\to i} \})}{{\cal N}^{i\to j}}\,  \langle Z^{i\to j}_m\rangle_{\rm hard}\, .
\label{eta_prelim}
\eea
A similar equation can formally be written for the soft fields 
\be 1 - q \eta^{i \to j} =
\frac{ 1-q w(\{ \eta^{k\to i} \})}{{\cal N}^{i\to j}}\, \langle Z^{i\to j}_m\rangle_{\rm soft}\, .
\label{eta_prelim2}
\ee
Writing explicitly the normalization ${\cal N}^{i \to j}$, we finally obtain the
generalized survey propagation equations:
\be
\label{hard_m}
 \eta^{i \to j} =
\frac{ w(\{ \eta^{k\to i} \})} {q
w(\{ \eta^{k\to i} \})
+\left[1-q
w(\{ \eta^{k\to i} \})
\right] r(m,\{ \eta^{k\to i} \})}\, ,
\ee
where $r$ is the ratio of average reweighting factors of the soft and hard fields
\be
 r(m,\{ \eta^{k\to i} \})=
\frac {\langle Z^{i\to j}_m\rangle_{\rm soft}}{\langle Z^{i\to j}_m\rangle_{\rm hard}}\, .
\ee 
In order to do this recursion, the only nontrivial information needed is the ratio $r$ between soft- and hard-field average reweightings, which depends on the full distribution of soft fields $P_{\rm soft}^{i\to j}(\psi^{i\to j})$. Eq.~(\ref{hard_m}) is easy to use in the population dynamics and allows to compute the fraction of frozen variables in typical clusters of a given size (for a given value $m$).  

There are two cases where eq.~(\ref{hard_m}) simplifies so that the 
hard-field recursion becomes {\it independent} from the soft-field 
distribution. The first case is, of course, $m=0$. Then $r=1$ independently 
of the edge $(ij)$, and the equation reduces to the original SP. The second case 
arises for $m=1$, where the eq.~(\ref{hard_m}) can be written as
the equation for the naive reconstruction\index{reconstruction!naive} (\ref{eq:hard}). 
The probability that a variables is frozen at $m=1$ is the same at the probability that leaves (far away variables) determine uniquely the root in the reconstruction problem, see sec.~\ref{trees}.

\paragraph{Frozen variables and minimal rearrangements ---}
Montanari and Semerjian~\cite{MontanariSemerjian05,Semerjian07} developed a very interesting connection between frozen variables and the so-called {\it minimal rearrangements}\index{minimal rearrangement}. Given a CSP instance, one of its solutions $\{s_i\}$ and a variable $i$, find the nearest solution to $\{s_i\}$ where the values of the variable $i$ is changed to $s'_i\neq s_i$. The set of variables on which these two solutions differ is called the {\it minimal rearrangement}. It was shown in~\cite{Semerjian07} that the size of the average (over variables $i$, the solution $\{s_i\}$, and the graph ensemble) minimal rearrangement diverges at the rigidity transition (when almost all the dominant clusters become frozen). Indeed, the cavity approach to minimal rearrangements leads to equations analogous to those for frozen variables. Part of the reasoning is the following \cite{SeitzAlava05}: Consider a solution of a $K$-SAT formula and a variable $i$ from its whitening core. By flipping the variable $i$ at least one neighbouring constraint $a$ is made unsatisfied, otherwise the variable would not be in the whitening core. All variables contained in $a$ are also in the whitening core, thus one of them has to be flipped in order to satisfy this constraint. There have to be a chain of flips which can be finished only by closing a loop. The length of the shortest loop going through a typical variable is of order $\log{N}$. Thus a diverging number of changes is needed to find another solution. Hence the connection between frozen variables and rearrangements is:
\begin{itemize}
    \item{If the variable $i$ is frozen in the cluster to which the solution $\{s_i\}$ belongs, then in order to change the value of $i$ one has to find a solution from a different cluster, thus at an extensive Hamming distance.}
    \item{If the variable $i$ is not frozen in the cluster to which the solution $\{s_i\}$ belongs, then the best rearrangement will probably also lie within that cluster and the Hamming distance is finite.}
\end{itemize}

Many more results about rearrangements can be found in \cite{Semerjian07}, they shed light on  the onset of frozen variables. An exciting possibility is that the cavity equations for rearrangements might be useful in incremental algorithms for CSPs, like the one of \cite{KrzakalaKurchan07}.

\subsection{The phase transitions: Rigidity and Freezing}
\label{sec:hard_sol}

A natural question is: ``In which clusters are the hard fields present?'' Or more in the terms of the 1RSB solutions: ``When does eq.~(\ref{hard_m}) have a nontrivial solution $\eta>0$?'' We answer this question in one of the simplest cases, that is for the coloring of random regular graphs of connectivity $c=k+1$. In tree-like regular graphs the neighbourhood of each node looks identical, thus also the distribution $P^{i\to j}(\psi^{i\to j})$ is the same for every edge $(ij)$. Moreover we search for a color-symmetric solution [\ZK], that is $\eta_s=\eta_r=\eta$ for all $s,r\in \{1,\dots, q\}$. The function $w(\{ \eta \})$ in the ensemble of random regular graphs simplifies to 
\be
       w(\eta)= \frac{ \sum_{l=0}^{q-1} (-1)^l {q-1 \choose l} 
         \left[1-(l+1)\eta \right]^{k}
         }{ \sum_{l=0}^{q-1} (-1)^l {q \choose l+1} 
         \left[ 1-(l+1)\eta \right]^{k} } \, . \label{SP_reg}
\ee
First notice that in order to constrain a variable into one color, i.e., create a hard field, one needs at
least $q-1$ incoming fields that forbids all the other colors. It means that the function $w(\{ \eta \})$ defined in eq.~(\ref{SP_reg}) is identically
zero for $k<q-1$ and might be non-zero only for $k \ge q-1$, where $k$ is the number of
incoming fields.

\begin{figure}[!ht]
 \resizebox{\linewidth}{!}{
\includegraphics{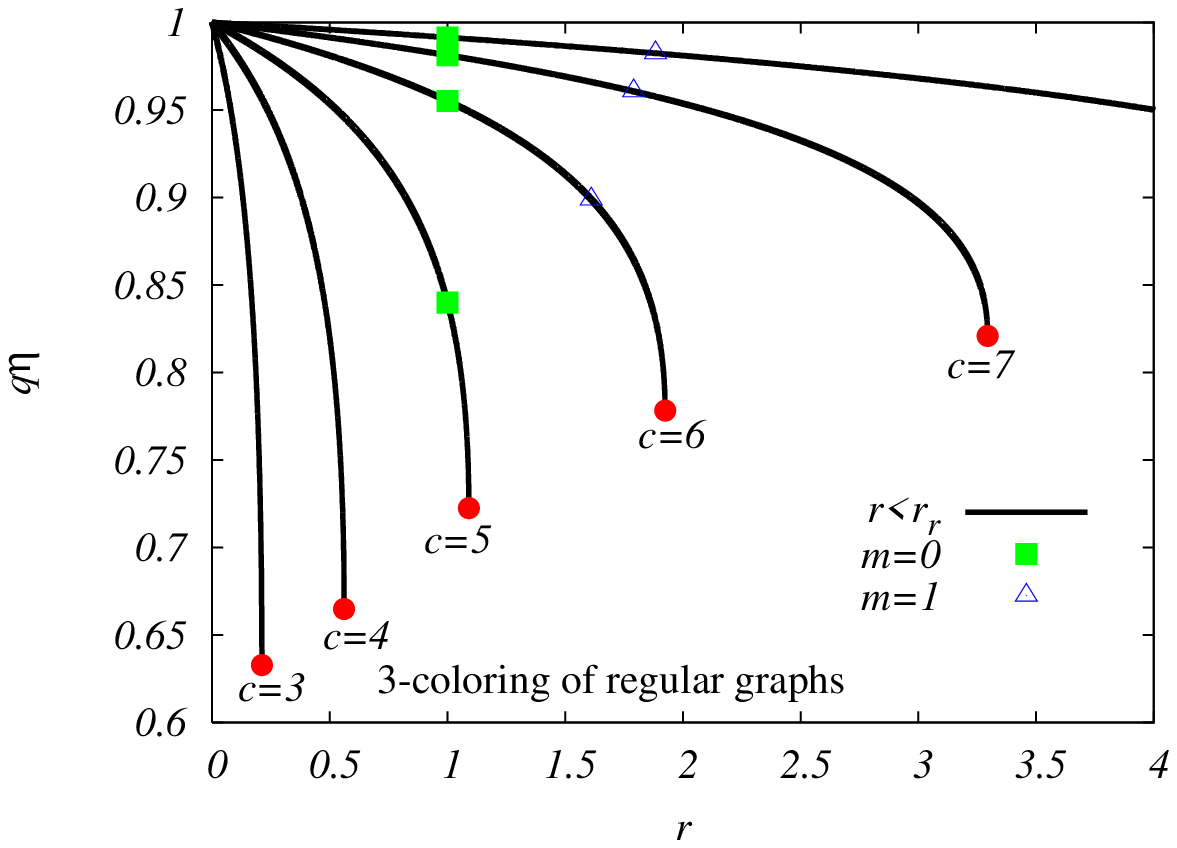}
\includegraphics{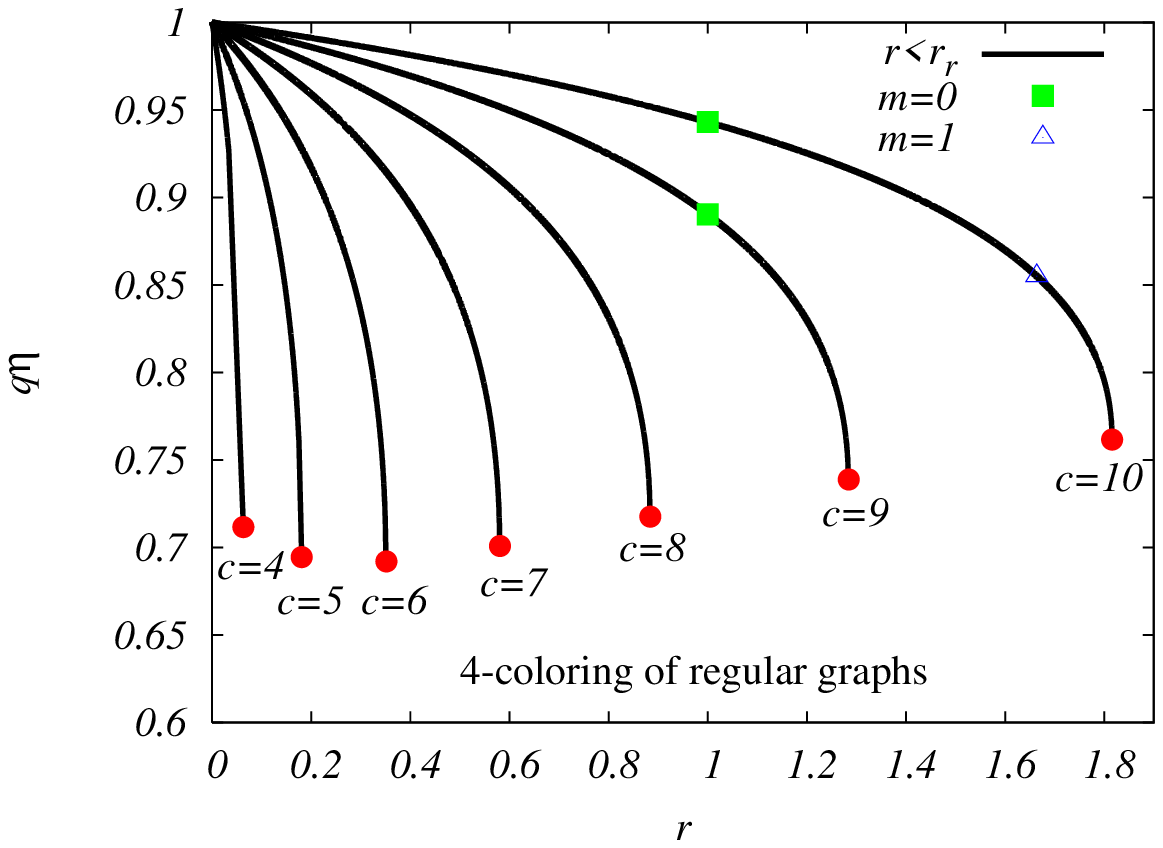}}
\caption{\label{fig_eta} The lines are solutions of eq.~(\ref{hard_m_reg}) and give the fraction $q\eta$ of hard fields for a given value of ratio $r=\langle Z^{i\to j}_m\rangle_{\rm soft}/\langle Z^{i\to j}_m\rangle_{\rm hard}$ for $q=3$ (left) and $q=4$ (right) in regular random graphs. There is a critical value of the ratio $r_r$ (red point) beyond which only the trivial solution $\eta=0$ exists. Note that the solutions corresponding to $m=0$ (green square) and $m=1$ (blue triangle) only exist for a connectivity large enough.}
\end{figure}

The equation (\ref{hard_m}) also simplifies on a regular graph and $\eta$ follows 
a self-consistent relation
\be
\label{hard_m_reg}\eta = w(\eta)\frac{1} {q w(\eta) +\left[1-q w(\eta) \right] r(m) } \, ,
\ee 
where $r(m)$ is the average of the reweighting of the soft fields divided by the average of the reweighting of the frozen fields (\ref{hard_m}). The function $r(m)$ is in general not easy to compute, the population dynamics is needed for that. Several properties are, however known:
\begin{subequations}
\label{eq:r_values}
\bea
    r &\to& 0 \quad {\rm when} \quad m \to -\infty\, , \\
    r &\to& \infty \quad {\rm when} \quad m \to \infty\, ,
\eea
\end{subequations}
and $r(m)$ is a monotonous function of $m$. Moreover, for the internal entropy of clusters $s(m)\to 0$ when $m\to -\infty$, and $s(m)\to \infty$ when $m\to \infty$, and $s(m)$ is also a monotonous function. We thus solve eq.~(\ref{hard_m_reg}) for every possible ratio $r$. For all $k \ge q-1$ we compute the solution $\eta(r)$.  
The result is shown in fig.~\ref{fig_eta} for the 3- and 4-coloring of random regular graphs. 

There is a discontinuous phase transition: For $r<r_r$ eq.~(\ref{hard_m_reg}) has a solution with a large fraction of frozen fields, $\eta>0$, whereas for $r<r_r$ the only solution is $\eta=0$. Note that the index $r$ stands for "rigidity". In terms of the parameter $m$, the critical value is $r(m_r)=r_r$. In terms of the internal entropy of clusters $s(m_r)=s_r$. The interpretation is the following:
\begin{itemize}
   \item{Clusters of internal entropy $s<s_r$ are almost all frozen, and the fraction of frozen variable's they contain is quite large.}
   \item{Clusters of internal entropy $s>s_r$ are almost all soft, meaning the fraction of frozen variables is zero.}
\end{itemize} 

\begin{figure}[!ht]
 \resizebox{\linewidth}{!}{
 \includegraphics{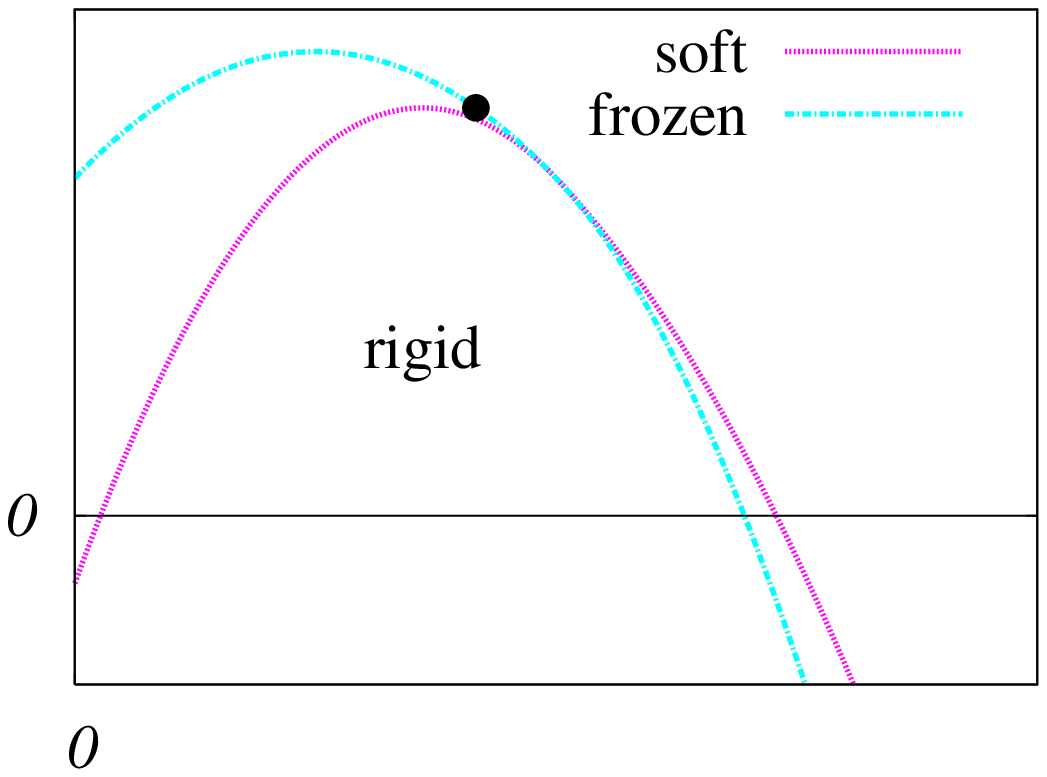}
 \includegraphics{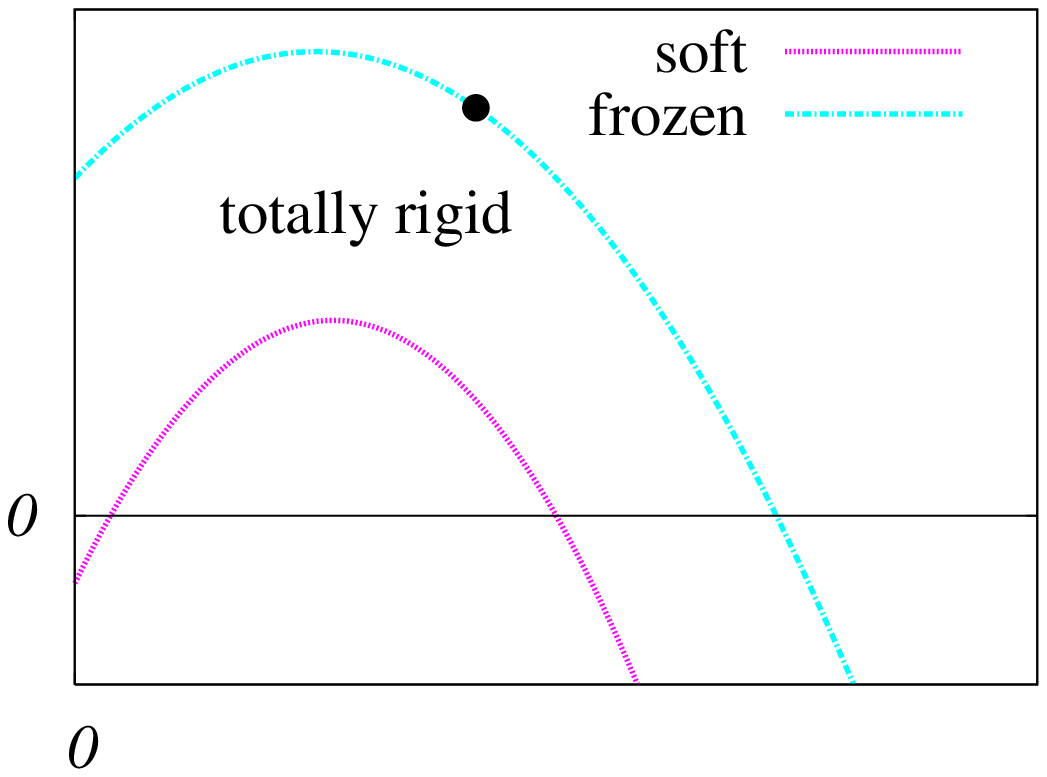}
 \includegraphics{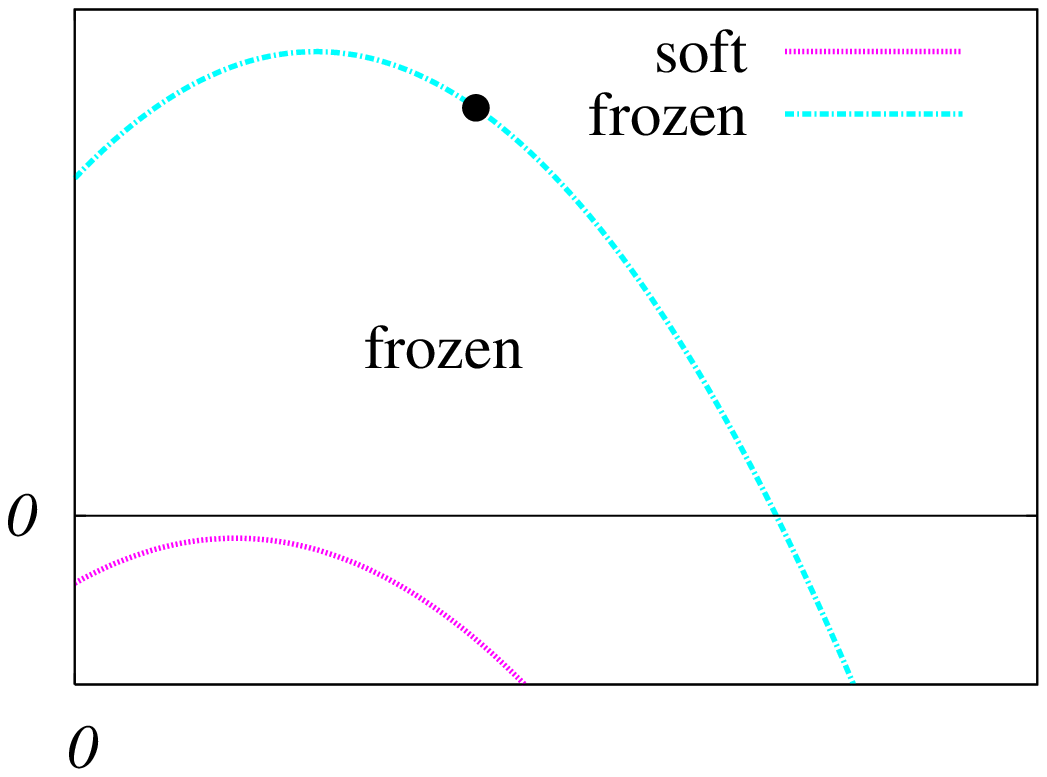}
 }
 \caption{\label{fig:phases} A pictorial sketch of the complexity function of clusters of a given size. Cyan-blue is the complexity of the frozen clusters, magenta of the soft clusters. The total complexity is the envelope, which can be calculated from the entropic 1RSB solution. The black point marks the dominating clusters. Left: In the rigid phase almost all the dominant clusters are frozen, but clusters corresponding to larger entropy might be mostly soft. Middle: In the totally rigid phase almost all clusters of all sizes are frozen, but there still might be exponentially many of soft clusters. Right: The frozen phase where soft clusters almost surely do not exist.}
\end{figure}

When we change the average constraint density there are at least three interesting phase transitions related to frozen variables. Fig.~\ref{fig:phases} sketches the difference between the phases they separate. Recall that $s^*$ is the internal entropy of the dominant clusters, and $s_{\rm max}$ the internal entropy of the largest clusters $\Sigma(s_{\rm max})=0$.
\begin{itemize}
    \item{The rigidity transition\index{phase transition!rigidity}\index{phase!rigid}, $c_r$, at which $s^*=s_r$, separates a phase where a typical dominant cluster is almost surely not frozen from a phase where a typical dominant cluster is almost surely frozen.}
    \item{The total rigidity transition\index{phase transition!total rigidity}, $c_{tr}$, at which $s_{\rm max}=s_r$, when almost all clusters of every size become frozen.}
    \item{The freezing transition\index{phase transition!freezing}, $c_{f}$, separates phase where exponentially many unfrozen cluster exists from a phase where such clusters almost surely do not exist\footnote{Note that what is called freezing transition in \cite{Semerjian07} or in sec.~IV.C of \cite{MontanariRicci08} is in fact what we define as the rigidity transition, in agreement with [\ZK].}. }
\end{itemize}

In general it have to be $c_r\le c_{tr} \le c_f$. The relation between the rigidity and total rigidity transition is easily obtained from the 1RSB solution. It is thus known that in the $q$-coloring of Erd\H{o}s-R\'enyi graphs $c_r=c_{tr}$ if and only if $q\le 8$, in $K$-SAT if and only if $K\le 5$. For larger $q$ or $K$ the rigidity transition is given by the onset of frozen variables in clusters corresponding to $m=1$, this is equivalent to the naive reconstruction~(\ref{eq:hard}).

The relation between the total rigidity transition and the freezing is less known. There are only few studies for the freezing transition in random $K$-SAT. The first one is the one of \cite{AchlioptasRicci06} where they prove that for every $K\ge 9$ the freezing transition is strictly smaller than the satisfiability one $c_f<c_s$. In the large $K$ limit they showed that the frozen phase covers a finite fraction (at least $20\%$) of the satisfiable region. The second study \cite{ManevaSinclair07} gives a rigorous upper bound on the freezing transition in 3-SAT $\alpha_f<4.453$, which is slightly better than the best known upper bound on the satisfiability transition in 3-SAT
\cite{DuboisBoufkhad00}. The third study is numerical [\AZ], presented in fig.~\ref{fig:rigidity}. It shows that in 3-SAT the frozen phase is tiny, about $0.3\%$ of the satisfiable region.

It is not known if the total rigidity transition coincides with the freezing transition. The entropic cavity method describes a typical but not every cluster of a given size. A~generalization of the 1RSB equations which would count only the number of soft cluster would answer this question. 

To summarize the description of the freezing of variables and clusters in the canonical constraint satisfaction problems, like $q$-coloring or $K$-satisfiability, is both numerically and conceptually involved task. 
Moreover in the experimentally feasible range of $q$ and $K$ the frozen phase is tiny. Thus conclusive statements about the connection between the freezing and the computational hardness are difficult to make. In the next section we introduce the so-called {\it locked} constraint satisfaction problems where the situation is much more transparent.

\section{Point like clusters: The locked problems}
\label{locked}

In order to get a better understanding of the frozen phase we introduce the so-called {\it locked}\index{locked problems} constraint satisfaction problems [\ZML]. 
In these problems the whole clustered phase is at the same time frozen, this is because in the locked problems all the clusters contain only one solution. 

\subsection{Definition}

A {\it locked} constraint satisfaction problem is made of $N$ variables and $M$ {\it locked} constraints\index{locked problems!locked constraint} in such a way that every variable is present in at least two constraints. 
A constraint consisting of $K>0$ variables is {\it locked} if and only if for every satisfying assignment of variables changing the value of any (but only one) variable makes the assignment unsatisfying.

A locked constraint of $K$ variables has the property that if $(K-1)$ variables are assigned then either the constraint cannot be satisfied by any value of the last variable or there is only one value of the last variable which makes the constraint satisfied. All the uniquely extendible constraints \cite{Connamacher04,ConnamacherMolloy04} are locked, XOR-SAT being the most common example. 1-in-K SAT (exact cover) constraint \cite{GareyJohnson79} is another common example. On the other hand, the most studied constraint satisfaction problems $K$-SAT or graph $q$-coloring ($q>2$) are not made of locked constraints. 

The second important part of the definition of locked constraint satisfaction problems is the requirement that every variable is present in at least two constraints, i.e., leaves are absent. An important property follows: In order to change a satisfying assignment into a different satisfying assignment at least a closed loop of variables have to be changed. If leaves would be allowed changing a path connecting two leaves might be sufficient. 

It seems to us that all the random locked constraint satisfaction problems should behave in the way we describe in the following. We, however, investigated in detail only a subclass of the locked problems called {\it locked occupation problems}\index{locked problems!locked occupation problems}\index{occupation problems!locked} (LOP). Occupation constraint satisfaction problem is  defined as a problem with binary variables (0-empty, 1-occupied) where each constraint containing $K$ variables is a function of how many of the $K$ variables are occupied. A constraint of the occupation CSP can thus be characterized via a $(K+1)$-component vector $A$, $A_i \in \{0,1\}, i \in 0,\dots,K$. A constraint is satisfied (resp. violated) if it contains $r$ occupied variables where $r$ is such that $A_r=1$ (resp. $A_r=0$). For example $A=(0,1,0,0)$ corresponds to the positive 1-in-3 SAT [\RS], $A=(0,1,1,0)$ is bicoloring \cite{CastellaniNapolano03}, $A=(0,1,0,1,0)$ is 4-odd parity check (4-XOR-SAT without negations) \cite{MezardRicci03}.  

An occupation problem is locked if all the variables are connected to at least two constraints and the vector $A$ is such that $A_i A_{i+1}=0$ for all $i=0,\dots,K-1$. We study the random ensembles of LOPs where all constraints are identical and the variable degree is either fixed of distributed according to a truncated Poissonian law (\ref{eq:Poiss}).

\subsection{The replica symmetric solution}

The replica symmetric cavity equations, belief propagation (\ref{eq:BP_1}-\ref{eq:BP_2}), for the occupation problems read
\begin{subequations}
\label{eq:BP_locked}
\bea
     \psi_{s_i}^{a\to i} &=& \frac{1}{Z^{a\to i}} \sum_{\{s_j\}} \delta(A_{s_i+\sum_{j} s_j} -1)\prod_{j\in \partial a-i} \chi_{s_j}^{j\to a} \, , \label{BP1} \\
     \chi_{s_j}^{j\to a} &=& \frac{1}{Z^{j\to a}} \prod_{b\in \partial j-a} \psi_{s_j}^{b\to j}\, , \label{BP2} 
\eea
\end{subequations}
where $\psi_{s_i}^{a\to i}$ is the probability that the constraint $a$ is satisfied conditioned that the value of the variable $i$ is $s_i$, and $\chi_{s_j}^{j\to a}$ is the probability that variable $j$ takes value $s_j$ conditioned that the constraint $a$ was removed from the graph. The normalizations $Z$ have the meaning of the partition function contributions. The replica symmetric entropy $s$ is a zero temperature limit of (\ref{free_given})
\be
   s=\frac{1}{N}\sum_a \log{(Z^{a+\partial a})} -  \frac{1}{N} \sum_i (l_i-1)  \log{(Z^i)}\, , \label{eq:entropy_l}
\ee
where the contributions $Z^{a+\partial a}$  (resp. $Z^i$) are the exponentials of the entropy shifts when the node $a$ and its neighbours (resp. the node $i$) is added (\ref{eq:Zaa}-\ref{eq:Zi})
\begin{subequations}
\label{eq:Z_locked}
\bea
     Z^{a+\partial a}&=& \sum_{\{s_i\}} \delta(A_{\sum_i s_i}-1) \prod_{i\in a} \left( \prod_{b\in i-a} \psi_{s_i}^{b\to i} \right)\, , \label{Za}\\
     Z^i&=&  \prod_{a\in i} \psi_{0}^{a\to i}+ \prod_{a\in i} \psi_{1}^{a\to i}\, . \label{Zi}
\eea 
\end{subequations}

Solving eqs.~(\ref{BP1}-\ref{BP2}) means finding their fixed points. A crucial property of the locked problems it that if $\{s_i\}$ is one of the solutions then 
\begin{subequations}
\bea
\psi^{a\to i}_{s_i} = 1\, , \quad   \psi^{a\to i}_{\neg s_i} = 0\, ,\\  
\chi^{i\to a}_{s_i} = 1\, , \quad   \chi^{i\to a}_{\neg s_i} = 0   
\eea
\end{subequations}
is a fixed point of eqs.~(\ref{BP1}-\ref{BP2}). The corresponding entropy is then zero, as $Z^i=Z^{a + \partial a}=1$ for all $i$, $a$. In the derivation of \cite{MezardMontanari07} fixed points of the belief propagation equations correspond to clusters. Thus in the locked problems every solution corresponds to a cluster. 

In the satisfiable phase there exist exponentially many solutions (i.e., clusters), thus the iterative fixed point of BP equations (\ref{BP1}-\ref{BP2}) obtained from a random initialization gives an asymptotically exact value for the total entropy. And the satisfiability threshold coincides with the condensation transition, described in chap.~\ref{condensation}. Furthermore, as each cluster contains only one solution the clustered phase is automatically 
frozen according to the definition in sec.~\ref{sec:hard_sol}.
Interestingly, part of the satisfiable phase is only "fake clustered" meaning that at infinitesimally small temperature there is a single fixed point of the BP equations. This has been discussed e.g. in the context of the perfect matchings in [\ZMM]. A general discussion and proper definition of the clustering transition in the locked problems follows in sec.~\ref{sec:sn_rec}. 

Iterative fixed point of eqs.~(\ref{BP1}-\ref{Zi}) averaged over the graph ensemble is in general found via the population dynamics technique, see appendix \ref{app:pop_dyn}. Note that the sum over $\{s_j\}$ in (\ref{BP1}) can be computed iteratively in $(K-1)^2$ steps instead of the naive $2^{K-1}$ steps. Moreover, on the regular graphs ensemble or for some of the symmetric locked problems, such that $A_i=A_{K-i}$ for all $i=0,\dots,K$, the solutions is {\it factorized}\index{solution!factorized}. In the factorized solution the messages $\chi^{i\to a}$, $\psi^{a \to i}$ are independent of the edge $(ia)$ and the population dynamics is thus not needed. 
\begin{itemize}
\item For the regular graph ensemble where each variable is present in $L$ constraints the factorized solution is 
\begin{subequations}
\label{eq:locked_reg}
\bea
 \psi_0&=& \frac{1}{Z^{\rm reg}}  \sum_{A_r=1}  {K-1 \choose r} \psi_1^{(L-1)r} \psi_0^{(L-1)(K-1-r)}\, , \label{RS1}\\
 \psi_1&=& \frac{1}{Z^{\rm reg}}  \sum_{A{r+1}=1}  {K-1 \choose r} \psi_1^{(L-1)r} \psi_0^{(L-1)(K-1-r)} \, ,\label{RS2}
\eea 
\end{subequations}
and the entropy is
\be
  s_{\rm reg}= \frac{L}{K} \log{\left[ \sum_{A_r=1}  {K \choose r} \psi_1^{(L-1)r} \psi_0^{(L-1)(K-r)}\right]} -
  (L-1) \log{\left[\psi_{0}^{L}+  \psi_{1}^{L}\right]}\, .
\ee
\item For the symmetric locked problems where the symmetry is not spontaneously broken the solution is also factorized. We call these the {\it balanced}\index{locked problems!balanced} locked problems. The BP solution is $\psi_1=\psi_0=1/2$ and the corresponding entropy 
\be
s_{\rm sym}(\overline l)=\log{2}+\frac{\overline l}{K}
\log{\left[2^{-K}\sum_{r=0}^K \delta(A_r-1) {K\choose r}\right] }\label{eq:s_sym}
\, ,        
\ee
where $\overline l$ is the average degree of variables. Notably, this result for the entropy can be proven rigorously by computing the first and second moment of the partition sum, i.e., $\langle Z \rangle, \langle Z^2 \rangle$, and using the Chebyshev's inequality. The exact value of the satisfiability threshold is then given by $s_{\rm sym}(l_{\rm s})=0$. This itself is a remarkable result, because so far the exact threshold was computed in only a handful of the sparse NP-complete CSPs. As far as we know only in the 1-in-$K$ SAT \cite{AchlioptasChtcherba01} and [\RS], the $2+p$-SAT \cite{MonassonZecchina99b,AchlioptasKirousis01} and the $(3,4)$-UE-CSP \cite{ConnamacherMolloy04}. We dedicate the appendix \ref{app:moments} to this computation. 
\end{itemize}

The replica symmetric solution might be incorrect if long range correlations are present in the system, as we discussed in detail in chap.~\ref{clustering}. A sufficient condition for its correctness is the decay of the point-to-set correlations, which we will discuss in the next section, again in context of the reconstruction problem.
A necessary condition for the RS solution to be correct is the non-divergence of the spin glass susceptibility, which can be investigated in several equivalent ways, as described in appendix \ref{app:RS_stab}. The result for all the locked problems we investigated is that the phase where the entropy (\ref{eq:entropy_l}) is positive is always RS stable, whereas part of the phase where the entropy  (\ref{eq:entropy_l}) is negative might be RS unstable (depending on the parameters and the vector $A$).

\subsection{Small noise reconstruction}
\label{sec:sn_rec}

It is immediate to observe that reconstruction as we defined it in sec.~\ref{trees} is always possible for the locked problems. Indeed, if we know $K-1$ out of $K$ variables around a constraint the last one is given uniquely (no contradiction is possible as we broadcasted a solution). This is related to the fact that at least one closed loop has to be flipped to go from one solution of a given instance of a locked problem to another solution. Typical
length of such a minimal loop is of order $\log{N}$. For very low connectivities, and at infinitesimally low temperature, the BP equations will have a unique fixed point, there the zero temperature $\log{N}$ clustering is "fake" and will not have a crucial influence on the dynamics and other properties of interest. 

Thus for the locked problem it is useful to modify the definition of the clustering transition presented in chap.~\ref{clustering}. In order to do that we need to introduce the {\it small noise (SN) reconstruction}\index{reconstruction!small noise reconstruction}. 
Construct an infinite tree hyper-graph, assign a value $1$ or $0$ to its root and iteratively assign its offsprings uniformly at random but in such a way that the constraints are satisfied (constraints play the role of noiseless channels). 
At the end of the procedure forget the values of all variables in the bulk but also of an infinitesimal fraction $\epsilon$ of leaves. If the remaining $1-\epsilon$ leaves contain some information about the original value on the root then we say that the small noise reconstruction is possible, if they do not the small noise reconstruction is not possible. The phase where the SN reconstruction is not possible is then only "fake clustered" and is more similar to the liquid phase. Whereas the phase where the SN reconstruction is possible has all the properties of the clustered phase, except that each of the clusters contains only one configuration\footnote{Note that a rigorous study of a related robust reconstruction exists \cite{JansonMossel04}. In robust reconstruction, however, one allows $\epsilon$ to be arbitrarily near to one.
}.

All the equations we derived in sec.~\ref{trees} for the reconstruction apply also for the SN reconstruction. Except the specification of the initial conditions (\ref{eq:init}) which for the SN reconstruction is instead
\be
    P^{\rm init}(\vec \psi) = \frac{1-\epsilon}{2} \big[ \delta(\vec \psi - \delta_0) +\delta(\vec \psi - \delta_1)  \big] + \epsilon\,  \delta\big(\psi_0 - \frac{1}{2}\big) \, \delta\big(\psi_1 - \frac{1}{2}\big)\, ,  \label{eq:init_sn}
\ee
where $\epsilon \ll 1$. The second term accounts for the fraction of leaves on which the value of the variable has been forgotten. The fixed point of the 1RSB equation (\ref{eq:1RSB_m1}) is then either trivial (corresponding to the replica symmetric solution) or nontrivial describing solutions as an ensemble of totally frozen clusters. This has several interesting consequences: The threshold for the naive SN reconstruction\index{reconstruction!naive} (i.e., the one taking into account only the frozen variables) coincide with the true threshold for SN reconstruction. The solution of the 1RSB equation (\ref{eq:1RSB_m1}) in the locked problem does not depend on the value of the parameter $m$. 

A general form of the 1RSB equations at $m=1$ for occupation problems is derived in appendix~\ref{app:m1}. First we consider only problems where the replica symmetric solution is factorized. We define $\mu_1$ (resp. $\mu_0$) as the probability that a variable which in the broadcasting had value $1$ (resp. $0$) is uniquely determined by the boundary conditions.  Based on the general eq.~(\ref{final_m1}), we derive self-consistent equations for $\mu_1$, $\mu_0$ on regular graphs ensemble of connectivity of variables $L$:
\begin{subequations}
\label{eq:mus}
\bea
   \mu_1 &=& \frac{1}{\psi_1 Z^{\rm reg}}\sum_{A_{r+1}=1,A_r=0} {k \choose r} (\psi_1)^{lr}  (\psi_0)^{l(k-r)} \sum_{s=0}^{s_1} {r \choose s}\nonumber \\  && \left[1-(1-\mu_0)^{l}\right]^{k-r} \left[1-(1-\mu_1)^{l}\right]^{r-s} (1-\mu_1)^{ls} \, ,  \label{eq:mu1}\\
    \mu_0 &=& \frac{1}{\psi_0 Z^{\rm reg}}\sum_{A_r=1,A_{r+1}=0} {k \choose r} (\psi_1)^{lr}  (\psi_0)^{l(k-r)}\sum_{s=0}^{s_0} {k-r \choose s} \nonumber \\ & &\left[1-(1-\mu_1)^{l}\right]^{r}  \left[1-(1-\mu_0)^{l}\right]^{k-r-s} (1-\mu_0)^{ls}\, ,\label{eq:mu0} 
\eea
\end{subequations}
where $l=L-1$, $k=K-1$. The indices $s_1,s_0$ in the second sum of both equations are the largest possible but such that $s_1\le r$, $s_0\le K-1-r$, and $\sum_{s=0}^{s_1}A_{r-s}=0$, $\sum_{s=0}^{s_0}A_{r+1+s}=0$.
The values $\psi_0$, $\psi_1$ are the fixed point of eqs.~(\ref{RS1}-\ref{RS2}), and $Z^{\rm reg}$ is the corresponding normalization.
These lengthy equations have in fact a simple meaning. The first sum is over the possible numbers of occupied variables on the descendants in the broadcasting. The sums over $s$ is over the number of variables which were not implied by at least one constraint but still such that the set of incoming implied variables implies the outcoming value. The term $1-(1-\mu)^l$ is the probability that at least one constraint implies the variable, $(1-\mu)^l$ is the probability that none of the constraints implies the variable. 

The second case where the BP equations are factorized are the {\it balanced}\index{locked problems!balanced} locked problems. That is LOPs with symmetric vector $A$ where the symmetry is not spontaneously broken. Then $\psi_0=\psi_1=1/2$ and thus also $\mu_0=\mu_1=\mu$. For the ensemble of graphs with truncated Poissonian degree distribution of coefficient $c$ we derive from (\ref{final_m1})
\be
       \mu = \frac{2}{g_A} \sum_{A_{r+1}=1} {k\choose r} \sum_{s=0}^{s_1} {r \choose s} \left( \frac{1 - e^{-c\mu}}{1-e^{-c}}\right)^{k-s} \left(\frac{e^{-c\mu}-e^{-c}}{1-e^{-c}}\right)^{s} \, , \label{eq:mu} 
\ee
where $k=K-1$, and $g_A =  \sum_{r,A_{r+1}=1} {k \choose r}+ \sum_{r,A_r=1} {k \choose r}$ and the value $s$ is, as before, the number of descendants which were not directly implied.

In both these cases, there are two solutions to eqs.~(\ref{eq:mu1}-\ref{eq:mu0}) and (\ref{eq:mu}). One is $\mu=0$ and the other $\mu=1$. The small noise reconstruction is investigated by the iterative stability of the solution $\mu=1$. If it is stable then the SN reconstruction is possible, all variables are almost surely directly implied. If it is not stable then the only other solution is $\mu=0$. Few observations are immediate, for example if $L\ge 3$ then the solution $\mu_1=\mu_0=1$ of (\ref{eq:mu1}-\ref{eq:mu0}) is always iteratively stable. Iterative stability of (\ref{eq:mu}) gives for the balanced locked problems, marked by $\ast$ in tab.~\ref{tab:locked}:
\be
         \frac{e^{c_d}-1}{c_d}=K-1 -\frac{\sum_{r=0}^{K-2} \delta(A_{r+1}-1)\, \delta(A_{r-1})\, \delta(A_{r})\,  {K-1 \choose r}}{\sum_{r=0}^{K-2} \delta(A_{r+1}-1) \, {K-1 \choose r}}\, .
\ee

\subsection{Clustering transition in the locked problems}

\begin{table}[!ht]
\begin{center}
\begin{tabular}{|l|l||l||l|l|l|l|} \hline
A       & name      &\, $L_{s}$\, &\, $c_{d}$&\, $c_{s}$&\, $ l_d$&\, $ l_s$\\ \hline 
0100    & 1-in-3 SAT       & 3   &0.685(3)\, &0.946(4)\, &2.256(3)\, &2.368(4)\,  \\ \hline 
01000   & 1-in-4 SAT       & 3   & 1.108(3)  & 1.541(4)  & 2.442(3)  & 2.657(4)   \\ \hline 
00100*  & 2-in-4 SAT       & 3   & 1.256     & 1.853     & 2.513     & 2.827      \\ \hline 
01010*  & 4-odd-PC        & 5   & 1.904     & 3.594     & 2.856     & 4          \\ \hline 
010000  & 1-in-5 SAT       & 3   & 1.419(3)  & 1.982(6)  & 2.594(3)  & 2.901(6)   \\ \hline 
001000  & 2-in-5 SAT       & 4   & 1.604(3)  & 2.439(6)  & 2.690(3)  & 3.180(6)   \\ \hline 
010100  & 1-or-3-in-5 SAT  & 5   & 2.261(3)  & 4.482(6)  & 3.068(3)  & 4.724(6)   \\ \hline 
010010  & 1-or-4-in-5 SAT  & 4   & 1.035(3)  & 2.399(6)  & 2.408(3)  & 3.155(6)   \\ \hline 
0100000 & 1-in-6 SAT       & 3   & 1.666(3)  & 2.332(4)  & 2.723(3)  & 3.113(4)   \\ \hline 
0101000 & 1-or-3-in-6 SAT  & 6   & 2.519(3)  & 5.123(6)  & 3.232(3)  & 5.285(6)   \\ \hline
0100100 & 1-or-4-in-6 SAT  & 4   & 1.646(3)  & 3.366(6)  & 2.712(3)  & 3.827(6)   \\ \hline
0100010 & 1-or-5-in-6 SAT  & 4   & 1.594(3)  & 2.404(6)  & 2.685(3)  & 3.158(6)   \\ \hline
0010000 & 2-in-6 SAT       & 4   & 1.868(3)  & 2.885(4)  & 2.835(3)  & 3.479(4)   \\ \hline 
0010100*& 2-or-4-in-6 SAT  & 6   & 2.561     & 5.349     & 3.260     & 5.489      \\ \hline
0001000*& 3-in-6 SAT       & 4   & 1.904     & 3.023     & 2.856     & 3.576      \\ \hline 
0101010*& 6-odd-PC        & 7   & 2.660     & 5.903     & 3.325     & 6          \\ \hline
\end{tabular}  
\caption{\label{tab:locked}The locked cases of the occupation CSPs for $K\le 6$ (cases with a trivial ferromagnetic solution are omitted). In the regular graphs ensemble the phase is clustered for $L \ge L_d=3$, and unsatisfiable for $L\ge L_s$. Values $c$ are the critical parameters of the truncated Poissonian ensemble (\ref{eq:Poiss}), the corresponding average connectivities $\overline l$ are given via eq.~(\ref{eq:l_aver}). All these problems are RS stable at least up to the satisfiability threshold. For the balanced cases, marked as *, the dynamical threshold follows from (\ref{eq:mu}), and the satisfiability threshold, which can be computed rigorously, app.~\ref{app:moments}, from (\ref{eq:s_sym}).}
\end{center}
\end{table}

In the locked problem where the replica symmetric solution is not factorized there is another equivalent way to locate the clustering transition, which is simpler than solving eq.~(\ref{final_m1}). It is the investigation of the iterative stability of the nontrivial fixed point of the survey propagation. In LOPs the survey propagation equations consist of eqs.~(\ref{SP_1}) and
\begin{subequations}
\label{eq:SP_locked}
\bea
     q_1^{a\to i} &=& \frac{1}{{\cal N}^{a\to i}}  \left[ \sum_{\{r_j\}} C_1(\{r_j\}) \prod_{j\in a-i} p_{r_j}^{j\to a}  \right]\, ,\\
 q_{-1}^{a\to i} &=& \frac{1}{{\cal N}^{a\to i}} \left[ \sum_{\{r_j\}} C_{-1}(\{r_j\}) \prod_{j\in a-i} p_{r_j}^{j\to a}  \right] \, , \\
 q_0^{a\to i} &=& \frac{1}{{\cal N}^{a\to i}}  \left[ \sum_{\{r_j\}} C_0(\{r_j\}) \prod_{j\in a-i} p_{r_j}^{j\to a}  \right]\, ,  \label{SP2}
\eea 
\end{subequations}
where the indexes $r_j\in \{1,-1,0\}$, ${\cal N}^{a\to i}$ is the normalization constant. The $C_1$/$C_{-1}$ (resp. $C_0$) takes values $1$ if and only if the incoming set of $\{r_j\}$ forces the variable $i$ to be occupied/empty (resp. let the variable $i$ free), in all other cases the $C$'s are zero. Let us call $s_1,s_{-1},s_0$ the number of indexes $1,-1,0$ in the set $\{r_j\}$ then 
\begin{itemize}
\item{$C_1=1$ if and only if $A_{s_1+s_0+1}=1$ and $A_{s_1+n}=0$ for all $n=0\dots s_0$;} 
\item{$C_{-1}=1$ if and only if $A_{s_1}=1$ and $A_{s_1+1+n}=0$ for all $n=0\dots s_0$; }
\item{$C_0=1$ if and only if there exists $m,n=0\dots s_0$ such that $A_{s_1+n}=A_{s_1+m+1}=1$.}
\end{itemize}
The SP equations in LOPs have two different fixed points:
\begin{itemize}
\item The trivial one: $q^{a\to i}_0=p^{i\to a}_0=1$, $q^{a\to i}_1=p^{i\to a}_1=q^{a\to i}_{-1}=p^{i\to a}_{-1}=0$ for all edges $(ai)$.
\item The BP-like one: $q^{a\to i}_0=p^{i\to a}_0=0$, $q^{a\to i}=\psi^{a\to i}$, $p^{i\to a}=\chi^{i\to a}$ for all edges $(ai)$, where $\psi$ and $\chi$ is the solution of the BP equations (\ref{BP1}-\ref{BP2}).
\end{itemize}
The small noise reconstruction is then investigated, using the population dynamics, from the iterative stability of the BP-like fixed point. If it is stable then the SN reconstruction is possible and the phase is clustered. If it is not stable then we are in the liquid phase. Of course, this approach gives the same critical connectivity $l_d$ as the previous one, because for the locked problems the solutions of the 1RSB equation (\ref{eq:1RSB_m1}) is independent of the parameter $m$.

We remind at this point that in a general CSP, where the sizes of clusters fluctuate, the SP equations are not related to the reconstruction problem, more technically said the 1RSB solutions at $m=0$ and at $m=1$ are different. The solution of the locked problems is sometimes called frozen 1RSB \cite{MartinMezard04,MartinMezard05}.

\section{Freezing - The reason for hardness?}
\label{sec:hardness}

We describe several strong evidences that it is hard to find frozen solutions.  We also give several arguments for why it is so. However, the precise mechanism stays an open question and strictly speaking the freezing of variables might just be going along with a true yet unknown reason. Or even there might be an algorithm which is able to find the frozen solutions efficiently waiting for a discovery. 
But in any case, we show that freezing of variables is an important new aspect in the search of the origin of the average computational hardness.  
  
\subsection{Always a trivial whitening core}

Several studies of the random 3-SAT problem~\cite{ManevaMossel05,BraunsteinZecchina04,SeitzAlava05}  showed that all known algorithms on large instances systematically find only solutions with a trivial whitening core (defined in sec.~\ref{sec:white}). On small instances of the problem solutions with a nontrivial whitening core can be found as observed by several authors, and studied systematically in sec.~\ref{sec:3SAT_num}.

For solutions found by the stochastic local search algorithms, see appendix \ref{app:alg}, this observation is reasonable, as argued already in \cite{SeitzAlava05}. Consider that a stochastic local search finds a configuration which is not a solution, but its whitening core is nontrivial. Then a diverging number of variables have to be rearranged in order to satisfy one of the unsatisfied constraints \cite{Semerjian07}. In the clusters with a trivial whitening core the rearrangements are finite  \cite{Semerjian07} and thus stochastic local dynamics might be able to find them more easily.  

The fact of finding only the "white" solutions is, however, quite surprising for the survey propagation algorithm. The SP equations compute probabilities (over clusters) that a variables is frozen in a certain value.  This information is then used in a decimation, reinforcement, etc. algorithms, see appendix \ref{app:alg}. Thus SP is explicitly exploring the information about nontrivial whitening cores and in spite of that it finishes finding solutions with trivial whitening cores. 

A related, and rather surprising, result was shown in \cite{DallAstaRamezanpour08}. The authors considered the random bi-coloring problem in the rigid, but not frozen, phase. That is a phase where most solutions are frozen, but rare unfrozen ones still exist. They showed that belief propagation reinforcement solver, see appendix \ref{app:alg}, is in some cases able to find these exponentially rare, but unfrozen, solutions. 

\begin{figure}[!ht]
 \begin{center}
 \resizebox{0.67\linewidth}{!}{
 \includegraphics{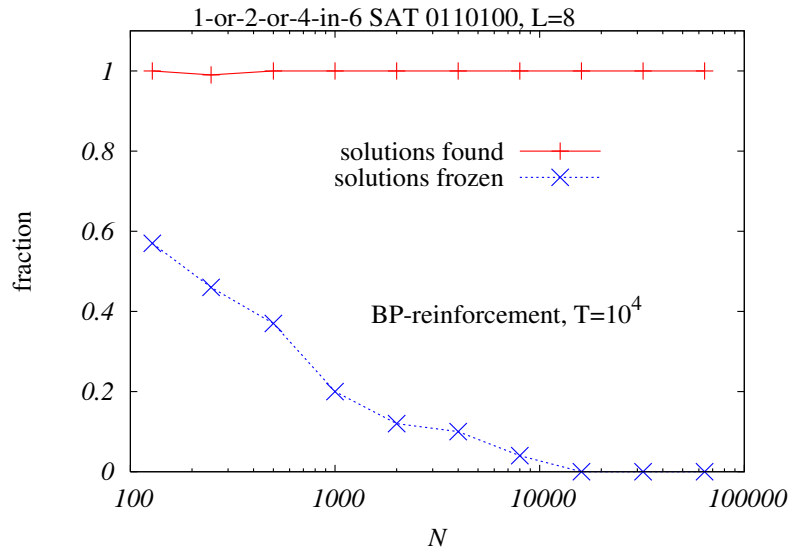}
 }
 \end{center}
 \caption{\label{fig:alg_nonlocked} Algorithmical performance in the rigid phase of the 1-or-2-or-4-in-6 SAT at $L=8$. In red is the rate of success of the belief propagation reinforcement algorithms as a function of system size (out of 100 trials). The algorithm basically always succeeds to find a solution. In blue is the fraction of solutions which were frozen (had a nontrivial whitening core). Almost all solutions are frozen in this problem, yet it is algorithmically easier to find the rare unfrozen solutions, in particular in instances of larger size.}
\end{figure}

We observed the same phenomena in one of the non-locked occupation problem $A=(0110100)$, that is 1-or-2-or-4-in-6 SAT. On regular factor graphs this problem is in the liquid phase for $L\le 6$, in the rigid phase for $7\le L \le 9$, where almost all the solutions are frozen, and it is unsatisfiable for $L\ge 10$. In fig.~\ref{fig:alg_nonlocked} we show that belief propagation reinforcement finds almost always solutions for $L=8$, but as the size of instances is growing the fraction of cases in which the solution is frozen goes to zero. 

We listed this paradox, that only the all-white solutions can be found, as one of the loose ends in sec.~\ref{sec:loose}. The resolution we suggest here, and substantiate in the following, is that every known algorithm is able to find efficiently (in polynomial - but more often in experiments we mean linear or quadratic - time) only the unfrozen solutions. The frozen solutions are intrinsically hard to find and all the known algorithms have to run for an exponential time to find them.

\subsection{Incremental algorithms}

Adopted from \cite{KrzakalaKurchan07}: Consider an instance of a constraint satisfaction problem of $N$ variables and $M$ constraints. Order randomly the set of constraints and remove all of them. Without constraints any configuration is a solutions. In each step: First, add back one of the constraints. Second, if needed rearrange the configuration in such a way that it satisfies the new and all the previous constraints. Repeat until there are some constraints left. We call such strategy the {\it incremental algorithm}\index{algorithms!incremental} for CSPs. And one can ask about its computational complexity. The way by which the rearrangement is found in the second step needs to be specified. But independently of this specification we know that if the new constraint connects frozen and contradictory variables then the size of the minimal rearrangement diverges \cite{Semerjian07}, thus in the frozen phase the incremental algorithm have to be at best super-linear. 

Another understanding of the situation is gained by imagining the space of solutions at a given constraint density. As we are adding the constraints some solutions are disappearing and none are appearing. At the clustering transition the space of solutions splits into exponentially many clusters. As more constraints are added the clusters are becoming smaller, they may split into several smaller ones and some may completely disappear. However, only the frozen clusters can disappear, if a constraint is added between two frozen and contradictory variables. Note also that each frozen cluster will almost surely disappear before an infinitesimally small fraction of constraints is added. An unfrozen cluster, on the other hand, may only become smaller or split. Indeed, if a constraint is added any solution belonging to an unfrozen cluster may be rearranged in a finite number of steps \cite{Semerjian07}. The incremental algorithm in this setting works as a non-intelligent animal would be escaping from a rising ocean on a Pacific hilly island \cite{KrzakalaKurchan07}. As the water starts to rise the animal would step away from it. As the water keeps rising at a point the animal would be blocked in one of the many smaller islands. This island will be getting smaller and smaller and it will disappear at a point and the animal will have to learn how to swim. But at this point there might still be many small higher island. All of them will disappear eventually. For sure the animal will be in trouble before all the clusters (island) start to contain frozen variables. 

Moreover, if the sequence of constraints to be added is not known in advance there is no way to choose the best cluster, because which cluster is the best depends completely on the constraints to be added. This proves that no incremental algorithm is able to work in linear time in the frozen phase. On the other hand it was shown experimentally in \cite{KrzakalaKurchan07} for the coloring problem that such algorithms work in linear time in part of the clustered (or even the condensed) phase.

\subsection{Freezing transition and the performance of SP in 3-SAT}

How does the freezing transition in 3-SAT, $\alpha_f=4.254\pm 0.009$ fig.~\ref{fig:rigidity}, compare to the performance of the best known random 3-SAT solver --- the survey propagation? We are aware of two studied where the performance of SP is investigated systematically and with a reasonable precision, \cite{Parisi03} and \cite{ChavasFurtlehner05}.

In \cite{Parisi03} the survey propagation decimation is studied. The SP fixed point is found on the decimated graph and the variable having the largest bias is fixed as long as the SP fixed point is nontrivial. When the SP fixed point becomes trivial the Walk-SAT algorithm finishes the search for a solutions. In \cite{Parisi03} the residual complexity is measured on the partially decimated graph. It is observed that if the residual complexity becomes negative then solutions are never found, if on the other hand the residual complexity is positive just before the survey propagation fixed point become trivial then solutions are found. The value of complexity in the last step before the fixed point becomes trivial is extrapolated, fig.~2 of \cite{Parisi03} for system size $N=3\cdot10^5$, to zero at a constraint density $\alpha=4.252\pm0.003$ (we estimated the error bar based on data from  \cite{Parisi03}).

In \cite{ChavasFurtlehner05} the survey propagation reinforcement is studied. The rate of success is plotted as a function of the complexity function. From fig.~8 of \cite{ChavasFurtlehner05} it is estimated that SP reinforcement (more precisely its implementation presented in  \cite{ChavasFurtlehner05}) finds solution in more than 50\% of trials if $\Sigma>0.0013$. The data do not really concentrate on this point, thus is is difficult to obtain a reliable error bar of this value, our educated guess is $0.0013\pm0.0003$ this would correspond to a constraint density $\alpha = 4.252\pm0.004$.

The striking agreement between our value for the freezing transition and the performance limit of the survey propagation supports the suggestion that the frozen phase is hard for any known algorithm. The trouble for a better study of the frozen phase in 3-SAT is its size, it covers only 0.3\% of the satisfiable phase. In $K$-SAT with large $K$ the frozen phase becomes wider, but as $K$ grows the constraint density of the satisfiability threshold grows like $2^K\log{K}$, empirical study thus becomes infeasible very fast. It is also not very easy to compute the freezing transition or to check if the 1RSB solution is correct in the frozen phase. Thus $K$-SAT (and $q$-coloring) are not very suitable problems for understanding better how exactly the freezing influences the search for a solution.

\subsection{Locked problems -- New extremely challenging CSPs}

We introduced the locked problems to challenge the suggestion about hardness of the frozen phase [\ZML]. It is rather easy to compute the freezing transition here, it coincides with the clustering transition $l_d$. Moreover, the frozen phase is wide, taking more than 50\% of the satisfiable phase for some of the locked problems, see table \ref{tab:locked}. As in the locked problems every cluster consists of one solution, all the variables are frozen. Consequently the replica symmetric approach describes correctly the phase diagram. From this point of view the locked problems seems extremely easy compared to $K$-SAT. 

On the other hand, experiments with the best known solvers of random CSPs show that the frozen phase of locked problems is very hard. And some of the very good solvers, e.g. the belief propagation based decimation, do not work at all even at the lowest connectivities (for an explanation see appendix \ref{app:alg}).

\begin{figure}[!ht]
 \resizebox{\linewidth}{!}{
 \includegraphics{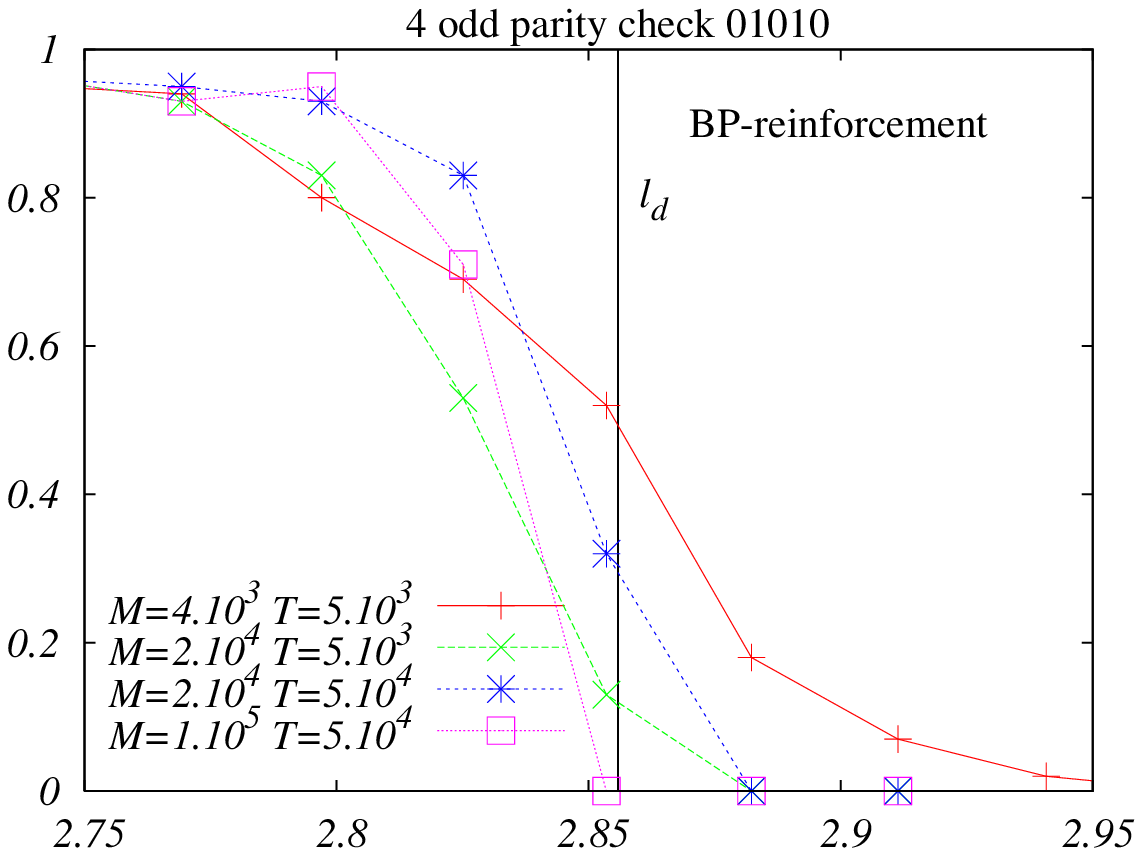}
 \includegraphics{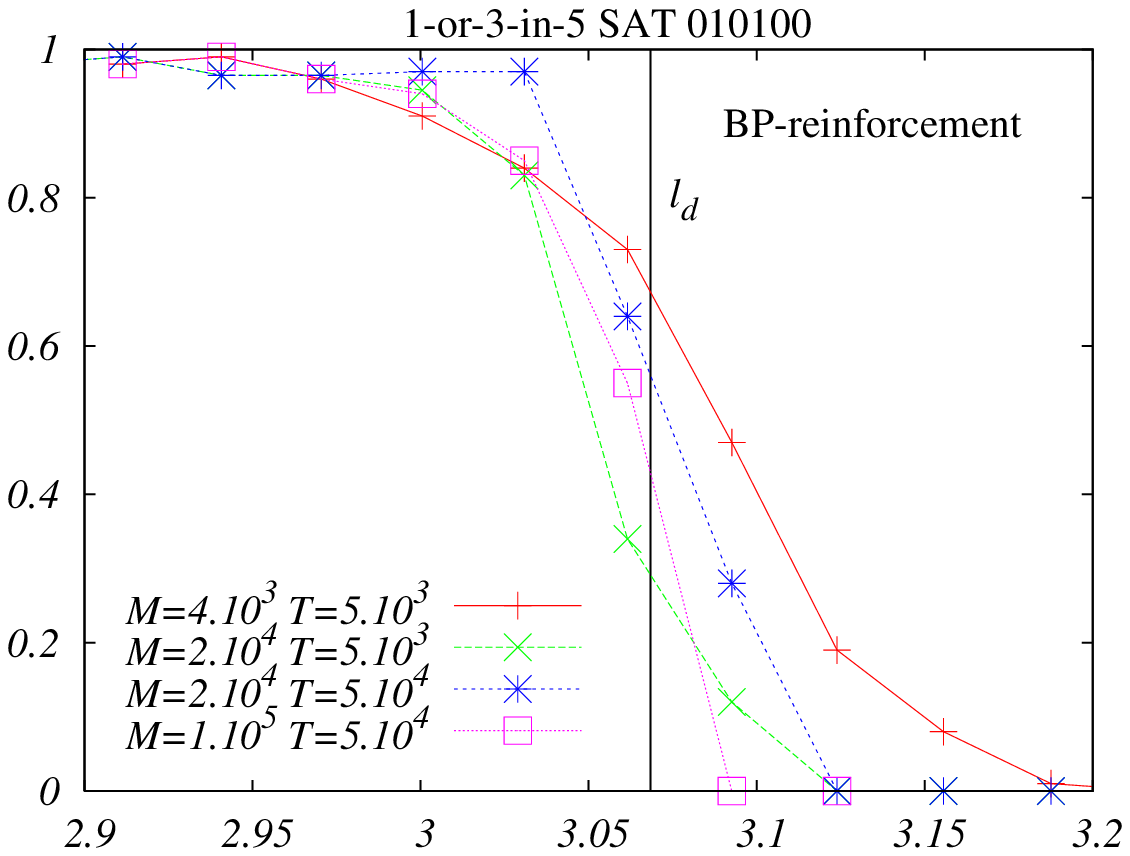}
 }
\resizebox{\linewidth}{!}{
 \includegraphics{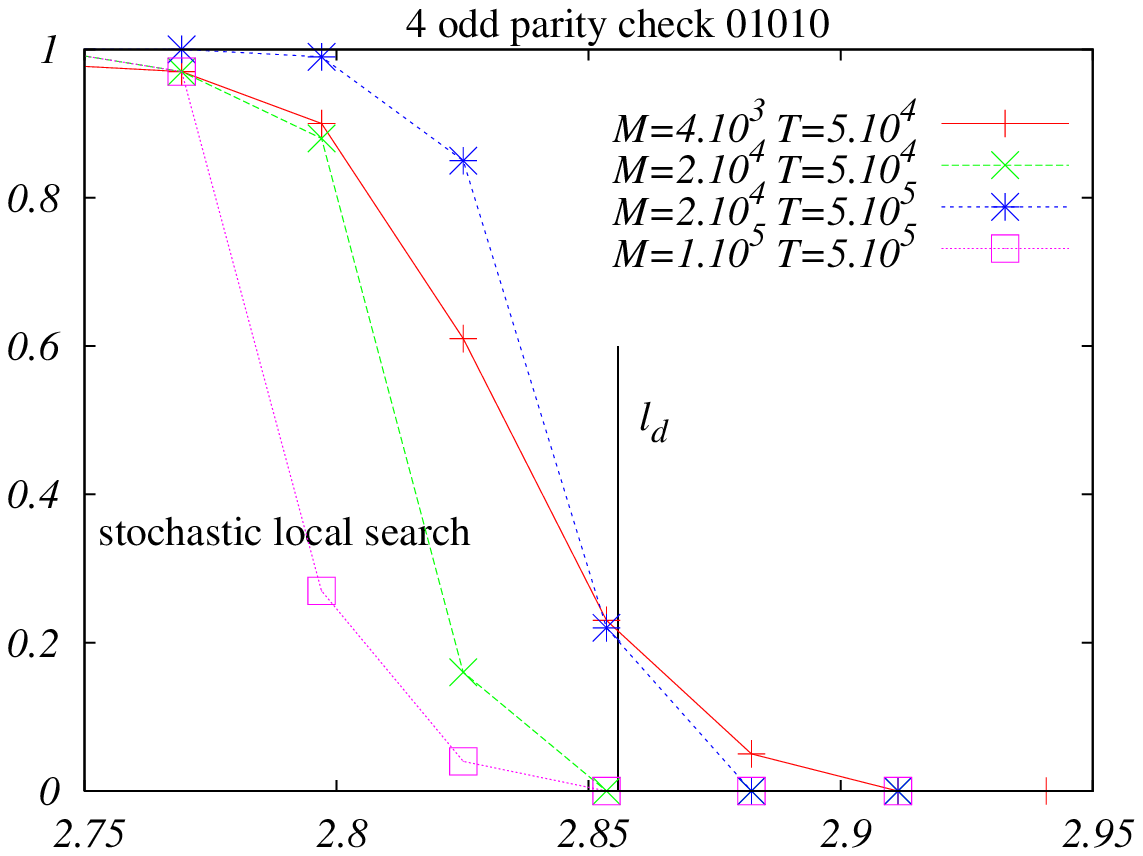}
 \includegraphics{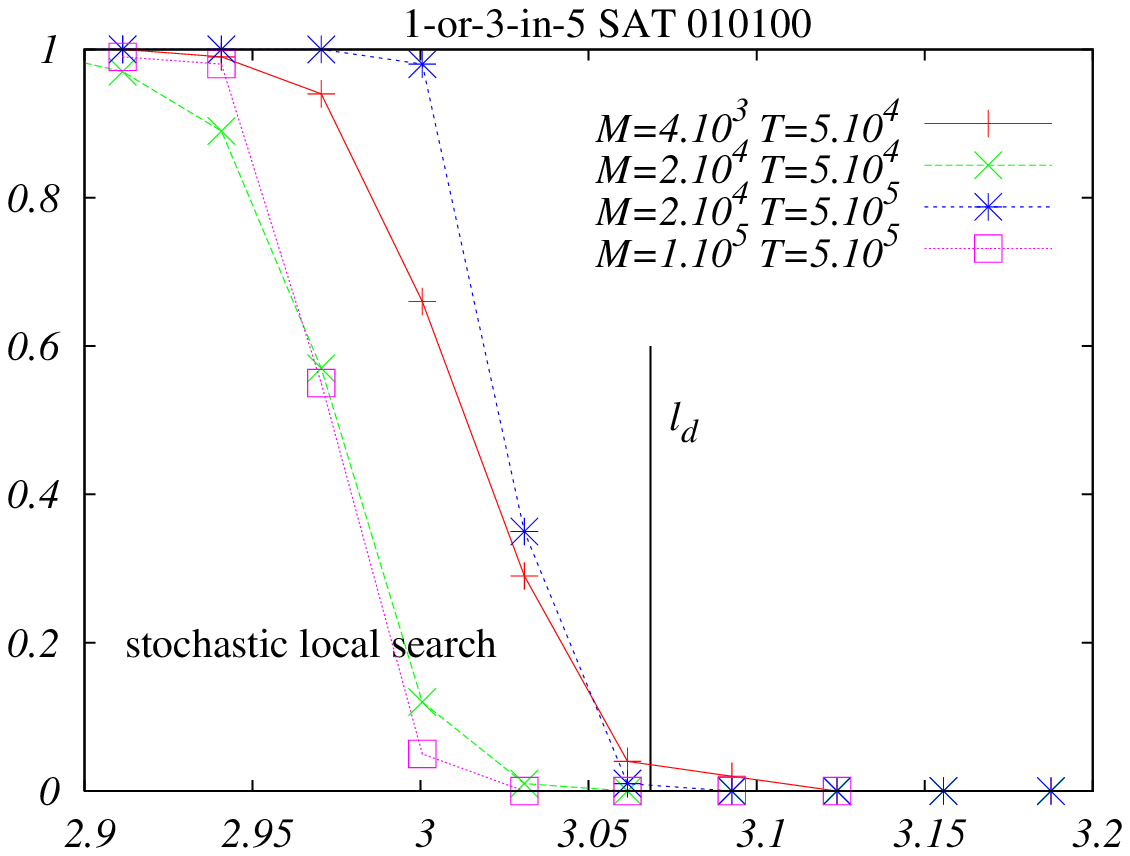}
 }
 \caption{\label{fig:alg_locked} The probability of success of the \proc{BP-reinforcement} (top) and the stochastic local search \proc{ASAT} (bottom) plotted against the average connectivity for two of the locked occupation problems. The clustering transition is marked by a vertical line, the satisfiability threshold is $l_s=4$ for the 4-odd parity checks, and $l_s=4.72$ for the 1-or-3-in-5 SAT. The challenging task is to design an algorithm which would work also in the clustered phase of the NP-complete locked problems.}
\end{figure}

In fig.~\ref{fig:alg_locked} we show the performance of the \proc{BP-reinforcement} and the stochastic local search \proc{ASAT} algorithms. Both the algorithms are described in appendix \ref{app:alg}, they are the best we were able to find for the locked CSPs. The greediness parameter in the stochastic local search \proc{ASAT} we evaluated as the most optimal is $p=5.10^{-5}$ for the 4-odd parity check, and $p=3.10^{-5}$ for the 1-or-3-in-5 SAT. In the \proc{BP-reinforcement} the optimal forcing parameter $\pi$ changes slightly with the connectivity. For the 1-or-3-in-5 SAT we used $\pi=0.42$ for $2.9\le\overline l< 3.0$ and $\pi=0.43$ for $3.0\le\overline l\le 3.2$. For the 4-odd parity checks we used $\pi=0.44$ for $2.75\le \overline l \le 2.95$. 

Of course, the parity check problem is an exceptional locked problem, as it is not NP-complete and can be solve via Gaussian elimination. However, our study shows that algorithms which do not use directly the linearity of the problem fail in the same way as they do in the NP-complete cases. Instances of the regular XOR-SAT indeed belong between the hardest benchmarks for all the best known satisfiability solvers which do not explore linearity of the problem, see e.g. \cite{HaanpaaJarvisalo06}. 

Fig.~\ref{fig:alg_locked} puts in the evidence that in all the random locked problems the best known algorithms stop to be able to find solutions (in linear time) at the clustering transition. This supports the conjecture about freezing being relevant for algorithmical hardness. The locked problems are thus (at least until they are "unlocked") the new benchmarks of hard constraint satisfaction problems.

\chapter{Coloring random graphs}
\label{coloring}

{\it In the previous three chapters we developed tools for describing the structure of solution and the phase diagram of random constraint satisfaction problems. These tools were applied to the problem of coloring random graphs in a series of works
[\KM, \ZK, \KZP, \KZJ]. In this section we summarize the results.}

\section{Setting}

\begin{figure}[!ht]
\begin{center}
  \resizebox{4cm}{!}{\includegraphics{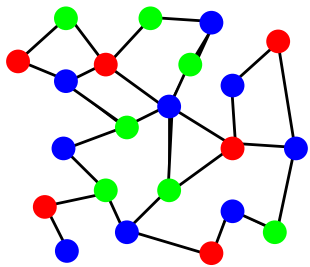}}
\end{center}
  \caption{ \label{fig:example}Example of a proper 3-coloring of a small graph.}
\end{figure}

Coloring of a graph is an assignment of colors to the vertices of the graph such that two adjacent vertices do not have the same color. The question is if on a given graph a coloring with $q$ colors exists. Fig.~\ref{fig:example} gives an example of 3-coloring of a graphs with $N=22$ vertices and $M=27$ edges, the average connectivity is $c=2M/N \approx 2.45$.

It is immediate to realize that the $q$-coloring problem is equivalent to the question of determining if the ground-state energy of a Potts anti-ferromagnet on a random graph is zero or not~\cite{KanterSompolinsky87}. Consider indeed a graph $G = ({ V,E})$ defined by its vertices ${V}=\{1,\dots,N\}$ and edges $(i,j)\in { E}$ which connect pairs of vertices $i,j\in { V}$; and the Hamiltonian
\be
{\cal H}(\{s\}) = \sum_{(i,j)} \delta(s_i,s_j)\, .
\label{Ham_col}
\ee 
With this choice there is no energy contribution for neighbours with different colors, but a positive contribution otherwise. The ground state energy is thus zero {\it if and only if} the graph is $q$-colorable. This transforms the coloring problem into a well-defined statistical physics model. 

Studies of coloring of sparse random graphs have a long history in mathematics and computer science, see [\ZK] for some references. From the statistical physics perspective it was first studied in \cite{MourikSaad02}, where the replica symmetric solution was worked out, and the replica symmetric stability was investigated numerically. Results were compared to Monte Carlo simulations and simulated annealing was used as a solver for coloring. The energetic 1RSB solution and the survey propagation algorithm for graph coloring were developed in \cite{MuletPagnani02,BraunsteinMulet03}. Subsequently \cite{KrzakalaPagnani04} studied the stability of the 1RSB solution and its large $q$ limit. The entropic 1RSB solution was studies in \cite{MezardPalassini05} for 3-coloring of Erd\H{o}s-R\'enyi graphs. The entropic 1RSB solution was, however, fully exploited only in [\KM, \ZK, \KZP, \KZJ] and the resulting phase diagram is discussed here.

\section{Phase diagram}

\begin{figure}[!ht]
  \resizebox{\linewidth}{!}{\includegraphics{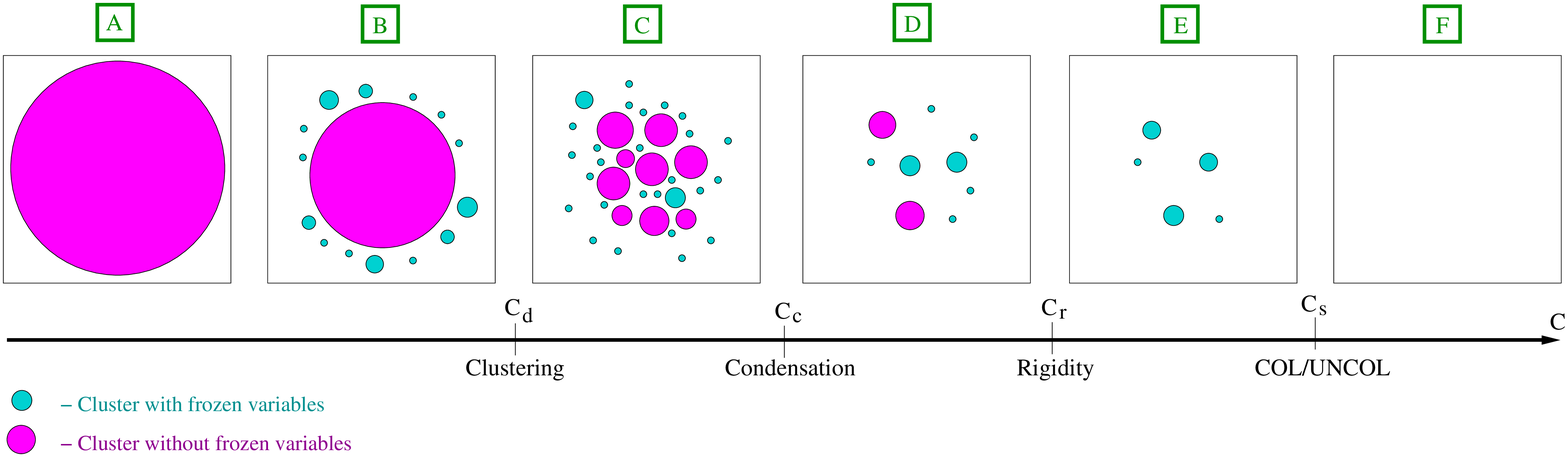}}
\begin{center}
  \resizebox{0.67\linewidth}{!}{\includegraphics{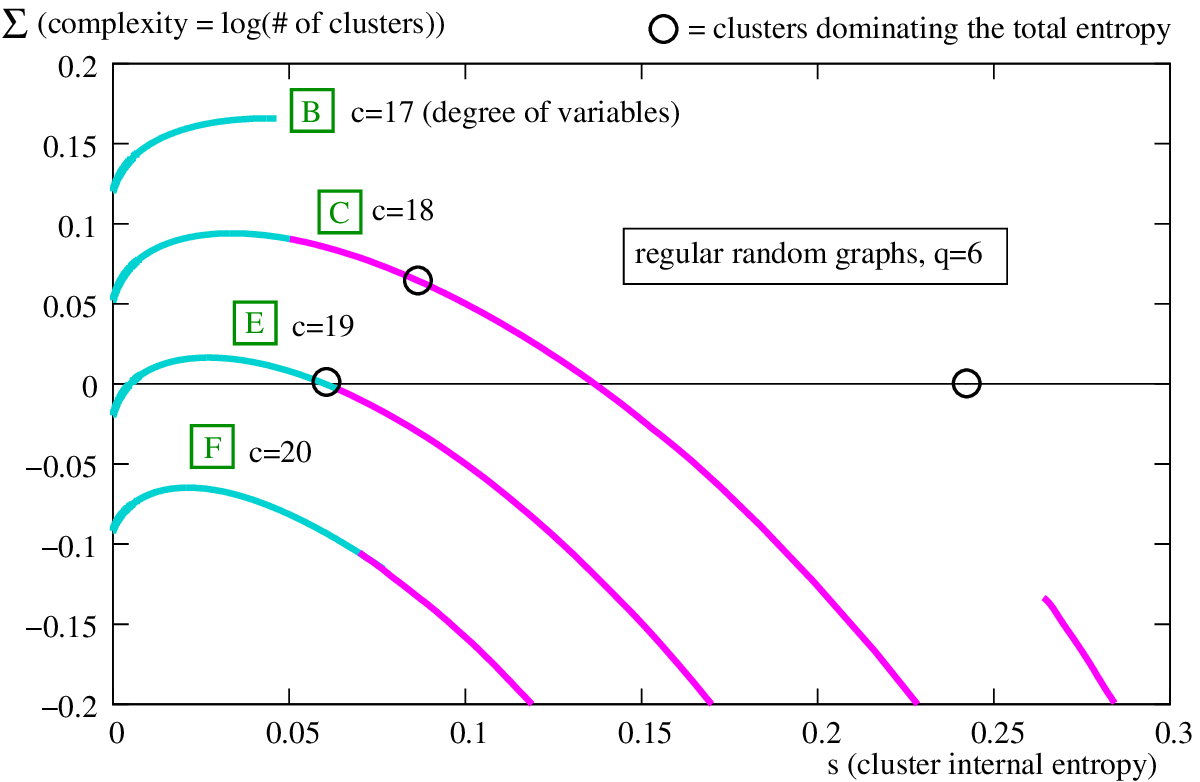}}
\end{center}
  \caption{ \label{fig:diag} Up: Sketch of the structure of solutions in the random coloring problem. The depicted phase transitions arrive in the above order on Erd\H{o}s-R\'enyi graphs for number of colors $4\le q \le 8$. Down: Complexity (log-number) of clusters of a given entropy, $\Sigma(s)$, for 6-coloring or random regular graphs. The circles mark the dominating clusters, i.e., those which cover almost all solutions.}
\end{figure}

Fig.~\ref{fig:diag} summarizes how does the structure of solutions of the coloring problem change when the average connectivity is increased, (A)$\to$(F). In fig.~\ref{fig:diag} up, each colored "pixel" corresponds to one solution, and each circle to one cluster. As the average connectivity is increased, some solutions disappear and the overall structure of clusters changes. This is depicted in the six snapshots (A)$\to$(F). The magenta clusters are the unfrozen ones, the cyan-blue clusters are the frozen ones. Fig.~\ref{fig:diag} down, the corresponding complexity (log-number) of clusters of a given entropy, $\Sigma(s)$, computed from the 1RSB approach (\ref{eq:1RSB_Sigma}) for the 6-coloring of random regular graphs. More detailed description of the different phases for $q$-coloring follows. 

\begin{itemize}
\item[(A)]{\bf A unique cluster exists}: For connectivities low enough, all the proper colorings are found in a single cluster, where it is easy to ``move'' from one solution to another. Only one possible ---and trivial--- fixed point of the BP equations exists at this stage (as can be proved rigorously in some cases \cite{BandyopadhyayGamarnik06}). The entropy can be computed and reads in the large graph size limit
\be 
s = \frac{\log{\cal{N}}_{\rm sol}} N = \log{q} + \frac {c}{2}
\log{\((1-\frac{1}{q}\))}\, .\label{S_RS} 
\ee

\item[(B)] {\bf Some (irrelevant) clusters appear}: As the connectivity is slightly increased, the phase space of solutions decomposes into a large (exponential) number of different clusters. It is tempting to identify that as the clustering transition. But in this phase all but one of these clusters contain relatively very few solutions, as compare to the whole set. Thus almost all proper colorings still belong to one single giant cluster, and the replica symmetric solution is correct, eq.~(\ref{S_RS}) gives the correct entropy.

\item[(C)] {\bf The clustered phase}: For larger connectivities, the large single cluster decomposes into an exponential number of smaller ones: this now defines the genuine clustering threshold $c_d$. Beyond this threshold, a local algorithm that tries to move in the space of solutions will remain prisoner of a cluster of solutions for a diverging time \cite{MontanariSemerjian06b}. Interestingly, it can be shown that the total number of solutions is still given by eq.~(\ref{S_RS}). Thus the free energy (entropy) has no singularity at the clustering transition, which is therefore not a phase transition in the sense of Ehrenfest. Only a diverging length scale (point-to-set correlation length) and time scale (the equilibration time) when $c_d$ is approached justify the name "phase transition".

\item[(D)] {\bf The condensed phase}: As the connectivity is  increased further, another phase transition arises at the condensation threshold, $c_c$, where most of the solutions are found in a finite number of the largest clusters. Total entropy in the condensed phase is strictly smaller than~(\ref{S_RS}). It has a non-analyticity at $c_c$ therefore this is a genuine static phase transition. The condensation transition can be observed from the two-point correlation functions or from the overlap distribution. 

\item[(E)] {\bf The rigid phase}: As explained in chapter \ref{freezing}, two different types of clusters exist. In the first type, the {\it unfrozen} ones, magenta in fig.~\ref{fig:diag}, all variables can take at least two different colors.  In the second type, {\it frozen} clusters, cyan in fig.~\ref{fig:diag}, a finite fraction of variables is allowed only one color within the cluster and is thus "frozen" into this color. In the rigid phase, a random proper coloring belongs almost surely to a frozen cluster. Depending on the value of $q$, this transition may arise before or after the condensation transition (see tab.~\ref{tab:results_complete}).

\item[(F)] {\bf The uncolorable phase}: Eventually, the connectivity $c_s$ is reached beyond which no more solutions exist. The ground state energy is zero for $c<c_{s}$ and then grows continuously for $c>c_{s}$.
\end{itemize}

In table \ref{tab:results_complete} we present all the critical values for coloring of Erd\H{o}s-R\'enyi graphs, in table \ref{tab:regular} for random regular graphs. Notice the special role of 3-coloring where the clustering and condensation transitions coincide and are given by the local stability of the replica symmetric solution, see app.~\ref{app:RS_stab}. Notice also that for $q\ge 9$ in Erd\H{o}s-R\'enyi graphs and $q\ge 8$ in regular graph the rigidity transition arrives before the condensation transition. 

\begin{table}[!ht]
\begin{center}
\begin{tabular}{|c||l|l|l|l| |l|l|}\hline
q & $c_d$ & $c_r$ &  $c_c$ & $c_s$ & $c_{\rm SP}$ & $c_{r(m=1)} $ \\
\hline \hline
3 & 4 & 4.66(1) & 4 & 4.687(2) & 4.42(1) & 4.911 \\
\hline
4 & 8.353(3) & 8.83(2) & 8.46(1) & 8.901(2) & 8.09(1) &  9.267 \\
\hline
5 & 12.837(3) & 13.55(2) & 13.23(1) & 13.669(2) & 12.11(2) &  14.036 \\
\hline
6 & 17.645(5) & 18.68(2) & 18.44(1) & 18.880(2) & 16.42(2) &  19.112 \\
\hline
7 & 22.705(5) & 24.16(2) & 24.01(1) & 24.455(5) & 20.97(2) &  24.435 \\
\hline
8 & 27.95(5) & 29.93(3) & 29.90(1) & 30.335(5) & 25.71(2) & 29.960 \\
\hline
9 & 33.45(5) & 35.658 & 36.08(5) & 36.490(5) & 30.62(2) & 35.658 \\
\hline
10 & 39.0(1) & 41.508 & 42.50(5) & 42.93(1) & 35.69(3) & 41.508 \\
\hline
\end{tabular}
\end{center}
\caption{\label{tab:results_complete} Critical connectivities $c_d$ (dynamical, clustering), $c_r$  (rigidity), $c_c$ (condensation, Kauzmann) and $c_s$ (colorability) for the phase transitions in the coloring problem on Erd\H{o}s-R\'enyi graphs. The connectivities  $c_{SP}$ (where the first non trivial solution of SP appears) and $c_{r(m=1)}$ (where hard fields appear at $m=1$) are also given. 
The error bars consist of the numerical precision on evaluation of the critical connectivities by the population dynamics technique, see appendix \ref{app:pop_dyn}.
}
\end{table}

\begin{table}[!ht]
\begin{center}
\begin{tabular}{|c|c|c|c|c|c|}
\hline
q & $c_{SP}$ & $c_d$& $c_r$ & $c_c$ & $c_s$ \\
\hline
3 & 5 & $5^+$ & - & 6 & 6 \\
\hline
4 & 9 & 9 & - & 10 & 10 \\
\hline
5 & 13 & 14  & 14 & 14 & 15 \\   
\hline
6 & 17 & 18 & 19 & 19 & 20  \\   
\hline
7 & 21 & 23 & - & 25 & 25  \\   
\hline
8 & 26 & 29 & 30 & 31 & 31  \\   
\hline
9 & 31 & 34 & 36 & 37 & 37  \\   
\hline
10 & 36 & 39 & 42 & 43 & 44  \\   
\hline
20 & 91 & 101 & 105 & 116 & 117  \\   
\hline 
\end{tabular}
\end{center}
\caption{ \label{tab:regular} The transition thresholds for regular random graphs: $c_{\rm SP}$ is the smallest connectivity with a nontrivial solution at $m=0$; the clustering threshold $c_d$  is the smallest connectivity with a nontrivial solution at $m=1$; the rigidity threshold $c_r$ is the smallest connectivity at which hard fields are present in the dominant states, the condensation $c_c$ is the smallest connectivity for which the complexity at $m=1$ is negative and $c_s$ the smallest uncolorable connectivity. Note that $3-$coloring of $5-$regular graphs is exactly critical for that $c_d=5^+$. The rigidity transition may not exist due to the discreteness of the connectivities.}
\end{table}

Few more words about the rigidity transition and the rigid phase in coloring. In sec.~\ref{sec:hard_sol}, next to the rigid phase, we also defined the {\it totally rigid} phase where almost all the clusters of every size become frozen. And the frozen phase where strictly all clusters become frozen. Note that in the random graph coloring the rigidity transition coincides with the total rigidity transition for $q\le 8$ for Erd\H{o}s-R\'enyi graphs and for $q\le 7$ for regular graphs. For larger values of $q$ the rigidity transition is given by the $m=1$ computation. We have not computed the total rigidity transition for larger $q$, but it is accessible from the present method. The freezing transition is, however, not accessible for the entropic 1RSB cavity approach. We cannot exclude that in the totally rigid phase there might still be some rare unfrozen clusters. 

Note also an interesting feature about the 1RSB entropic solution; in fig.~\ref{fig:diag} down, for the connectivity $c=17$ the function $\Sigma(s)$ consists of two branches. The low-entropy branch with frozen clusters, and the high-entropy branch with soft clusters. Note that the soft branch may also exist for positive values of complexity, e.g. in 4-coloring of Erd\H{o}s-R\'enyi graphs. We interpreted the gap as the nonexistence of clusters of the corresponding size. The gap might, however, be an artifact of the 1RSB approximation which most likely does not describe correctly clusters of the corresponding size. For the discussion of correctness of the 1RSB solutions see appendix \ref{app:1RSB_stab}. 

\begin{figure}[!ht]
\begin{center}
\resizebox{0.67\linewidth}{!}{\includegraphics{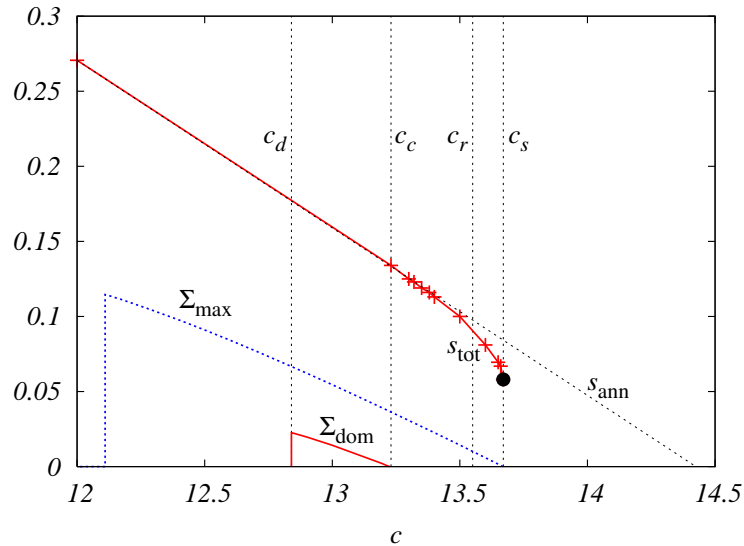}}
\caption{\label{fig_col} Entropies and complexities as a function of the average connectivity for the 5-coloring of Erd\H{o}s-R\'enyi graphs. The replica symmetric entropy is in dashed black, the total entropy in red. The complexity of dominant clusters in red. The total complexity, computed from the survey propagation, is in dashed blue.}
\end{center}
\end{figure}

To make the picture complete we plot the important complexities and entropies as a function of the average connectivity, for 5-coloring of Erd\H{o}s-R\'enyi graphs see fig.~\ref{fig_col}. We plotted in dashed black the replica symmetric entropy (\ref{S_RS}), which in coloring is equal to the annealed one $s_{\rm ann}$. The correct total entropy $s_{\rm tot}$ (in red) differs from the replica symmetric one in the condensed and uncolorable phase. The complexity of the dominating clusters (those covering almost all solutions)  $\Sigma_{\rm dom}$ (in red, computed at $m=1$) is non-zero between the clustering and the condensation transition. The total complexity $\Sigma_{\rm max}$ (in blue), maximum of the curves $\Sigma(s)$, can be computed in the region where survey propagation gives a nontrivial result. The colorability threshold corresponds to $\Sigma_{\rm max}=0$. We call $c_{\rm SP}$ the smallest connectivity at which survey propagation gives a nontrivial result, i.e., the part of the curve $\Sigma(s)$ with a zero slope exists. Clusters exists also for $c<c_{\rm SP}$, but computing their total complexity is more involved and we have not done it. The rigidity transition $c_r$ cannot be determined from these quantities.  

\begin{figure}[!ht]
\begin{center}
  \resizebox{0.67\linewidth}{!}{\includegraphics{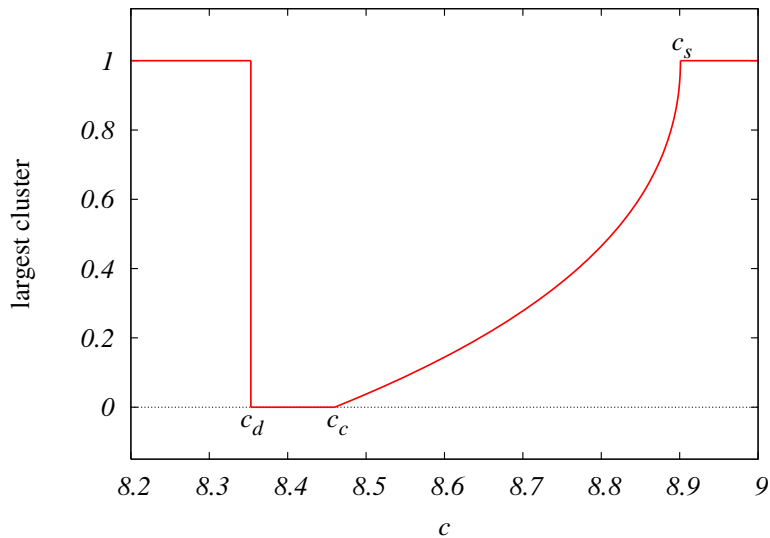}}
\end{center}
\caption{ \label{fig:largest} The fraction of solutions covered by the largest cluster as a function of the average connectivity for 4-coloring of Erd\H{o}s-R\'enyi graphs. In the condensed phase the fraction covered by the largest cluster is not self-averaging and is determined by the Poisson-Dirichlet process with parameter $m^*$.}
\end{figure}

In fig.~\ref{fig:largest} we sketch what fraction of solutions is covered by the largest cluster as the average connectivity increases  for 4-coloring of Erd\H{o}s-R\'enyi graphs. 
In the replica symmetric phase $c<c_d$ the largest cluster covers almost all solutions. In the dynamical 1RSB phase the largest cluster covers an exponentially small fraction of solutions. In the condensed  phase the largest state covers fraction of about $1-m^*$ of solutions\footnote{More precisely from the properties of the Poisson-Dirichlet process, described in sec.~\ref{sec:PD}, if the fraction of solutions covered by the largest state is $w$ then $1-m^*=1/\mathbb{E}(1/w)$.}, but this part of the curve in not self-averaging. In the uncolorable phase there are no clusters of solutions, the ground state is made from one cluster. 

\section{Large $q$ limit}

The coloring of random graphs in the limit of large number of colors might seem a very unpractical and artificial problem. However, it allows many simplifications in the statistical description (rigorous or not) and a lot of insight can be obtained from this limit. 

It is known from the cavity method, but also from a rigorous lower \cite{AchlioptasNaor05} and upper \cite{Luczak91} bound that the colorability threshold for large number of colors scales like $2q\log{q}$. At the same time a very naive algorithm: Pick at random an uncolored vertex and assign it at random a color which is not assigned to any of its neighbours, was shown to work in polynomial (linear) time up to a connectivity scaling as $q\log{q}$. In other words this algorithm uses about twice as many colors than needed. Such a performance is not very surprising, a very naive algorithm performs half as good as possible. The surprise comes with the fact that it is an open problem if there is a polynomial algorithm which would work at connectivity $(1+\epsilon)q\log{q}$ for an arbitrarily small positive $\epsilon$.

\subsection{The $2q\log{q}$ regime: colorability and condensation}
\label{sec:largeq}

The complexity function $\Sigma(s)$ at connectivity 
\be
     c=2q\log{q}-\log{q} +\gamma
\ee
where $\gamma=\Theta(1)$ was computed in [\ZK] and reads
\be
    \Sigma(s)  = \frac{s}{\log{2}} \left[1-\log{\frac{s}{\varepsilon \log{2}}}\right] -\varepsilon(2+\gamma) + o(\varepsilon).  \label{eq:lq_S}
\ee
where $\varepsilon = 1/2q$. From this expression it is easy to see that the coloring threshold corresponds to 
\be
        \gamma_s =-1. 
\ee
and the condensation transition
\be
        \gamma_c = -2\log{2}\, .  
\ee

Notice, as in [\MZ], that the complexity of the random subcubes model (\ref{sigma}), sec.~\ref{sec:subcubes}, gives exactly the expression (\ref{eq:lq_S}) if we take the parameters of the random subcubes model as~\footnote{We remind that in the section \ref{sec:subcubes} entropies were logarithms of base 2 whereas everywhere else they are natural logarithms.}
\be
    p=1-\varepsilon\, , \quad \alpha = 1+ \varepsilon \frac{1+\gamma}{\log 2}\, .
\ee
This is a striking property of the coloring problem in the limit of large number of colors near to the colorability threshold. The $1-\varepsilon$ is a fraction of frozen variables in each cluster. Almost all the soft variables can take only one of two colors. The expression (\ref{eq:lq_S}) means that the soft variables are mutually almost independent and the clusters have shape of small hypercubes. And the other way around, this property makes the random subcubes model more than just a pedagogical example of the condensation transition.

\subsection{The $q\log{q}$ regime: clustering and rigidity}

Another interesting scaling regime is defined as
\be
      c = q(\log{q}+\log{\log{q}}+\alpha) \, ,  \label{eq:lq_r}
\ee
where $\alpha=\Theta(1)$ is of order one. 
\begin{figure}[!ht]
\begin{center}
\resizebox{0.67\linewidth}{!}{
\includegraphics{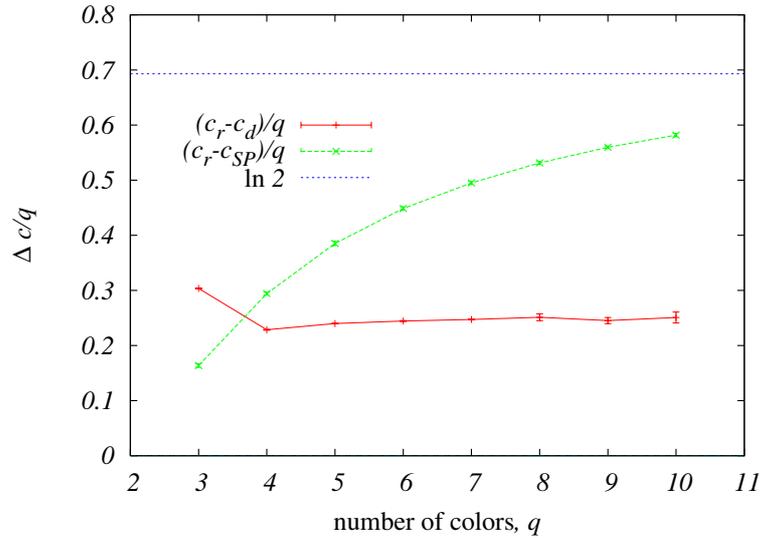}
}
\end{center}
\caption{\label{fig_largeq_cd} We plotted the difference $(c_r-c_d)/q=\alpha_r-\alpha_d$ and $(c_r-c_{\rm SP})/q=\alpha_r-\alpha_{\rm SP}$. The data are takes from table \ref{tab:results_complete}. The difference $\alpha_r-\alpha_{\rm SP}$ indeed seems to converge to the theoretical $\log{2}$, the difference $\alpha_r-\alpha_{\rm SP}$ seems to converge to around $1/4$.}
\end{figure}
The large $q$ scaling of the rigidity transition ($m=1$) is easily expressed from (\ref{eq:hard}):
\be
          \alpha_r = 1\, .   
\ee
This was originally computed in [\KM, \ZK] and \cite{Semerjian07}.
The onset of a nontrivial solution for the survey propagation corresponds to the rigidity transition at $m=0$ and reads \cite{KrzakalaPagnani04} 
\be
     \alpha_{\rm SP} = 1- \log 2\, .
\ee 
An empirical observation is that for $q=3$ the threshold for survey propagation is smaller than the rigidity at $m=1$, but for $q\ge 4$ the order changes and the distances between the two threshold grows with $q$. Based on this observation we conjectured that the clustering transition is
\be
     1-\log 2 \ge \alpha_d \ge 1 \, .
\ee
Note that recently the dynamical transition was proved to be $1-\log 2 \ge \alpha_d$ \cite{Sly08}. Fig.~\ref{fig_largeq_cd} actually suggest that $\alpha_d \approx 1/4$. Its precise location is actually an interesting problem because it could shed light on the way soft fields converge to hard fields in the cavity approach.

Concerning the total rigidity transition, where almost all the clusters of all sizes become frozen, we have not manage to compute it in the large $q$ limit. It is not even clear if the relevant scaling is as (\ref{eq:lq_r}). The same is true for the even more interesting freezing transition, where all the clusters become frozen.

\section{Finite temperature}
\label{sec:finiteT}

It is interesting to study how does the antiferromagnetic Potts model, coloring at zero temperature, behave at finite temperature. In particular which of the zero temperature phase transitions survive to positive temperatures and what do they correspond to in the phenomenology of glasses. This has been done in [\KZP] and we summarize the main results here.  

The belief propagation equation for coloring (\ref{eq:col_BP}) generalizes at finite temperature to
\begin{equation}
  \psi^{i\to j}_{s_i}= \frac{1}{Z^{i\to j}} \prod_{k\in \partial i -j} \left[1- \left(1-e^{-\beta}\right) \psi_{s_i}^{k\to i} \right] \equiv {\cal F}_{s_i}(\{ \psi^{k\to i}\})\, . \label{update_T}
\end{equation}
The distributional 1RSB equation (\ref{eq:1RSB_m1}) is the same. 

\begin{itemize}
   \item {\bf The clustering transition} --- becomes the dynamical phase transition $T_d$ at positive temperature. The notion of reconstruction on trees, introduced in sec.~\ref{trees}, generalizes to positive temperatures. Constraints then play the role of noisy channels in the broadcasting. The dynamical temperature $T_d$ is then defined via divergence of the point-to-set correlations (\ref{eq:pts}). Or equivalently via the onset of a nontrivial solution of the 1RSB equations at $m=1$. At the dynamical transition the point-to-set correlation length and the equilibration time diverge. There is however no non-analyticity in the free energy, Ehrenfest might thus not call it a phase transition.   
   \item {\bf The condensation transition} --- becomes the Kauzmann phase transition $T_K$ at positive temperature. The point at which the complexity function at $m=1$ (structural entropy) becomes negative defines the Kauzmann temperature \cite{Kauzmann48}. At the Kauzmann temperature the free energy has a discontinuity in the second derivative. This corresponds to the discontinuity in the specific heat. Kauzmann transition is thus genuine even in the sense of Ehrenfest. 
   \item {\bf The rigidity transition} --- is a purely zero temperature phase transition. At positive temperature the fields $\psi^{i\to j}_{s_i}$ (\ref{update_T}) cannot be hard.
   \item {\bf The colorability transition} --- is a purely zero temperature phase transition. At the colorability threshold the ground state energy becomes positive (it has discontinuity in the first derivative). At a finite temperature, however, there is no corresponding non-analyticity. 
\end{itemize}

\begin{figure}[!ht]
\resizebox{\linewidth}{!}{
\includegraphics{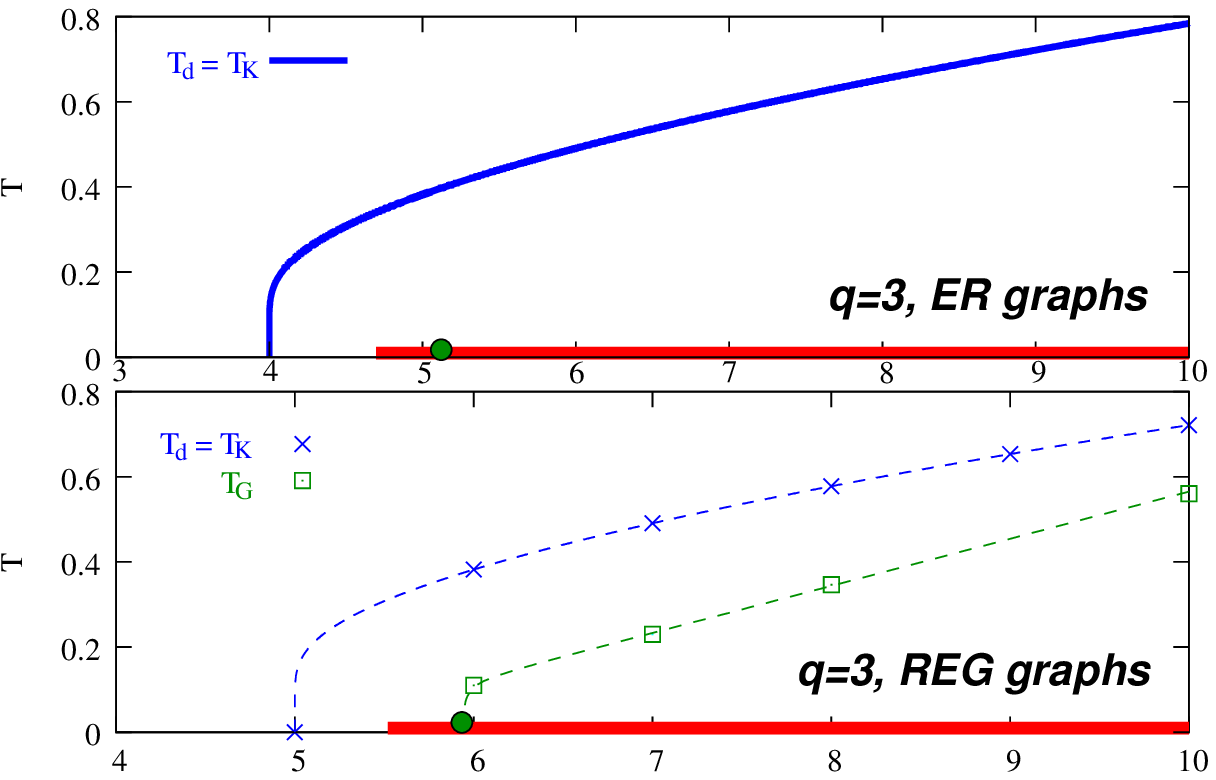}
\includegraphics{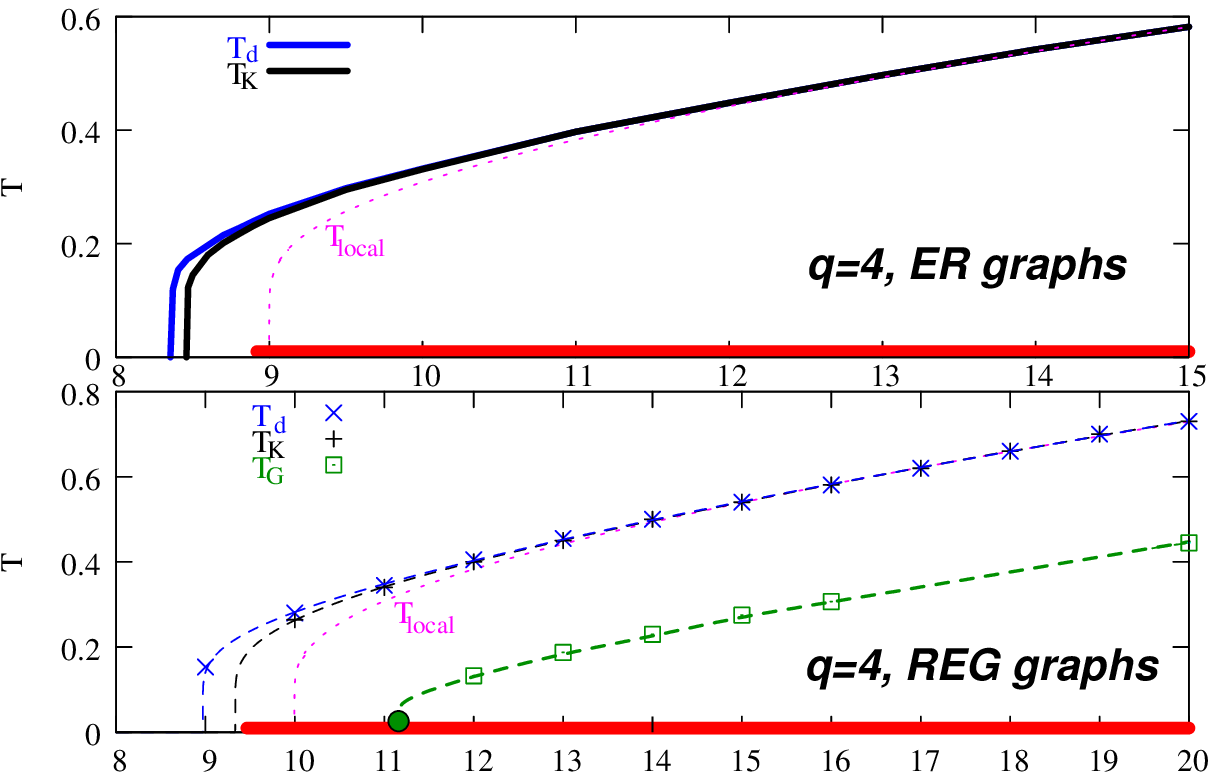}}
\caption{\label{fig_col_T} Phase diagrams for the 3-state (left) and 4-state (right) anti-ferromagnetic Potts glass on Erd\H{o}s-R\'enyi graphs of average degree $c$ (top) and regular graphs of degree $c$ (bottom). For $q=3$ the transition is continuous $T_d=T_K=T_{\rm local}$. For $q=4$, we find that $T_d > T_K > T_{\rm local}$, while for larger connectivities these three critical temperatures become almost equal. The Gardner temperature $T_G$ for regular graphs is also shown (green), bellow $T_G$ the 1RSB solution is not correct anymore (for Erd\H{o}s-R\'enyi graph we expect this curve to look similar).  The bold (red) lines at zero temperature represent the uncolorable connectivities $c>c_s$.}
\end{figure}

Fig.~\ref{fig_col_T} shows the temperature phase diagram of 3- (left) and 4-coloring (right) on both Erd\H{o}s-R\'enyi (up) and regular (down) random graphs. The dynamical temperature is in blue, the Kauzmann temperature in black. 

The temperature at which the replica symmetric solution becomes locally unstable, see appendix \ref{app:RS_stab}, is called $T_{\rm local}$. In the terms of reconstruction on trees this is the Kesten-Stigum bound \cite{KestenStigum66,KestenStigum66b}. This temperature is a lower bound on the dynamical temperature $T_d$, but also on the Kauzmann temperature $T_K$. This is because bellow $T_{\rm local}$ the two-point correlations do not decay, which is possible only bellow $T_K$. Note that in the 3-coloring $T_d=T_K=T_{\rm local}$ and this phase transition is continuous in the order parameter $P^{i\to j}( \psi^{i\to j})$ (\ref{eq:1RSB_m1}). For $q\ge 4$ colors we find instead $T_d>T_K>T_{\rm local}$ and the dynamical transition is discontinuous. At large connectivity, however, the three temperatures are very close, see fig.~\ref{fig_col_T} where the $T_{\rm local}$ is in pink.

\paragraph{Correctness of the 1RSB solution ---}
The last question concerns correctness of the 1RSB solutions itself. The local stabilities of the 1RSB solution are discussed in appendix~\ref{app:1RSB_stab}. The temperature at which the 1RSB solutions becomes type II locally unstable, see appendix \ref{app:1RSB_stab},  is called the Gardner temperature $T_G$ \cite{Gardner85}. We computed it only on the ensemble of random regular graphs, see fig.~\ref{fig_col_T}, the $T_G$ is in green.  
We do not know how to compute the stability of the type I, but we argued that the corresponding critical temperature should be smaller than the $T_{\rm local}$. An important consequence is that in the colorable region the 1RSB solution is stable for $q\ge 4$ coloring. 

Coloring with three colors is a bit special, as $T_{\rm local}=T_d=T_K$. However, at small temperatures the stability of type I can be investigated from the energetic approach, again discussed in app. \ref{app:1RSB_stab}. It follows that at least in interval $c\in (c_s,c_G)=(4.69,5.08)$ the 1RSB solution is stable at low temperature. For $c>c_G$ on contrary the Gardner temperature is strictly positive. We cannot exclude that part of the colorable phase is unstable, but in such a case the unstable region would have a sort of re-entrant behaviour. Moreover the ferromagnetic fully connected 3-state Potts model has also a continuous dynamical transition $T_d=T_{\rm local}$ yet it is 1RSB stable near to $T_d$ \cite{GrossKanter85}. We thus find more likely that also the colorable phase of 3-coloring is 1RSB stable. 

Finally, the local stability is only a necessary condition. The full correctness of the 1RSB approach have to be investigated from the 2RSB approach. We implemented the 2RSB on the regular coloring, the results are not conclusive, as the numerics is involved. but we have not found any sign for a nontrivial 2RSB solution in the colorable region.

\chapter*{Conclusions and perspectives}
\label{conclusions}
\addcontentsline{toc}{chapter}{Conclusions and perspectives}
\markboth{CONCLUSIONS AND PERSPECTIVES}{CONCLUSIONS AND PERSPECTIVES}

{\it In this final section we highlight the, in our view, most interesting results of this thesis. More complete overview of the original contributions is presented in sec.~\ref{sec:summary}. Scientific research is such that every answered question raises a number of new questions to be answered. We thus bring up a list of open problems which we find particularly intrinsic. Finally we give a brief personal view on the perspective applications of the results obtained in this work. }

\section*{Key results}
\addcontentsline{toc}{section}{Key results}

The main question underlying this study is: How to recognize if an NP-complete problem is typically hard and what are the main reasons for this? 

In order to approach the answer we studied the structure of solutions in random constraint satisfaction problem - mainly in the graph coloring. We did not neglect the entropic contributions, as was common in previous studies, and this led to much more complete description of the phase diagram and associated phase transitions, see summarizing fig.~\ref{fig:diag}.

The most interesting concept in these new findings was the freezing of variables. We pursued its study and investigated its relation to the average computational hardness. We introduced the {\it locked} constraint satisfaction, where the statistical description is easily solvable and the clustered phase is automatically frozen. We indeed observed empirically that these problems are much harder than the canonical K-satisfiability. They should thus become a new challenge for algorithmical development. As we mention in the perspectives, we also anticipate that the locked constraint satisfaction problems are of a more general interest.

\section*{Some open problems}
\addcontentsline{toc}{section}{Some open problems}

\paragraph{(A) Clusters and their counting  on trees ---} 
In sec.~\ref{clustering} we derived the 1RSB equations on purely tree graphs. Our derivation was, however, not complete as it is not straightforward why the complexity function should be counting the clusters as we defined them on trees. More physically founded derivations are for example the original one \cite{MezardParisi00}. And also the one presented in \cite{MezardMontanari07} where the complexity is shown to count the fixed points of the belief propagation. We are, however, persuaded that the purely tree approach is more appealing from the probabilistic point of view, as treating correlations in the boundary conditions on trees is easier than treating the random graphs directly, for a recent progress see e.g. \cite{Sly08,GerschenfeldMontanari07,DemboMontanari08}. This is why we chose to present this derivation despite its incompleteness. 

In general we should say that creating better mathematical grounds for the replica symmetry breaking approach is a very important and challenging task.

\paragraph{(B) What is the meaning of the gap in the $\Sigma(s)$ function ---} We computed the number of clusters of a given entropy via the 1RSB method. For some intervals of parameters there is no solution corresponding to certain intermediate sizes. In other words there is a gap in the 1RSB function $\Sigma(s)$. See e.g. fig.~\ref{fig:diag}, we observed such a gap in many other cases. Does this gap mean that there are truly no clusters of corresponding sizes or does it mean that the 1RSB method is wrong in that region or is there another explanation?

\paragraph{(C) Analysis of dynamical processes ---} In this thesis we described in quite a detail the static (equilibrium) properties of the constraint satisfaction problems. Very little is known about the dynamical properties -- here we mean both the physical dynamics (with detailed balance) and the dynamics of algorithms. Focusing on results described here: the dynamics of the random subcubes model can be solved [\MZ], and the uniform belief propagation decimation can be analyzed \cite{MontanariRicci07}, see also appendix \ref{sec:BP_anal}. However in general even the performance of simulated annealing as a solver is not known. And the understanding of why the survey propagation decimation works so well in 3-SAT and not that well in other problems, e.g. the locked problems or for larger $K$, is also very pure. 

The most exciting conjecture of this work is the connection between the algorithmical hardness and freezing of variables. Several indirect arguments and empirical results were explained in sec.~\ref{sec:hardness} to support this conjecture. It is, however, not very clear what is the detailed origin of the connection between presence of frozen variables in solutions and the fact that dynamics (of a solver) does not seem to be able to find them.

\paragraph{(D) Beyond random graphs and the thermodynamical limit ---} 

For practical application the perhaps most important point is to understand what is the relevance of our results for instances which are not random or not infinite. For example fig.~\ref{fig:cmplx} suggests that even on small random instances the clustering can be observed and is thus probably relevant. We also observed that the solutions-related quantities seems to have stronger finite size effects than the clusters-related properties, compare e.g. fig.~\ref{fig:sat_threshold} with fig.~\ref{fig:rigidity}. This is an interesting point and it should be pursued. 

\section*{Perspectives}
\addcontentsline{toc}{section}{Perspectives}

This work should have a practical impact on the design of new solvers of constraint satisfaction problems. Instances with only frozen solutions should be used as new benchmarks for SAT solvers. At the same time where the design allows such instances should be avoided. 

More concretely, the belief propagation algorithm is used as a standard approximative inference technique in artificial intelligence and information theory. One of the important problems with applications of the belief propagation is the fact that in many cases it does not converge. Many converging modifications were introduced. In might be interesting to investigate in this context the reinforced belief propagation, see appendix \ref{sec:rein}, which sometimes converges towards a fixed point when the standard belief propagation does not. As the reinforcement algorithm seems to be very efficient, robust and is not theoretically well understood different variants of the implementation should be studied empirically. It would be interesting to see if this algorithm performs well on non-random graphs, or if it can provide information useful for the practical solvers. Several other concepts enhanced in this thesis might show up useful in algorithmic applications. We feel that the whitening of solutions might be one of them. 

We introduced the locked constraint satisfaction problems as a new algorithmical challenge. Moreover the simplicity of their statistical description makes accessible several quantities which are difficult to compute in the $K$-SAT problem. For example the weight enumerator function or the $x$-satisfiability threshold. But these new models are exciting from many other points of view. Their hardness might be appealing for noise tolerant cryptographic applications. Planted ensemble of the locked problems might be a very good one-way functions. The fact that the solutions of the locked problems are well separated makes them excellent candidates for nonlinear error correcting codes. It will be interesting to investigate if they can be advantageous over the standard linear low-density-parity-check codes \cite{Gallager62,Gallager68,MacKayNeal95,Montanari01}. 

Clusters of solutions come up naturally in the pattern recognition and machine learning problems. There each cluster corresponds to a pattern which should be learned or recognized. Similarly the different phenotypes of a cell might be viewed as clusters of fixed points of the corresponding gene regulation network. The methods developed in this thesis might thus have impact also in these exciting fields.

\begin{appendices}
\appendixpage
\addappheadtotoc


\chapter{1RSB cavity equations at $m=1$}
\label{app:m1}

Here we derive how the 1RSB equation (\ref{eq:1RSB_m1}) simplifies at $m=1$ for the problems where the replica symmetric solution is not factorized. We restrict to the occupation models, but a generalization to other models is straightforward. Advantage of these equations is that the unknown object is not a functional of functionals but only a single functional. Moreover, the final self-consistent equation does not contain the reweighting term. This simplification makes implementation of the population dynamics at $m=1$ much simpler, and thus the computation of the clustering and condensation transitions easier. 
Derivation of the corresponding equations for the $K$-SAT problem can be found in \cite{MontanariRicci08}. 

Write the RS equation (\ref{eq:BP_one}) for the occupation problems in the form
\be
   \psi_{s_i}^{a\to i} = \frac{1}{Z^{j\to i}} \sum_{\{s_j\}} C_a(\{s_j\},s_i) \prod_{j\in \partial a -i}\left( \prod_{b\in \partial j-a} \psi_{s_j}^{b\to j} \right) \equiv {\cal F}_{s_i}(\{\psi^{b\to j}\})\, , \label{eq:m1_RS}
\ee
where the constraints $C_a(\{s_j\},s_i)=1$ if $\sum_j s_j +s_i \in A$, and $0$ otherwise. Let ${\cal P}_{\rm RS}(\psi)$ be the distribution of RS fields over the graph.

The 1RSB equations (\ref{eq:1RSB_m1}) at $m=1$ are 
\bea
     P^{a\to i}(\psi^{a\to i}) &=& \frac{1}{{\cal Z}^{j\to i}} \int \prod_{j \in \partial a -i} \prod_{b \in \partial j -a} \left[{\rm d}\psi^{b\to j} P^{b\to j}(\psi^{b\to j})\right] \nonumber \\
& & Z^{j\to i}(\{\psi^{b\to j}\})\,  \delta\left[\psi^{a\to i} - {\cal F}(\{\psi^{b\to j}\})\right]\equiv {\cal F}_2(\{P^{b\to j}\}) \, .\label{1RSB_link}
\eea
The averages over states
\be
   \overline \psi_{s_i}^{a\to i} = \int {\rm d}\psi_{s_i}^{a\to i} P^{a\to i}(\psi_{s_i}^{a\to i}) \, \psi_{s_i}^{a\to i} \label{eq:psi_mean}
\ee
satisfy the RS equation (\ref{eq:m1_RS}). And consequently the RS and 1RSB normalizations are equal ${\cal Z}^{j\to i}=Z^{j\to i}$. The full order parameter is the probability distribution of $P$'s over the graph, it follow the self-consistent equation
\bea
       {\cal P}_{\rm 1RSB}[P(\psi)] &=& \sum_{l_1,\dots l_{K-1}} q(l_1, \dots, l_{K-1}) \nonumber \\ &&\int \prod_{i=1}^{K-1} \prod_{j=1}^{l_i} \Big\{ {\rm d} P^j(\psi^j) \,  {\cal P}^j_{\rm 1RSB}[P^j(\psi^j)] \Big\} \, \delta[P(\psi) - {\cal F}_2(\{P^j\})]\, . \label{1RSB_gr}
\eea

We define the average distribution $\overline P(\psi|\overline \psi)$ on those edges where the RS field is equal to a given value $\overline \psi$
\be
   \overline P(\psi|\overline \psi) {\cal P}_{\rm RS}(\overline \psi) \equiv \int {\rm d}P(\psi)\,  {\cal P}_{\rm 1RSB}[P(\psi)] \,  P(\psi)\,  \delta\left[\overline \psi - \int {\rm d}\psi P(\psi) \psi\right]  \, . \label{new_P} 
\ee
Now we rewrite all the terms on the right hand side using the incoming fields and distributions, i.e., using first eq.~(\ref{1RSB_gr}) and then (\ref{1RSB_link}). 
\bea
   \overline P(\psi|\overline \psi) {\cal P}_{\rm RS}(\overline \psi)&=& \sum_{\{l\}} q(\{l\})  \int \prod_{i=1}^{K-1} \prod_{j=1}^{l_i} \Big\{ {\rm d} P^j(\psi^j)  {\cal P}^j_{\rm 1RSB}[P^j(\psi^j)] \Big\} \nonumber \\ && {\cal F}_2(\{P^j\}) \,  \delta\left[\overline \psi - \int {\rm d}\psi\,  {\cal F}_2(\{P^j\})\,  \psi\right]\nonumber \\
&=& \sum_{\{l\}} q(\{l\})  \int \prod_{i=1}^{K-1} \prod_{j=1}^{l_i} \Big\{ {\rm d} P^j(\psi^j)  {\cal P}^j_{\rm 1RSB}[P^j(\psi^j)] \Big\}  \nonumber 
\\ && \int \prod_{i=1}^{K-1} \prod_{j=1}^{l_i} \left[{\rm d}\psi^j P^{j}(\psi^j)\right]  \frac{Z(\{\psi^{j}\})}{{\cal Z}  } \,  \delta\left[\psi - {\cal F}(\{\psi^{j}\})\right]
\delta\left[\overline \psi - {\cal F}(\{\overline \psi^j\}) \right] \nonumber \\
&=& \sum_{\{l\}} q(\{l\}) \int \prod_{i=1}^{K-1} \prod_{j=1}^{l_i} \left[ {\rm d} \overline \psi^j {\cal P}_{\rm RS}(\overline \psi^j) \right]   \delta\left[\overline \psi - {\cal F}(\{\overline \psi^j\}) \right]\nonumber
\\ && \int \prod_{i=1}^{K-1} \prod_{j=1}^{l_i} \left[{\rm d}\psi^j \overline P^{j}(\psi^j|\overline \psi^j)\right]  \frac{Z(\{\psi^{j}\})}{Z(\{\overline \psi^{j}\})  } \,  \delta\left[\psi - {\cal F}(\{\psi^{j}\})\right]\, ,
\eea
where the original Dirac function was rewritten using
\bea
    \int {\rm d}\psi\,  {\cal F}_2(\{P^j\}) \, \psi &=&\frac{1}{{\cal Z}} \int \prod_{i=1}^{K-1} \prod_{j=1}^{l_i} \left[{\rm d}\psi^j P^{j}(\psi^j\right]  Z(\{\psi^j\})\,  \int {\rm d}\psi \, \psi \, \delta\left[\psi - {\cal F}(\{\psi^j\})\right]\nonumber\\
   &=& \frac{1}{{\cal Z}} \int \prod_{i=1}^{K-1} \prod_{j=1}^{l_i} \left[{\rm d}\psi^j P^{j}(\psi^j\right]  Z(\{\psi^j\})\,   {\cal F}(\{\psi^j\} = {\cal F}(\{\overline \psi^j\}) \, ,
\eea
and in last equality was obtained using the integral of eq.~(\ref{new_P})
\be
      \int {\rm d}\overline \psi\,  \overline P(\psi|\overline \psi)\,  {\cal P}_{\rm RS}(\overline \psi) = \int {\rm d}P(\psi)\,  {\cal P}_{\rm 1RSB}[P(\psi)] \, P(\psi)\, .
\ee

To simplify the equations further, in particular to get rid of the reweighting term $Z(\{\psi^{j}\})$, we define a distribution $\overline P_s$ 
\be
       \overline \psi_s \overline P_s(\psi|\overline \psi) \equiv \psi_s \overline P(\psi|\overline \psi)  \quad  \Rightarrow \quad \overline P(\psi|\overline \psi) = \sum_s  \overline \psi_s \overline P_s(\psi|\overline \psi)   \, ,
\ee
then by factorizing the sum over components $s$ we get
\bea
    \overline \psi_s \overline P_s(\psi|\overline \psi) {\cal P}_{\rm RS}(\overline \psi) &=& 
\sum_{\{l\}} q(\{l\})  \int \prod_{i=1}^{K-1} \prod_{j=1}^{l_i} \left[ {\rm d} \overline \psi^j {\cal P}_{\rm RS}(\overline \psi^j) \right] \,  \delta\left[\overline \psi - {\cal F}(\{\overline \psi^j\})\right] \nonumber \\ && \sum_{\{s_i\}} C(\{s_i\},s)\,  \frac{\prod_{i=1}^{K-1} \prod_{j=1}^{l_i} \overline \psi^j_{s_i}}{Z(\{\overline \psi^{j}\})} \nonumber
 \\ && \int \prod_{i=1}^{K-1} \prod_{j=1}^{l_i} \left[{\rm d}\psi^j \overline P_{s_i}^{j}(\psi^j|\overline \psi^j)\right]  \,  \delta\left[\psi - {\cal F}(\{\psi^{j}\})\right]\, . \label{final_m1}
\eea
This final equation might look more complicated than the original one, but, in fact, it is much easier to solve. It could seem that we need a population of populations to represent the distribution  $\overline P_s(\psi|\overline \psi) {\cal P}_{\rm RS}(\overline \psi)$. But keeping in mind that the proper initial conditions are
\be
  \overline P_1(\psi_1=1|\overline \psi)=1\, ,  \quad \quad  \overline P_0(\psi_0=1|\overline \psi)=1\, ,   \label{component}  
\ee
independently of the RS field $\overline \psi$ we see that the probability 
distribution $\overline P_s(\psi|\overline \psi) {\cal P}_{\rm RS}(\overline \psi)$ 
may be represented by a population of triplets of fields - the first one corresponding to the RS field $\overline \psi$ and the other two corresponding to the two components (\ref{component}).

In the population dynamics we first equilibrate the RS distribution ${\cal P}_{\rm RS}(\overline \psi)$ and then initialize the other two components according to (\ref{component}). In every step of the update we first fix randomly the set of indexes $\{j\}$ and compute the new $\overline \psi$, then given the value $s$ we choose the set of indexes $\{s_i\}$ according to a probability law given by the first line of eq.~(\ref{final_m1}), then we compute the new $\psi$ for $s=0$ and $s=1$ and change a random triplet in the population for the new values.
In summary, eq.~(\ref{final_m1}) allows to reduce the double-functional equations at $m=1$ into a simple-functional form, which is much easier to solve. 

The internal entropy $s=s_{RS}-\Sigma$, and thus also the complexity function, may be computed by making very similar manipulations as
\bea
     s&=& \alpha \sum_{\{l\}} q(\{l\})  \int \prod_{i=1}^{K} \prod_{j=1}^{l_i} \left[ {\rm d} \overline \psi^j {\cal P}_{\rm RS}(\overline \psi^j) \right] \frac{\sum_{\{s_i\}} C(\{s_i\}) \prod_{i=1}^K \prod_{j=1}^{l_i} \overline \psi^{j}_{s_i}}{Z^{a+\partial a}(\{\overline \psi^{j}\})}\nonumber\\ && \int\prod_{i=1}^K \prod_{j=1}^{l_i} \left[ {\rm d}\psi^{j} P_{s_i}^{j}(\psi^{j}|\overline \psi^{j}) \right] \log{Z^{a+\partial a}(\{\psi^{j}\})}\nonumber\\ &-& \sum_{l} {\cal Q}(l) (l-1)   \int \prod_{i=1}^{l} \left[ {\rm d} \overline \psi^i {\cal P}_{\rm RS}(\overline \psi^i) \right]  \frac{ \sum_{s_i} \prod_{i=1}^l \overline \psi^i_{s_i} }{Z^{i}(\{\overline \psi ^{i}\})}\nonumber \\ &&\int \prod_{i=1}^l \left[ {\rm d}\psi^{i} P_{s_i}^{i}(\psi^{i}|\overline \psi^{i}) \right] \log{Z^{i}(\{\psi^{i}\})}\, .
\eea

We can also express other quantities, e.g. the inter $q_0=q_{RS}$ and intra $q_1$ state overlaps. 
\be
    q_1 = \int {\rm d}P(\psi) {\cal P}_{\rm 1RSB} \int {\rm d}\psi P(\psi) \sum_{\sigma} \psi_{\sigma} = \sum_{\sigma,s} \int {\rm d}\overline \psi\,   {\cal P}_{\rm RS}(\overline \psi) \, \overline \psi_s \int {\rm d}\psi \, \overline P_s(\psi|\overline \psi) \, \psi_{\sigma}^2 \, .
\ee

\paragraph{Factorized RS solution ---}

Several times,  see e.g. sec.~\ref{sec:sn_rec}, we used the equations at $m=1$ for problems with factorized RS solution, ${\cal P}_{\rm RS}(\psi) = \delta(\psi - \overline \psi)$. The derivation is straightforward from (\ref{final_m1})
\be
    \overline P_s(\psi)=\sum_{\{l\}} q(\{l\}) \frac{1}{\overline \psi_s Z}  \sum_{\{s_i\}} C(\{s_i\},s)\prod_{i=1}^{K-1}
\prod_{j=1}^{l_i} \overline \psi^{j}_{s_i}  \int \prod_{i=1}^{K-1}\prod_{j=1}^{l_i} {\rm d}\overline{P}_{s_i}(\psi^j)  \,  \delta{(\psi-{\cal F}(\psi^j))} \, . \label{m1_1RSB}
\ee
Proper initial conditions for the population dynamics resolution of (\ref{m1_1RSB}) is $ \overline P_s(\psi_s=1)=1$.

At zero temperature the distributions can be written as the sum of the frozen and soft part
\begin{subequations}
\bea
      \overline P_1(\psi) = \mu_1 \delta{(\psi-{1 \choose 0})} + (1-\mu_1) \tilde{P_1}(\psi) \, ,\\
      \overline P_0(\psi) = \mu_0 \delta{(\psi-{0 \choose 1})} + (1-\mu_0) \tilde{P_0}(\psi) \, .
\eea
\end{subequations}
Self-consistent equations for the fractions of hard fields $\mu_1$, $\mu_0$ (\ref{eq:mu1}-\ref{eq:mu0}) follow from (\ref{m1_1RSB}).

\chapter{Exact entropy for the balanced LOPs}
\label{app:moments}

Rigorous results about the entropy and the satisfiability threshold can be obtain comparing the first and second moment of the number of solutions, that is: If a number of solution on a graph $G$ is ${\cal N}_G$ then the first moment is average over the graph ensemble:
\be
      \langle {\cal N}_G \rangle =  \sum_{\{\sigma\}} {\rm Prob}\left(\{\sigma\}\,  {\rm is}\,  {\rm SAT}\right)\, .
\ee
The second moment is 
\be
     \langle {\cal N}^2_G \rangle =  \sum_{\{\sigma_1\},\{\sigma_2\}} {\rm Prob}\left(\{\sigma_1\}\,  {\rm and} \,  \{\sigma_2\} \,  {\rm are} \,  {\rm both} \, {\rm SAT}\right) \, .  \label{2nd}
\ee
The Markov inequality then gives an upper bound on the entropy and the  satisfiability threshold 
\be
      {\rm Prob}({\cal N}_G>0) \le   \langle {\cal N}_G \rangle \, .
\ee  
The Chebyshev's inequality gives a lower bound via
\be
      {\rm Prob}({\cal N}_G>0) \ge   \frac{ \langle {\cal N}^2_G \rangle }{ \langle {\cal N}_G \rangle^2}  \, .
\ee

\section{The $1^{\rm st}$ moment for occupation models}

Let us remind that the occupation models are defined via a $(K+1)$-component vector $A$, such that $A_i=1$ if and only if there can be $i$ occupied particles around a constraint of $K$ variables. We consider by default $A_0=A_K=0$, i.e., that everybody full of empty is not a solution. We also consider all the $M$ constraints are the same. We have $Q(l)N$ variables of connectivity $l$, where $\sum_{l=0}^\infty Q(l) = 1$ and $\overline l = \sum_{i=0}^\infty l Q(l) = KM/N$.

In order to compute the first moment we divide variables into groups according to their connectivity and in each groups we choose fraction $t_l$ of occupied variables. Number of ways in which this is possible is then multiplied by a probability that such a configuration satisfies simultaneously all the constraints. 
\bea
    \langle {\cal N}_G \rangle = \int_0^1 {\rm d}t \sum_{\{t_l\}} \prod_l {Q(l) N \choose t_l Q(l) N} \sum_{r_1,\dots,r_M=1}^K \prod_{a=1}^M   \delta(A_{r_a}-1)  { N\sum_l l (1-t_l)Q(l) \choose (K- r_1) \dots (K-r_M) }\nonumber \\ {N\sum_l  l \, t_l Q(l)\choose r_1 \dots r_M}
\left[ { \overline l N \choose K \dots K} \right]^{-1}   \delta\left(\sum_{a=1}^M r_a - \overline l t N\right)\,  \delta\left( t\, \overline l N - \sum_l l\,  t_l Q(l) N\right) \, ,\label{1st_a}
\eea
where $t$ is the total fraction of occupied variables, this variable might seem ambiguous, as it can be integrated out, but we will appreciate its usefulness later, $r_a$ is a number of occupied variables in a constraint $a$.

We develop expression (\ref{1st_a}) in the exponential order. In order to do so we exchange the last two delta functions by their Fourier transforms, introducing two complex Lagrange parameters $\log{x}$ and $\log{u}$. 
\bea
    \langle {\cal N}_G \rangle &\approx& \int {\rm d}t \int \prod_l {\rm d}t_l \int {\rm d}x \int {\rm d}u  \exp N \Bigg\{ -\sum_l Q(l) \left[t_l \log{t_l} + (1-t_l)\log{(1-t_l)}\right]  \nonumber \\ &+& \overline l \left[t \log{t}+ (1-t)\log{(1-t)}\right]   + \log{u} \left[\sum_l l t_l Q(l)-t\overline l\right]  \nonumber \\  &+& \frac{\overline l}{K} \log{\left[\sum_{r=1}^K \delta(A_r-1) {K\choose r} x^r\right]} -t\overline l \log{x}\Bigg\} \, .\label{1st_b}
\eea
Saddle point with respect to parameters $t_l$ gives us 
\be
     t_l = \frac{u^l}{1+t^l}\, ,
\ee
and we call $p_A(x)=\sum_{r=1}^K \delta(A_r-1) {K\choose r} x^r$. Using this we have 
\bea
    \langle {\cal N}_G \rangle &\approx& \int   {\rm d}t \, {\rm d}x \, {\rm d}u \,  \exp N \Bigg\{  \frac{\overline l}{K} \log{p_A(x)} -t\, \overline l \log{x} \nonumber \\  &+& \sum_l Q(l) \log{(1+u^l)} - t\, \overline l \log{u} + \overline l \left[t \log{t}+ (1-t)\log{(1-t)}\right]   \Bigg\}  \, .  \label{1st_c}
\eea
The saddle point equations read
\begin{subequations}
\label{eq:saddle}
\bea
 \partial_u: \quad   t&=&\frac{1}{\overline l} \sum_l l\,  Q(l) \frac{u^l}{1+u^l}\, , \label{saddle_u}\\
 \partial_x: \quad    t&=& \frac{x \partial_x p_A(x)}{K p_A(x)}\, , \label{saddle_x}\\
 \partial_t: \quad   t&=& \frac{xu}{1+xu}\, ,\label{saddle_t}
\eea
\end{subequations}

As the parameter $t$ is the only physically meaningful from the three, the goal is to express the annealed entropy as a function of $t$ and find its maxima. We do that by inverting numerically (\ref{saddle_u}) and plugging (\ref{saddle_t}) in (\ref{1st_c}). Eq.~(\ref{saddle_t}) then express the saddle point with respect to the parameter $t$. We can write 
\be
    s_{\rm ann}(t) = \sum_l Q(l) \log{[1+u(t)^l]} + \frac{\overline l}{K} \log{p_A(t)}\, , \label{1st_d}
\ee
where
\be 
p_A(t) = \sum_{r=1}^K \delta(A_r-1) {K \choose r}  \left(\frac{t}{u(t)}\right)^r (1-t)^{K-r} \, ,
\ee
where $u(t)$ is an inverse of (\ref{saddle_u}). 

For the regular graphs $Q(l)=\delta(l-L)$ the inverse of (\ref{saddle_u}) is explicit $u=[t/(1-t)]^{1/L}$ and thus 
\be
  s_{\rm ann \, reg }(t) =  \frac{L}{K} \log{\left\{ \sum_{r=1}^K \delta(A_r-1) {K \choose r}  \left[ t^r (1-t)^{K-r} \right]^{\frac{L-1}{L}}  \right\} }\, .
\ee

\section{The $2^{\rm nd}$ moment for occupation models}

The second moment is computed in a similar manner. First we fix that in a fraction $t_{x,l}$ of nodes of connectivity $l$ the variable is occupied in both the solutions $\sigma_1,\sigma_2$ in (\ref{2nd}). In a fraction $t_{y,l}$ the variable is occupied in $\sigma_1$ and empty in $\sigma_2$ and the other way round for $t_{z,l}$. We sum over all possible combinations of $0\le t_{x,l}, t_{y,l}, t_{z,l}$ such that $\sum_{w=x,y,z} t_{w,l}\le 1$. All this is multiplied by the probability that such two configurations $\sigma_1, \sigma_2$ both satisfy all the constraints. 
\bea
   && \langle {\cal N}^2_G \rangle= \int {\rm d}t_x {\rm d}t_y {\rm d}t_z  \sum_{\{t_{x,l}\},\{t_{y,l}\},\{t_{z,l}\}} \prod_l {Q(l) N \choose (t_{x,l} Q(l) N) \, (t_{y,l} Q(l) N) \, (t_{z,l} Q(l) N) } \nonumber \\ & &\sum_{r_{x,1},\dots,r_{x,M}} \sum_{r_{y,1},\dots,r_{y,M}} \sum_{r_{z,1},\dots,r_{z,M}} \prod_{a=1}^M  \delta(A_{r_{x,a}+r_{y,a}}-1) \delta(A_{r_{x,a}+r_{z,a}}-1)  \nonumber \\ & &  { N\sum_l l (1-\sum_{w=x,y,z} t_{w,l})Q(l) \choose (K- \sum_{w=x,y,z} r_{w,1}) \dots (K-\sum_{w=x,y,z}r_{w,M}) } \prod_{w=x,y,z}  {N\sum_l  l \, t_{w,l} Q(l)\choose r_{w,1} \dots r_{w,M}}  \nonumber \\ & &  \left[ { \overline l N \choose K \dots K} \right]^{-1}  
\prod_{w=x,y,z} 
  \delta\left(\sum_{a=1}^M r_{w,a} - \overline l t_w N\right)\,  \delta\left( t_w\, \overline l N - \sum_l l\,  t_{w,l} Q(l) N\right)\, .
\label{2nd_a}
\eea
We introduce Fourier transforms at a place of both the Dirac functions, the conjugated parameters are $\log{x},\log{y},\log{z}$ for the first Dirac function, and $\log{u_x},\log{u_y},\log{u_z}$ for the second one. After that we suppress the parameters $t_{w,l}$ in the same manner as we did for the first moment. We obtain for the second moment entropy 
\bea
  s_{\rm 2nd} &=&  \overline l \left[ t_x\log{t_x}+ t_y\log{t_y} + t_z\log{t_z} + (1-t_x-t_y-t_z)\log{(1-t_x-t_y-t_z)} \right] \nonumber \\ & &  - \overline l (t_x \log{x} + t_y\log{y} + t_z\log{z}) + \frac{\overline l }{K} \log{p_A{(x,y,z)}}\nonumber \\ & &  + \sum_l Q(l) \log{(1+u_x^l +u_y^l +u_z^l)} - \overline l ( t_x \log{u_x} + t_y\log{u_y} + t_z\log{u_z} )\, , \label{2nd_b}
\eea
where
\be
  p_A{(x,y,z)} = \sum_{r_1,r_2=0}^K  \delta{(A_{r_1}A_{r_2}-1)}  \sum_{s=\max{(0,r_1+r_2-K)}}^{\min{(r_1,r_2)}} {K \choose (r_1-s) (r_2-s) \, s} x^s y^{(r_1-s)} z^{(r_2-s)}\, ,
\ee
and the saddle point with respect to $t_w$, $w$ and $u_w$ ($w=x,y,z$) is
\begin{subequations}
\label{eq:saddle_2}
\bea
  \partial_{t_w}: \quad t_w &=& \frac{1}{\overline l} \sum_l l\, Q(l) \frac{u_w^l}{1+ u_x^l +u_y^l+u_z^l}\, ,  \quad \quad  w=x,y,z \, , \label{saddle_tw}\\ 
  \partial_{w}:  \quad t_w&=&\frac{w \partial_w p_A(x,y,z)}{Kp_A(x,y,z)}\, ,  \quad \quad  w=x,y,z \, ,\label{saddle_w}\\
  \partial_{u_w}: \quad w u_w &=& \frac{t_w}{1-t_x-t_y-t_z}\, ,  \quad \quad  w=x,y,z \, .\label{saddle_uw}
\eea
\end{subequations}

Once again the parameters $t_w$ are physically meaningful, so we want to express $s_{\rm 2nd}$ as a function of these. We thus need to inverse (\ref{saddle_tw}), note that such an inverse is well defined, and using (\ref{saddle_uw}) we obtain
\be
   s_{\rm 2nd}(t_x,t_y,t_z) = \frac{\overline l}{K} \log{p_A(t_x,t_y,t_z)}+ \sum_l Q(l) \log{\left\{1+ \sum_{w\in \{x,y,z\}} [u_w(t_x,t_y,t_z)]^l\right\}}  \, , \label{2nd_d}
\ee
where 
\bea
   && p_A(t_x,t_y,t_z) = \sum_{r_1,r_2=0}^K  \delta{(A_{r_1}A_{r_2}-1)}  \sum_{s=\max{(0,r_1+r_2-K)}}^{\min{(r_1,r_2)}} {K \choose (r_1-s) (r_2-s) \, s} \left( \frac{t_x}{u_x(t_x,t_y,t_z)} \right)^s \nonumber \\ & &\left( \frac{t_y}{u_y(t_x,t_y,t_z)} \right)^{(r_1-s)} \left( \frac{t_z}{u_z(t_x,t_y,t_z)} \right)^{(r_2-s)}  (1-t_x-t_y-t_z)^{(K-r_1-r_2+s)}\, . \label{pttt}
\eea
The global maximum with respect to $t_x,t_y,t_z$ needs to be found. 

For the regular ensemble $Q(l)=\delta(l-L)$ the function (\ref{saddle_tw}) is explicitly reversible and the final expression for the second moment entropy simplifies significantly
\bea
   s_{\rm 2nd,reg}(t_x,t_y,t_z)& =& \frac{L}{K} \log \Bigg\{ \sum_{r_1,r_2,s}  \frac{ K! \delta{(A_{r_1}-1)} \delta{(A_{r_2}-1)}  }{ (r_1-s)! \,  (r_2-s)! \, s! \, (K-r_1-r_2+s)!} \nonumber \\ & &
\left[ t_x^{s} t_y^{(r_1-s)} t_z^{(r_2-s)}  (1-\sum_w t_w)^{(K-r_1-r_2+s)}  \right]^{\frac{L-1}{L}} \Bigg\}\, ,
\eea
where the range of summations is the same as in (\ref{pttt}).

\section{The results}

The main result is that for some of the symmetric ($A_{K-r}=A_r$ for all $r=0,\dots,K$) and locked occupation problems ($Q(0)=Q(1)=0$) the first and second moments computation leads the exact entropy of solutions (\ref{eq:s_sym}). And thus also the exact satisfiability threshold. The cases where this statement holds are marked by a $\ast$ in tab.~\ref{tab:locked}, and we call them {\it balanced} LOPs. We observed that some of the balanced problems $A$ are created iteratively starting from $010$ or $01010$ and adding
\be
    A^{K+2} = 0A^K0\, ,   \quad \quad A^{K+4}= 01A^K10\, . \label{eq:balanced}
\ee
We, however, found also other balanced cases than (\ref{eq:balanced}). The simplest example of symmetric locked problem which is not balanced is $A=010010$, and many others of higher $K$.

Let us now show this result. For all the symmetric occupation problems: 
\begin{itemize}
  \item{The annealed entropy (\ref{1st_d}) has a stationary point at $t=1/2$ ($u=1$, $x=1$). At this stationary the entropy evaluates to (\ref{eq:s_sym}).}
  \item{The second moments entropy (\ref{2nd_d}) has a stationary point at $t_x=t_y=t_z=1/4$ ($u_x=u_y=u_z=1$, $x=y=z=1$). At this stationary point the second moment entropy evaluates to twice the (\ref{eq:s_sym}). To prove this statement observe that for the symmetric problems $p_A(1/4,1/4,1/4)=[p_A(1/2)]^2$. This last identity can be derived from the Vandermonde's combinatorial identity
\be
       {K \choose r_2} = \sum_{s=0}^{r_1} {r_1 \choose s} {K-r_1 \choose r_2 -s}\, .
\ee}
\item{The second moment entropy has another stationary point at $t_x=1/2, t_y=t_z=0$ or $t_x=0, t_y=t_z=1/2$. This stationary point is equal to the first moment entropy at $t=1/2$.}
\end{itemize}

In the problems where one of the above stationary points is the global maximum the annealed entropy is exact and the satisfiability threshold easily calculable from (\ref{eq:s_sym}). 

In the symmetric problems with leaves (${\cal Q}(1)>0$), or those which are not locked (e.g. $0110$) or not balanced (e.g. $010010$) another competing maximum of the second moment entropy appears before the annealed entropy goes to zero. 

We investigated numerically that this does not happen for the balanced problems described by the recursion (\ref{eq:balanced}). So far we were not able to prove this last point analytically. This is, however, a technical problem, much simpler that the original one. 

The main message of this analysis is what are the ingredients of the model which make the satisfiability threshold accessible to the second moment computations. Here we showed that it is on one hand the (unbroken) symmetry of the problem and on the other hand the point-like clusters. Such a general result might be surprising because otherwise the satisfiability threshold is known exactly in only a handful of the NP-complete problems \cite{AchlioptasChtcherba01,MonassonZecchina99b,AchlioptasKirousis01,ConnamacherMolloy04}.

\chapter{Stability of the RS solution}
\label{app:RS_stab}

In chapter \ref{clustering} we argued in detail that the replica symmetric solution is correct if and only if the point-to-set correlations decay to zero, or equivalently if the reconstruction is not possible. Failure of the RS solution may (but does not have to) manifest itself via the divergence of the spin glass susceptibility. In a system with Ising variables $s_i \in \{-1,+1\}$ this is defined as
\be
       \chi_{\rm SG} =\frac{1}{N} \sum_{i,j} \langle s_i s_j \rangle^2_c \, ,\label{eq:chi_sg}
\ee 
where $\langle \cdot \rangle_c$ is the connected expectation with respect to the Boltzmann measure.\index{stability!replica symmetric} 

Originally the replica symmetric instability was investigated from the spectrum of the Hessian matrix in a celebrated paper by de Almeida and Thouless \cite{AlmeidaThouless78}. Equivalence between the RS stability and the convergence of the belief propagation equations on a single large graph is also often stated. In the reconstruction on tress this corresponds to the Kesten-Stigum condition \cite{KestenStigum66,KestenStigum66b}. It is not straightforward to see that all these statements are equivalent. We thus try to put a bit of order to the different ways of expressing the stability of the RS solution\footnote{This overview has been worked out in collaboration with F. Krzakala and F. Ricci-Tersenghi.}. 

\section{Several equivalent methods for RS stability}
\label{sec:BP_stab}

\paragraph{Susceptibility chains ---}
Perhaps the most direct way how to investigate the divergence of the spin glass susceptibility (\ref{eq:chi_sg}) is to write
\be
    \chi_{\rm SG} \approx \sum_i {\mathbb E}( \langle s_i s_0 \rangle^2_c ) \approx    \sum_d \gamma^d  {\mathbb E}(\langle s_d s_0 \rangle^2_c) \, , \label{eq:chi_dist}
\ee
where $s_0$ is a typical variable (the origin), $s_d$ is a variable at distance $d$ from $s_0$, and $\gamma^d$ is the typical number of variables at distance $d$ from $s_0$ ($\gamma = \overline{l^2}/\overline l -1$). The average $ {\mathbb E}(\cdot)$ is over the randomness of the graph. The spin glass susceptibility diverges if and only if $\lambda>1$ where
\be
   \lambda = \gamma \lim_{d\to \infty} \Big[ {\mathbb E}(\langle s_d s_0 \rangle^2_c ) \Big ]^{\frac{1}{d}}   \label{eq:chain}
\ee
Using the fluctuation dissipation theorem we can rewrite
\be
  {\mathbb{E}}(\langle s_0 s_d \rangle^2_c) \approx        
  {\mathbb{E}}  \left[\left( \frac{\partial 
    h_0 } {\partial h_d}  \right)^2\right] =   {\mathbb{E}} 
    \left[  \prod_{i=1}^d \left( \frac{\partial h_{i-1} } 
    {\partial h_{i}} \right)^2\right] ,
    \label{eq:fdt} 
\ee
where $h_0,\dots,h_d$ is a sequence of cavity fields (\ref{def_hu}) on the shortest path from $s_0$ to $s_d$. The dependence of the cavity field $h_i$ on $h_{i-1}$ is given by the belief propagation equations. This method to investigate the RS stability was used e.g. in \cite{MartinMezard05} or [\ZMM]. It is numerically involved and not very precise as in practice $d$ can be taken only at maximum $10-20$. 

\paragraph{Noise propagation ---} 
Call $v^0_d$ the contribution to the spin glass susceptibility from the layer of variables at a distance $d$ from $0$
\be
   v^0_d = \sum_{k, |k,0|=d} \Big( \frac{\partial 
    h_0 } {\partial h_k}  \Big)^2 = \sum_{i\in \partial 0}  \Big( \frac{\partial h_0 } {\partial h_i}  \Big)^2 \sum_{k, |k,i|=d-1} \Big( \frac{\partial h_i } {\partial h_k}  \Big)^2 = \sum_{i\in \partial 0}  \Big( \frac{\partial h_0 } {\partial h_i}  \Big)^2 v_{d-1}^i \, , 
\label{eq:stab_var}
\ee
where $h_k$ are cavity fields at distance $d$ from $h_0$, and the sum is over all the cavity fields needed to compute $h_0$. The spin glass susceptibility diverges if and only if the numbers $v_d$ are on average growing with the distance $d$. 

The evolution of numbers $v$ can be followed via the population dynamics method. Next to the population of fields $h$ we keep also a population of positive numbers $v$. When a field $h_0$ is updated according to the belief propagation equations, we update also the number $v^0$ according to (\ref{eq:stab_var}). The RS solution is stable if and only if the overall sum $\sum_i v^i$ is decreasing during the population dynamics updates. This method was implemented e.g. in \cite{MarinariSemerjian06} or [\RS]. It is simple and numerically very precise. 

\paragraph{Deviation of two replicas ---}
Consider a general form of the belief propagation equations $h=f(\{h_i\})$. After averaging over the graph ensemble we obtain distributional equations (\ref{eq:aver_P}-\ref{eq:aver_O}) which are solved via the population dynamics technique. Consider now two replicas of the resulting population, each element $i$ differs by $\delta h_i$. Keep running the population dynamics on both these replicas and record how the differences $\delta h_i$ are changing
\be
      \delta h_0 = \sum_{i\in \partial 0} \frac{\partial h_0}{\partial h_i} \delta h_i  \, . \label{eq:stab_lin}
\ee
The differences $\delta h$ can be negative and positive. Take $v = (\delta h)^2$ then 
\be
    v_0 = \left(\sum_{i\in \partial 0} \frac{\partial h_0}{\partial h_i} \delta h_i\right)^2 = \sum_{i\in \partial 0}\left( \frac{\partial h_0}{\partial h_i} \right)^2 v_i + \sum_{i\neq j} \frac{\partial h_0}{\partial h_i} \frac{\partial h_0}{\partial h_j} \delta h_i  \delta h_j\, .
\ee
The second term can be neglected because the terms $\delta h_i$ and $\delta h_j$ are independent. This brings us back to the equation (\ref{eq:stab_var}). 

Thus the replica symmetric solutions is stable if and only if the two infinitesimally different replicas do not deviate one from another. This method is very fast to implement and is thus useful for preliminary checks of the RS stability.

\paragraph{Convergence of the belief propagation ---}

The stability of replica symmetric solutions is equivalent to the convergence of the belief propagation equations on a large random graph. This fact follows directly from the previous paragraph. 
Eq.~(\ref{eq:stab_lin}) gives the rate of convergence (divergence) of two nearby trajectories of the dynamical map defined by the BP iterative equations.

\paragraph{Variance propagation ---}

Often a "variance" formulation of the stability if described. 
Assume that instead of a value $h_i$ on every link, there is a narrow distribution of values $g(h_i)$ parameterized by a mean
$\overline{h_i}$ and a small variance $v_i$. How does $\overline h$ and $v$ evolve? We have now
\bea
\overline{h} &=&\int {\rm d}h\,  g(h)\,  h = \int \prod_i \left[ {\rm d}h_i\,  g_i(h_i) \right] f(\{h_i\})\, , \\
v &=&\int {\rm d}h\,  g(h)\,  (h-\overline h)^2  = \int \prod_i \left[ {\rm d}h_i \, g_i(h_i) \right] f^2(\{h_i\}) - (\overline h )^2\, ,
\eea
where $h=f(\{h_i\})$ is the belief propagation equation.
However, since the variance is infinitesimal, the variation of $h_i$
around $\overline{h_i}$ is very small, so that
\be 
f(\{h_i\}) = f(\{\overline{h_i}\}) + \sum_i \((
h_i-\overline{h_i}\)) \frac{\partial f(\{h_i\})}{\partial h_i}\Big|_{\overline{h_i}}  \, ,
\ee
and therefore one obtains $\overline{h} = f(\{\overline{h_i}\}) $ and 
\be
v = \sum_i v_i \((\frac{\partial f(\{h_i\})}{\partial h_i}\Big|_{\overline{h_i}}\))^2\, , \label{eq:variance}
\ee
which is nothing else then equations  (\ref{eq:stab_var}). 

\paragraph{Numerical instability towards  the 1RSB solution ---}

The RS stability can also be investigated from the numerical stability of the trivial solution of the 1RSB equations. Indeed if the distribution of fields over states is regarded the probability distribution of a small variance $g(h)$ then the 1RSB equation (\ref{eq:1RSB_m1}) gives for a $p^{\rm th}$ moment of $g(h)$
\be
      \overline{ h^{p} } = \frac{1}{{\cal Z}} \int  \prod_i \left[ {\rm d}h_i\,  g_i(h_i) \right] Z^m(\{h_i\}) f^p(\{h_i\}) \, ,
\ee
where $Z$ is the normalization of the BP equations and its $m^{\rm th}$ power is the reweighting factor. Expansion gives
\be
   Z^m(\{h_i\}) =  Z^m(\{\overline {h_i}\}) + m Z^{m-1}(\{h_i\}) \sum_i \((
h_i-\overline{h_i}\)) \frac{\partial Z(\{h_i\})}{\partial h_i}\Big|_{\overline{h_i}} \, .  \label{eq:z_var}  
\ee
The equations for the variances (\ref{eq:variance}) does not depend on the second term from (\ref{eq:z_var}), as this is of a smaller order. 
As a consequence the condition for stability is independent of the parameter $m$.

It is quite remarkable fact that the divergence of the spin glass
susceptibility corresponds to the appearance of a nontrivial solution
of the 1RSB equation at {\it all} the values of~$m$. In particular because we observed that when the instability is not present the onset of a nontrivial 1RSB solution is $m$ dependent, see e.g. fig.~\ref{fig:slope_stab}. 

\paragraph{The eigenvalues of the Hessian ---}
The replica symmetric solution is a minimum of the Gibbs free energy. This is often investigated from the spectra of the matrix of second derivatives called Hessian. The equivalence between this approach and the divergence of the spin glass susceptibility is a classical result, see e.g. the book of Fischer and Hertz \cite{FischerHertz91}, page 98-100.

\section{Stability of the warning propagation}
\label{sec:WP_stab}

At zero temperature the necessary (but not sufficient) condition for the replica symmetric solution to be stable is the convergence of the warning propagation on a single graph. Obviously if the warning propagation does not converge then BP does not either, and convergence of the BP is equivalent to the replica symmetric stability. Advantage of the investigation of the warning propagation convergence is that it can be treated analytically, without using the population dynamics method.  

Consider a model with Ising spins where the warnings $u$ (\ref{BP_0}) can take only three possible values $u \in \{-1,0,1\}$. Consider warning $u$ and one of the warnings on which $u$ depends, say $u_0$. Except $u_0$ the warning $u$ depends also on $u_1,\dots,u_k$, where $k$ is distributed according to $\tilde {\cal Q}(k)$. The degree distribution conditioned on the presence of two edges is 
\be
        \tilde {\cal Q}(k-2) = \frac{k(k-1) }{\overline{k^2}-\overline k} {\cal Q}(k) \, . \label{eq:Q2}
\ee   
Call $P(a\to b| c\to d)$ the probability that the warning $u$ changes from value $a$ to value $b$ provided that the warning $u_0$ was changed from value $c$ to value $d$. This probability can be always computed from the probabilities $p_-,p_0,p_+$ that a warning $u=-1,0,+1$
\be
      P(a\to b| c\to d) = \sum_{k} \tilde {\cal Q}(k) \, P_k(p_-,p_0,p_+;a\to b| c\to d)\, ,
\ee
where the function $P_k$ depends on the model in consideration. This probability describes a proliferation of a "bug" in the warning propagation.
We define a {\it bug proliferation} matrix~$P_{ij}$ of dimension $6$, $i\equiv a\to b$, $j \equiv c\to d$. The stability of the warning propagation is then governed by the largest (in absolute value) eigenvalue of this matrix $\lambda_{\rm max}$. The warning propagation is stable if and only if 
\be
      \gamma \lambda_{\rm max} <1\, ,
\ee
where $\gamma = \overline{k^2}/\overline k -1$ is the growth rate of the tree ($\gamma^d$ is the typical number of vertices at distance $d$ from the root). This analysis is often called bug proliferation  \cite{KrzakalaPagnani04,MertensMezard06} (mostly in the context of the 1RSB stability). This investigation of the warning propagation stability was used e.g. in [\ZMM] or \cite{CastellaniKrzakala05}.

An example where the warning propagation is stable, however, the belief propagation is not, can be found in [\RS] for the 1-in-K SAT problem. In 1-in-K SAT the warning propagation stability threshold corresponds to the unit clause propagation upper bound [\RS].

\chapter{1RSB stability}
\label{app:1RSB_stab}

Concerning the correctness of the 1RSB solution: the Boltzmann measure is split into clusters. This leads to an exact description of the system if and only if both the following conditions are satisfied.\index{solution!1RSB!correctness}
\begin{itemize}
    \item{Condition of type I --- the point-to-set correlation with respect to the measure over clusters decay to zero. The statistics over clusters may be described on the replica symmetric (tree) level. Clusters do not tend to aggregate.}
    \item{Condition of type II --- the point-to-set correlations within the dominating clusters decay to zero. The interior of these clusters may be described on the replica symmetric (tree) level. Clusters do not tend to fragment into smaller ones.}
\end{itemize}  

Within the cavity approach these conditions can be checked from the 2RSB equation
\be
    P^{i\to j}_2\big[ P^{i\to j} \big]= \frac{1}{{\cal Z}_2^{i\to j}} \int \prod_{k\in \partial i -j} {\rm d} P^{i\to j}_2\big[P^{k\to i}\big] \,   ({\cal Z}^{i\to j})^{m_2} \,   \delta\big[P^{i\to j} -{\cal F}_2(\{P^{k\to i}\})\big]\, , \label{eq:2RSB_m}
\ee
where the functional ${\cal F}_2$ is given by the 1RSB equation (\ref{eq:1RSB_m1}). 
We call the solution of (\ref{eq:2RSB_m}) trivial if either $ P^{i\to j}_2\big[ P^{i\to j} \big] = \delta[P^{i\to j} ]$ or  each $ P^{i\to j}(  \psi^{i\to j})  = \delta(  \psi^{i\to j} -  \overline \psi^{i\to j})$, where the $P^{i\to j}$ is the solution of (\ref{eq:1RSB_m1}).
If and only if the (population dynamics) solution of the 2RSB equation at $m=m^*,\, m_2=1$ and at $m=1,\, m_2=m^*$ is trivial then the two conditions are satisfied, and the 1RSB solution at $m^*$ is correct.

Solving the 2RSB equation is, however, numerically involved. Even on random regular graphs the population dynamics of populations is needed, see app.~\ref{sec:pop_pop}. Moreover the reweighting taking in account the term $ ({\cal Z}^{i\to j})^{m_2}$ is costly. It is thus extremely useful to check the local stability of the 1RSB solution in the lines of the appendix \ref{app:RS_stab}. The two types of local stability follow. 
\begin{itemize}
    \item{Stability of type I\index{stability!1RSB type I} --- the inter-cluster spin glass susceptibility does not diverge.
\be
 \chi^{\rm inter}_{\rm SG} = \frac{1}{N}\sum_{i,j} \big( \overline{ \langle s_i \rangle \langle s_j \rangle} - \overline{\langle s_i \rangle} \, ,\overline{ \langle s_j \rangle } \big)^2 \, ,\label{eq:inter}
\ee
where the overline denotes an average over clusters 
\be
\overline{ x(  \psi^{i\to j})} = \int x(  \psi^{i\to j}) \, {\rm d}P^{i\to j}(  \psi^{i\to j})\, .
\ee}
    \item{Stability of type II\index{stability!1RSB type II} --- the intra-cluster spin glass susceptibility does not diverge.
\be
    \chi^{\rm intra}_{\rm SG} = \frac{1}{N}\sum_{i,j} \langle s_i s_j \rangle^2_c \, .\label{eq:intra}
\ee
The instability of second type is sometimes called the Gardner instability due to \cite{Gardner85}.}
\end{itemize} 
Again, there are several equivalent ways how to investigate the 1RSB stability. This time we first describe the zero temperature - frozen fields - version before turning to the general formalism. 

\section{Stability of the energetic 1RSB solution}

In the energetic zero temperature limit the 1RSB distribution $P^{i\to j}(  \psi^{i\to j})$ can be split into the frozen and soft part as in (\ref{eq:frozen}). Moreover the self-consistency equations on the weights of the frozen fields, called the SP-$y$ equations,  do not depend on the details of the soft part. The methods for stability investigation of the SP-$y$ equations were developed in \cite{Parisi02c,MontanariRicci03,MontanariParisi04,RivoireBiroli04}.

\paragraph{Type I ---  SP-$y$ convergence ---}

The divergence of the inter-cluster spin glass susceptibility is in general equivalent to the non-convergence of the 1RSB equations (\ref{eq:1RSB_m1}) on a single graph. The reason is exactly the same as for the equivalence of the non-divergence of the spin glass susceptibility and the convergence of the belief propagation equations, which we explained in app.~\ref{sec:BP_stab}. In the energetic zero temperature limit the convergence of the general 1RSB equations becomes convergence of the SP-$y$ equations on a single graph. All the methods described in app.~\ref{sec:BP_stab} for the stability of the belief propagation equations can be used directly.

Remark in particular that the chain method (\ref{eq:chain}), used e.g. in \cite{RivoireBiroli04,KrzakalaPagnani04}, is not the simplest choice. The chains of length $d\to \infty$ have to be considered numerically, and the treatable  values are only $d\approx 10-20$. This leads to an imprecision for a relatively large numerical effort. It is much more precise to use for example the noise propagation (\ref{eq:stab_var}) as e.g. in [\RS].

\paragraph{Type II --- Bug proliferation ---}

The intra-state susceptibility is investigated in exactly the same manner as the replica symmetric stability. The only difference is that the average over clusters have to be taken properly. The energetic 1RSB solution is based on the warning propagation equations averaged properly over the clusters. Thus the 1RSB stability of the type II leads to the bug proliferation, as in app.~\ref{sec:WP_stab}, averaged over the clusters. 

Roughly explained, if we consider a model with Ising spins, we have the three components surveys $  p=(p_-,p_0,p_+)$ on each edge. Where $p_s$ is the probability over clusters that the warning on this edge takes the value $s$. Consider, as in app.~\ref{sec:WP_stab}, a warning $u$ and one of the incoming warnings $u_0$, the remaining incoming warnings are indexed by $i=1,\dots,k$ where $k$ is distributed according to $\tilde {\cal Q}(k)$ (\ref{eq:Q2}). Define $P_k(a\to b| c\to d)$  as the probability, over clusters, that the warning $u$ changes from a value $a$ to a value $b$ provided that the warning $u_0$ was changed from a value $c$ to a value $d$. Consider $P_k(a\to b| c\to d)$ as a matrix of dimension $6$. And consider a chain of edges of length $d$. The proliferation of an instability "bug" is given by the product of matrices $P_k$ along this chain. The product is averaged over the realizations of disorder (in degrees, etc.). We define the stability parameter as
\be
     \lambda_{II}(d) = \gamma \left( {\rm Tr}\langle P^1_{k_1}\dots P^d_{k_d}\rangle \right)^{\frac{1}{d}}\, .
\ee
The SP-$y$ is 1RSB stable if and only if $\lim_{d\to \infty}  \lambda_{II}(d)<1$. For more detailed presentation of the 1RSB bug proliferation method or concrete examples see e.g. \cite{RivoireBiroli04,MertensMezard06,KrzakalaPagnani04} and [\RS].
In all the implementations of this method the chain of $d\to \infty$ edges was used. Unlike in the type I stability, it is not know if this can be avoided in general.

\paragraph{Some results ---}

\begin{figure}[!ht]
\resizebox{\linewidth}{!}{
\includegraphics{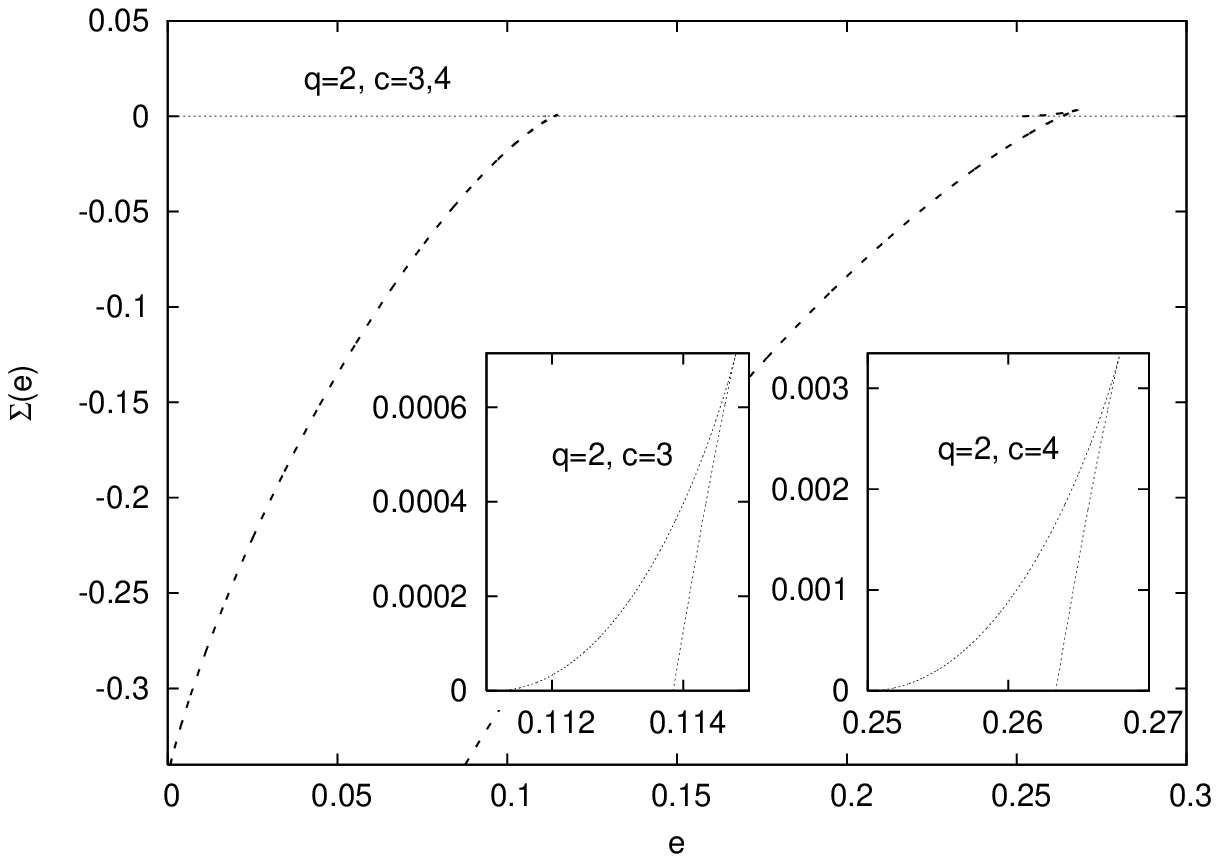}
\includegraphics{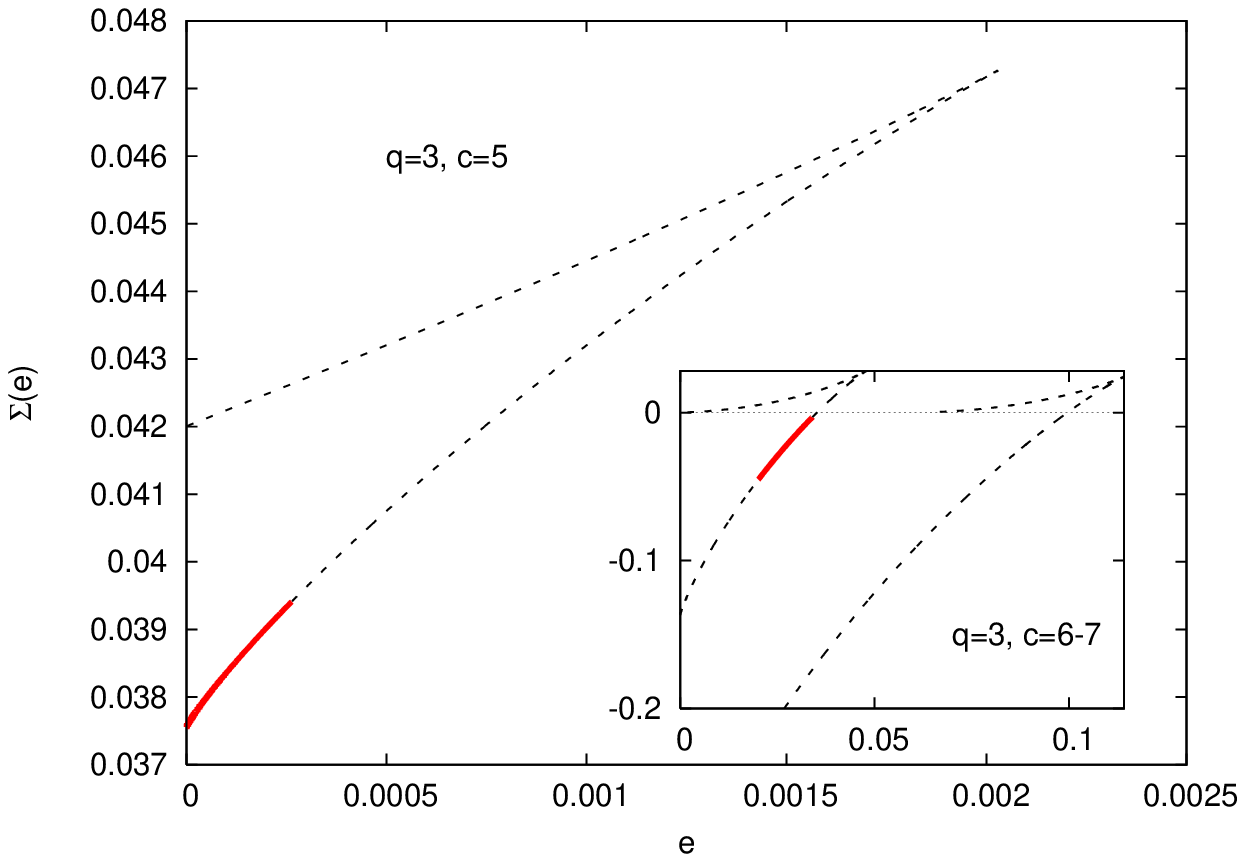}}
\resizebox{\linewidth}{!}{
\includegraphics{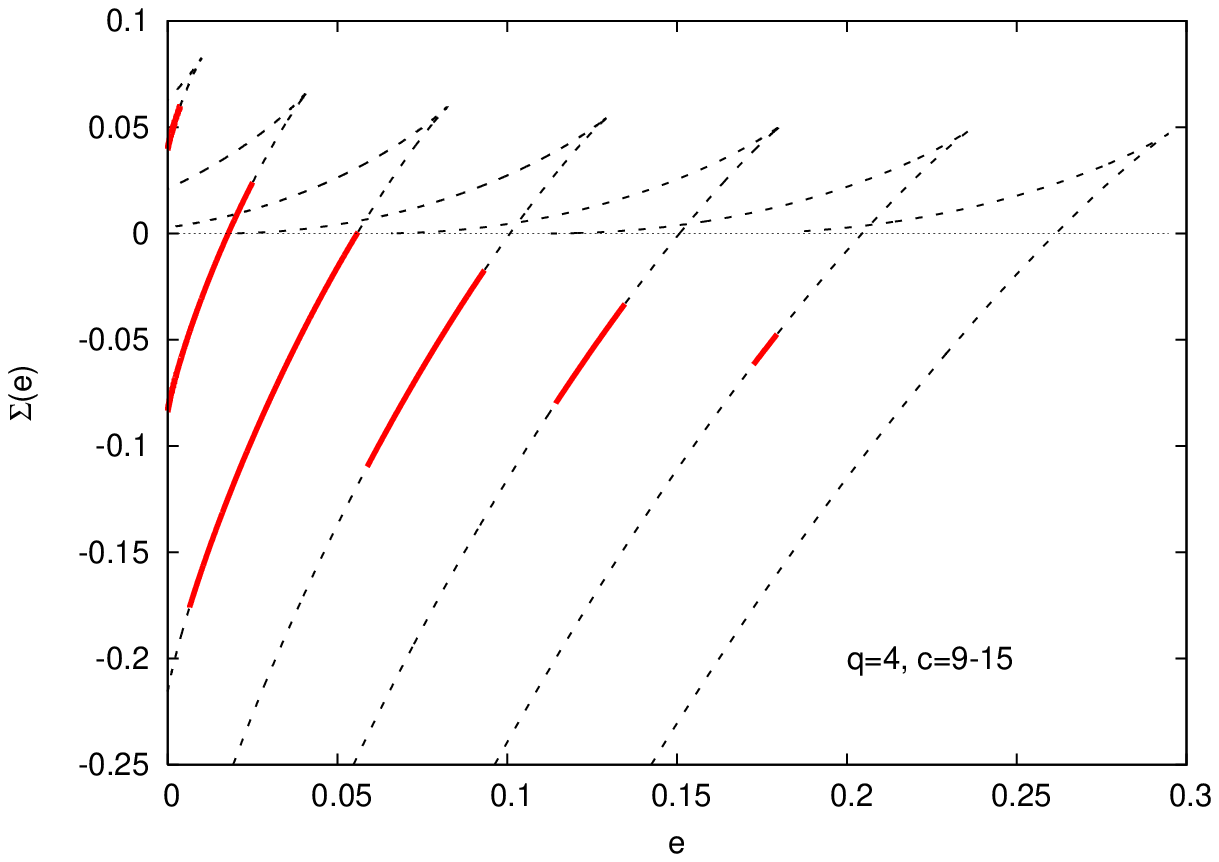}
\includegraphics{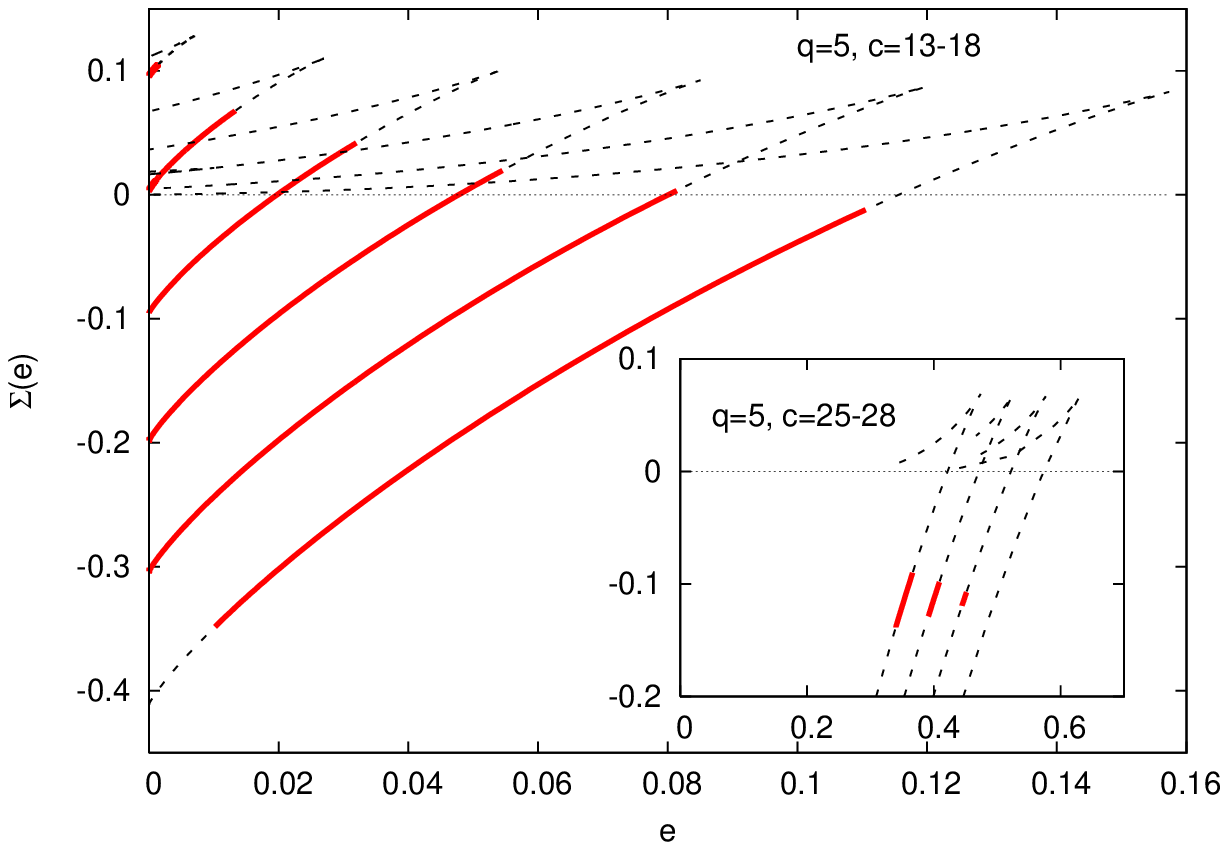}}
\caption{\label{fig_1rsb_stab} The complexity as a function of energy for the coloring of random regular graphs. The 1RSB stable parts of the curves are in bold red.}
\end{figure}

The investigation of the 1RSB stability as we just described can be very simply incorporated to the population dynamics method used to solve the survey propagation equations. This means that on random regular graphs the stability equations become algebraic, as the values of surveys do not depend on the index of the edge. In fig.~\ref{fig_1rsb_stab} we present the result for coloring of random regular graphs. 

On all the parts of fig.~\ref{fig_1rsb_stab} the complexity function is plotted against energy, $\Sigma(e)$ (\ref{eq:Legendre_ener}). This function is the main output of the 1RSB energetic method, the SP-$y$ equations. The parameter $y$ corresponds to the slope of the complexity function $y = \partial \Sigma(e)/ \partial e$. Note that only the concave parts of the curves are physical. 

The red parts of the $\Sigma(e)$ curves are the 1RSB stable parts. It seems to be a general fact that the instability of type I happens first for large values of $y$, and the instability of the type II for small values of $y$. The unphysical (convex) branch is always type II instable.
The instability of type I is sometimes completely absent. 

An important observation is that the stability of the 1RSB energetic solution does not guarantee the stability of the full 1RSB solution.
Differently said, the soft fields can destabilize the full solution. On the other hand also the opposite is true --- the instability of the clusters corresponding to $m=0$ does not imply the instability of the dominating clusters at $m^*$. We thus want to stress that the results of \cite{MontanariParisi04,RivoireBiroli04,MertensMezard06,KrzakalaPagnani04} and others have to be taken with these two facts in mind.

\section{1RSB stability at general $m$ and $T$}

The stability of the full 1RSB equations at a general value of the parameter $m$ and of the temperature $T$ is a more difficult task. 
We are not aware of any study where this would be practically considered for models on random graphs, apart from [\KZP]. We review shortly the main findings and difficulties. 

\paragraph{Type I ---} Divergence of the inter-cluster spin glass susceptibility (\ref{eq:inter}) is equivalent to the non-convergence of the probability distributions $P^{i\to j}(  \psi^{i\to j})$ (\ref{eq:1RSB_m1}). But here arrives the biggest problem, how to judge if a probability distribution converges? The probability distribution $P^{i\to j}(  \psi^{i\to j})$ is represented by a population of random elements picked from this distribution. How to decouple the randomness coming from this sampling and the one coming from the eventual non-convergence? Of course, provided that the numerical difficulty does not rise to the level of directly solving the 2RSB equations. This is not known in general and it is a technical but important open problem in the subject. 

One interesting observation can be made, however: If the RS solution is instable then the 1RSB solution at $m=1$ is type I instable.  Indeed, if the mean value of the probability distribution does not converge then the 1RSB solution is type I instable. At the value $m=1$ the mean (\ref{eq:psi_mean}) satisfies the simple belief propagation equations, as explained in app.~\ref{app:m1}. 

\paragraph{Type II ---} 

Divergence of the intra-cluster spin glass susceptibility (\ref{eq:intra}) is much easier to investigate on a general level.
It is equivalent to checking if the 1RSB iteration are stable against small changes in the probabilities $  \psi$. Arguably the simplest way to do so is the {\it deviation of two replicas} method, described for the RS stability in app.~\ref{sec:BP_stab}. We first find a fixed point of the 1RSB equations (\ref{eq:1RSB_m1}) using the population dynamics method. Then we create a second copy of the populations representing the distributions $P^{i\to j}(  \psi^{i\to j})$. We perturb infinitesimally every of its elements $  \psi^{i\to j}$. The 1RSB is type II stable if and only if the two copies converge to the same point. The noise propagation and other methods from \ref{sec:BP_stab} can be used equivalently.

\begin{figure}[!ht]
\resizebox{\linewidth}{!}{
\includegraphics{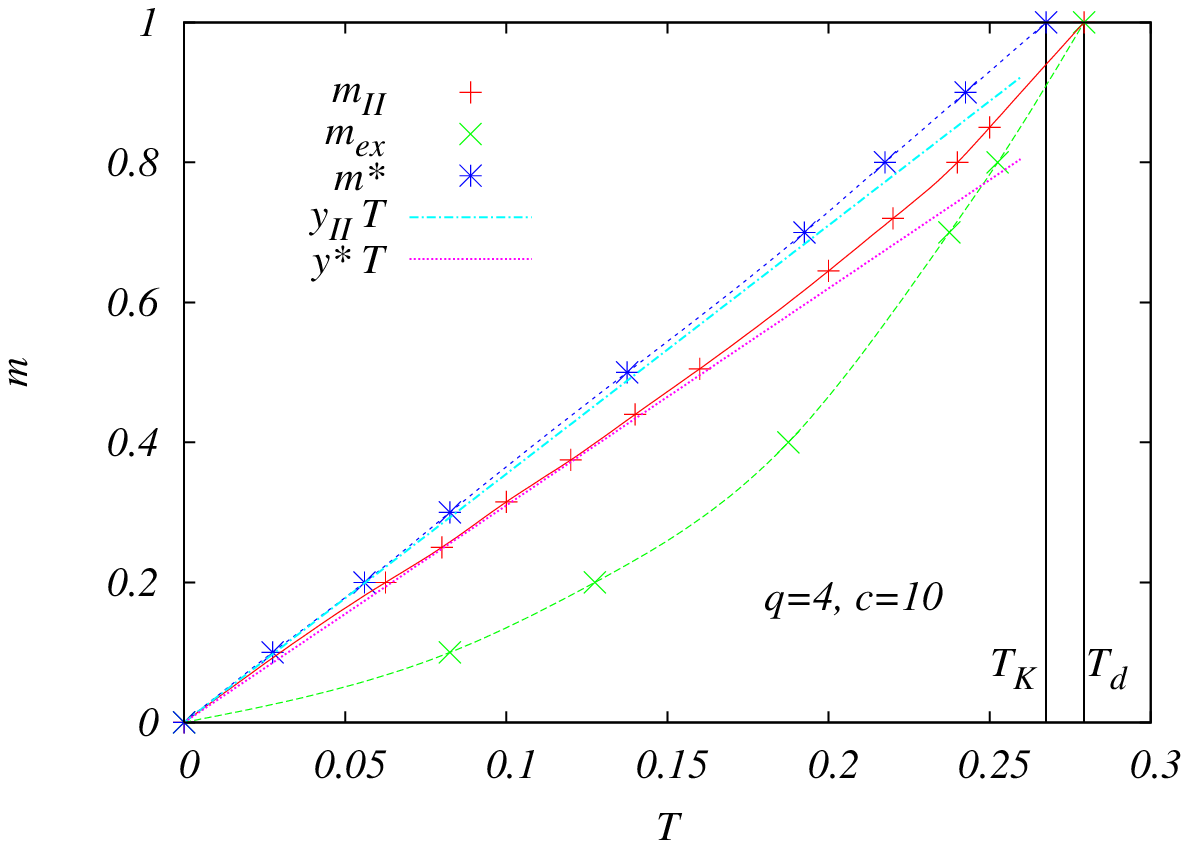}
\includegraphics{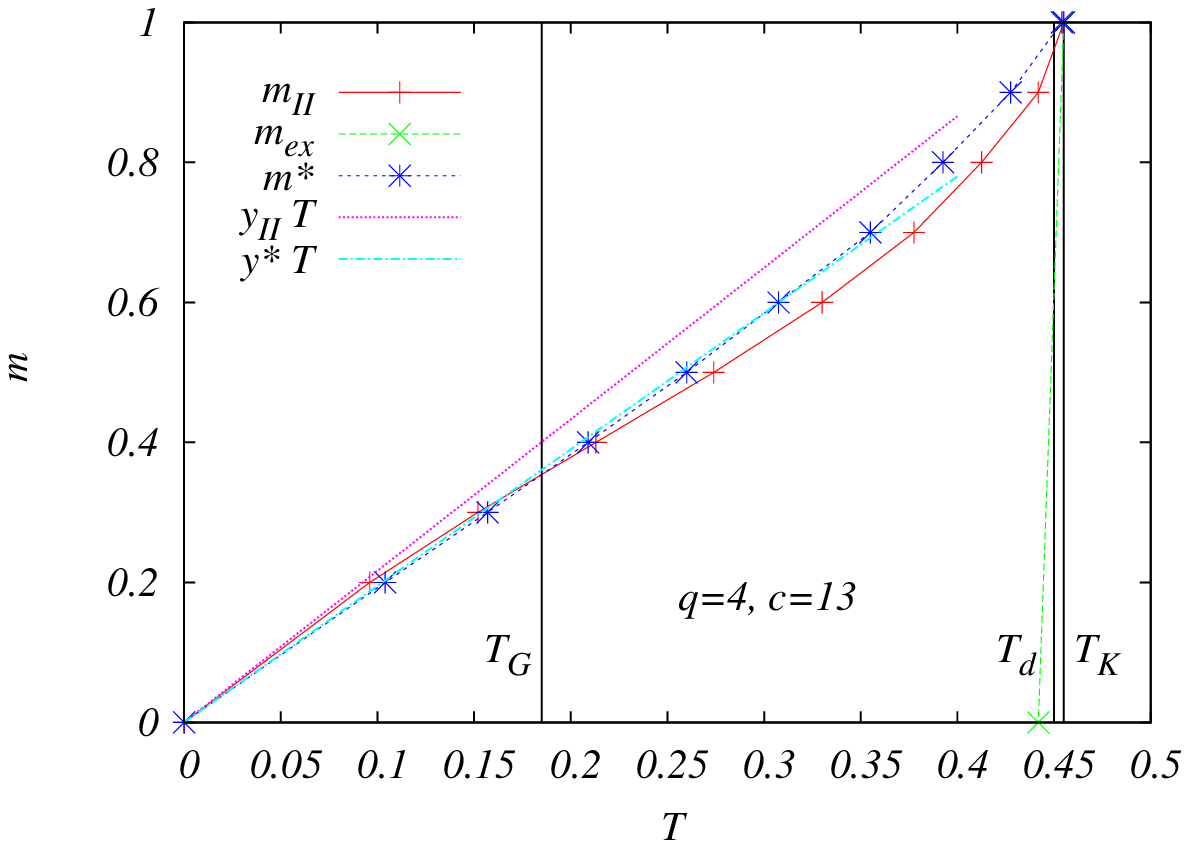}}
\caption{\label{fig:slope_stab} Example of $m$-$T$ diagrams for the $4$-state anti-ferromagnetic Potts model on $c$-regular random graphs (left $c=10$, right $c=13$). A nontrivial solution of the 1RSB eq.~(\ref{eq:1RSB_m1}) exists above the curve $m_{\rm ex}$ (green). The curves in blue $m^*$ represent the thermodynamic value of the parameter $m$. The red curve $m_{\rm II}$ is the lower border of the type II stable region. The straight lines ($y_{\rm II}T$ and $y^* T$) represent the slopes corresponding to the energetic 1RSB solution $yT=m$ in $T \to 0$. The energetic 1RSB solution is type II stable for $y > y_{\rm II}$. The line $y_{\rm II}T$ seems to give correctly the slope of $m_{\rm II}$. This suggest that the stability of frozen variables is equivalent to the full stability for small $m$ and $T$. Other examples of diagrams of this type are presented in [\KZP].}
\end{figure}

\paragraph{Some results and connection to the SP-$y$ stability ---}
Fig.~\ref{fig:slope_stab} depicts the results for the stability of type II in the space of the parameters $m$ and temperature $T$. The 1RSB solution is type II stable above the red curve $m_{\rm II}$. 

It is interesting to state the connection between the general $m$, $T$ stability and the energetic zero temperature limit. The parameter $m=yT$ when $T\to 0$, thus when the stability of the frozen fields is relevant for the full stability the parameter $y_{\rm II}T$ gives the slope of $m_{\rm II}(T)$ near to zero $T$. This indeed seems to be the case, as shown in fig.~\ref{fig:slope_stab}.

Based on the arguments above, it seems reasonable that the following assumptions are correct:
\begin{itemize}
   \item[(i)]{The stability of the energetic method gives the full stability for small $m$ and $T$.}
   \item[(ii)]{If the RS solutions is stable then the 1RSB is stable type I at $m=1$.}
   \item[(iii)]{If the 1RSB at a given temperature is type I (II resp.) stable at a given $m$, then it is type I (II resp.) stable for all smaller (larger resp.) $m$.}
\end{itemize}

Assuming as above, the stability of the 1RSB solution in the region where the RS solution is stable is given by the type II (Gardner) stability, which we know how to investigate. The result is depicted e.g. in fig.~\ref{fig_col_T}. This would mean that the stability of type II is always more important for the thermodynamical solution. And in particular that in the random coloring problem for $q\ge 4$ the 1RSB solution is stable in all the colorable phase. 

The situation for 3-coloring is more subtle as 3-coloring is not RS stable for $c\ge c_d$. However, from assumption (i) follows that the interval of connectivities $(c_s,c_G)=(4.69,5.08)$ is 1RSB stable at small temperatures. Thus we expect also all the colorable phase to be 1RSB stable (otherwise the phase diagram at fig.~\ref{fig_col_T} would have to present a sort of re-entrant behaviour). This would also be in agreement with the situation in the fully connected ferromagnetic 3-state Potts model \cite{GrossKanter85}
\footnote{This is a contra-example to the common claim that in the systems with continuous dynamical transition ($T_d=T_{\rm local}$) the 1RSB solution is not stable.}.

\chapter{Populations dynamics}
\label{app:pop_dyn}

Population dynamics is a numerical method to solve efficiently distributional equations of type (\ref{eq:aver_m}) or (\ref{eq:1RSB_m1}) and compute observables of type (\ref{free_en}) or (\ref{eq:1RSB_entr}). In this context it was developed in \cite{MezardParisi01}. As the form of the 1RSB equations was more or less known before, and they were solved approximatively using various forms of the variational ansatz, see e.g. \cite{BiroliMonasson00}, it may be argued that the population dynamics technique was the crucial ingredient which made the spin glass models on random graphs solvable. Recently rigorous versions of this method were developed to analyze the performance of decoding algorithms \cite{RichardsonUrbanke01}, the name {\it density evolution} is often used in this context. 

The main idea is to represent the probability distribution by a {\it population} (sample) of $N$ elements drawn independently at random from this distribution. The algorithm starts from a random list and it mimics $T$ iterations of the distributional equations and (hopefully) converges to a good representation of the desired fixed point. Several generalizations or subtleties are encountered and we describe some of them in the following. Consider the a random constraint satisfaction model specified by degree distribution ${\cal R}(k)$ of constraints, and ${\cal Q}(l)$ of variables, the excess degree distributions $r(k)$ and $q(l)$ are given by (\ref{eq:excess}).

\section{Population dynamics for belief  propagation}

The simplest version of the population dynamics is used to solve 
\begin{itemize}
\item Belief propagation distributional equations (\ref{eq:aver_P}-\ref{eq:aver_O}) and compute the corresponding average free energy (\ref{free_given}), entropy, etc. The complete replica symmetric solution is obtained this way.
\item Survey propagation distributional equations, obtained from (\ref{SP_1}-\ref{SP_2}), and compute the average complexity function (\ref{Sigma_0}). The satisfiability transition is obtained this way. 
\end{itemize}

The pseudocode for the procedures \proc{Population-Dynamics} and \proc{One-Measurement} follows.
To compute the observable $\Phi$ (free energy, entropy, complexity, etc.) we first call procedure \proc{Population-Dynamics} with $T=T_{\rm equil}$ (equilibration time) and sufficiently large $N$. After we repeat \proc{One-Measurement} plus \proc{Population-Dynamics} with $T=T_{\rm rand}$ (randomization time) and $M$ sufficiently large, but smaller than $N$. And finally we compute averages and error bars of these measurements.  

In some problems the constraints are themselves random (negations in $K$-SAT, interactions in a spin glass etc.). The choice of this quenched randomness is then done at line~\ref{line_q} of \proc{Population-Dynamics}, and at line~\ref{line_q2} of \proc{One-Measurement}. 

The population $\{\psi\}$ is randomly initialized to a random assignment at line \ref{line_init} of \proc{Population-Dynamics}. That is all the zero components of the surveys (\ref{SP_1}-\ref{SP_2}) are zero, and the beliefs are completely biased, i.e., either $(1,0)$ or $(0,1)$. Such a choice is justified from the analogy with the reconstruction on trees where the proper initial condition is given by (\ref{eq:init}).

Satisfactory results are usually obtained with the population sizes and times of order $N\approx 10^4-10^5$, $T_{\rm equil}\approx 10^3-10^4$, $T_{\rm rand}\approx 10$, $M\approx N$. But these values may change problem from problem and a special care have to be taken about the numerics every time as basically no convergence theorems are known for a general case. 

\begin{codebox}
\Procname{$\proc{Population-Dynamics}(r(k),q(l),N,T)$}
\li Initialize randomly $N$-component array $\{\psi\}$; \label{line_init}
\li \For $t=1,\dots,T$:
\li     \Do
           \For $i=1,\dots,N$:
\li        \Do
             Draw an integer $k$ from the distribution  $r(k)$;
\li          \For $d=1,\dots,k$:
\li             \Do
                Draw an integer $l$ from the distribution  $q(l)$;
\li             Draw indexes $j_1,\dots,j_l$ uniformly in $\{1,\dots,N\}$;
\li             Compute $\chi_d$ from $\{\psi_{j_1},\dots,\psi_{j_l}\}$ according to eq. (\ref{eq:BP_2}); 
                \End
\li          Compute $\psi_{\rm new}$ from  $\{\chi_{1},\dots,\chi_{k}\}$ according to eq. (\ref{eq:BP_1});   \label{line_q} 
\li           $\psi_i \gets \psi_{\rm new}$;
            \End
         \End 
\li \Return array $\{\psi\}$;    
\end{codebox}

\begin{codebox}
\Procname{$\proc{One-Measurement}({\cal R}(k),{\cal Q}(l),q(l),N,M)$}
\li Initialize $\Phi_{\rm constraint}=0$; $\Phi_{\rm variable}=0$;
\li     \For $i=1,\dots,M$: \Comment Compute the constraint part.
\li        \Do
             Draw an integer $k$ from the distribution  ${\cal R}(k)$;
\li          \For $d=1,\dots,k$:
\li             \Do
                Draw an integer $l$ from the distribution  $q(l)$;
\li             Draw indexes $j_1,\dots,j_l$ uniformly in $\{1,\dots,N\}$;
\li             Compute $\chi_d=\prod_{n=1}^l \psi_{j_n}$; 
                \End
\li        Compute $Z_{\rm new}$ from  $\{\chi_{1},\dots,\chi_{k}\}$ according to eq. (\ref{eq:Zaa});    \label{line_q2}
\li         $\Phi_{\rm constraint}\gets \Phi_{\rm constraint}+\log{Z_{\rm new}}$;
         \End 
\li     \For $i=1,\dots,M$: \Comment Compute the variable part.
\li        \Do
           Draw an integer $l$ from the distribution  ${\cal Q}(l)$;
\li        Draw indexes $j_1,\dots,j_l$ uniformly in $\{1,\dots,N\}$;
\li        Compute $Z_{\rm new}$ from  $\{\psi_{j_1},\dots,\psi_{j_l}\}$ according to eq. (\ref{eq:Zi});    
\li         $\Phi_{\rm variable}\gets \Phi_{\rm variable}+(l-1)\log{Z_{\rm new}}$;
         \End 
\li \Return $(\alpha \Phi_{\rm constraint}-\Phi_{\rm variable})/M$;   
\end{codebox}

\section{Population dynamics to solve 1RSB at $m=1$}

The general 1RSB equations for general random graph ensemble require a population dynamics with population of populations. We will explain this in sec.~\ref{sec:pop_pop}. Treating the population of populations requires a lot of CPU time and it is not very precise, thus anytime we have the opportunity to avoid this we have to take it. One such opportunity is the simplification of the 1RSB equations at $m=1$ explained in appendix \ref{app:m1}. Conveniently, both the clustering and the condensation transitions are obtained this way.  

The population dynamics method have to be adapted to solve eq.~(\ref{final_m1}) and to measure the entropy of states (\ref{component}). We give the $m=1$ generalization of the procedure \proc{Population-Dynamics}, the changes in \proc{One-Measurement} are then straightforward. Note that lines \ref{line_s} and \ref{line_r} take in general $2^{k}$ steps as we need to compute probability of every combination of the set $\{s_1,\dots,s_k\}$.

\begin{codebox}
\Procname{$\proc{PD-$(m=1)$-Generalization}(r(k),q(l),N,T)$}
\li $\{\psi^{\rm RS}\} \gets \proc{Population-Dynamics}(r(k),q(l),N,T)$;
\li Initialize $N$-component arrays $\{\psi^1\gets 1\}$ and $\{\psi^0 \gets 0\}$;
\li \For $t=1,\dots,T$:
\li \For $i=1,\dots,N$:
\li        \Do
             Draw an integer $k$ from the distribution  $r(k)$;
\li          \For $d=1,\dots,k$:
\li             \Do
                Draw an integer $l_d$ from the distribution  $q(l)$;
\li             Draw indexes $j(d,1),\dots,j(d,l_d)$ uniformly in $\{1,\dots,N\}$;
\li             Compute $\chi^{\rm RS}_d$ from $\{\psi^{\rm RS}_{j(d,1)},\dots,\psi^{\rm RS}_{j(d,l_d)}\}$ according to eq.~(\ref{eq:BP_2}); 
                \End   
\li          $s \gets 1$;
\li          Choose $\{s_1,\dots,s_k\}$ with prob. given by the 2$^{\rm nd}$ line of eq.~(\ref{final_m1}); \label{line_s}
\li          $s \gets 0$;
\li          Choose $\{r_1,\dots,r_k\}$ with prob. given by the 2$^{\rm nd}$ line of eq.~(\ref{final_m1}); \label{line_r}
\li          \For $d=1,\dots,k$:
\li             \Do
                Compute $\chi^{1}_d$ from $\{\psi^{s_d}_{j(d,1)},\dots,\psi^{s_d}_{j(d,l_d)}\}$ according to eq.~(\ref{eq:BP_2});
\li             Compute $\chi^{0}_d$ from $\{\psi^{r_d}_{j(d,1)},\dots,\psi^{r_d}_{j(d,l_d)}\}$ according to eq.~(\ref{eq:BP_2}); 
                \End                             
\li          Compute $\psi^{\rm RS}_{\rm new}$ from  $\{\chi^{\rm RS}_{1},\dots,\chi^{\rm RS}_{k}\}$ according to eq.~(\ref{eq:BP_1}); 
\li          Compute $\psi^{1}_{\rm new}$ from  $\{\chi^{1}_{1},\dots,\chi^{1}_{k}\}$ according to eq.~(\ref{eq:BP_1}); 
\li          Compute $\psi^{0}_{\rm new}$ from  $\{\chi^{0}_{1},\dots,\chi^{0}_{k}\}$ according to eq.~(\ref{eq:BP_1});  
\li           $\psi^{\rm RS}_i \gets \psi^{\rm RS}_{\rm new}$;
\li           $\psi^{1}_i \gets \psi^{1}_{\rm new}$;
\li           $\psi^{0}_i \gets \psi^{0}_{\rm new}$;
            \End
\li \Return arrays $\{\psi^{\rm RS}\}$, $\{\psi^1\}$, $\{\psi^0\}$;    
\end{codebox}

\vspace{-0.5cm}

\section{Population dynamics with reweighting}
A simplification of the 1RSB equations (\ref{eq:1RSB_m1}) arises for the ensemble of random regular graphs, there the distribution ${\cal P}^{i\to j}(\psi^{i\to j})$ over clusters is the same for every edge $(ij)$. In the corresponding population dynamics a special care have to be taken about the reweighting term $\big (Z^{i\to j}\big)^m$. 

We describe two different strategies to deal with the reweighting. In the first one \proc{Reweighting-Faster} the elements of the population have all the same weight and thus in each sweep the population needs to be re-sampled and some less probable elements might be lost. In the second strategy \proc{Regular-Reweighting-Precise} each element has its own weight, no re-sampling is needed, but the search of a random element, at the line \ref{line_search}, takes $\log{N}$ steps. Thus the first strategy is faster, the second one is slightly more precise. Which one is eventually better seems to be problem specific. 

Consider a population $\{\psi\}$ where each element $\psi_i$ has weight $w_i$. The weights are computed from the BP update (\ref{eq:BP_1}-\ref{eq:BP_1}) as
$
      w_i = \Big( Z^{a \to i} \prod_{j\in \partial a -i} Z^{j\to a}\Big)^m 
$. 

\begin{codebox}
\Procname{$\proc{Reweighting-Faster}(N,\{\psi\},\{w\})$}
\li  
           $w_{\rm tot}\gets 0$; 
\li        \For $i=1,\dots,N$:
\li        \Do
               $w_{\rm tot}\gets w_{\rm tot}+w_i$;
            \End
\li      \Comment $z_i$ is the cumulative distribution of indexes $i$; \label{line_rewf}
\li      $z_0=0$;
\li      \For $i=1,\dots,N$:
\li         \Do
            $z_i \gets z_{i-1} + w_i/w_{\rm tot}$ 
            \End
\li       \Comment Trick to make a list of ordered random numbers $n_i$ in $O(N)$ steps.
\li       $G\gets 0$; 
\li       \For $i=1,\dots,N$:
\li         \Do
              $n_i \gets -\log{\proc{Rand}}$;
\li           \Comment \proc{Rand} outputs a random number in the interval $(0,1)$.  
\li           $G \gets G + n_i$;   
            \End
\li       $G \gets G-\log{\proc{Rand}}$; 
\li       $n_1\gets n_1/G$; 
\li       \For $i=2,\dots,N$:
\li         \Do
                $n_i\gets n_i/G$;
\li             $n_i\gets n_i+n_{i-1}$;
            \End 
\li       \Comment Finally making the new population.
\li       $p\gets 0$;    
\li       \For $i=1,\dots,N$
\li         \Do
              \While ($n_i>z_{p}$) $p \gets p+1$;
\li           $\psi^{\rm new}_i\gets \psi_{p}$; \label{line_rewl}
            \End   
\li \Return array $\{\psi^{\rm new}\}$;    
\end{codebox}

\begin{codebox}
\Procname{$\proc{Regular-Reweighting-Precise}(r(k),q(l),N,T,m)$}
\li Initialize randomly $N$-component arrays $\{\psi\}$ and $\{w\}$;
\li \For $t=1,\dots,T$:
\li \For $i=1,\dots,N$:
\li        \Do
             Draw an integer $k$ from the distribution  $r(k)$;
\li          $Z_{\rm new} \gets 1$;         
\li          \For $d=1,\dots,k$:
\li             \Do
                Draw an integer $l$ from the distribution  $q(l)$;
\li             \For $n=1,\dots,l$:
\li             \Do 
                   Create cumulative probability distribution from weights $\{w\}$;     
\li                Draw index $j_n$ from this cumulative distribution; \label{line_search}
                \End
\li             Compute $\chi_d$ from $\{\psi_{j_1},\dots,\psi_{j_l}\}$ according to eq.~(\ref{eq:BP_2}); 
\li              $Z_{\rm new} \gets Z_{\rm new}\cdot Z_d$, where $Z_d$ is the norm. from eq.~(\ref{eq:BP_2}); 
                \End
\li          Compute $\psi_{\rm new}$ from  $\{\chi_{1},\dots,\chi_{k}\}$ according to eq.~(\ref{eq:BP_1});  
\li              $Z_{\rm new} \gets Z_{\rm new}\cdot Z_d$, where $Z_d$ is the norm. from eq.~(\ref{eq:BP_1}); 
\li           $\psi_i \gets \psi_{\rm new}$;
\li           $w_i \gets (Z_{\rm new})^m$; 
            \End 
\li \Return array $\{\psi\}$, weights $\{w\}$;    
\end{codebox}

\section{Population dynamics with hard and soft fields}

Fraction of frozen variables (again on random regular graphs for simplicity) can be obtained by solving equation (\ref{hard_m_reg}). To compute the value $r(m)$ a population needs to be kept for the soft part of the distribution $P_{\rm soft}$, eq.~(\ref{eq:frozen}). 
It is important to stress that when evaluating the \If conditions on lines \ref{line_if1},\ref{line_if2} and \ref{line_if3} we consider as frozen only the incoming fields created at line \ref{line_fr}.

\begin{codebox}
\Procname{$\proc{PD-Hard-Soft}(r(k),q(l),N,T,m)$}
\li Initialize randomly $N$-component array $\{\psi\gets \proc{Rand}\}$;
\li $\eta \gets 1$;
\li \For $t=1,\dots,T$:
\li \Do
           $i\gets 1$; 
\li        $h\gets 0$; $Z_{\rm hard} \gets 0$; $Z_{\rm soft} \gets 0$; 
\li        \While $i\le N$:
\li        \Do
             Draw an integer $k$ from the distribution  $r(k)$; \label{line_Z}   
\li          $Z_{\rm new} \gets 1$;     
\li          \For $d=1,\dots,k$:
\li             \Do
                Draw an integer $l$ from the distribution  $q(l)$;
\li             \For $r=1,\dots,l$:
\li             \Do
                \If $\proc{Rand} < \eta$
\li                  \Then Set $\psi_r$ to be a frozen field;  \label{line_fr}
\li                  \Else Draw $\psi_r$ uniformly from $\{\psi\}$;
                     \End
                \End  
\li             \If No contradiction between the frozen fields in  $\{\psi_{1},\dots,\psi_{l}\}$ \label{line_if1}
\li                 \Then Compute $\chi_d$ from $\{\psi_{1},\dots,\psi_{l}\}$ using eq.~(\ref{eq:BP_2}); 
\li              $Z_{\rm new} \gets Z_{\rm new}\cdot Z_d$, $Z_d$ is the norm. from (\ref{eq:BP_2}); 
\li                 \Else \Goto line \ref{line_Z};
                    \End                 
                \End
\li             \If No contradiction between the frozen fields in $\{\chi_{1},\dots,\chi_{k}\}$ \label{line_if2}
\li                \Then Compute $\psi_{\rm new}$ from  $\{\chi_{1},\dots,\chi_{k}\}$ according to eq.~(\ref{eq:BP_1});  
\li              $Z_{\rm new} \gets Z_{\rm new}\cdot Z_d$, where $Z_d$ is the norm. from eq.~(\ref{eq:BP_1}); 
\li                 \Else \Goto line \ref{line_Z};
                 \End
\li           \If $\psi_{\rm new}$ is a frozen field \label{line_if3}
\li              \Then $Z_{\rm hard}\gets Z_{\rm hard} + \big( Z_{\rm new}\big)^m$;
\li                    $h\gets h +1$;
\li              \Else $Z_{\rm soft}\gets Z_{\rm soft} + \big( Z_{\rm new}\big)^m$;
\li                    $\psi_i \gets \psi_{\rm new}$;
\li                    $w_i \gets (Z_{\rm new})^m$; 
\li                    $i\gets i +1$;                
            \End
    \End
\li      $r\gets (Z_{\rm soft} h )/(Z_{\rm hard} N)$;
\li      Update $\eta$ according to eq.~(\ref{hard_m_reg});  
\li      $\{\psi\}\gets \proc{Reweighting-Faster}(N,\{\psi\},\{w\})$;  
    \End  
\li \Return array $\{\psi\}$, $\eta$;    
\end{codebox}

\section{The population of populations}
\label{sec:pop_pop}

The general 1RSB equations take form (\ref{eq:1RSB_aver}), the order parameter ${\cal P}[P(\psi)]$ is a distribution (over the graph ensemble) of distributions (over the clusters). It can be represented by a population $\{\{\psi\}\}$ of $N$-component populations $\{\psi\}_i$, where $i=1,\dots,M$.
We sketch here the corresponding population dynamics of populations. Again this has been first described in \cite{MezardParisi01}.

\begin{codebox} 
\Procname{$\proc{Population-of-Populations}(r(k),q(l),N,M,T,m)$}
\li Initialize randomly $M\times N$-component array $\{\{\psi\}\}$; 
\li \For $t=1,\dots,T$:
\li     \Do
           \For $i=1,\dots,M$:
\li        \Do
             Draw an integer $k$ from the distribution  $r(k)$;
\li          \For $d=1,\dots,k$:
\li             \Do
                Draw an integer $l_d$ from the distribution  $q(l)$;
\li             Draw indexes $i(d,1),\dots,i(d,l_d)$ uniformly in $\{1,\dots,M\}$;
                \End
\li          $\{\psi\}_{\rm new}\gets \proc{One-Step}(\{\{\psi\}\},\{i(1,1),\dots,i(k,l_k)\},\{l\},k,N,m)$; 
\li           $\{\psi\}_i \gets \{\psi\}_{\rm new}$;
            \End 
         \End
\li \Return array $\{\{\psi\}\}$;    
\end{codebox} 

\begin{codebox} 
\Procname{$\proc{One-Step}(\{\{\psi\}\},\{i(1,1),\dots,i(k,l_k)\},\{l\},k,N,m)$}
\li \For $j=1,\dots,N$:
\li \Do
     $Z_{\rm new} \gets 1$; 
\li     \For $d=1,\dots,k$:
\li     \Do
            Draw indexes $j(d,1),\dots,j(d,l_d)$ uniformly in $\{1,\dots,N\}$;  
\li         Compute $\chi_d$ from $\{\psi_{i(d,1),j(d,1)},\dots,\psi_{i(d,l_d),j(d,l_d)}\}$ using (\ref{eq:BP_2}); 
\li         $Z_{\rm new} \gets Z_{\rm new}\cdot Z_d$, $Z_d$ is the norm. from (\ref{eq:BP_2}); 
        \End
\li     Compute $\psi_{\rm new}$ from  $\{\chi_{1},\dots,\chi_{k}\}$ according to eq.~(\ref{eq:BP_1});  
\li     $Z_{\rm new} \gets Z_{\rm new}\cdot Z_d$, $Z_d$ is the norm from (\ref{eq:BP_1});
\li     $w_j\gets \big(Z_{\rm new}\big)^m$;  
\li     $\psi_j \gets \psi_{\rm new}$;
    \End
\li $\{\psi\} \gets \proc{Reweighting-Faster}(N,\{\psi\},\{w\})$; 
\li \Return array $\{\psi\}$;    
\end{codebox} 

Depending on the problem we are about to solve the population of populations might also be combined with the reweighting of populations or the separation of the frozen and soft fields, see e.g. appendix D of [\ZK]. 

\section{How many populations needed?}

We make a summary of which level of the population dynamics technique is needed depending on the problem. References are just examples and are biased towards works presented in this thesis. 

\begin{itemize}
\item{ Analytical solution
  \begin{itemize}
  \item Belief propagation on regular graphs [\ZMM, \ZK, \ZML].
  \item General warning propagation with integer warnings [\ZMM, \RS].
  \item Frozen variables at $m=1$ [\ZK, \ZML].
  \item Survey propagation on regular graphs (frozen variables at $m=0$, energetic cavity) \cite{KrzakalaPagnani04} or [\ZK, \ZML].
  \end{itemize}
}
\item{ Single population
  \begin{itemize}
  \item General belief propagation in models with discrete variables [\ZMM, \ZK, \ZML].
  \item General survey propagation (1RSB at $m=0$, energetic cavity) on model with integer warnings [\RS, \ZML], or very precise numerics in \cite{MertensMezard06}.
  \item 1RSB at $m=1$ \cite{MezardMontanari06,MontanariRicci08} or [\KM, \ZK].
  \item 1RSB on random regular graphs [\KM, \ZK].
  \item 2RSB at $m=0$ (energetic cavity) on regular graphs \cite{RivoireThese}.
  \end{itemize}
}
\item{ Population of populations
  \begin{itemize}
  \item General 1RSB (also finite temperature)  \cite{MezardParisi01,MezardPalassini05,MontanariRicci08} or [\KM, \ZK, \KZP].
  \item 2RSB of random regular graphs [\KZP].
  \item 2RSB at $m=0$ (energetic cavity).
  \item 3RSB at $m=0$ (energetic cavity) on regular graphs.
  \end{itemize}
}
\end{itemize}

We are not aware on any work where the last two points would be implemented. 
More levels of replica symmetry breaking would require more levels of populations. We are not aware of any work where more than population of populations would be treated. Rather than pushing the numerics in this direction new theoretical works are needed for models where the 1RSB solution is not correct.

\chapter{Algorithms}
\label{app:alg}

Here we do not aim to provide a complete summary of algorithms used to solve the random constraint satisfaction problems. We just define and briefly discuss algorithms which were used, generalized or tested in the context of this thesis. Strictly speaking we are almost always dealing with incomplete solvers, that is algorithms which might find a solution but never provide a certificate of unsatisfiability. It is an open and interesting questions if the methods presented in this thesis can imply something for certification of unsatisfiability.

\section{Decimation based solvers}

A large class of algorithms for CSPs is based on the following iterative scheme:

\begin{codebox}
\Procname{$\proc{Decimation}$}
\li \Repeat Choose a variable $i$ \label{line_var};
\li         Choose a value $s_i$ \label{line_val};
\li         Assign $i$ the value $s_i$ and simplify the formula;
\li \Until  Solution or contradiction is found;
\end{codebox}

The nontrivial part is how to choose a variable in step \ref{line_var} and how to choose its value in step \ref{line_val}. In the following we describe several more or less sophisticated or efficient strategies.

Note that all these strategies can be improved by {\it backtracking}, that is if a contradiction was found we return to the last variable where another value than the one we chose was possible and make this choice instead. 

\subsection{Unit Clause propagation}

One of the simplest (and obvious) strategies is to choose and assign a variable which is present in a constraint which is compatible with only one value of that variable. In K-SAT this is equivalent to assigning variables belonging to clauses which contain only this variable, hence the name {\it unit clause}. If no such variable exists one possibility (the random heuristics) is to choose an arbitrary  
variable and assign it a random value from the available ones. The unit clause propagation combined with the random heuristics (without backtracking) is not very efficient solver of K-SAT. But the situation is more fortunate for some other constraint satisfaction problems. The most interesting example being perhaps the 1-in-K SAT \cite{AchlioptasChtcherba01} and [\RS]. The random 1-in-K SAT exhibits a sharp satisfiability phase transition for $K\ge 3$. Moreover, if the probability of negation of variables lies in the interval $(0.2726,0.7274)$ (for $K=3$) then: 
\begin{itemize}
   \item In the satisfiable phase the unit clause propagation combined with the random heuristics finds a solution with finite probability in every run.
   \item In the unsatisfiable phase every run of the unit clause propagation leads to a contradiction with finite probability after the assignment of the very first variable. 
\end{itemize}
Hence, with random restarts the random 1-in-3 SAT is almost surely solvable in polynomial time in the whole phase space (given the probability of a negation is as specified above). At the same time the 1-in-3 SAT is an NP-complete problem, it thus provides a rare example of an on average easy NP-complete problem with a satisfiability phase transition.

Unit clause propagation is the main element of all the exact solvers of constraint satisfaction problems. The most studied example being the Davis-Putnam-Logemann-Loveland (DPLL) algorithm \cite{DavisPutnam60,DavisLogemann62}\index{algorithms!DPLL} for K-SAT which combines the unit clause propagation with the pure literal elimination (pure literal appears either only negated or non-negated) with backtracking. It was mostly this algorithm which was used when the connection between the algorithmical hardness and phase transitions was being discovered \cite{MitchellSelman92,CheesemanKanefsky91}. Moreover, all the modern complete solvers of the satisfiability problem follow a similar, more elaborated, path.

\subsection{Belief propagation based decimation}
\label{sec:BP_dec}

Belief propagation \cite{Pearl82} computes, or on general graphs approximates, marginal probabilities. These can then be used to find an actual solution. In some problems the marginals give the solution directly, e.g. in the error correcting codes \cite{Gallager68}, in the matching \cite{BayatiShah05,BayatiShah06}, or the random field Ising model at zero temperature \cite{KolmogorovWainwright05,Chertkov08} etc. In constraint satisfaction problems, typically, marginals do not give a direct information about a solution. For example in coloring of random graphs, the BP equations always converge to all marginals being equal to $1/q$. 
Belief propagation based decimation strategies have been studied recently. 

In every cycle of the algorithm \proc{Decimation}, the belief propagation equations are updated until they converge or a maximal number of updates per variable $T_{\rm max}$ is reached. At least two strategies how to choose the decimated variable and its value were tested and studied, see e.g. [\KM] and \cite{MontanariRicci07}:
\begin{itemize}
   \item{Uniform BP decimation\index{algorithms!belief propagation!decimation}\index{belief propagation!decimation} -- Choose a variable at random and assign its value according to the marginal probability estimated by BP. }
   \item{Maximal BP decimation -- Find the variable with the most biased BP marginal and assign it the most probable value.}
\end{itemize}
The other two combinations where a random variables is assigned its most probable value or when the most biased variable is assigned random value according to its marginal probability can be think of. 
The BP decimation, as described above, runs in quadratic time. In eventual practical implementations a small fraction of variables should be decimated at each step, thus reducing the computational complexity to linear (or log-linear if the maximum convergence time increases as $\log{N}$).

The empirically best strategy is the maximal BP decimation. This can be understood from the fact that this strategy aims to destroy the smallest possible number of solutions in every step, as argued on a more quantitative level in \cite{Parisi03}. We gave as an example the performance of the maximal BP decimation in the 3- and 4-coloring of random Erd\H{o}s-R\'enyi graphs [\ZK] in fig.~\ref{fig:BP_col}.

The uniform BP decimation is less successful, because it aims not only to find a solution but also to sample solutions uniformly at random. Indeed, if an exact calculation of marginal probabilities would be used instead of the BP estimates the uniform {\it exact} decimation would lead to a perfect sampling. The uniform exact decimation is a process which can be analyzed using the cavity method. The result then sheds light on the limitations of the BP decimation. This analysis was developed in \cite{MontanariRicci07}, and we give an example for the factorized occupation problems in the following.  

\subsubsection{Maximal BP decimation on the random coloring}
\label{sec:BP_col}

We implemented the maximal BP decimation algorithm on the random graph coloring. We chose $T_{\max}=10$, if a solutions is not found we restart with $T_{\max}=20$ and eventually once again with $T_{\max}=40$. The fraction of successful runs is plotted in fig.~\ref{fig:BP_col} and we see that this algorithm works even in condensed phase where the BP marginals are not asymptotically correct, or in a phase where the equations do not even converge. The non-convergence of the belief propagation equations is ignored (in 3-coloring from the beginning, in 4-coloring after a small fraction, typically around 10\%, of variables was fixed). It thus seems that in coloring the BP decimation is a very robust algorithm.

What is the reason for the failure of the maximal BP decimation at higher connectivities? A straightforward suggestion would be that is should not work in the condensed phase where the BP marginals are not asymptotically correct. But we do not observe anything particular in the performance curves at the condensation transition. A second natural suggestion would be that BP should converge in order that the algorithm works, this also does not seem to be the case, as BP does not converge in the 3-coloring for connectivity $c>4$ and yet the algorithm is perfectly able to find solutions. Moreover, even in 4-coloring where the BP equations converge on large formulas in all the satisfiable phase, after a certain (rather small) fraction of variables is decimated the convergence is lost. As we argued in appendix \ref{app:RS_stab} the non-convergence of BP is equivalent to the local instability of the replica symmetric solution. It thus seems that the reduced problem, after a certain fraction of variable was fixed, is even harder from the statistical physics perspective than the original problem. Yet, this does not seem to be fatal for the finding of solutions. Finally, in the region where the BP decimation algorithm really does not succeed we observed that a precursor of the failure exists. The normalizations in the BP equations (\ref{eq:BP_1}-\ref{eq:BP_2}) gradually decreases to zero, meaning that the incoming beliefs become almost contradictory.

\subsubsection{Analysis of the uniform exact decimation}
\label{sec:BP_anal}

The uniform exact decimation after $\theta N$ steps is equivalent to taking a solution uniformly at random and fixing its first $\theta N$ variables. Such a procedure can be analyzed \cite{MontanariRicci07} and conclusions made about the influence of small errors in the BP estimates of marginals. 

Given an instance of the CSP, consider a solution $\{s\}$ taken uniformly at random and reveal the value of each variable with probability $\theta$. Denote $\Phi$ the fraction of variables which were either revealed or are directly implied by the revealed ones. To compute $\Phi(\theta)$ we derive the cavity equations on a tree. Denote $\Phi^{i\to b}_s$ the probability that a variable $i$ is fixed conditioned on the value $s$ of the variable $i$ and on the absence of the edge $(ib)$:
\be
  \Phi^{i\to b}_s = \theta + (1-\theta) \left[ 1 - \prod_{a\in \partial i - b}(1-q_s^{a \to i}  )  \right]\, .
\ee 
Meaning that the variable $i$ was either revealed or not, and if not it is implied if at least one of the incoming constraints implies it. The $q_s^{a \to i}$ is a probability that constraint $a$ implies variable $i$ to be $s$ conditioned on: 1) variable $i$ taking the value $s\in \{s\}$ in the solution we chose, 2) variable $i$ was not  revealed directly and 3)  the edge $(ai)$ is absent.

We write the expression for $q_s^{a \to i}$ only for random occupation CSPs on random regular graphs where the replica symmetric equation is factorized. Then also $q_s^{a \to i}$ and $\Phi^{i\to b}_s$ are factorized, that is independent of $a,b,i$. The conditioned probability $q_s$ is the ratio of the probability that variable $i$ takes the value $s$ and is implied by the constraint $a$ and probability that variable $i$ takes the value $s$:
\begin{subequations}
\label{eq:qs}
\bea
   q_1 &=& \frac{1}{\psi_1 Z^{\rm reg}}\sum_{A_{r+1}=1,A_r=0} {k \choose r} (\psi_1)^{lr}  (\psi_0)^{l(k-r)} \sum_{s=0}^{s_1} {r \choose s} \Phi_0^{k-r} \Phi_1^{r-s} (1-\Phi_1)^{s} \, ,  \label{eq:q1}\\
   q_0 &=& \frac{1}{\psi_0 Z^{\rm reg}}\sum_{A_r=1,A_{r+1}=0} {k \choose r} (\psi_1)^{lr}  (\psi_0)^{l(k-r)}\sum_{s=0}^{s_0} {k-r \choose s} \Phi_1^{r} \Phi_0^{k-r-s} (1-\Phi_0)^{s}\, ,\label{eq:q0} 
\eea
\end{subequations}
where $l=L-1$, $k=K-1$. The indexes $s_1,s_0$ in the second sum of both equations are the largest possible but such that $s_1\le r$, $s_0\le K-1-r$, and $\sum_{s=0}^{s_1}A_{r-s}=0$, $\sum_{s=0}^{s_0}A_{r+1+s}=0$. The terms $\Phi^r_1 \Phi^{k-r-s}_0 (1-\Phi_0)^s$ and $\Phi^{r-s}_1 \Phi^{K-r-1}_0(1-\Phi_1)^s$ are the probabilities that a sufficient number of incoming variables was revealed such that the out-coming variable is implied (not conditioned on its value). The first sum goes over all the possible numbers of $1$'s being assigned on the incoming variables, $r$. The term $\psi_1^{lr} \psi_0^{l(k-r)}$ is then the probability that such a configuration took place. The cavity probabilities that the corresponding variable takes value $0/1$, $\psi_0,\psi_1$ are taken from the BP equations (\ref{RS1}-\ref{RS2}), $Z^{\rm reg}$ is the normalization in (\ref{RS1}-\ref{RS2}). The first condition on $r$  takes care about the values of the incoming neighbours being compatible with the value of the variable $i$ on which is conditioned, the second condition on $r$ is satisfied  if and only if the value of the variable $i$ is implied by the incoming configuration. 

Once a solution for $q_s$ is found (from initial conditions $\Phi=\theta$) the total probability that a variable is fixed is computed as
\be
      \Phi(\theta) = \theta + (1-\theta) \left\{ \mu_1 [1- (1- q_1)^L ]    + \mu_0  [1- (1- q_0)^L ] \right\}\, , \label{eq:phi_th}
\ee
where $\mu_0,\mu_1$ are the total BP marginals, $\mu_s=\psi_s^L/(\psi^L_0+\psi^L_1)$.

Notice the complete analogy between eqs.~(\ref{eq:q0}-\ref{eq:q1}) and the equations for hard fields at $m=1$ (\ref{eq:mu0}-\ref{eq:mu1}). To compute the function $\Phi(\theta)$ for a general CSP on a general graph ensemble a derivation in the lines of app.~\ref{app:m1} have to be adapted, see also \cite{MontanariRicci07}. Finally note that as the probabilities $\psi_1$, $\psi_0$ are taken from the belief propagation equations the form (\ref{eq:q0}-\ref{eq:q1}) is not correct in the condensed phase (but in the locked problems the satisfiable phase is never condensed).

\subsubsection{The Failure of Decimation in the Locked problems}

In the locked problems, see sec.~\ref{locked}, the BP decimation algorithm does not succeed to find a satisfying assignment even at the lowest possible connectivity. To give an example in the 1-or-3-in-5 SAT on truncated Poissonian graphs the maximal BP decimation succeeds to find a solution in only about 25\% at the lowest average connectivity $\overline l=2$, and this fraction drops down to less than 5\% at already $\overline l = 2.3$ (to be compared with the clustering threshold $l_d=3.07$, or the satisfiability threshold $l_s=4.72$).  

Interestingly, the precursors of the failure of the BP decimation algorithm observed in the graph coloring are not present in the locked problems. In particular the BP equations converge during all the process and the normalizations in the BP equations (\ref{eq:BP_1}-\ref{eq:BP_2}) stays finite. However, the above analysis of the function $\Phi(\theta)$ sheds light on the origin of the failure. 

\begin{figure}[!ht]
  \resizebox{\linewidth}{!}{
  \includegraphics{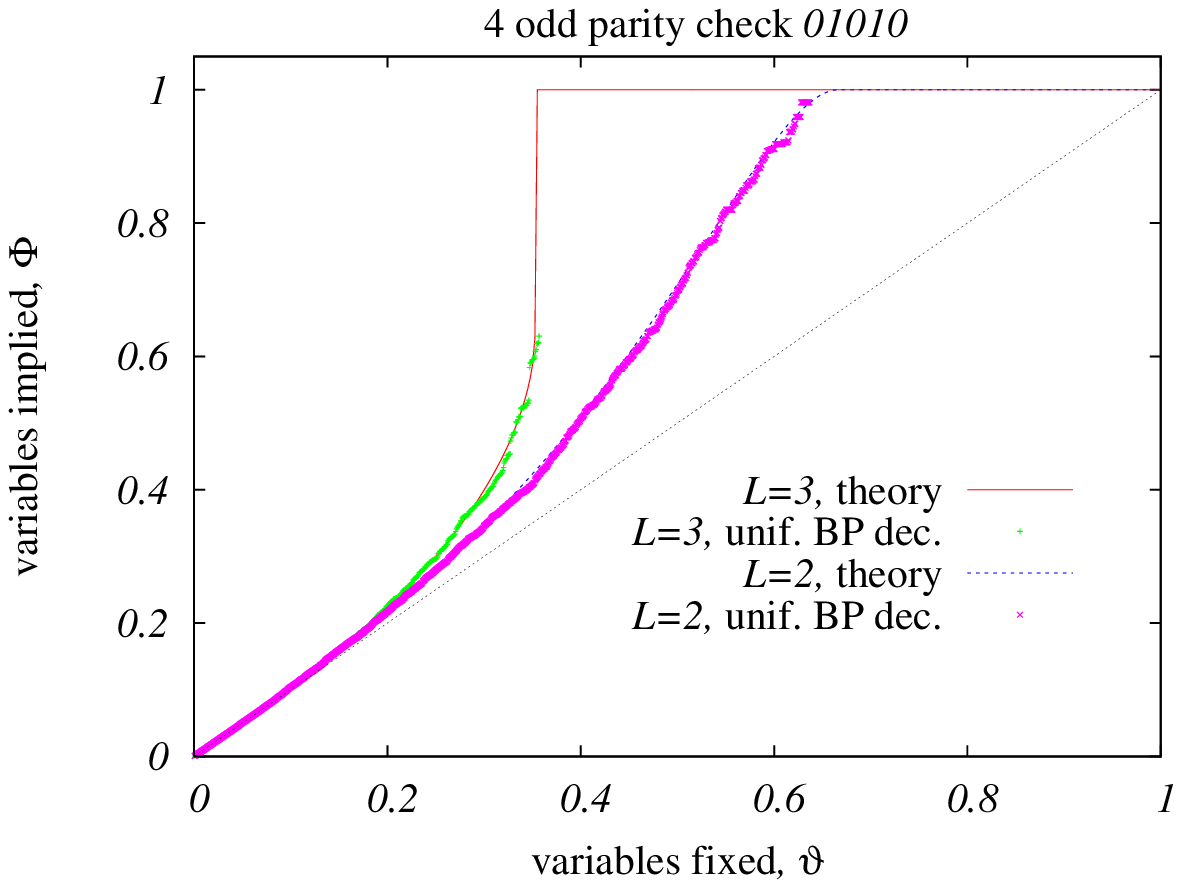}
  \includegraphics{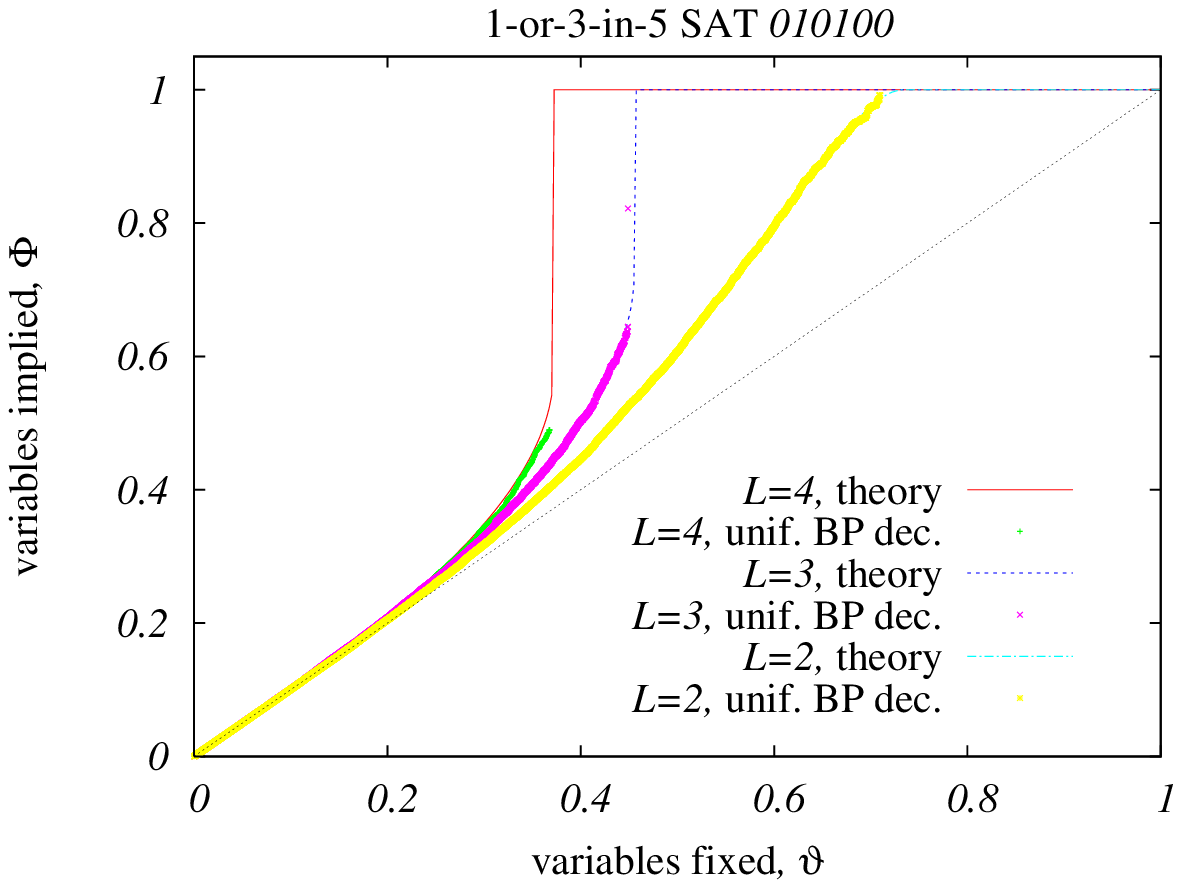}}
  \caption{\label{Fig:BP_anal} Analytical analysis of the BP inspired uniform decimation. Number of variables directly implied $\Phi(\theta)$ plotted against number of variables fixed $\theta$.
}
\end{figure}

\begin{figure}[!ht]
  \resizebox{\linewidth}{!}{
  \includegraphics{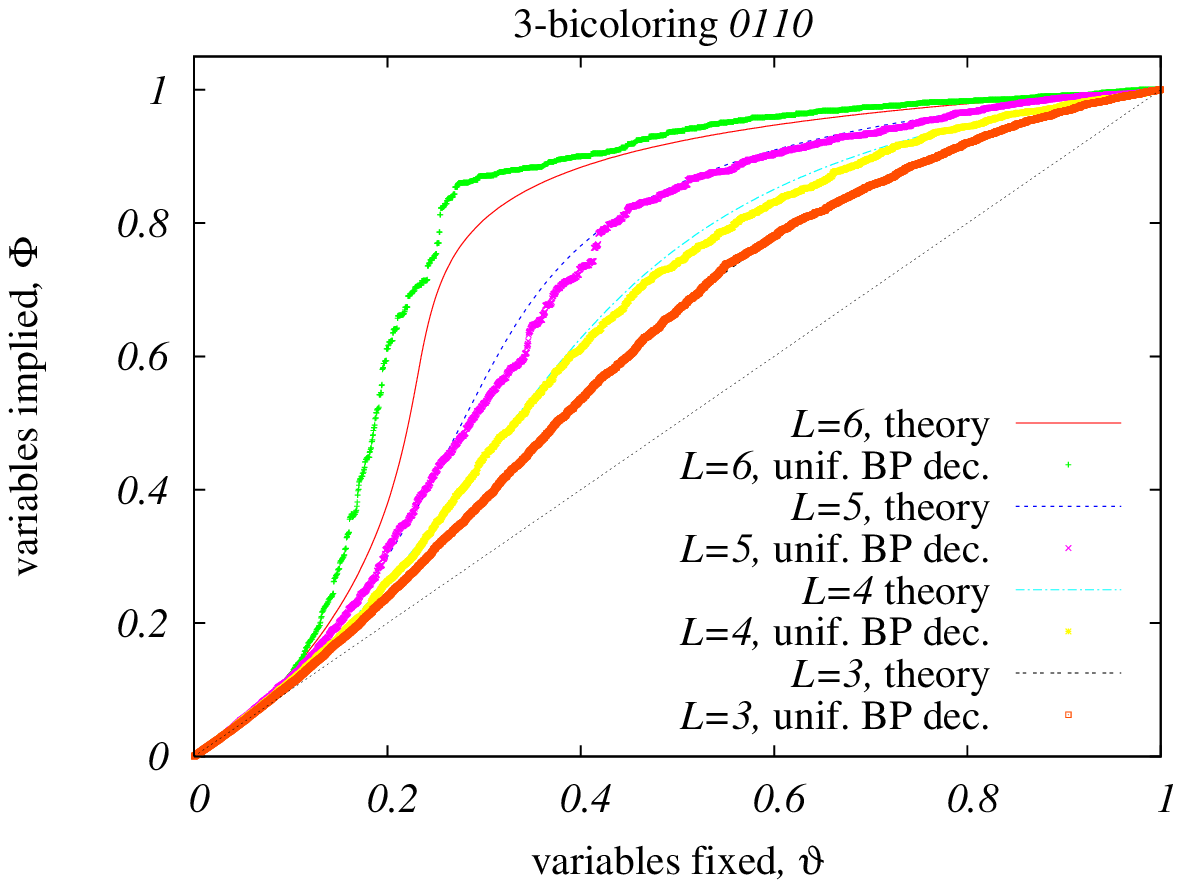}
  \includegraphics{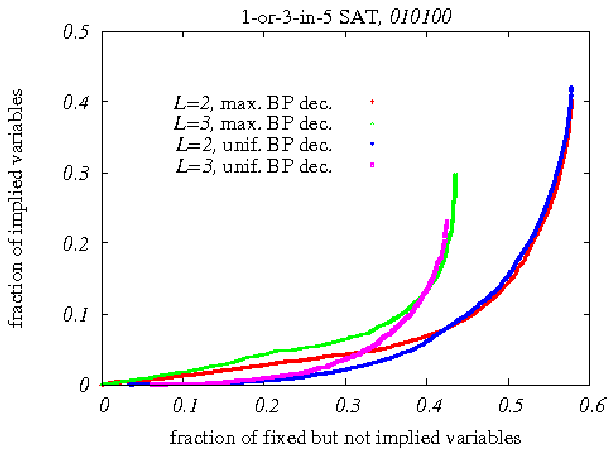}}
  \caption{\label{Fig:BP_dec} Left: For comparison, the BP uniform decimation works well on the non-locked problems, the example is for bicoloring. Right: Comparison of the maximal and uniform decimation. Number of directly implied variables is plotted against number of variables which were free just before being fixed. Behaviour of the two decimation strategies is similar.}
\end{figure}

In fig.~\ref{Fig:BP_anal} we compare the function $\Phi(\theta)$ (\ref{eq:phi_th}) with the experimental performance of the uniform BP decimation. Before the failure of the algorithm (when a contradiction is encountered) the two curves collapse perfectly. The reason why the algorithm fails to find solutions is now transparent. 
\begin{itemize}
   \item{Avalanche of direct implications -- In some cases the function $\Phi(\theta)$ has a discontinuity at a certain spinodal point $\theta_s$ ($\theta_s\approx 0.46$ at $L=3$ of the 1-or-3-in-5 SAT). Before $\theta_s$ after fixing one variable there is a finite number of direct implications. As the loops are of order $\log{N}$ these implications never lead to a contradiction. At the spinodal point $\theta_s$ after fixing one more variable and extensive avalanche of direct implications follows. Small (order $1/N$) errors in the previously used BP marginals may thus lead to a contradiction. This indeed happens in almost all the runs we have done. For more detailed discussion see \cite{MontanariRicci07}.}
   \item{No more free variables -- The second reason for the failure is specific to the locked problems, more precisely to the problems where $\Phi=1$ is a solutions of (\ref{eq:q1}-\ref{eq:q0}). In these cases function $\Phi(\theta)\to 1$ at some $\theta_1<1$ ($\theta_1\approx 0.73$ at $L=4$ of 1-or-3-in-5 SAT). In other words if we reveal a fraction $\theta>\theta_1$ of variables from a random solution, the reduced problem will be compatible with only that given solution. 
Again a little error in the previously fixed variables and the BP uniform decimation ends up in a contradiction. 
If on the contrary the function $\Phi(\theta)$ reaches value 1 only for $\theta=1$ then the residual entropy is positive and there should everytime be some space to correct previous small errors, demonstrated on a non-locked problem in fig.~\ref{Fig:BP_dec}.  
}
\end{itemize}

These two reasons of failure of the BP uniform decimation seems quite different. But they have one property in common. As the point of failure is approached we observe a divergence of the ratio between the number of variables which were not directly implied before being fixed and the number of directly implied variables, see fig.~\ref{Fig:BP_dec}. This ratio can also be computed for the maximal BP decimation and no quantitative difference is observed for the locked problems, thus the two reasons above explain also the failure of the, otherwise more efficient, maximal BP decimation.  

\subsection{Survey propagation based decimation}

The seminal works \cite{MezardParisi02,MezardZecchina02} not only derived the survey propagation equations, but also suggested it 
as a base for a decimation algorithm for random 3-SAT. The performance is spectacular, near to the satisfiability threshold on large random 3-SAT formulas it works faster than any other known algorithm. SP based decimation seem to be able to find solutions in $O(N\log{N})$ time up to the connectivity $\alpha=4.252$ in 3-SAT \cite{Parisi03} (to be compared with the satisfiability threshold $\alpha_s=4.267$).

Survey propagation equations (\ref{SP_1}-\ref{SP_2}) aim to compute the probability (over clusters) that a certain variables is frozen to take a certain value. This information can then be used to design a strategy for the \proc{Decimation} algorithm. In particular, as long as the result of survey propagation is nontrivial (not all $p_0^{i\to a}=1$) the variable with the largest bias $|p_+^i-p_-^i|$ is chosen and is assigned the more probable value. After a certain fraction of variables is decimated the fixed point of the survey propagation on the reduced formula is trivial. The suggestion of  \cite{MezardParisi02,MezardZecchina02} is that such a reduced formula is easily satisfiable and some of the well known heuristic algorithms may be used to solve it (Walk-SAT, see the next section \ref{sec:SLS}, was used in the original implementation). Note also that the original implementation of \cite{MezardParisi02,MezardZecchina02} decimated a fraction of variables at each \proc{Decimation} step, thus reducing significantly the computational time. \index{algorithms!survey propagation!decimation}\index{survey propagation!decimation}

Originally, the success of the survey propagation based algorithm was contributed to the fact that survey propagation equations take into account the clustering of solutions. This was, however, put in doubt since. To give an example, in the locked problems, see sec.~\ref{locked}, the survey propagation equations give an identical fixed point as the belief propagation and as we argued in the previous section \ref{sec:BP_dec} the maximal BP decimation fails to find solutions in the locked problem in the whole range of connectivities. 

The true reason for the high performance of survey propagation in 3-SAT thus stays an open problem. For example, and unlike with BP, there are usually no problems with SP convergence during the decimation. Two very interesting observations were made in \cite{KrocSabharwal08} for SP the decimation algorithm on $K$-SAT. First, the SP decimation indeed makes the formula gradually simpler for local search algorithms, see sec.~\ref{sec:SLS}, again in contrast with BP decimation. Second, the SP decimation on $K$-SAT does not create any (or a very small number) of direct implications (unit clauses) during the process. Given that creation of direct implication makes the decimation fail in the locked problems, as we just showed, this might be a promising direction for a new understanding.

\section{Search of improvement based solvers}

Here we describe another large class of CSPs solvers, the {\it search of improvement algorithms}. All these algorithms start with a random assignment of variables. Then different rules are adopted to gradually improve this assignment and eventually to find a solution. The most typical example of that strategy is the simulated annealing \cite{KirkpatrickGelatt83} or stochastic local search algorithms like Walk-SAT \cite{SelmanLevesque92,SelmanKautz94}.

\subsection{Simulated annealing}

In physics simulated annealing is a popular and very universal solver of optimization problems. It is based on running the Metropolis \cite{MetropolisRosenbluth53} (or other Monte Carlo) algorithm and gradually decreasing the temperature-like parameter. Simulated annealing algorithm respects the detailed balance condition, after large time it thus converges to the equilibrium state, and it is thus guaranteed to find the optimal state in a finite time for a finite system size. In general, the time can of course depend exponentially on the system size, and in such a case it is not really of practical interest. 

We argued in chap.~\ref{clustering} that at the clustering (dynamical) transition the equilibration time of a detailed balance local dynamics diverges. However, the clusters which appear at the dynamical energy $E_d>0$ have bottom at an energy $E_{\rm bottom}\le E_d$ and numerical performance of the simulated annealing in the 3-coloring of random graphs \cite{MourikSaad02} suggests that $E_{\rm bottom}$ might be zero even if $E_d$ is positive. More precise numerical investigation of this point is, however, needed.

\subsection{Stochastic local search}
\label{sec:SLS}

Solving $K$-SAT by a pure random walk was suggested in \cite{Papadimitriou91}:

\begin{codebox}
\Procname{$\proc{Pure-Random-Walk-SAT}(T_{\rm max})$}
\li Draw a random assignment of variables;
\li $T\gets 0$;
\li \Repeat Draw a random unsatisfied constraint $a$;
\li         Flip a random variable $i$ belonging to $a$;
\li         $T\gets T+1$;
\li \Until  Solution is found or $T>NT_{\rm max}$; 
\end{codebox}

In random 3-SAT this simple strategy seems to work in linear time up to $\alpha_{\rm RW}\approx 2.7$ \cite{SemerjianMonasson03}. Improvements of the \proc{Pure-Random-Walk-SAT} have led to a large class of so-called stochastic local search algorithms. All are based on a random walk in the configurational space with more complicated rules about which variables would be flipped. The version called \proc{WalkSAT} introduced in  \cite{SelmanKautz94,SelmanKautz96} became, next to the DPLL-based exact solvers, the most widely used solver of practical SAT instances. In random 3-SAT the Walk-SAT with $p=0.5$ was shown to work in linear time up to about $\alpha_{\rm WS}=4.15$ \cite{AurellGordon04}. \index{algorithms!stochastic local search}\index{algorithms!Walk-SAT} 

\begin{codebox}
\Procname{$\proc{WalkSAT}(T_{\rm max},p)$}
\li Draw a random assignment of variables;
\li $T\gets 0$;
\li \Repeat Pick a random unsatisfied constraint $a$;
\li         \If Exists a variable $i$ in $a$ that is not necessary in any other constraint;
\li         \Then Flip this variable $i$;
\li         \Else \If $\proc{Rand}<p$;
\li               \Then  Flip a random variable $i$ belonging to $a$;
\li               \Else  Flip $i$ (from $a$) that minimizes the $\#$ of unsat. constraints;
                   \End 
            \End      
\li         $T\gets T+1$;
\li \Until  Solution is found or $T>NT_{\rm max}$; 
\end{codebox}

Several other variants of stochastic local search on random 3-SAT were studied in \cite{SeitzAlava05} showing that with a proper tuning of parameters like $p$ the linear performance can be extended up to at least $\alpha \approx 4.20$. Finally a version of the stochastic local search called \proc{ASAT} was introduced in \cite{ArdeliusAurell06}. In random 3-SAT \proc{ASAT} works in a linear time at least up to $\alpha=4.21$  \cite{ArdeliusAurell06}. We adapted the implementation of \proc{ASAT} and studied its performance in coloring and on the occupation CSPs.\index{algorithms!ASAT}

\begin{codebox}
\Procname{$\proc{ASAT}(T_{\rm max},p)$}
\li Draw a random assignment of variables;
\li $T\gets 0$;
\li Create the list $\{v\}$ of variables which are present in unsatisfied constraints.
\li \Repeat Pick a random variable $i$ from the list $\{v\}$;
\li         Compute the change of energy $\Delta E$ if the value of $i$ is flipped. \label{line_flip}
\li         \If $\Delta E \le 0$;
\li         \Then Flip $i$;
\li         \Else \If $\proc{Rand}<p$;
\li               \Then  Flip $i$;
\li               \Else  Do nothing;
               \End
            \End  
\li         Update list $\{v\}$ of variables which are present in unsatisfied constraints.   
\li         $T\gets T+1$;
\li \Until  Solution is found or $T>NT_{\rm max}$; 
\end{codebox}

In the coloring problem where variables take one from more than two possible values, the only modification of \proc{ASAT} is that we choose a random value into which the variable is flipped on line \ref{line_flip}. The performance for the 4-coloring of Erd\H{o}s-R\'enyi graphs was sketched in fig.~\ref{fig:WalkCOL}.

There are two free parameters in the \proc{ASAT} algorithm, the maximal number of steps per variable $T_{\rm max}$ and, more importantly, the greediness (temperature-like) parameter $p$, which need to be optimized. In \cite{ArdeliusAurell06} and [\ZK] it was observed that in the random K-SAT and random coloring problems the optimal value of $p$ does not depend on the system size~$N$, neither very strongly on the constraint density $\alpha$. But these observation might be model dependent, as it indeed seems to be the case for the locked problems.

\subsection{Belief propagation reinforcement}
\label{sec:rein}

A "search of improvement" solver can also be based on the belief propagation equations. The idea of the {\it belief propagation reinforcement}\index{algorithms!belief propagation!reinforcement}\index{belief propagation!reinforcement}, introduced in \cite{ChavasFurtlehner05}~\footnote{Strictly speaking the reinforcement strategy was fist introduced for the survey propagation equations, but the concept is the same for belief propagation.}, is to write belief propagation equations with an external "magnetic" field (site potential) $\mu^i_{s_i}$
\begin{subequations}
\bea
     \psi_{s_i}^{a\to i} &=& \frac{1}{Z^{a\to i}} \sum_{A_{s_i+\sum s_j}=1} \prod_{j\in \partial a-i} \chi_{s_j}^{j\to a}\, , \label{BP1_r}  \\
     \chi_{s_i}^{i\to a} &=& \frac{1}{Z^{i\to a}} \, \mu^i_{s_i} \prod_{b\in \partial i-a} \psi_{s_i}^{b\to i}\, , \label{BP2_r} 
\eea
\end{subequations}
and then iteratively update this field in order to make the procedure converge to a solution given by the direction of the external field $r_i={\rm argmax}_{s_i} \mu^i_{s_i}$. At every step the configuration given by the direction of the external field is regarded as the current configuration which is being improved. 

The question is how to update the external field. The basic idea is to choose the local potential $\mu^i_{s_i}$ in some way proportional to the current value of the total marginal probability $\chi^i_{s_i}$, which is computed without the external fields as 
\be
   \chi_{s_i}^{i} = \frac{1}{Z^{i}} \prod_{b\in \partial i} \psi_{s_i}^{b\to i} \label{BPf_r} \, .
\ee

How exactly, and how often should the value of local potential be updated is open to many different implementations, some of them can be found in \cite{BraunsteinZecchina06,DallAstaRamezanpour08}. The same as in the local search algorithm it is not well understood, beyond a purely experimental level, how the details of the implementation influence the final performance. We tried several ways and the best performing seemed to be the following 
\begin{subequations}
\bea
   \mu^i_1 = (\pi)^{l_i-1},  \quad  \mu^i_0 = (1-\pi)^{l_i-1},  \quad {\rm if} \quad \xi^i_{0}>\xi^i_{1} \, , \label{rein1}\\
   \mu^i_1 = (1-\pi)^{l_i-1},  \quad  \mu^i_0 = (\pi)^{l_i-1},  \quad {\rm if} \quad \xi^i_{0}\le \xi^i_{1} \, , \label{rein2}
\eea
\end{subequations}
where $0\le \pi\le 1/2$, $l_i$ is the degree of variable $i$ and the auxiliary variable $\xi^i_{s_i}$ is computed before updating the field $\mu^i$
\be
     \xi^i_{s_i} = (\mu^i_{s_i})^{\frac{1}{l_i-1}}\,  \chi_{s_i}^{i}\, .
\ee

\begin{codebox}
\Procname{$\proc{BP-Reincorcement}(T_{\rm max},n,\pi)$}
\li Initialize $\mu^i_{s_i}$ and $\psi_{s_i}^{a\to i}$ randomly;
\li $T\gets 0$;
\li Compute the current configuration $r_i={\rm argmax}_{s_i} \mu^i_{s_i}$;
\li \Repeat Make $n$ sweeps of the BP iterations (\ref{BP1_r}-\ref{BP2_r}); \label{line_BP}
\li         Update all the local fields $\mu_{s_i}^i$ according to (\ref{rein1}-\ref{rein2});
\li         Update $r_i={\rm argmax}_{s_i} \mu^i_{s_i}$;
\li         $T\gets T+1$;
\li \Until  $\{r\}$ is a solution or $T>T_{\rm max}$; 
\end{codebox}

How should the strength of the forcing $\pi$ be chosen? Empirically we observed three different regimes:
\begin{itemize}
   \item[a)]{$\pi_{\rm BP-like}<\pi<0.5 $: When the forcing is weak the \proc{BP-Reinforcement} converges very fast to a BP-like fixed point, the values of the local fields do not point towards any solution. On contrary many constraints are violated by the final configuration $\{r_i\}$.}
   \item[b)]{$\pi_{\rm conv}< \pi < \pi_{\rm BP-like} $: The \proc{BP-Reinforcement} converges to a solution $\{r_i\}$.}
   \item[c)]{$0 < \pi < \pi_{\rm conv}$: When the forcing is too strong the \proc{BP-Reinforcement} does not converge. And many constraints are violated by the configuration $\{r_i\}$ which is reached after $T_{\rm max}$ steps.}
\end{itemize}
When the constraint density in the CSP is large the regime b) disappears and $\pi_{\rm conv}= \pi_{\rm BP-like}$. For an obvious reason our goal is to find $\pi_{\rm conv}<\pi<\pi_{\rm BP-like}$. The point $\pi_{\rm BP-like}$ is very easy to find, because for larger $\pi$ the convergence of \proc{BP-Reinforcement} to a BP-like fixed point happens in just several sweeps. Thus in all the runs we chose $\pi$ to be just bellow $\pi_{\rm BP-like}$, that is to hit the possible gap between $\pi_{\rm BP-like}$ and $\pi_{\rm conv}$. The value of $\pi$ chosen in this way does not seem to depend on the size of the system, it, however, depends slightly on the constraint density. 

Experimentally it seems that the optimal number of BP sweeps on line \ref{line_BP} of \proc{BP-Reinforcement} is very small, typically $n=2$, in agreement with \cite{ChavasFurtlehner05}. We observed with a surprise that when $n$ is much larger not only the total running time is larger but the overall performance of the algorithm is worse.  

In the regime where the \proc{BP-reinforcement} algorithm performs well the median running time $T$ seems to be independent of the size, leading to an overall linear time complexity. The total CPU time is comparable to the time achieved by the stochastic local search \proc{ASAT}.

There is an imperfection of our implementation of the \proc{BP-reinforcement}, because in small fraction of cases, for all connectivities, the algorithm is blocked in a configuration with only 1-3 violated constraints. If this happens we reinforce stronger the problematic variables which sometimes shifts the problem to a different part of the graph, where it might be resolved. Also a restart leads to a solution. 

We tested the \proc{BP-Reinforcement} algorithm mainly in the occupation CSPs, the results are shown in sec.~\ref{locked}. Survey propagation reinforcement can be implemented in a similar way, as was done originally in \cite{ChavasFurtlehner05}.\index{algorithms!survey propagation!reinforcement}

\end{appendices}

\newpage
\chapter*{\hspace{2.5cm} }
\vspace{4cm}
\begin{center}
{\huge \bf Reprints of Publications}
\end{center}

\pagestyle{myheadings}
\markboth{REPRINTS OF PUBLICATIONS}{REPRINTS OF PUBLICATIONS}

\addcontentsline{toc}{chapter}{Reprints of Publications}

\addcontentsline{toc}{section}{{\bf [ZDEB-1]} LZ, M. M\'ezard, "The number of matchings in random graphs", \\ {\it J. Stat. Mech.} (2006) P05003}
\chapter*{[ZDEB-1]}

\markboth{REPRINT OF PUBLICATION [ZDEB-1]}{REPRINT OF PUBLICATION [ZDEB-1]}

\begin{center}

{\LARGE\bf "The number of matchings in random graphs"}

\vspace{1cm}

{\large L. Zdeborov\'a, M. M\'ezard, {\it J. Stat. Mech}, P05003 (2006).}

\vspace{1cm}

{\large arXiv:cond-mat/0603350v2}

\vspace{2cm}

\end{center}

This article develops a way how to count matchings in random graphs. We used this as an example to introduce the replica symmetric method in sec.~\ref{sec:matching}. The main result of this work is that the belief propagation estimates asymptotically correctly the entropy of matchings, this was partially proven on a rigorous level in \cite{BayatiNair06}.

\addcontentsline{toc}{section}{{\bf [ZDEB-2]} E. Maneva, T. Meltzer, J. Raymond, A. Sportiello, LZ, "A Hike in the Phases of the 1-in-3 Satisfiability," \\Lecture notes of the Les Houches Summer School 2006, Session LXXXV, Complex Systems, Volume 85, 491-498, Elsevier 2007}
\chapter*{[ZDEB-2]}

\markboth{REPRINT OF PUBLICATION [ZDEB-2]}{REPRINT OF PUBLICATION [ZDEB-2]}

\begin{center}

{\LARGE \bf "A Hike in the Phases of \\ 
\vspace{0.2cm}
the 1-in-3 Satisfiability"}

\vspace{1cm}

{\large E. Maneva, T. Meltzer, J. Raymond, A. Sportiello, L. Zdeborov\'a, In proceedings of the Les Houches Summer School, Session LXXXV 2006 on Complex Systems.}

\vspace{1cm}

{\large arXiv:cond-mat/0702421v1}

\vspace{2cm}

\end{center}

This work on the 1-in-$K$ SAT problem started as a student project on the summer school in Les Houches 2006: Complex Systems, organized by M. M\'ezard and J.-P. Bouchaud. This short note contains a non-technical overview of our findings, and appeared in the collection of lecture notes from the school.

\addcontentsline{toc}{section}{{\bf [ZDEB-3]} J. Raymond, A. Sportiello, LZ, "The Phase Diagram of 1-in-3 Satisfiability,"  {\it Phys. Rev. E} {\bf 76} (2007) 011101}
\chapter*{[ZDEB-3]}

\markboth{REPRINT OF PUBLICATION [ZDEB-3]}{REPRINT OF PUBLICATION [ZDEB-3]}

\begin{center}

{\LARGE\bf "The Phase Diagram of 1-in-3 Satisfiability"}

\vspace{1cm}

{\large J. Raymond, A. Sportiello, L. Zdeborov\'a, {\it Phys. Rev. E} {\bf 76}, 011101 (2007)}

\vspace{1cm}

{\large arXiv:cond-mat/0702610v2}

\vspace{2cm}

\end{center}

In this article we present in detail the energetic 1RSB solution of the 1-in-3 SAT problem. We also analyze the performance of the unit clause propagation algorithms. We show how the phase diagram changes from an on average easy to K-SAT like when the probability of negating a variable is varied. An interesting point is the existence of a region where the replica symmetric solution is unstable, yet the unit clause propagation provably finds solutions in a randomized polynomial time. This work is a continuation of [\MM]. We used the 1-in-K SAT to present the energetic 1RSB solution in sec.~\ref{sec:1in3}. Note that 1-in-K SAT on factor graphs without leaves is one of the locked problems, this article however studies the Poissonian graph ensemble.

\addcontentsline{toc}{section}{{\bf [ZDEB-4]} F. Krzakala, A. Montanari, F. Ricci-Tersenghi, G. Semerjian, LZ, "Gibbs States and the Set of Solutions of Random Constraint Satisfaction Problems," {\it Proc. Natl. Acad. Sci.} {\bf 104} (2007) 10318}
\chapter*{[ZDEB-4]}

\markboth{REPRINT OF PUBLICATION [ZDEB-4]}{REPRINT OF PUBLICATION [ZDEB-4]}

\begin{center}

{\LARGE\bf "Gibbs States and the Set of \\
\vspace{-0.2cm}
Solutions of Random Constraint \\
\vspace{0.2cm}
Satisfaction Problems"}

\vspace{1cm}

{\large F. Krzakala, A. Montanari, F. Ricci-Tersenghi, G. Semerjian, L. Zdeborov\'a, {\it Proc. Natl. Acad. Sci.} {\bf 104}, 10318 (2007).}

\vspace{1cm}

{\large	arXiv:cond-mat/0612365v2}

\vspace{2cm}

\end{center}

In this article the clustering transition was defined via the extremality of the uniform measure over solutions, or equivalently via the onset of a nontrivial solution of the 1RSB equations at $m=1$. The derivation of the 1RSB equations on trees is sketched. This is a basis of our chapter \ref{clustering}. The condensation transition in constraint satisfaction problems, different from the clustering one, was discovered here. This is a basis of our chapter \ref{condensation}. The use of the belief propagation as a solver in the clustered but non-condensed phase was suggested here and studied. 
The results of this short article are developed in greater detail in [\ZK] for the graph coloring, and in \cite{MontanariRicci08} for the $K$-SAT problem. 
This article is addressed mainly to a mathematical and computer science audience.

\addcontentsline{toc}{section}{{\bf [ZDEB-5]}  LZ, F. Krzakala, "Phase transition in the Coloring of Random Graphs," {\it Phys. Rev. E} {\bf 76} (2007) 031131}
\chapter*{[ZDEB-5]}

\markboth{REPRINT OF PUBLICATION [ZDEB-5]}{REPRINT OF PUBLICATION [ZDEB-5]}

\begin{center}

{\LARGE\bf "Phase transition in the Coloring of Random Graphs"}

\vspace{1cm}

{\large L. Zdeborov\'a, F. Krzakala, {\it Phys. Rev. E} {\bf 76}, 031131 (2007).}

\vspace{1cm}

{\large	arXiv:0704.1269v2}

\vspace{2cm}

\end{center}

This is a detailed article about the phase diagram of the random coloring problem, summarized in chapter \ref{coloring}. We give an overview of the entropic 1RSB solution of the problem. We derive many results about the clustering and condensation transitions. The cavity method study of the frozen variables, as presented in \ref{sec:freezing_cav}, is developed here. The conjecture about freezing of variables being relevant for the computational hardness, which we discuss in \ref{sec:hardness}, is made here.

\addcontentsline{toc}{section}{{\bf [ZDEB-6]} F. Krzakala, LZ, "Potts Glass on Random Graphs,"\\ {\it Eur. Phys. Lett.} {\bf 81} (2008) 57005}
\chapter*{[ZDEB-6]}

\markboth{REPRINT OF PUBLICATION [ZDEB-6]}{REPRINT OF PUBLICATION [ZDEB-6]}

\begin{center}

{\LARGE\bf "Potts Glass on Random Graphs"}

\vspace{1cm}

{\large F. Krzakala, L. Zdeborov\'a, {\it Eur. Phys. Lett.} {\bf 81} (2008) 57005.}

\vspace{1cm}

{\large	arXiv:0710.3336v2}

\vspace{2cm}

\end{center}

In this letter we present the finite temperature phase diagram of the coloring problem, or in other words the antiferromagnetic Potts model on random graphs. We showed the phase diagram in sec.~\ref{sec:finiteT}.  We also analyze the stability of the 1RSB solution and in particular show that the colorable phase is 1RSB stable (at least for $q\ge 4$). This is reviewed in more detail in appendix \ref{app:1RSB_stab}.

\addcontentsline{toc}{section}{{\bf [ZDEB-7]} F. Krzakala, LZ, "Phase Transitions and Computational Difficulty in 
Random Constraint Satisfaction Problems,"\\ {\it J. Phys.: Conf. Ser.} {\bf 95} (2008) 012012}
\chapter*{[ZDEB-7]}

\markboth{REPRINT OF PUBLICATION [ZDEB-7]}{REPRINT OF PUBLICATION [ZDEB-7]}

\begin{center}

{\LARGE\bf "Phase Transitions and Computational Difficulty in 
Random Constraint Satisfaction Problems"}

\vspace{1cm}

{\large F. Krzakala, L. Zdeborov\'a, {\it J. Phys.: Conf. Ser.} {\bf 95} (2008) 012012.}

\vspace{1cm}

{\large	arXiv:0711.0110v1}

\vspace{2cm}

\end{center}

In this article we present in a accessible and non-technical way the main new results for the phase diagram of the random coloring.
This might be a good reading for uninitialized audience. 
We also summarize the present ideas about the origin of the average computational hardness. This article was presented in the Proceedings of the International Workshop on Statistical-Mechanical Informatics, Kyoto 2007. Chapter \ref{coloring} is largely inspired by this colloquial presentation.

\addcontentsline{toc}{section}{{\bf [ZDEB-8]} T. Mora, LZ, "Random subcubes as a toy model for constraint satisfaction problems," {\it J. Stat. Phys.}  {\bf 131} n.6 (2008) 1121}
\chapter*{[ZDEB-8]}

\markboth{REPRINT OF PUBLICATION [ZDEB-8]}{REPRINT OF PUBLICATION [ZDEB-8]}

\begin{center}

{\LARGE\bf "Random subcubes as a toy model for constraint satisfaction problems"}

\vspace{1cm}

{\large T. Mora, L. Zdeborov\'a, {\it J. Stat. Phys.}  {\bf 131} n.6  (2008) 1121-1138.}

\vspace{1cm}

{\large	arXiv:0710.3804v2}

\vspace{2cm}

\end{center}

In this article we introduced the random subcubes model. It plays the same role for constraint satisfaction problems as the random energy model played for spin glasses.  It is an exactly solvable toy model which reproduces the series of phase transitions studied in CSPs. The condensation transition comes out very naturally in this simple model, see sec.~\ref{sec:subcubes}. The space of solutions in the random subcubes model compares even quantitatively to the space of solutions in the $q$-coloring and $K$-SAT in the limit of large $q$ and $K$ near to the satisfiability threshold, as explained in sec.~\ref{sec:largeq}. We also introduced an energy landscape and showed that the glassy dynamics in this model can be understood purely from the static solution.

\addcontentsline{toc}{section}{{\bf [ZDEB-9]} LZ, M. M\'ezard, "Hard constraint satisfaction problems,"\\ preprint arXiv:0803.2955v1}
\chapter*{[ZDEB-9]}

\markboth{REPRINT OF PUBLICATION [ZDEB-9]}{REPRINT OF PUBLICATION [ZDEB-9]}

\begin{center}

{\LARGE\bf "Locked constraint satisfaction problems"}

\vspace{1cm}

{\large L. Zdeborov\'a, M. M\'ezard, to be accepted in {\it Phys. Rev. Lett.}}

\vspace{1cm}

{\large	arXiv:0803.2955v1}

\vspace{2cm}

\end{center}

In this letter we introduce the locked constraint satisfaction problems, presented in sec.~\ref{locked}. The space of solutions of these problems have an extremely easy statistical description, as illustrated e.g. by the second moment computation of the entropy in app.~\ref{app:moments}. On the other hand these problems are algorithmically very challenging, none of the algorithms we tried is able to find solutions in the clustered phase. Some classical algorithms do not work at all in these problems, for example the belief propagation decimation analyzed in app.~\ref{app:alg}. A more detailed version of this article is in preparation.

\addcontentsline{toc}{section}{{\bf [ZDEB-10]} J. Ardelius, LZ, "Exhaustive enumeration unveils clustering and freezing in random 3-SAT," preprint arXiv:0804.0362v1.}
\chapter*{[ZDEB-10]}

\markboth{REPRINT OF PUBLICATION [ZDEB-10]}{REPRINT OF PUBLICATION [ZDEB-10]}

\begin{center}

{\LARGE\bf "Exhaustive enumeration unveils clustering and freezing in random 3-SAT"}

\vspace{1cm}

{\large J. Ardelius, L. Zdeborov\'a, submitted to {\it Phys. Rev.}}

\vspace{1cm}

{\large	arXiv:0804.0362v2}

\vspace{2cm}

\end{center}

In this letter we study via an exhaustive enumeration the phase space in the random 3-SAT. The main question we addressed here is the relevance of the asymptotic predictions to instances of moderate size. We show that the complexity of clusters compares strikingly well to the analytical prediction. We also locate for a first time the freezing transition and show that it practically coincides with the performance limit of the survey propagation based algorithms. Results of this work appear on several places of the thesis, mainly figs.~\ref{fig:sat_threshold}, \ref{fig:cmplx}, and \ref{fig:rigidity}.

\backmatter

\addcontentsline{toc}{chapter}{Bibliography}

\markboth{BIBLIOGRAPHY}{BIBLIOGRAPHY}
\bibliographystyle{alpha}
\bibliography{myentries}

\newpage

\addcontentsline{toc}{chapter}{Index}
\markboth{INDEX}{INDEX}
\printindex

\end{document}